\documentclass[12pt,a4paper]{report}
\usepackage[paperwidth=22.4cm, paperheight=31cm]{geometry}
\usepackage[T1]{fontenc}
\usepackage[latin1]{inputenc}
\usepackage{titlesec,epstopdf,tocloft,verbatim,pdfpages,epstopdf,epsfig,graphicx,amsthm,amsmath,amssymb,amsthm,amsfonts,titling,url,array,subfig,makeidx}
\usepackage[nottoc]{tocbibind}
\usepackage{pdfpages}
\newcommand{\bl}{\boldsymbol}

\renewcommand{\bibname}{\centerline{References}\global\def\bibname{\hspace*{.64cm}References}}
\renewcommand{\indexname}{\centerline{Index}\global\def\indexname{\hspace*{.64cm}Index}}
\renewcommand{\tableofcontents}{\centerline{Contents}}
\makeindex

\begin{document}

\begin{titlepage}
%minipage cria subambientes ideias para usar figuras ao lado de textos.
\begin{figure}[h]
\begin{minipage}[c]{0.3\linewidth}
\flushleft{\includegraphics[width=3cm]{{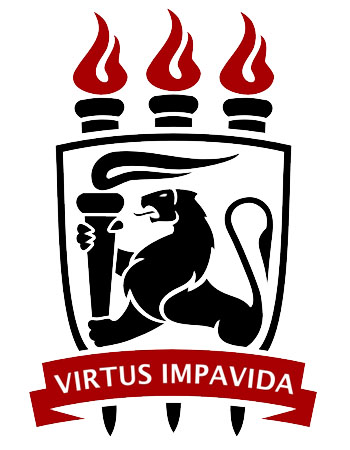}}}
   \end{minipage}
\hfill
\begin{minipage}[c]{0.7\linewidth}
\flushleft{\textbf{UNIVERSIDADE FEDERAL DE PERNAMBUCO}\\
\textbf{DEPARTAMENTO DE FÍSICA - CCEN}\\
\textbf{PROGRAMA DE PÓS-GRADUAÇÃO EM FÍSICA}} 
\end{minipage}
\end{figure}

\begin{center}
\vspace{2.3cm}
\textbf{JOÁS DA SILVA VENÂNCIO}
\vspace{2.3cm}
\end{center}

\vspace{2.3cm}
\begin{center}
\resizebox{!}{0,35cm}{\textbf{ON THE QUASINORMAL MODES IN GENERALIZED}} 
\resizebox{!}{0,35cm}{\textbf{NARIAI SPACETIMES}}
\end{center}
\vspace{2.3cm}

\vspace{4cm}

\begin{center}
Recife\\
2021.
\end{center}

%\begin{center}
%\begin{figure}[h]
%	\centering
%		\includegraphics[width=1cm]{{Figuras/ufpe-logo2.jpg}}
%	\quad\quad\quad\quad
 %  \includegraphics[width=1cm]{{Figuras/cnpq-logo.jpg}}
%	\quad\quad\quad\quad
%	  \includegraphics[width=1cm]{{Figuras/brasil-logo.png}}
%\quad\quad\quad\quad
 %   \includegraphics[width=1cm]{{Figuras/fisica-logo.jpeg}}
%\end{figure}
%\end{center}

\end{titlepage}

%%%%%%%%%%%%%%%%%%%%%%%%%%%%%%%%%%%%%%%%%%%%%%%%%%%%%%%%%%%%%%%%%%%
%%%%%%%%%%%%%%%%%%%%%%%%%%%%%%%%%%%%%%%%%%%%%%%%%%%%%%%%%%%%%%%%%%%

%%%%%%%%%%%%%%%%%%%%%%%%%%%%%%%%%%%%%%%%%%%%%%%%%%%%%%%%%%%%%%%%%%%
%%%%%%%%%%%%%%%%%%%%%%%%%%%%%%%%%%%%%%%%%%%%%%%%%%%%%%%%%%%%%%%%%%%

\begin{titlepage}

\begin{center}
\textbf{JOÁS DA SILVA VENÂNCIO}
\end{center}

\vspace{3cm}
\begin{center}
\textbf{ON THE QUASINORMAL MODES IN GENERALIZED NARIAI SPACETIMES}\\ 
\end{center}
\quad
\vspace{1cm}

\small{
\begin{flushright}
\parbox{3in}{Thesis submitted to the graduation program of the Physics Department of Federal University of Pernambuco in partial fulfilment of the requirements for the award of the degree of Doctor of Philosophy in Physics.
\\
\\
\\
\textbf{Concentration Area}: Theoretical and Computational Physics
\\
\\
\\
\textbf{Supervisor}: Prof. Dr. Carlos Alberto Batista da Silva Filho}
\end{flushright}}

\normalsize

\vspace{6.5cm}

\begin{center}
Recife\\
2021.
\end{center}

\end{titlepage}

%%%%%%%%%%%%%%%%%%%%%%%%%%%%%%%%%%%%%%%%%%%%%%%%%%%%%%%%%%%%%%%%%%%
%\newpage
\includepdf{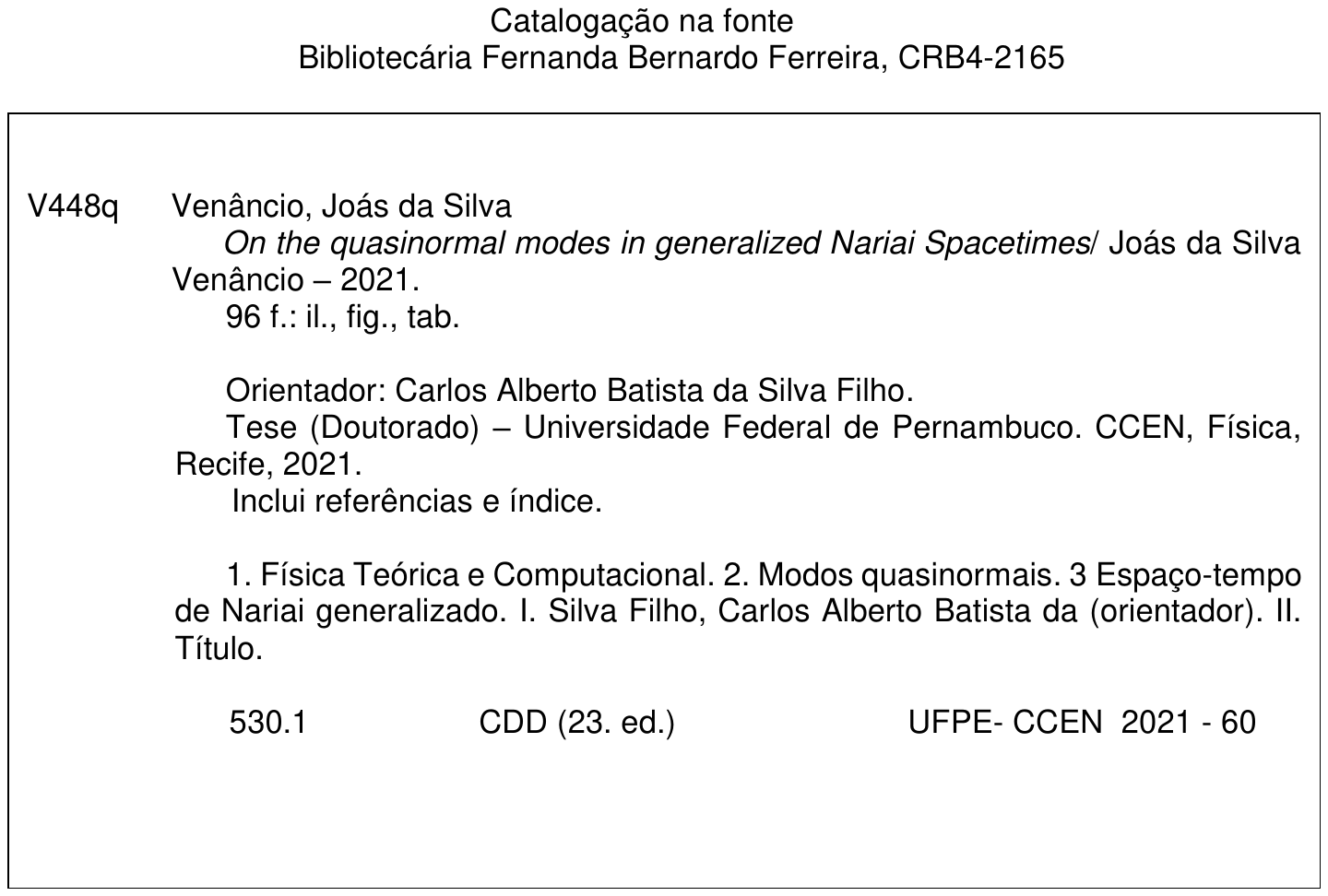}

%%%%%%%%%%%%%%%%%%%%%%%%%%%%%%%%%%%%%%%%%%%%%%%%%%%%%%%%%%%%%%%%%%%
%\pagebreak
%\includepdf{Catalogacao}
%%%%%%%%%%%%%%%%%%%%%%%%%%%%%%%%%%%%%%%%%%%%%%%%%%%%%%%%%%%%%%%%%%%
%\newpage\ \newpage

\begin{titlepage}

\begin{center}
\textbf{JOÁS DA SILVA VENÂNCIO}
\end{center}

\vspace{.5cm}
\begin{center}
\textbf{ON THE QUASINORMAL MODES IN GENERALIZED NARIAI SPACETIMES}\\ 
\end{center}
\quad
\vspace{.4cm}

\small{
\begin{flushright}
\parbox{3in}{Thesis submitted to the graduation program of the Physics Department of Federal University of Pernambuco in partial fulfilment of the requirements for the award of the degree of Doctor of Philosophy in Physics.}
\end{flushright}}
\normalsize

\vspace{.4cm}

\begin{flushleft}
  Accepted on: 25/02/2021.\\
\end{flushleft}

\vspace{.4cm}

\begin{center}
  \textbf{EXAMINING BOARD:}\\
	\vspace{.5cm}
	$\overline{ \qquad\qquad\qquad\qquad\qquad\qquad\qquad\qquad\qquad\qquad\qquad\qquad\qquad }$ \\
	 Prof. Dr. Carlos Alberto Batista da Silva Filho \\
	 Supervisor \\
   Federal University of Pernambuco (PD-UFPE, Brazil) \\
\vspace{.5cm}
	$\overline{ \qquad\qquad\qquad\qquad\qquad\qquad\qquad\qquad\qquad\qquad\qquad\qquad\qquad }$\\
   Prof. Dr. Azadeh Mohammadi \\
	 Internal Examiner \\
   Federal University of Pernambuco (PD-UFPE, Brazil) \\
\vspace{.5cm}
$\overline{ \qquad\qquad\qquad\qquad\qquad\qquad\qquad\qquad\qquad\qquad\qquad\qquad\qquad }$\\
   Prof. Dr. Fernando Jorge Sampaio Moraes \\
	 Internal Examiner \\
   Federal Rural University of Pernambuco (PD-UFRPE, Brazil) \\
\vspace{.5cm}		
$\overline{ \qquad\qquad\qquad\qquad\qquad\qquad\qquad\qquad\qquad\qquad\qquad\qquad\qquad }$\\
   Prof. Dr. Riccardo Sturani \\
	 External Examiner \\
   Federal University of Rio Grande do Norte (IIP-UFRN, Brazil) \\
   \vspace{.5cm}	
   $\overline{ \qquad\qquad\qquad\qquad\qquad\qquad\qquad\qquad\qquad\qquad\qquad\qquad\qquad }$\\
   Prof. Dr. Andr\'{e}s Fernando Anabal\'{o}n Dupuy \\
	 External Examiner \\
   Universidad Adolfo Iba\~{n}ez (Vin\~{a} del Mar, Chile)\\
   \vspace{.5cm}

\end{center}
\end{titlepage}

%%%%%%%%%%%%%%%%%%%%%%%%%%%%%%%%%%%%%%%%%%%%%%%%%%%%%%%%%%%%%%%%%%%%%%%%%%%%%%%%%%%%%%%%%%%%%%%%%%%%%%%%%%%%%%%%%%%%%%%%%%%%%%%%%%%%%%%%%

\begin{titlepage}

\begin{center}
 \Huge{\textbf{{Abstract}}}\\
 %\textbf{\uppercase{Abstract}}\\
\end{center}
   \normalsize
  \vspace{0.3cm}
  \quad

Quasinormal modes are eigenmodes of dissipative systems. For instance, if a spacetime with an event or cosmological horizon is perturbed from its equilibrium state, quasinormal modes arise as damped oscillations with a spectrum of complex frequencies, called quasinormal frequencies\index{Quasinormal frequencies}, that does not depend on the details of the excitation. In fact, these frequencies depend just on the charges which define the geometry of the spacetime in which the perturbation is propagating, such as the mass, electric charge, and angular momentum. Quasinormal modes have been studied for a long time and the interest in this topic has been renewed by the recent detection of gravitational waves, inasmuch as these are the configurations that are generally measured by experiments. Mathematically, this discrete spectrum of quasinormal modes stems from the fact that certain boundary conditions must be imposed to the physical fields propagating in such a spacetime. In this book, we shall consider a higher-dimensional generalization of the charged Nariai spacetime that is comprised of the direct product of the two-dimensional de Sitter space, $dS_{2}$, with an arbitrary number of two-spheres, $S^{2}$, and investigate the dynamics of spin-$s$ field perturbations\index{Spin-$s$ field perturbations} for $s = 0 ,1/2, 1$ and $2$. As a first step, we shall attain the separability\index{Separability} of the equations of motion for each perturbation type in such a geometry and its reduction into a Schr\"{o}dinger-like differential equation\index{Schr\"{o}dinger-like differential equation} whose potential is contained in the Rosen-Morse\index{Rosen-Morse class of potentials} class of integrable potentials, which has the so-called P\"{o}schl-Teller potential\index{P\"{o}schl-Teller potential} as a particular case. A key step in order to attain this separability\index{Separability} is to use a suitable basis for the angular functions depending on the rank of the tensorial degree of freedom that one needs to describe. Here we define such a basis, which is a generalization of the tensor spherical harmonics that is suited for spaces that are the product of several spaces of constant curvature. Finally, with the integration of the Schr\"{o}dinger-like differential equation\index{Schr\"{o}dinger-like differential equation} at hand, the boundary conditions\index{Boundary conditions} leading to quasinormal modes are analyzed and the quasinormal frequencies\index{Quasinormal frequencies} are analytically obtained.   

\end{titlepage}

%%%%%%%%%%%%%%%%%%%%%%%%%%%%%%%%%%%%%%%%%%%%%%%%%%%%%%%%%%%%%%%%%%%%%%%%%%%%%%%%%%%%%%%%%%%%%%%%%%%%%%%%%%%%%%%%%%%%%%%%%%%%%%%%%%%%%%%%%

\begin{titlepage}

\begin{center}
 \Huge{\textbf{{Resumo}}}\\
%\textbf{\uppercase{Resumo}}\\
     \end{center}
   \normalsize
  \vspace{0.3cm}
  \quad

Modos quasinormais s\~ao modos pr\'{o}prios de sistemas dissipativos. Por exemplo, se um espa\c{c}o-tempo com um horizonte de evento ou horizonte cosmol\'{o}gico \'{e} perturbado de seu estado de equil\'{i}brio, os modos quasinormais surgem como oscila\c{c}\~oes amortecidas com um espectro de frequ\^{e}ncias complexas, chamadas frequ\^{e}ncias quasinormais, que n\~ao dependem dos detalhes da excita\c{c}\~ao. De fato, estas frequ\^{e}ncies dependem apenas das cargas que definem a geometria do espa\c{c}o-tempo no qual a perturba\c{c}\~ao est\'{a} se propagando, tais como: massa, carga el\'{e}trica e momento angular. Os modos quasinormais v\^{e}m sendo estudados h\'{a} muito tempo e o interesse nesse tema tem sido renovado pela recente detec\c{c}\~ao de ondas gravitacionais, visto que essas s\~{a}o as configura\c{c}\~oes que s\~{a}o geralmente medidas por experimentos. Matematicamente, esse espectro discreto de modos quasinormais decorre do fato de que certas condi\c{c}\~oes de contorno devem ser impostas aos campos f\'{i}sicos que se propagam em um tal espa\c{c}o-tempo. Neste livro, devemos considerar uma generaliza\c{c}\~ao em dimens\~{o}es superiores do espa\c{c}o-tempo de Nariai carregado que \'{e} formado do produto direto do espa\c{c}o de Sitter bidimensional, $dS_{2}$, com v\'{a}rias esferas bidimensionais, $S^{2}$, e investigar a din\^{a}mica das perturba\c{c}\~oes de campos de spin $s$ para $s = 0,1/2,1$ e $2$. Como um primeiro passo, devemos atingir a separabilidade das equa\c{c}\~oes de movimento para cada tipo de perturba\c{c}\~ao em tal geometria e, em seguida, a redu\c{c}\~ao em uma equa\c{c}\~ao diferencial tipo Schr\"{o}dinger cujo potencial est\'{a} contido na classe de Rosen-Morse\index{Rosen-Morse class of potentials} de potenciais integr\'{a}veis, que tem o chamado potencial P\"{o}schl-Teller\index{P\"{o}schl-Teller potential} como um caso particular. Um passo fundamental para atingir essa separabilidade \'{e} usar uma base adequada para as fun\c{c}\~oes angulares, dependendo do rank do grau de liberdade tensorial que se precisa descrever. Aqui definimos tal base, que \'{e} uma generaliza\c{c}\~ao dos harm\^{o}nicos esf\'{e}ricos tensoriais. Tal base tamb\'{e}m \'{e} adequada para quaisquer espa\c{c}os que s\~ao o produto de v\'{a}rios espa\c{c}os de curvatura constante. Finalmente, com a integra\c{c}\~ao da equa\c{c}\~ao diferencial tipo Schr\"{o}dinger em m\~aos, as condi\c{c}\~oes de contorno que conduzem aos modos quasinormais s\~{a}o analisadas e as frequ\^{e}ncias quasinormais s\~ao obtidas analiticamente.
\\
\\
\\
\small{\textbf{Palavras chaves:} Modos quasinormais. Espa\c{c}o-tempo de Nariai generalizado. Separabilidade. Condi\c{c}\~oes de contorno.}

\end{titlepage}

%%%%%%%%%%%%%%%%%%%%%%%%%%%%%%%%%%%%%%%%%%%%%%%%%%%%%%%%%%%%%%%%%%%%%%%%%%%%%%%%%%%%%%%%%%%%%%%%%%%%%%%%%%%%%%%%%%%%%%%%%%%%%%%%%%%%%%%%%

\begin{titlepage}

\begin{center}
\Huge{\textbf{{List of Publications}}}\\
%\textbf{\uppercase{List of Publications}}\\
\end{center}
   \normalsize
  \vspace{0.6cm}
\large{\textbf{Published articles}}
 \vspace{0.3cm}

$\bullet$\quad J. Ven\^{a}ncio and C. Batista, \textit{Separability of the Dirac equation on backgrounds that are the direct product of bidimensional spaces}, Phys. Rev. D \textbf{95} (2017), 084022. \\
\\
\\

$\bullet$\quad J. Ven\^{a}ncio and C. Batista, \textit{Quasinormal modes in generalized Nariai spacetimes}, Phys. Rev. D \textbf{97} (2018), 105025.\\
\\
\\

$\bullet$\quad  J. Ven\^{a}ncio and C. Batista, \textit{Spin-$2$ quasinormal modes in generalized Nariai spacetimes}, Phys. Rev. D \textbf{101} (2020), 084037..
\\
\\
\\
\large{\textbf{Published book}}
 \vspace{0.3cm}

$\bullet$\quad J. Ven\^{a}ncio, \textit{The Spinorial formalism: An Introduction to 
the  Spinorial  Formalism  with  Applications  in  Physics},  	Lambert  Academic  Publishing,  (2019). \\
\\
\\
\\
\large{\textbf{Submitted articles}}
\vspace{0.3cm}

$\bullet$\quad  J. Ven\^{a}ncio and C. Batista, \textit{Two-component spinorial formalism using quaternions for six-dimensional spacetimes}, (2020). arXiv:2007.04296v2

\end{titlepage}

%%%%%%%%%%%%%%%%%%%%%%%%%%%%%%%%%%%%%%%%%%%%%%%%%%%%%%%%%%%%%%%%%%%
%%%%%%%%%%%%%%%%%%%%%%%%%%%%%%%%%%%%%%%%%%%%%%%%%%%%%%%%%%%%%%%%%%%

%%%%%%%%%%%%%%%%%%%%%%%%%%%%%%%%%%%%%%%%%%%%%%%%%%%%%%%%%%%%%%%%%%%
%%%%%%%%%%%%%%%%%%%%%%%%%%%  List of Symbols
%%%%%%%%%%%%%%%%%%%%%%%%%%%%%%%%%%%%%%%%%%%%%%%%%%%%%%%%%%%%%%%%%%%

\begin{titlepage}

\begin{center}
\Huge{\textbf{
{List of Symbols}}}
%\textbf{\uppercase{List of Symbols}}
\end{center}
\vspace{0.5cm}
%\noindent
\begin{eqnarray}
\text{QNMs}&       \quad\quad& \textrm{Quasinormal modes: } \small{\textsf{  Page \pageref{QNM}.}} \nonumber\\
\text{QNFs}&       \quad\quad& \textrm{Quasinormal frequencies: } \small{\textsf{  Page \pageref{QNF}.}} \nonumber\\
\text{AdS}&       \quad\quad& \textrm{Anti-de Sitter space: } \small{\textsf{  Page \pageref{AdS}.}} \nonumber\\
\text{CFT}&       \quad\quad& \textrm{Conformal Field Theory: } \small{\textsf{  Page \pageref{CFT}.}} \nonumber\\
\mathcal{R}&       \quad\quad& \textrm{Ricci scalar: } \small{\textsf{  Page \pageref{RS}.}} \nonumber\\
\mathcal{R}_{\mu\nu}&       \quad\quad& \textrm{Ricci tensor: } \small{\textsf{  Page \pageref{RT}.}} \nonumber\\
\mathcal{T}_{\mu\nu}&       \quad\quad& \textrm{Stress-energy tensor: } \small{\textsf{  Page \pageref{ST}.}} \nonumber\\
\partial_{\mu}&    \quad\quad& \textrm{Differential operators } \frac{\partial}{\partial x^\mu}:  \small{\textsf{  Page \pageref{DO}.}}  \nonumber\\
\text{EP}&    \quad\quad& \textrm{Effective potential}:  \small{\textsf{  Page \pageref{EPF}.}}  \nonumber\\
\text{S},  g^{S},f_{S}&       \quad\quad& \textrm{Schwarzschild: } \small{\textsf{  Page \pageref{SB}.}} \nonumber\\
\text{RW},  V^{RW}_{s=2}(r)&       \quad\quad& \textrm{Regge-Wheeler: } \small{\textsf{  Page \pageref{RWL}.}} \nonumber\\
\text{Z},  V^{Z}_{s=2}(r)&       \quad\quad& \textrm{Zerilli: } \small{\textsf{  Page \pageref{ZPL}.}} \nonumber\\
\star, H^{\star}&      \quad\quad&   \textrm{Complex conjugation:}   \small{\textsf{  Page \pageref{CC}.}}  \nonumber\\
\text{GN},  g^{GN} &       \quad\quad& \textrm{Generalized Nariai: } \small{\textsf{  Page \pageref{GN}.}} \nonumber\\
\text{N},  g^{N} &       \quad\quad& \textrm{Nariai: } \small{\textsf{  Page \pageref{N}.}} \nonumber\\
\text{PT},  V^{PT}(x) &       \quad\quad& \textrm{P\"{o}schl-Teller: } \small{\textsf{  Page \pageref{PTPL}.}} \nonumber\\
P_l, \overset{P_l\,\,}{\rightarrow }&   \quad\quad& \textrm{Parity transformation: }  \small{\textsf{  Page \pageref{PT}.}}\nonumber\\
F(a,b,c;y)& \quad\quad& \textrm{Hypergeometric function: }  \small{\textsf{  Page \pageref{HG}.}}\nonumber\\
\pounds_{\boldsymbol{K}}&       \quad\quad&   \textrm{Lie derivative along of $\boldsymbol{K}$:}   \small{\textsf{  Page \pageref{LD}.}}\nonumber\\
\mathbb{V}&       \quad\quad& \textrm{Two-dimensional vector space: } \small{\textsf{  Page \pageref{TVS}.}} \nonumber\\
\mathbb{S}&       \quad\quad& \textrm{Two-dimensional spinor space: } \small{\textsf{  Page \pageref{SP}.}} \nonumber\\
%\mathbb{R}&       \quad\quad& \textrm{Field of the real numbers: } \small{\textsf{  Page \pageref{Real Field}.}}  \nonumber\\
\mathcal{C}\ell(\mathbb{V},\hat{\bl{g}})&      \quad\quad& \textrm{Clifford algebra of $\mathbb{V}$ endowed with $\hat{\bl{g}}$: }   \small{\textsf{  Page \pageref{CA}.}}  \nonumber\\	
\wedge, \bl{V}\wedge \bl{U}&   \quad\quad& \textrm{Exterior product: }   \small{\textsf{  Page \pageref{EP}.}}  \nonumber\\
\mathbb{C}&        \quad\quad& \textrm{Field of the complex numbers: } \small{ \textsf{  Page \pageref{CF}}.} \nonumber\\
\bl{R}_{\zeta}&    \quad\quad& \textrm{Rotation operator: }  \small{\textsf{  Page \pageref{RO}.}}\nonumber\\
SPin(\mathbb{V}) &     \quad\quad& \textrm{Spin group of $\mathbb{V}$: }   \small{\textsf{  Page \pageref{SG}.}}  \nonumber\\
\omega_{ab}^{\phantom{ab}c}\,,\; \omega_{abc}&    \quad\quad&    \textrm{Components of the spin connection:} \small{\textsf{  Page \pageref{CSC}.}} \nonumber\\
{_{s}Y_{j,m}(\theta,\phi)}&   \quad\quad& \textrm{Spin-$s$ spherical harmonics: }  \small{\textsf{  Page \pageref{SSSH}.}}\nonumber\\	
\boldsymbol{g}&     \quad\quad&   \textrm{The metric of the manifold:}   \small{\textsf{  Page \pageref{Metric}.}} \nonumber
\end{eqnarray}

\end{titlepage}

%%%%%%%%%%%%%%%%%%%%%%%%%%%%%%%%%%%%%%%%%%%%%%%%%%%%%%%%%%%%%%%%%%%
%%%%%%%%%%%%%%%%%%%%%%%%%%%  List of Figures
%%%%%%%%%%%%%%%%%%%%%%%%%%%%%%%%%%%%%%%%%%%%%%%%%%%%%%%%%%%%%%%%%%%

%\listoffigures
%\pagebreak 

%%%%%%%%%%%%%%%%%%%%%%%%%%%%%%%%%%%%%%%%%%%%%%%%%%%%%%%%%%%%%%%%%%%
%%%%%%%%%%%%%%%%%%%%%%%%%%%  List of Figures
%%%%%%%%%%%%%%%%%%%%%%%%%%%%%%%%%%%%%%%%%%%%%%%%%%%%%%%%%%%%%%%%%%%

%%%%%%%%%%%%%%%%%%%%%%%%%%%%%%%%%%%%%%%%%%%%%%%%%%%%%%%%%%%%%%%%%%%
%%%%%%%%%%%%%%%%%%%%%%%%%%%  List of Tables
%%%%%%%%%%%%%%%%%%%%%%%%%%%%%%%%%%%%%%%%%%%%%%%%%%%%%%%%%%%%%%%%%%%

%\listoftables
%\pagebreak 

%%%%%%%%%%%%%%%%%%%%%%%%%%%%%%%%%%%%%%%%%%%%%%%%%%%%%%%%%%%%%%%%%%%
%%%%%%%%%%%%%%%%%%%%%%%%%%%  List of Tables
%%%%%%%%%%%%%%%%%%%%%%%%%%%%%%%%%%%%%%%%%%%%%%%%%%%%%%%%%%%%%%%%%%%

\pagenumbering{gobble}{\tableofcontents}

\newpage
\pagenumbering{arabic}
\setcounter{page}{10}

\setcounter{chapter}{0}

%%%%%%%%%%%%%%%%%%%%%%%%%%%%%%%%%%%%%%%%%%%%%%%%%%%%%%%%%%%%%%%%%%%
%%%%%%%%%%%%%%%%%%%%%%%%%%% B: Part I
%%%%%%%%%%%%%%%%%%%%%%%%%%%%%%%%%%%%%%%%%%%%%%%%%%%%%%%%%%%%%%%%%%%
%\part{\uppercase{Quasinormal Modes and Some Classical Results}}\label{Part-Rev}
\part{Quasinormal Modes and Some Classical Results}\label{Part-Rev}

%%%%%%%%%%%%%%%%%%%%%%%%%%%%%%%%%%%%%%%%%%%%%%%%%%%%%%%%%%%%%%%%%%%
%%%%%%%%%%%%%%%%%%%%%%%%%%% E: Part I
%%%%%%%%%%%%%%%%%%%%%%%%%%%%%%%%%%%%%%%%%%%%%%%%%%%%%%%%%%%%%%%%%%%

%the wise man said, but he walked behind?
%%%%%%%%%%%%%%%%%%%%%%%%%%%%%%%%%%%%%%%%%%%%%%%%%%%%%%%%%%%%%%%%%%%
%%%%%%%%%%%%%%%%%%%%%%%%%%%  Motivation and Outline
%%%%%%%%%%%%%%%%%%%%%%%%%%%%%%%%%%%%%%%%%%%%%%%%%%%%%%%%%%%%%%%%%%%
%\chapter{\uppercase{Motivation and Outline}}
\chapter{Motivation and Outline}
%\chapter*{1. \quad Motivation and Outline}
%\thispagestyle{headings}
%\addcontentsline{toc}{chapter}{1.\qquad Motivation and Outline}

It is well-known that a string of a guitar produces a characteristic sound when someone
hits it. This characteristic sound is the natural way the system finds to respond to
the external excitation. Interestingly, similar phenomena are ubiquitous in dynamical
systems that are in equilibrium states. These systems typically respond to a perturbation by oscillating around the equilibrium configuration with a set of natural frequencies, known as the normal frequencies. In particular, when some specific frequency is selected we say that the system is in a normal mode. Now, do black holes have a characteristic `sound' as well? The answer is yes. Studying scattering in Schwarzschild geometry, Vishveshwara found that the evolution of perturbations is given by damped oscillations with natural frequencies that do not depend on the details of the excitation \cite{Vishveshwara1970}. Since these perturbations decay exponentially in time, they are characterized by complex frequencies. Hence, they are called quasinormal frequencies\index{Quasinormal frequencies} (QNFs)\label{QNF}\index{Quasinormal Modes}, and the configurations with a single frequency are the quasinormal modes (QNMs)\label{QNM} \cite{Press1971, Ferrari1970}. The qualifier ``quasi'' is used to indicate that these modes are similar to, but not exactly equal to normal modes. The real part of a QNF is associated with the oscillation frequency of the perturbation, while the imaginary part is related to its decay rate. This damping stems from the existence of an event horizon\index{Event horizon}, which prevents incoming signals to be reflected back, yielding dissipation. The interesting fact is that these frequencies depend on the charges of the black hole, such as the mass, electric charge, and angular momentum. Therefore, the measurement of QNFs can be used to obtain the charges of astrophysical black holes \cite{Vishveshwara1970, Andersson1988}. This has incited a wide effort to find the QNFs of several gravitational configurations, with several numerical and analytical techniques being devised \cite{Kokkotas1999,Berti2009,CardosoThesis,Nollert1999}. The interest in QNMs has been renewed by the recent detection of gravitational waves \cite{Abbott2016}, since now the QNFs are closer of being experimentally accessible. Another reason for studying QNMs is that we would expect, in light of Bohr's correspondence principle, that they should give some hint about quantum nature of gravity \cite{Hod1998}. Indeed, a connection between QNFs and the quantization of the event horizon\index{Event horizon}
area has been put forward \cite{Dreyer2003,Maggiore2008,Domagala2004}. 

From the theoretical point of view, most of the recent works featuring QNMs are concerned with higher-dimensional spacetimes \cite{Konoplya2001,Frolov2018,Zhidenko2006}. For instance, QNMs are used to test the stability of certain solutions, this is particularly useful in dimensions greater than four, in which case there is no uniqueness theorem for black holes, so that the stability may be the criteria to select physical configurations among several gravitational solutions \cite{Zhidenko2009,Liu2008}. There are several motivations for studying gravitational configurations in dimensions greater than four. For example, string theory, which intends to describe the fundamental interactions of nature in a unified scheme, requires the spacetime to have 10 dimensions \cite{Mukhi2011}. Actually, there are many other theories that seek to explain our Universe through the use of higher-dimensional theories, for reviews see \cite{Emparan2008,Csaki2005}. Another source of interest in higher-dimensional spacetimes is the anti-de Sitter/conformal field theory (AdS/CFT)\label{AdS} correspondence, which provides tools to tackle field theories living in $d$ dimensions by means of studying gravitational solutions in $d+1$ dimensions \cite{Maldacena1999,Horowitz2009,Hubeny2015}. Through AdS/CFT correspondence\label{CFT}, QNFs can be associated to the thermalization of perturbations in finite temperature field theories \cite{Horowitz2000,Birmingham2002,Chirenti2018,Nunez2003,Keranen2016,David2015}. 

With the above motivations in mind, in the present book we shall consider a higher-dimensional generalization of the charged Nariai spacetime\index{Generalized Nariai spacetimes} \cite{Batista2016} and investigate the dynamics of perturbations of test fields with spins $0, 1/2, 1$ and $2$. In particular, we investigate the boundary conditions\index{Boundary conditions} that lead to QNMs and analytically obtain the spectrum of QNFs. The background used here is the direct product of two-dimensional spacetimes of constant curvature, $dS_{2} \times S^{2} \times \ldots \times S^{2}$ , while the most known higher-dimensional generalization of Nariai spacetime is given by $dS_{2} \times S^{D-2}$ \cite{Kodama2004,Cardoso2004}. One interesting feature of the spacetime considered here is that it supports magnetic charges besides the electric charge \cite{Batista2016}, which lead to a richer physics. Moreover, spaces that are the direct product of two-dimensional spaces can also be of relevance to model internal spaces in string theory compactifications \cite{Brown2014}.

%%%%%%%%%%%%%%%%%%%%%%%%%%%%%%%%%%%%%%%%%%%%%%%%%%%%%%%%%%%%%%%%%%%
%%%%%%%%%%%%%%%%%%%%%%%%%%%  End: Introduction
%%%%%%%%%%%%%%%%%%%%%%%%%%%%%%%%%%%%%%%%%%%%%%%%%%%%%%%%%%%%%%%%%%%

%%%%%%%%%%%%%%%%%%%%%%%%%%%%%%%%%%%%%%%%%%%%%%%%%%%%%%%%%%%%%%%%%%%
%%%%%%%%%%%%%%%%%%%%%%%%%%% B: Chapter 1. Quasinormal Modes: An Introduction
%%%%%%%%%%%%%%%%%%%%%%%%%%%%%%%%%%%%%%%%%%%%%%%%%%%%%%%%%%%%%%%%%%%

%\chapter{\uppercase{Quasinormal Modes: An Introduction}}\label{ChapPTP}
\chapter{Quasinormal Modes: An Introduction}\label{ChapPTP}
%\chapter*{2.\quad Quasinormal Modes: An Introduction}\label{ChapPTP}
%\pagestyle{empty}
%\setcounter{chapter}{1} 
%\addcontentsline{toc}{chapter}{2.\qquad Quasinormal Modes: An Introduction}

The study of perturbations is of central importance in almost all branches of physics, since often the physical systems are in a stable configuration and the changes are all due to small disturbances that do not build up as time passes by. The perturbation formalism is even more necessary when the mathematical equations that describe the dynamics of a system are nonlinear, since the effect of perturbations can generally be handled by means of linear equations, providing thus a great deal of simplification.
Einstein's General Relativity theory is an important example of this, since its field equation in $D$ dimensions, Einstein's equation, is a coupled set of $D(D+1)/2$ nonlinear (in all orders) partial differential equations that is impossible to solve analytically in the generic case, that is, without assuming the existence of special symmetries. This nonlinearity makes the study of perturbations of metric and matter fields in general relativity a non-trivial problem, once the matter fields appear in Einstein's equation through its energy-momentum tensor, which is typically quadratic or of higher order in the matter fields. To overcome this difficulty it is a standard procedure to work with linear perturbation theory in which one assumes the weak regime solution, in the sense that the energy-momentum tensor is small enough in order to allow us to neglect it. Let us see now such a procedure in more details as well as how to define quasinormal modes.

\vspace{.5cm}
%\section{\uppercase{Linear Perturbation Theory}}\label{Sec.LPT}
\section{Linear Perturbation Theory}\label{Sec.LPT}
\vspace{.5cm}
%\section{2.1 \quad Linear Perturbation Theory}\label{Sec.LPT}
%\section*{2.1 \quad Linear Perturbation Theory}\label{Sec.LPT}
%\addcontentsline{toc}{section}{2.1\qquad Linear Perturbation Theory}

In $D$ dimensions, the dynamics of general relativity in curved spacetimes with cosmological constant $\Lambda$ is described by the following version of the Einstein-Hilbert action
\begin{equation}
S\,=\,\frac{1}{16\pi} \int d^{D}x\sqrt{\left | g \right |}\left[ \mathcal{R}-(D-2)\Lambda \right] \,+\,S_{m},
\end{equation}
with $|g|$ being the determinant of the metric $g_{\mu\nu}$, $\mathcal{R}$ being the Ricci scalar\label{RS} and $S_{m}$ being the action of the matter fields $\{\Phi_i\}$ coupled to gravity. The least action principle allows us to find the equations of motion for the fields $g_{\mu\nu}$ and $\Phi_{i}$ which are given, respectively, by
\begin{align}
 &\mathcal{R}_{\mu\nu} + \frac{1}{2}\left[ \Lambda(D-2)  -\mathcal{R} \right]\,g_{\mu\nu} = 8\pi \mathcal{T}_{\mu\nu} \,, \label{EE}\\
 & \frac{\delta S_{m}}{\delta\Phi_{i}}\,=\,0 \label{FE}\,,
\end{align}
where $\mathcal{R_{\mu\nu}}$ is the Ricci tensor\label{RT}, and the symmetric tensor $\mathcal{T}_{\mu\nu}$, defined by the equation\label{ST}
\begin{equation}
\mathcal{T}^{\mu\nu}\,=\,\frac{2}{\sqrt{\left | g \right |}} \frac{\delta S_{m}}{\delta g_{\mu\nu}}\,,
\end{equation}
is the stress-energy tensor associated to the matter fields. Now, let the pair $g^{0}_{\mu\nu}$ and $\Phi^{0}_{i}$ be a solution for the equations of motion \eqref{EE} and \eqref{FE}. Then, in order to study the perturbations around this solution, we write our fields as a sum of the unperturbed
fields $g^{(0)}_{\mu\nu}$ and $\Phi^{(0)}_{i}$ and the small perturbations $h_{\mu\nu}$ and $\phi_{i}$
\begin{equation}\label{Pertubation}
g_{\mu\nu}\,=\,g^{(0)}_{\mu\nu}\,+\,h_{\mu\nu} \quad , \quad  \Phi_{i} \,=\,\Phi^{(0)}_{i}\,+\,\phi_{i} \,,
\end{equation}
where $h_{\mu\nu}$ and $\phi_{i}$ are assumed to be small in comparison with $g^{0}_{\mu\nu}$ and $\Phi^{0}_{i}$, respectively.  In such a case, we can drop terms of order $\mathcal{O}(h_{\mu\nu}^{2})$, $\mathcal{O}(\phi_{i}^{2})$ and $\mathcal{O}(h_{\mu\nu} \phi_{i})$ and higher in all equations and get consequently the linearized version of general relativity. Indeed, inserting the above ansatz \eqref{Pertubation} into \eqref{EE} and \eqref{FE} and neglecting quadratic and higher order powers of the perturbation fields, we are left with a set of linear equations satisfied by $h_{\mu\nu}$ and $\phi_{i}$. In general, these equations are coupled, namely $\phi_{i}$ is a source for $h_{\mu\nu}$ and vice versa.
However, around the particular background fields $g^{(0)}_{\mu\nu}$ and $\Phi^{(0)}_{i}=0$, the equations governing the perturbed fields $\phi_{i}$ can be decoupled from the metric perturbation $h_{\mu\nu}$ and vice versa. 
The reason why this happens is because, when $\Phi^{(0)}_{i}=0$, the stress-energy tensor $\mathcal{T}_{\mu\nu}$ can be set to zero at first order in the perturbation, since it is typically quadratic or of higher order in the matter fields and therefore can be neglected. In such a case, the dynamics of generic small perturbations of the matter fields is equivalent to studying the test fields $\phi_{i}$ in the fixed background $g^{(0)}_{\mu\nu}$.

\vspace{.5cm}
%\section{\uppercase{Effective Potential and Quasinormal Modes}}\label{EPQNMs}
\section{Effective Potential and Quasinormal Modes}\label{EPQNMs}
\vspace{.5cm}
%\section*{2.2\quad Effective Potential and Quasinormal Modes}\label{EPQNMs}
%\addcontentsline{toc}{section}{2.2\qquad Effective Potential and Quasinormal Modes}

In order to solve the perturbation equation for a given test field $\phi_{i}$ propagating in a given background $g^{(0)}_{\mu\nu}$, the first step is to separate the degrees of freedom of $\phi_{i}$ by
carefully choosing an angular basis\index{Angular basis} that allows us to decouple the angular variables in
the perturbation equation. Once the variables are decoupled, most of the problems concerning solving the perturbation equation, for instance, for the scalar field\index{Scalar field} (spin-$0$), the Dirac field\index{Dirac field} (spin-$1/2$), the Maxwell field\index{Maxwell field} (spin-$1$) and the gravitational field (spin-$2$), can be reduced to a second order partial differential equation for radial and time variables of the form\label{DO}

\begin{equation}\label{WEQ}
\left[ \partial_{x}^{2} - \partial_{t}^{2} - V(x) \right] Q(t, x) = 0\,, \quad \text{where} \quad \partial_{\mu} = \dfrac{\partial}{\partial x^{\mu}} \,,
\end{equation}
with $Q$ being a field related to the radial function of the perturbation, $x$ being the
tortoise coordinate\index{Tortoise coordinate} whose domain is the entire real line and $V(x)$ being an effective potential (EP)\label{EPF} that depends on the perturbation. However, the reduction process is not always so easy. Actually, the variables in perturbation equations cannot even be decoupled for perturbations of an arbitrary background once an arbitrary background possesses no symmetry. Indeed, the choice of a suitable angular basis\index{Angular basis} depends directly on the symmetries of the background. So, for the reduction process to be possible, the
background must possess sufficient symmetries. Such a symmetry is expressed by the
existence of Killing vectors and of other tensor associated with symmetries, such as
Killing tensors, Killing-Yano tensors and its conformal versions, and conformal Killing
tensors \cite{Carter1968A,Hughston1973,Frolov2008,Batista2014}. In this scenario, the choice of an appropriate spin basis\index{Spin basis} for the angular functions plays a central role for the reduction process \cite{Konoplya2000,Carter1968B,Bagrov}.

Before proceeding let us work out some examples in four dimensions in which we present a detailed derivation for the effective potential in Schwarzschild's background for various field perturbations. In order to perform this, the key point is the choice of an appropriate spin basis\index{Spin basis} for the angular functions which in its turn takes into account the spherical symmetry of the Schwarzschild background.

\vspace{.5cm}
\subsection{Example 1: Spin-0 Field Perturbations Around the Schwarzschild Background}\label{EX1}
\vspace{.5cm}

As a simple example, let us study the dynamics of spin-0 field perturbations on the $4$-dimensional Schwarzschild (S)\label{SB} background. The Schwarzschild line element for a spherical object of mass $M$ is given by
\begin{equation}\label{SBM}
g_{\mu\nu}^{S}dx^{\mu}dx^{\nu} = -f_{S}(r)dt^{2} + \frac{1}{f_{S}(r)}dr^{2}+ r^{2} \left(d\theta^{2} + \text{sin}^{2}\theta \, d\phi^{2} \right) \,,
\end{equation}
where the function $f_{S} = f_{S}(r)$ is given by
\begin{equation}
f_{S} = 1 - \dfrac{2M}{r} \,.
\end{equation}
As the background is spherically symmetric, it is useful to expand the angular dependence of a scalar field\index{Scalar field} $\boldsymbol{\Phi}$ in terms of scalar spherical harmonics\index{Scalar spherical harmonics} $Y_{\ell_{l}}^{m}(\theta, \phi)$, that is
\begin{equation}\label{ASFE1}
\boldsymbol{\Phi} = \sum_{\ell, m}R_{\ell,m}(t, r) \,Y_{\ell,m}(\theta, \phi).
\end{equation}
For a given value of $\ell \geq 0$ and integer  $m$  with $0-\ell  \leq m \leq \ell$, scalar spherical harmonics\index{Scalar spherical harmonics} are the only regular functions on the sphere satisfying the eigenvalue equation
\begin{equation}
\Delta_{S^{2}}Y_{\ell,m}(\theta, \phi) = -\ell(\ell + 1)\,Y_{\ell,m}(\theta, \phi) \,,
\end{equation}
with $\Delta_{S^{2}}$  being the Laplace-Beltrami operator on the unit sphere $S^{2}$, namely
\begin{equation}
\Delta_{S^{2}} \equiv \frac{1}{\text{sin}\theta} \partial_{\theta}(\text{sin}\theta\,\partial_{\theta}) + \frac{1}{\text{sin}^{2}\theta} \,\partial^{2}_{\phi} \,.
\end{equation}

Now, a scalar field\index{Scalar field} $\boldsymbol{\Phi}$ of mass $\mu$ is governed by the Klein-Gordon equation that, in curved spacetime, is given by
\begin{equation}\label{KGEE1}
\frac{1}{\sqrt{\left | g^{S} \right |}} \,\partial_{\mu} \left( g^{S\,\mu\nu}\sqrt{ \left | g^{S} \right |} \,\partial_{\nu} \right)\boldsymbol{\Phi} \,=\,\mu^{2} \boldsymbol{\Phi} \,.
\end{equation}
%\begin{equation}
%\frac{1}{\sqrt{\left | g^{S} \right |}} \,\partial_{\mu} \left( g^{SB\,\mu\nu}\sqrt{ \left | g^{S} %\right |} \,\partial_{\nu} \right)\boldsymbol{\Phi} \,=\,0 \,.
%\end{equation}
The advantage of using scalar spherical harmonics\index{Scalar spherical harmonics} as angular basis\index{Angular basis}, namely \eqref{ASFE1}, is that in the above equation the angular dependence automatically factors out as a global multiplicative term, so that we end up with an equation that depends just on the coordinate $r$. Indeed, plugging the ansatz \eqref{ASFE1} into the Eq. \eqref{KGEE1}, we find a second order partial differential equation for the field $R_{\ell,m}$:
\begin{equation}\label{DESFE1}
f_{S}^{2}\,\partial_{r}^{2} (r R_{\ell,m}) + f_{S}f'_{S} \,\partial_{r}(r R_{\ell,m}) - \partial_{t}^{2}(r R_{\ell,m}) - f_{S}\left[\dfrac{\ell(\ell + 1)}{r^{2}} + \mu^{2}  \right] (r R_{\ell,m}) = 0 \,.
\end{equation}
A very common and useful trick is to change to tortoise coordinate\index{Tortoise coordinate} $x$ defined by the
equation
\begin{equation}\label{TCSB}
dx = \dfrac{1}{f_{S}}\, dr \quad \Rightarrow \quad x = r + 2M \text{ln}(r - 2M) \,.
\end{equation}
Indeed, in addition to this change of variable, if we make the field redefinition
\begin{equation}
R_{\ell,m}(t, r)  = \dfrac{\phi_{\ell,m}(t, r)}{r} \,,
\end{equation}
%\begin{equation}
%\boldsymbol{\Phi} = \int d\omega \sum_{\ell, m}\dfrac{e^{i\omega t} \phi^{\omega}_{\ell,m}(r)}{r} %\,Y_{\ell,m}(\theta, \phi) \,.%
%\end{equation}
from \eqref{DESFE1}, we are left with the following one-dimensional wave-like equation\index{Wave-like equation} for the field $\phi_{\ell,m}$:
\begin{equation}
\left[ \partial_{x}^{2} - \partial_{t}^{2} - V_{s=0}(r) \right]\phi_{\ell,m}(t, r) = 0 \,,
\end{equation}
where the effective potential $V_{s=0}$ has the form
\begin{equation}
V_{s=0}(r) = f_{S} \left[ \dfrac{\ell(\ell +1)}{r^{2}} + \dfrac{2M}{r^{3}} + \mu^{2}  \right] \,.
\end{equation}
%\begin{equation}
%V(r) = f_{S} \left[ \dfrac{\ell(\ell +1)}{r^{2}} + \dfrac{2M}{r^{3}}\right] \,.
%\end{equation}
The $s=0$ label stands for the spin of the scalar field\index{Scalar field}.

\hfill\(\Box\)
\\

%%%%%%%%%%%%%%%%%%%%%%%%%%%%%%%%%%%%%%%%%%%%%%%%%%%%%%%%%%%%%%%%%%%%%%%%%%%%%%%%%%%
%%%%%%%%%%%%%%%%%%%%%%%%%%% E: Example 1
%%%%%%%%%%%%%%%%%%%%%%%%%%%%%%%%%%%%%%%%%%%%%%%%%%%%%%%%%%%%%%%%%%%%%%%%%%%%%%%%%%%

%%%%%%%%%%%%%%%%%%%%%%%%%%%%%%%%%%%%%%%%%%%%%%%%%%%%%%%%%%%%%%%%%%%%%%%%%%%%%%%%%%%
%%%%%%%%%%%%%%%%%%%%%%%%%%% B: Example 2
%%%%%%%%%%%%%%%%%%%%%%%%%%%%%%%%%%%%%%%%%%%%%%%%%%%%%%%%%%%%%%%%%%%%%%%%%%%%%%%%%%%

\vspace{.5cm}
\subsection{Example 2: Spin-1 Field Perturbations Around the Schwarzschild Background}\label{EX2}
\vspace{.5cm}
%Here and in the rest of this thesis, for notational simplicity, we usually omit the integral over frequency in the Fourier transform

Consider a massless, uncharged, spin-$1$ field $\mathcal{\boldsymbol{A}}$, propagating in a background described by the metric $g_{\mu\nu}^{S}$, namely \eqref{SBM}. To separate the angular dependence, once the background is spherically symmetric and the field has spin-$1$, a suitable angular basis\index{Angular basis} for the angular functions is provided by the vector spherical harmonics\index{Vector spherical harmonics}, denoted here by $\boldsymbol{\mathcal{E}}_{a, \ell m}$, where $ a \in \{1, 2, 3 \}$.
The latter objects are given by
\begin{equation}
  \left\{
     \begin{array}{ll}
       \boldsymbol{\mathcal{E}}_{1,\ell m} = \frac{1}{r}\,\boldsymbol{r}\,\, Y_{\ell,m}(\theta,\phi) \,, \\
       \boldsymbol{\mathcal{E}}_{2,\ell m} = \boldsymbol{r} \times \bl{\nabla} Y_{\ell,m}(\theta,\phi) \,, \\
        \boldsymbol{\mathcal{E}}_{3,\ell m} = r\, \boldsymbol{\nabla} Y_{\ell,m} (\theta,\phi)   \,,
     \end{array}
   \right.
\end{equation}
where $\{r,\theta,\phi\}$ is a spherical coordinate system in $\mathbb{R}^3$. Since these three vector fields are orthogonal to each other, it follows that they are linearly independent and, therefore, form a frame for the space of vector fields in  $\mathbb{R}^3$. Thus, in a spherically symmetric problem it is natural to expand vector fields $\boldsymbol{A}$ in terms of the basis $\{\boldsymbol{\mathcal{E}}_{a,\ell m}\}$ as
\begin{equation}\label{A}
  \boldsymbol{A} = A^{a}(r) \,\boldsymbol{\mathcal{E}}_{a,\ell m}\,,
\end{equation}
where the sum over the indices $\ell$ and $m$ of $Y_{\ell,m}$ have been omitted for simplicity. Using the expression for the gradient in spherical coordinates, it follows that the vector spherical harmonics\index{Vector spherical harmonics} are given by
\begin{align}
 \boldsymbol{\mathcal{E}}_{1,\ell m} &=  Y_{\ell,m}(\theta,\phi)\,\hat{\boldsymbol{e}}_r  \;,\;\; \nonumber \\
          \boldsymbol{\mathcal{E}}_{2,\ell m} &= \frac{-\,1}{\sin\theta}\partial_\phi Y_{\ell,m}(\theta,\phi)\, \,\hat{\boldsymbol{e}}_\theta
          + \partial_\theta Y_{\ell,m}(\theta,\phi)\, \,\hat{\boldsymbol{e}}_\phi \,, \;\;\;  \nonumber \\
           \boldsymbol{\mathcal{E}}_{3,\ell m} &=  \partial_\theta Y_{\ell,m}(\theta,\phi)\, \,\hat{\boldsymbol{e}}_\theta
          + \frac{1}{\sin\theta}\partial_\phi Y_{\ell,m}(\theta,\phi)\, \,\hat{\boldsymbol{e}}_\phi     \nonumber
\end{align}
where $\{\hat{\boldsymbol{e}}_r,\hat{\boldsymbol{e}}_\theta,\hat{\boldsymbol{e}}_\phi\}$ is the orthonormal frame associated to the spherical coordinates $\{r,\theta,\phi\}$. More precisely, their connection with coordinate frame $\{\partial_{r}, \partial_{\theta}, \partial_{\phi} \}$ is the following:
\begin{equation}
  \hat{\boldsymbol{e}}_r = \partial_r  \quad , \quad  \hat{\boldsymbol{e}}_\theta = \frac{1}{r}\partial_\theta  \quad , \quad \hat{\boldsymbol{e}}_\phi = \frac{1}{r \sin\theta}\partial_\phi \,.
\end{equation}
Thus, the generic vector field $\boldsymbol{A}$ of Eq. (\ref{A}) is written as
\begin{align}
  \boldsymbol{A} =& A^1\,Y_{\ell,m}\,\hat{\boldsymbol{e}}_r +  \left( A^3 \partial_\theta Y_{\ell,m} - A^2\frac{\partial_\phi Y_{\ell,m}}{\sin\theta} \right)\hat{\boldsymbol{e}}_\theta \nonumber\\
&+ \left( A^3  \frac{\partial_\phi Y_{\ell,m}}{\sin\theta} + A^2\partial_\theta Y_{\ell,m} \right) \hat{\boldsymbol{e}}_\phi\,.
\end{align}

Likewise, in a spherically symmetric problem, a $1$-form $\boldsymbol{\tilde{A}}$ is conveniently expanded in the following way
\begin{align}
  \boldsymbol{\tilde{A}} = & A_1\,Y_{\ell,m}\,\hat{\boldsymbol{e}}^r +
\left( A_3 \partial_\theta Y_{\ell,m} - A_2\frac{\partial_\phi Y_{\ell,m} }{\sin\theta} \right)\hat{\boldsymbol{e}}^\theta \nonumber\\
& + \left( A_3  \frac{\partial_\phi Y_{\ell,m} }{\sin\theta} + A_2\partial_\theta Y_{\ell,m}  \right) \hat{\boldsymbol{e}}^\phi\,,
\end{align}
where $\{ \hat{\boldsymbol{e}}^r,\hat{\boldsymbol{e}}^\theta,\hat{\boldsymbol{e}}^\phi\}$ stands for the frame of $1$-forms that is dual to the frame of vector fields $\{\hat{\boldsymbol{e}}_r,\hat{\boldsymbol{e}}_\theta,\hat{\boldsymbol{e}}_\phi\}$, namely $\hat{\boldsymbol{e}}^{a}(\hat{\boldsymbol{e}}_{b})=\delta^a_b$. This frame is related to the coordinate frame $\{dr,d\theta,d\phi\}$ as follows:
\begin{equation}
 \hat{\boldsymbol{e}}^r = dr\,,\; \hat{\boldsymbol{e}}^\theta=r d\theta \,,\; \hat{\boldsymbol{e}}^\phi=r \sin\theta d\phi \,,
\end{equation}
so that the line element of $\mathbb{R}^3$ is written, in spherical coordinates, as $ds^2 =  (\hat{\boldsymbol{e}}^r)^2 + (\hat{\boldsymbol{e}}^\theta)^2 +(\hat{\boldsymbol{e}}^\phi)^2$.  Thus, the most general decomposition of the $1$-form field in a problem with spherical symmetry is:
\begin{align}
  \boldsymbol{\tilde{A}} =   A^{+}_{1} Y_{\ell,m}\,dr   + A^{+}\, V_{\ell,m}^{+} + A^{-} \,V_{\ell,m}^{-}\,,
\end{align}
where $A_{1}^{+} = A_{1}, A^{+} = r A_{3}, A^{-} = -r A_{2}$ and $V_{\ell,m}^{\pm}$ being the ideal basis for the angular dependence once the background has spherical symmetry
\begin{equation}
V_{\ell,m}^{+} = \partial_\theta Y_{\ell,m}d\theta +  \partial_\phi Y_{\ell,m}d\phi \quad \text{and} \quad 
V_{\ell,m}^{-} = \dfrac{1}{\sin\theta }\, \partial_\phi Y_{\ell,m} \, d\theta -\sin\theta \, \partial_\theta Y_{\ell,m} \, d\phi \,.
\end{equation}
Taking into account such a spherical symmetry of the Schwarzschild background, the ansatz for the gauge field $\boldsymbol{\mathcal{A}} = \mathcal{A}_{\mu} dx^{\mu}$ which is in agreement with such symmetries is given by
\begin{align}
\boldsymbol{\mathcal{A}} =  \left( A^{+}_{0} \,dt + A^{+}_{1}\,dr \right) Y_{\ell,m}  + A^{+}\, V_{\ell,m}^{+} + A^{-} \,V_{\ell,m}^{-} \,,
\end{align}
where now  $A^{+}_{0} = A^{+}_{0}(t, r),A^{+}_{1} = A^{+}_{1}(t, r) $ and $A^{\pm} = A^{\pm}(t, r)$.  Notice, however, that the above expression can be rewritten in the following form
\begin{align}
\boldsymbol{\mathcal{A}} =  \left( A^{+}_{0} \,dt + \tilde{A}^{+}_{1}\,dr \right) Y_{\ell,m} + A^{-} \,V_{\ell,m}^{-} + d\tilde{A}^{+} \,,
\end{align}
where $\tilde{A}^{+}_{1} = A^{+}_{1} - \partial_{r} A^{+}$ and $\tilde{A}^{+} = A^{+} Y_{\ell,m}$. In a $U(1)$ gauge field theory, we can ignore the degree of freedom $\tilde{A}^{+}$ in the previous equation, since an exact differential can be eliminated by a gauge transformation\index{Gauge transformation}. Thus, dropping the tildes, we can say that a natural ansatz for a $1$-form gauge field in Schwarzschild background which is a problem with spherical symmetry, is:
\begin{align}
\boldsymbol{\mathcal{A}} =  \sum_{\ell , m}\left[ A^{+}_{0, \ell m}(t, r) \,dt + A^{+}_{1, \ell m}(t, r)\,dr \right] Y_{\ell,m} + A^{-}_{\ell, m}(t, r) \,V_{\ell,m}^{-} \,,
\end{align}
after we recover the sum over the indices $\ell, m$.

Now, spin-$1$ field perturbations are governed by Maxwell's equations
\begin{equation}\label{MEE2}
\nabla_{\mu}\mathcal{F}^{\mu\nu}\,=\,0 \,, \quad \text{with} \quad \mathcal{F}_{\mu\nu}\,=\,\partial_{\mu}\mathcal{A}_{\nu}\,-\,\partial_{\nu}\mathcal{A}_{\mu}\,,
\end{equation}
where $\mathcal{F}^{\mu\nu}$ is the Maxwell tensor and $\mathcal{A}^{\mu}$ are the components of the gauge field $\boldsymbol{\mathcal{A}}$. It is worth mentioning that the decomposition of $\boldsymbol{\mathcal{A}}$ in the basis $\{ Y_{\ell,m},V_{\ell,m}^{+}, V_{\ell,m}^{-} \}$ is a crucial
factor in order to attain the separation process of the Maxwell equation, since its angular
dependence becomes a global multiplicative factor, so that we end up with an equation
depending just on the coordinate $r$. Besides that, another important advantage of using
such a basis comes from the behavior of $\boldsymbol{\mathcal{A}}$ under a parity transformation\index{Parity transformation} as described in the following. A parity transformation is a noncontinuous operation $P$ such that

\begin{equation}\label{PTE2}
P\,:\,\theta \rightarrow \theta - \pi \quad \text{and} \quad \phi \rightarrow \phi + \pi \,.
\end{equation}
By noncontinuous, we mean that the operation cannot be decomposed in infinitesimal operations and thus such operations have no generator. Since the operation $P$ applied to itself is the identity operation, eigenvalues of the operator $P$ can be only $\pm 1$. Thus when acting on angular dependence of the gauge field $\boldsymbol{\mathcal{A}}$ via \eqref{PTE2} the parity transformation\index{Parity transformation} $P$ splits it into a sum of two distinct classes of fields. In order to see this, notice that the fields $A^{+}_{0, \ell m}, A^{+}_{1, \ell m}$ and $A^{\pm}_{\ell,m}$ remain unchanged when a parity transformation\index{Parity transformation} is applied, so that the parity of $\boldsymbol{\mathcal{A}}$ is completely determined from its angular part, the angular basis\index{Angular basis} $\{ Y_{\ell,m},V_{\ell,m}^{+}, V_{\ell,m}^{-} \}$. Under parity transformation \eqref{PTE2}, scalar spherical harmonic transforms as

\begin{equation}
Y_{\ell,m} \,\overset{P\,}{\rightarrow}\,(-1)^{\ell} \,Y_{\ell,m} \,,
\end{equation}
and using this, it is easy matter to see that vectorial spherical harmonic transforms as $V_{\ell,m}^{\pm} \overset{P\,}{\rightarrow}\,\pm (-1)^{\ell} \,V_{\ell,m}^{\pm}$. It follows that we can write $\boldsymbol{\mathcal{A}}$ as
\begin{equation}
\boldsymbol{\mathcal{A}} = \boldsymbol{\mathcal{A}}^{+} + \boldsymbol{\mathcal{A}}^{-} \,,
\end{equation}
where the objects $\boldsymbol{\mathcal{A}}^{\pm}$ defined by
\begin{align}
\boldsymbol{\mathcal{A}}^{+} &=\,\,\sum_{\ell , m}\left[ A^{+}_{0, \ell m}(t, r) \,dt + A^{+}_{1, \ell m}(t, r)\,dr \right] Y_{\ell,m} \,, \nonumber\\
\\
\boldsymbol{\mathcal{A}}^{-} &=\,\, \sum_{\ell , m}A^{-}_{\ell, m}(t, r) \,V_{\ell,m}^{-} \,,\nonumber
\end{align}
transform as $\boldsymbol{\mathcal{A}}^{\pm} \overset{P\,}{\rightarrow}\,\pm (-1)^{\ell} \,\boldsymbol{\mathcal{A}}^{\pm}$ under parity transformation\index{Parity transformation}. The fields corresponding to eigenvalue $+1$ will be dubbed \textit{even}, while the fields corresponding to eigenvalue $-1$ will be dubbed \textit{odd}, the reason why $\pm$ has been employed in order to label the fields. In particular, even fields have parity $(-1)^{\ell}$, while odd fields have parity $(-1)^{\ell+1}$. Finally, since the Schwarzschild background metric does not change when a parity transformation\index{Parity transformation} is applied, we expect that the perturbation equations will not mix $(-1)^{\ell}$ and $(-1)^{\ell+1}$ parities. So, we can, without loss of generality, separate the perturbation into its $\boldsymbol{\mathcal{A}}^{+}$ and $\boldsymbol{\mathcal{A}}^{-}$ parts and study them separately.

\subsubsection{Even Perturbation (spin-1)}

By an even perturbation\index{Even perturbation} we mean the most general perturbation for a given $\ell, m$ and
parity $(-1)^{\ell}$, namely
\begin{equation}
\boldsymbol{\mathcal{A}}^{+} = \sum_{\ell , m}\left[ A^{+}_{0, \ell m}(t, r) \,dt + A^{+}_{1, \ell m}(t, r)\,dr \right] Y_{\ell,m} \,.
\end{equation}
Inserting this ansatz for the massless spin-$1$ field perturbations into the Maxwell equation, we are eventually led to the following equations:
\begin{align}
E_{t}^{+} &\equiv\,\, \partial_{x} \left[r^{2}\left(\partial_{r} A^{+}_{0, \ell m} - \partial_{t} A^{+}_{1, \ell m} \right ) \right ] - \ell (\ell +1)A^{+}_{0, \ell m} = 0 \,, \\
E_{r}^{+} &\equiv\,\, \partial_{t} \left[r^{2}\left(\partial_{r} A^{+}_{0, \ell m} - \partial_{t} A^{+}_{1, \ell m} \right ) \right ] - \ell (\ell +1)f_{S}  A^{+}_{1, \ell m} = 0 \,, \\
E_{\theta}^{+} &\equiv\,\, \partial_{t}E_{t}^{+} -\partial_{x}E_{r}^{+} = 0 \,,\\
E_{\phi}^{+} &\equiv\,\, E_{\theta}^{+}  = 0 \,. 
\end{align}
Assuming that $E_{t}^{+} = 0$ and $E_{r}^{+} = 0$, it follows that $E_{\theta}^{+} = 0$ and hence $E_{\phi}^{+}$ are trivially satisfied. Defining, now, a new field $A^{\ell m}_{01} = A^{\ell m}_{01}(t, r)$ as
\begin{equation}
A^{\ell m}_{01} := r^{2}\left(\partial_{r} A^{+}_{0, \ell m} - \partial_{t} A^{+}_{1, \ell m} \right )\,,
\end{equation}
and assuming the field equations $E_{t}^{+} = 0$ and $E_{r}^{+} = 0$, it follows immediately from the
relation
\begin{equation}
\partial_{x}E_{t}^{+} -\partial_{t}E_{r}^{+} = 0 
\end{equation}
that $A^{\ell m}_{01}$ obeys the one-dimensional wave-like equation\index{Wave-like equation}
\begin{equation}
\left[ \partial_{x}^{2} - \partial_{t}^{2} - V_{s=1}(r) \right]A^{\ell m}_{01}(t, r) = 0 \,,
\end{equation}
where the effective potential $V_{s=1}$ has the form
\begin{equation}\label{EPMFE2}
V_{s=1}(r) = f_{S} \left[\dfrac{\ell(\ell +1)}{r^{2}}\right] \,,
\end{equation}
with the $s=1$ label indicating the spin of the Maxwell field\index{Maxwell field}. In particular, the fields $ A^{+}_{0, \ell m}$ and $ A^{+}_{1, \ell m}$ of the Maxwell perturbation are related to $A^{\ell m}_{01}$ by
\begin{equation}
A_{0, \ell m}^{+} = \dfrac{1}{\ell(\ell +1)} \, \partial_{x} A^{\ell m}_{01} \quad \text{and} \quad  A_{1, \ell m}^{+} = \dfrac{1}{\ell(\ell +1)} \, \dfrac{\partial_{x} A^{\ell m}_{01}}{f_{S}} \,,
\end{equation}
where, as in the previous example, we make use of the tortoise coordinate\index{Tortoise coordinate} defined in Eq. \eqref{TCSB}.

\subsubsection{Odd Perturbation (spin-1)}

By an odd perturbation\index{Odd perturbation} we mean the most general perturbation for a given $\ell, m$ and
parity $(-1)^{\ell+1}$, namely
\begin{equation}
\boldsymbol{\mathcal{A}}^{-} =  \sum_{\ell , m}A^{-}_{\ell, m}(t, r) \,V_{\ell,m}^{-} \,.
\end{equation}
Inserting the above ansatz into the Maxwell equation, we are left with the following equations:
\begin{align}
E_{\theta}^{-} &\equiv\,\, \left[ \partial_{x}^{2} - \partial_{t}^{2} - V_{s=1}(r) \right]A^{-}_{\ell, m} = 0 \,,\nonumber\\
E_{\phi}^{-} &\equiv\,\, E_{\theta}^{-}  = 0 \,, \nonumber
\end{align}
where $V_{s=1}(r)$ is the potential defined in Eq. \eqref{EPMFE2}. Hence, assuming that $E_{\theta}^{-} = 0$, we have that $E_{\phi}^{-} = 0$ and that $A^{-}_{\ell, m}$ obeys the same Schr\"{o}dinger-like differential equation\index{Schr\"{o}dinger-like differential equation} as $A^{\ell m}_{01}$.

\hfill\(\Box\)
\\

%%%%%%%%%%%%%%%%%%%%%%%%%%%%%%%%%%%%%%%%%%%%%%%%%%%%%%%%%%%%%%%%%%%%%%%%%%%%%%%%%%%
%%%%%%%%%%%%%%%%%%%%%%%%%%% E: Example 2
%%%%%%%%%%%%%%%%%%%%%%%%%%%%%%%%%%%%%%%%%%%%%%%%%%%%%%%%%%%%%%%%%%%%%%%%%%%%%%%%%%%

%%%%%%%%%%%%%%%%%%%%%%%%%%%%%%%%%%%%%%%%%%%%%%%%%%%%%%%%%%%%%%%%%%%%%%%%%%%%%%%%%%%
%%%%%%%%%%%%%%%%%%%%%%%%%%% B: Example 3
%%%%%%%%%%%%%%%%%%%%%%%%%%%%%%%%%%%%%%%%%%%%%%%%%%%%%%%%%%%%%%%%%%%%%%%%%%%%%%%%%%%

\vspace{.5cm}
\subsection{Example 3: Spin-2 Field Perturbations Around the Schwarzschild Background}\label{EX3}
\vspace{.5cm}

Let us consider spin-$2$ field perturbations in the Schwarzschild Background, a spherically symmetric vacuum solution of Einstein's equations, $\mathcal{R}_{\mu\nu} = 0$. In order to perform this, consider a small perturbation $h_{\mu\nu}$ in $g_{\mu\nu}^{S}$ such that the perturbed metric can be taken as the sum of unperturbed background metric and perturbation,
\begin{equation}
g_{\mu\nu} = g_{\mu\nu}^{S} + h_{\mu\nu} \,,
\end{equation}
with $h_{\mu\nu}$ being very small compared with $g_{\mu\nu}^{S}$. In order to build the ansatz for the spin-$2$ perturbation $h_{\mu\nu}$, we must note that its $10$ degrees of freedom transform differently under rotations on the sphere $S^{2}$. Indeed, by decomposing the perturbation $\boldsymbol{h} = h_{\mu\nu}dx^{\mu}dx^{\nu}$ as 
\begin{align}
\boldsymbol{h} &=\,\, h_{tt}\,dt^{2} +  2 h_{tr}\,dt dr +  h_{rr}\,dr^{2} + (h_{ta}dx^{a})dt + (h_{ra}dx^{a})dr + \nonumber\\
 &+\,\, h_{ab}\,dx^{a}dx^{b} \quad \text{where} \quad a,b \in \{\theta, \phi \} \,,
\end{align}
we see that, under such a rotation, the perturbation $h_{\mu\nu}$ comprises $3$ fields which transform as scalar fields\index{Scalar field}, $2$ fields transforming as the components of $1$-forms with respect to the $S^{2}$ and finally $1$ field transforming as the components of a second order tensor in $S^{2}$.  Each type of field should be expanded in terms of an angular basis\index{Angular basis} that has the same nature. The scalar fields\index{Scalar field} $h_{tt}, h_{tr}$ and $h_{rr}$ are naturally expanded in terms of scalar spherical harmonics\index{Scalar spherical harmonics}, $Y_{\ell,m}$, and the $1$-forms $h_{ta}dx^{a}$ and $h_{ra}dx^{a}$ in terms of vector spherical harmonics\index{Vector spherical harmonics}, $V^{\pm}_{\ell, m}$, as seen in the examples 1 and 2. Now, for the second order tensor $h_{ab}\,dx^{a}dx^{b}$,
there are three fundamental types of elements, they are: $T^{\oplus}_{\ell,m} = Y_{\ell,m}\, \hat{g}_{ab}\,dx^{a}dx^{b}, T^{+}_{\ell,m} = \hat{\nabla}_{a} \hat{\nabla}_{b} Y_{\ell,m}dx^{a}dx^{b}$ and $T^{-}_{\ell,m} = (\hat{\epsilon}_{ac}  \hat{\nabla}_{b} \hat{\nabla}^{c} Y_{\ell,m} +
\hat{\epsilon}_{bc}   \hat{\nabla}_{a} \hat{\nabla}^{c} Y_{\ell,m})dx^{a}dx^{b}$, where $\hat{g}_{ab}$ is the metric tensor on the sphere, $\hat{g}_{\theta\theta} = 1, \hat{g}_{\theta\phi} = 0, \hat{g}_{\phi\phi} = \sin^{2}\theta$, whereas $\hat{\epsilon}_{ab}$ is the volume form in
the sphere. Explicitly, we have
\begin{align}
T^{\oplus}_{\ell,m} &=\,\, Y_{\ell,m}d\theta^2 + \sin^2\theta \,  Y_{\ell,m} d\phi^2 \\
\nonumber\\
T^{+}_{\ell, m} &=\,\, \partial_\theta^2 Y_{\ell,m}d\theta^2 +
2\left( \partial_\theta \partial_\phi Y_{\ell,m}- \cot\theta\, \partial_\phi Y_{\ell,m} \right) \, d\theta d\phi \\
&+\,\,
\left( \partial_\phi^2 Y_{\ell,m} + \cos\theta\,\sin\theta\, \partial_\theta Y_{\ell,m} \right) \, d\phi^2 \nonumber\\
\nonumber\\
T^{-}_{\ell, m} &= 2\csc\theta \left(\partial_\theta  \partial_\phi  Y_{\ell,m} - \cot\theta\, \partial_\phi Y_{\ell,m} \right)  d\theta^2 +\nonumber\\
&+\,\,\,2 \left( \cos\theta\, \partial_\phi Y_{\ell,m} -  \sin\theta\, \partial_\theta  \partial_\phi  Y_{\ell,m} \right) \, d\phi^2 \\
 &+\,\, 4\left(\csc\theta \, \partial_\phi^2 Y_{\ell,m} +  \cos\theta \, \partial_\theta Y_{\ell,m} -
 \sin\theta \, \partial_\theta^2  Y_{\ell,m} \right) \, d\theta d\phi \nonumber \,.
\end{align}
Thus, it follows that the most suitable way to expand the perturbation $\boldsymbol{h}$ is
\begin{align}
 \boldsymbol{h} &=\,\, \left[ H_{tt}(t,r)\, dt^2 + H_{rr}(t,r) \,dr^2 + 2H_{tr}(t,r) \, dtdr \right] Y_{\ell,m} \nonumber\\
&+\,\, \left[ H^{+}_{t}(t,r)dt  +  H^{-}_{r}(t,r)dr \right] V^{+}_{\ell,m} + \left[H^{-}_{t}(t,r)dt  +  H^{-}_{r}(t,r)dr \right] V^{-}_{\ell,m} \\
&+\,\, H^{\oplus}(t,r)\,T^{\oplus}_{\ell,m} + H^{+}(t,r)\,T^{+}_{\ell,m} +  H^{-}(t,r)\,T^{-}_{\ell,m}\,\,.\nonumber
\end{align}

Spin-$2$ field perturbations are governed by the linearized version of Einstein's equation
\begin{align}\label{PES2E3}
 \delta \mathcal{R}_{\mu\nu} = 0 \quad \Rightarrow \quad  2\nabla^{\sigma}\nabla_{(\mu}h_{\nu)\sigma}\,-\,\Box h_{\mu\nu} \,-\, \nabla_{\mu}\nabla_{\nu}\,h  = 0\,,
\end{align}
where
\begin{equation}
\Box = g^{SB\,\mu\nu}\nabla_\mu\nabla_\nu \quad \text{and} \quad h = g^{SB\,\mu\nu}h_{\mu\nu} \,.
\end{equation}
In spite of the fact that the Eq. \eqref{PES2E3} is much simpler to solve than the full Einstein's equation, even in the simplest cases like perturbations on the Schwarzschild background, proved to be challenging. Regge and Wheeler were the first to decompose the spin-$2$ perturbations in Schwarzschild background in the angular basis\index{Angular basis} $\{Y_{\ell,m}, V^{\pm}_{\ell, m}, T^{\oplus}_{\ell,m}, T^{\pm}_{\ell,m}\}$ \cite{Regge1957}. The great advantage of using such a basis is that, in terms of it, they were able
to classify spin-$2$ perturbations into two types: odd, which have parity $(-1)^{\ell + 1}$ and
even, which have parity $(-1)^{\ell}$, that is
\begin{equation}
\boldsymbol{h} = \boldsymbol{h}^{+} + \boldsymbol{h}^{-} \,,
\end{equation}
where $\boldsymbol{h}^{\pm}$ are given by
\begin{align}
\boldsymbol{h}^{+} &=\,\, \left[ H_{tt}(t,r)\, dt^2 + H_{rr}(t,r) \,dr^2 + 2H_{tr}(t,r) \, dtdr \right] Y_{\ell,m}\nonumber\\
&+\,\, \left[ H^{+}_{t}(t,r)dt  +  H^{+}_{r}(t,r)dx \right] V^{+}_{\ell,m} \\
&+\,\,H^{\oplus}(t,r)\,T^{\oplus}_{\ell,m} + H^{+}(t,r)\,T^{+}_{\ell,m} \,,\nonumber\\
\nonumber\\
\boldsymbol{h}^{-} &=\,\,\left[H^{-}_{t}(t,r)dt  +  H^{-}_{x}(t,r)dx \right]\,V^{-}_{\ell,m}  +  H^{-}(t,r)\,T^{-}_{\ell,m} \,.
\end{align}
The labels $\pm$ in $\boldsymbol{h}^{\pm}$ means that, under parity transformation\index{Parity transformation} \eqref{PTE2}, $\boldsymbol{h}^{\pm}$ is multiplied by $\pm (-1)^{\ell}$. For this reason, object $\boldsymbol{h}^{+}$ is said to have even parity, $(-1)^{\ell}$, while $\boldsymbol{h}^{-}$ is said to have odd parity, $(-1)^{\ell+1}$. Since the Schwarzschild background metric does not change when a parity transformation\index{Parity transformation} is applied, we expect that the perturbation equations will not mix $(-1)^{\ell}$ and $(-1)^{\ell+1}$ parities. So, we can, without loss of generality, separate the perturbation into its $\boldsymbol{h}^{+}$ and $\boldsymbol{h}^{+}$ parts and study them separately.

\vspace{.5cm}
\subsubsection{{Odd Perturbation (spin-2)}}
\vspace{.5cm}

Regge and Wheeler showed that the equations for $\boldsymbol{h}^{-}$ can be put into the form
of a Schr\"{o}dinger-like differential equation\index{Schr\"{o}dinger-like differential equation} with a nonintegrable potential. However, there were some minor errors in the equations given by Regge and Wheeler. Indeed, Manasse pointed out that the equations appearing in the literature contained mistakes and were inconsistent with Einstein's field equation \cite{Manasse1963}. Brill and Hartle rederived the odd parity equations which once again contained some errors as published \cite{Brill1964}. Let us now obtain the correct differential equations for perturbations on the Schwarzschild metric for odd parity which have been given by Vishveshwara \cite{VishveshwaraThesis}, displayed for both parities in Appendix of his doctoral thesis and published later in Ref. \cite{Edelstein1970}. In order to perform this, we can use the freedom of choosing a gauge to simplify the general form of the perturbations. Let us work with the classical Regge-Wheeler\label{RWL} (RW) gauge in which the canonical form for the odd perturbations\index{Odd perturbation} is \cite{ Regge1957}
\begin{equation}
\boldsymbol{h}^{-}_{RW} \,=\,\left[H^{-}_{t}(t,r)dt  +  H^{-}_{x}(t,r)dx \right]V^{-}_{\ell,m} \,.
\end{equation}
Inserting this ansatz into the equation \eqref{PES2E3}, we find that, out of the $10$ Einstein equations\index{Einstein equation}, only $3$ are independent. They are:
\begin{align}
E_{t\phi}^{-} &\equiv\,\, \partial_{r}^{2}H^{-}_{t} -  \partial_{r}\partial_{t} H^{-}_{r} - \dfrac{2}{r}\,\partial_{t}H^{-}_{r} - \left[\dfrac{\ell (\ell +1)}{r^{2}}- \dfrac{4M}{r^{3}}\right ]\dfrac{H_{r}^{-}}{f_{S}}  = 0 \,,\\
\nonumber\\
E_{r\phi}^{-} &\equiv\,\, \partial_{t}^{2}H^{-}_{r} + \left(\dfrac{2}{r} - \partial_{r}\right )  \partial_{t} H^{-}_{t} + \left[\dfrac{\ell (\ell +1)}{r}- \dfrac{2}{r}\right ]\dfrac{f_{S}}{r}\,H_{r}^{-}  = 0 \,,\\
\nonumber\\
E_{\theta\phi}^{-} &\equiv\,\, \partial_{t}H^{-}_{t} - f_{S}\,\partial_{r}(f_{SB}H^{-}_{r}) = 0 \,.
\end{align}
It appears that we have $3$ differential equations for only $2$ fields, $H^{-}_{t}$ and $H^{-}_{r}$. However,
it is not hard to verify that the first equation is a consequence of the other two. Defining the field
\begin{equation}
Q^{-}_{RW}(t, r) := \dfrac{f_{SB}}{r}\, H_{r}^{-}(t, r) \,,
\end{equation}
it follows immediately from the equation
\begin{equation}
\left(\dfrac{2}{r} - \partial_{r}\right ) E^{-}_{\theta\phi} - E^{-}_{r\phi} = 0 \,,
\end{equation}
which is a direct consequence of the fact that $ E^{-}_{\theta\phi}=0$ and $E^{-}_{r\phi}=0$, that $Q^{-}_{RW}$ satisfies the one-dimensional wave-like equation\index{Wave-like equation}:
\begin{equation}\label{RWE}
\left[ \partial_{x}^{2} - \partial_{t}^{2} - V_{s=2}^{RW}(r) \right]Q^{-}_{RW}(t, r) = 0 \,,
\end{equation}
where $x$ is the usual tortoise coordinate\index{Tortoise coordinate}, namely \eqref{TCSB}, and the effective potential $V_{s=2}^{RW}$ is given by
\begin{equation}\label{RWP}
V_{s=2}^{RW}(r) = f_{S} \left[ \dfrac{\ell(\ell +1)}{r^{2}} - \dfrac{6M}{r^{3}} \right] \,,
\end{equation}
with the $s=2$ label standing for the spin of the gravitational field. The equation \eqref{RWE}
became known as the Regge-Wheeler equation and the effective potential \eqref{RWP} became
known as the Regge-Wheeler potential.

\vspace{.5cm}
\subsubsection{{Even Perturbation (spin-2)}}
\vspace{.5cm}

In a similar way to the odd perturbation\index{Odd perturbation} case, we can use the gauge freedom in order to simplify the general form of the perturbation. In the case of even perturbations\index{Even perturbation}, in which it is considerably more complicated due to the larger number of fields involved, judicious gauge field fixing can simplify calculations immensely. Here, taking into account the fact that $t$ is a cyclic coordinate in the Schwarzschild metric, it is useful to
decompose the temporal dependence of the perturbation as $\boldsymbol{h}^{+} = e^{-i\omega t} \tilde{\boldsymbol{h}}^{+}$, so that $t$ appears in the perturbation equation just through the Killing vector $\partial_{t}$. In particular, in the Regge-Wheeler gauge, the form for $\tilde{\boldsymbol{h}}^{+}$ is
%\begin{equation}
%\boldsymbol{h}^{+}_{RW} =\,\, \left[ f_{S}\,H_{0}(t,r)\, dt^2 + \dfrac{H_{2}(t,r)}{f_{S}} \,dr^2 + 2H_{1}(t,r) \, dtdr \right] Y_{\ell,m} + r^{2} K(t,r) \,T^{\oplus}_{\ell,m} \,,
%\end{equation}
\begin{equation}
\tilde{\boldsymbol{h}}^{+}_{RW} =\,\, \left[ f_{S}\,H_{0}(r)\, dt^2 + \dfrac{H_{2}(r)}{f_{S}} \,dr^2 + 2H_{1}(r) \, dtdr \right] Y_{\ell,m} + r^{2} K(r) \,T^{\oplus}_{\ell,m} \,,
\end{equation}
with the fields $H_{0}, H_{1}, H_{2}$ and $K$ given by
\begin{align}
H_{0}(r) = \dfrac{\tilde{H}_{tt}(r)}{f_{S}} \quad &,\, \quad H_{2}(r) = f_{S} \tilde{H}_{tt}(r) \,, \nonumber\\
H_{1}(r) = \tilde{H}_{tr}(r) \quad &,\, \quad K(r) = \dfrac{\tilde{H}^{\oplus}(r)}{r^{2}} \,.
\end{align}
Inserting this ansatz into the equation \eqref{PES2E3}, we end up with only $7$ independent equations for $\ell > 1$ out of the $10$ components of the linearized Einstein field equations. It is, however, possible to avoid the task of solving these coupled equations by combining them into Schrödinger-like differential equation\index{Schr\"{o}dinger-like differential equation}, just as the odd perturbations\index{Odd perturbation}. Regge and Wheeler could not reduce them as far as those for odd perturbations, but Zerilli\label{ZPL} (Z) much later has found that the even perturbation\index{Even perturbation} equations can also be put into a Schr\"{o}dinger-like equation with a more complicated form for the potential \cite{Zerilli1970}. In order to perform this, he defined a new field $Q_{Z}$ implicitly through the equations
\begin{align}
K &=\,\, \left[\dfrac{\lambda(\lambda + 1)r^{2} + 3\lambda M r + 6 M^{2}}{r^{2}(\lambda r +3M)^{2}} \right] Q_{Z} + \partial_{x} Q_{Z} \,,\nonumber\\
\\
H_{1} &=\,\, -i \omega \left[\dfrac{\lambda r^{2} - 3\lambda M r - 3 M^{2}}{r^{2}(\lambda r +3M)^{2}} \right] \dfrac{r}{f_{S}} \, Q_{Z} + \partial_{x} Q_{Z} - \dfrac{ i\omega r}{f_{S}} \, Q_{Z} + \partial_{x} Q_{Z} \,,\nonumber
\end{align}
where $\lambda = (\ell -1)(\ell + 2)/2$, and $H_{0}$ obtained from the algebraic relation
\begin{align}
\left[(\ell -1)(\ell +2) + \dfrac{6M}{r}   \right] H_{0} - \left[ \dfrac{\ell (\ell + 1)M}{i\omega r^{2}} + 2i\omega r  \right ]H_{1} \,+\nonumber\\
 \,-\,  \left[ (\ell -1)(\ell +2) - \dfrac{2\omega^{2}r^{2}}{f_{S}^{2}} -\dfrac{2M (M - r f_{S})}{r} \right ] K = 0 \,.
\end{align}
Then Einstein's equations for even parity perturbations can be put into a Schr\"{o}dinger-like equation for the field $Q_{Z}$ with effective potential $V^{Z}_{s=2}$ given by
\begin{equation}
V_{s=2}^{Z} = f_{S} \left[\dfrac{2\lambda^{2}(\lambda + 1)r^{3} + 6\lambda^{2}M r^{2} + 18\lambda M^{2} r + 18M^{3}}{r^{3}(\lambda r +3M)^{2}}   \right]\,.
\end{equation}
Such a potential became known as Zerilli's potential and has nearly identical properties to the Regge-Wheeler potential. A Schr\"{o}dinger-like differential equation\index{Schr\"{o}dinger-like differential equation} with the above potential became known as the Zerilli equation. Zerilli's equation yields an enormous simplification in the analysis of such perturbations and his work was of great significance in the study of gravitational radiation formed from an asymmetric gravitational collapse. It is also worth mentioning the contribution of Fackerell on the analysis of the solutions to Zerilli's equation \cite{Fackerell1971}. The Regge-Wheeler formalism was later extended to other static black holes in four dimensions \cite{Zerilli1970, Moncrief1974A,Moncrief1974B} and in higher dimensions \cite{Takahashi2010, Kodama2004}. Rotating black holes in four dimensions were tackled in the seminal works of Teukolsky \cite{Teukolsky1972,Teukolsky1974}. Recently, some techniques based on monodromy calculations have been put forward to obtain analytical expressions for the quasinormal spectrum of perturbations in five-dimensional Kerr background \cite{Amado2017}. However, the latter spectrum is written in terms of transcendental equations whose solutions must be found numerically \cite{Amado2019}.

\hfill\(\Box\)
\\

%%%%%%%%%%%%%%%%%%%%%%%%%%%%%%%%%%%%%%%%%%%%%%%%%%%%%%%%%%%%%%%%%%%%%%%%%%%%%%%%%%%
%%%%%%%%%%%%%%%%%%%%%%%%%%% E: Example 3
%%%%%%%%%%%%%%%%%%%%%%%%%%%%%%%%%%%%%%%%%%%%%%%%%%%%%%%%%%%%%%%%%%%%%%%%%%%%%%%%%%%

\begin{figure}[ht!]
  \centering
  \includegraphics[width=14cm]{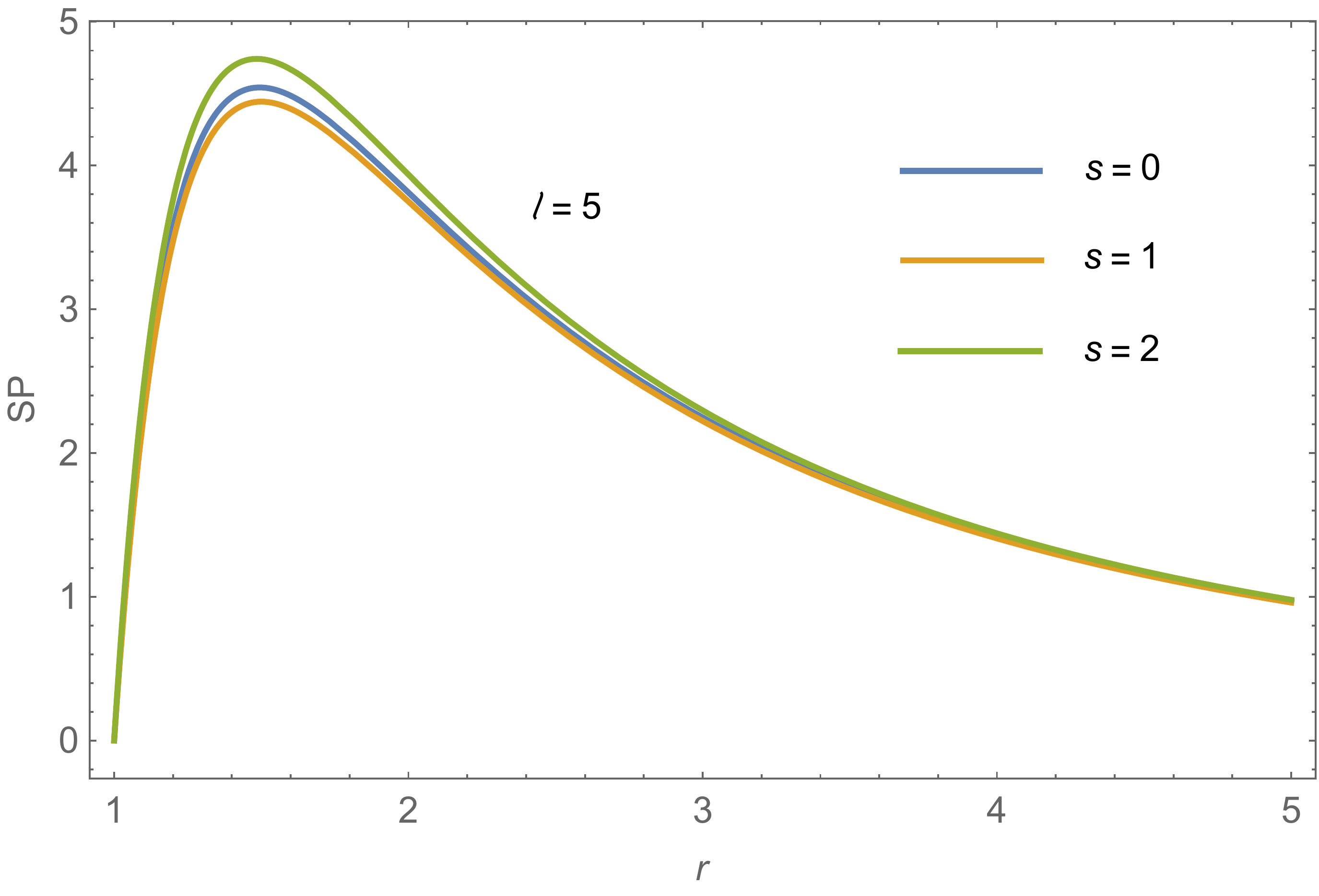}
  \caption{Plots of the effective potential \eqref{SPSpinS} in Schwarzschild spacetime for different values of $s= 0,1,2$ and $\ell=5$, in units $2M =1$.}\label{FSPSpinS}
\end{figure}

To summarize this set of examples, we have seen that in four-dimensional Schwarzschild background, spin-$s$ field perturbations\index{Spin-$s$ field perturbations} can be described by a Schr\"{o}dinger-like differential equation\index{Schr\"{o}dinger-like differential equation} with the Scwarzschild potential (SP)
\begin{equation}\label{SPSpinS}
V_{s}(r) = f_{S} \left[ \dfrac{\ell(\ell +1)}{r^{2}} + (1-s^{2})\left(\dfrac{2M}{r^{3}} +\dfrac{ (4-s^{2})\mu^{2}}{4}\right )  \right] \,,
\end{equation}
where the $s$ label standing for the spin of the perturbation, being
%\begin{equation}
%s = \left\{\begin{matrix}
%0 &\text{scalar\,\, perturbation}\\ 
%1 &\text{Maxwell\,\, perturbation}\\ 
%2 &\text{odd\,\,gravitational\,\, perturbation}
%\end{matrix}\right.
%\end{equation}
$s=0$ for scalar field\index{Scalar field}, $s=1$ for Maxwell field\index{Maxwell field} and $s=2$ for the odd part of the gravitational perturbation. For plots of the potentials \eqref{SPSpinS} for different values of $s \in \{0,1,2\}$, see Fig. \ref{FSPSpinS}.

\begin{figure}[ht!]
  \centering
  \includegraphics[width=14cm]{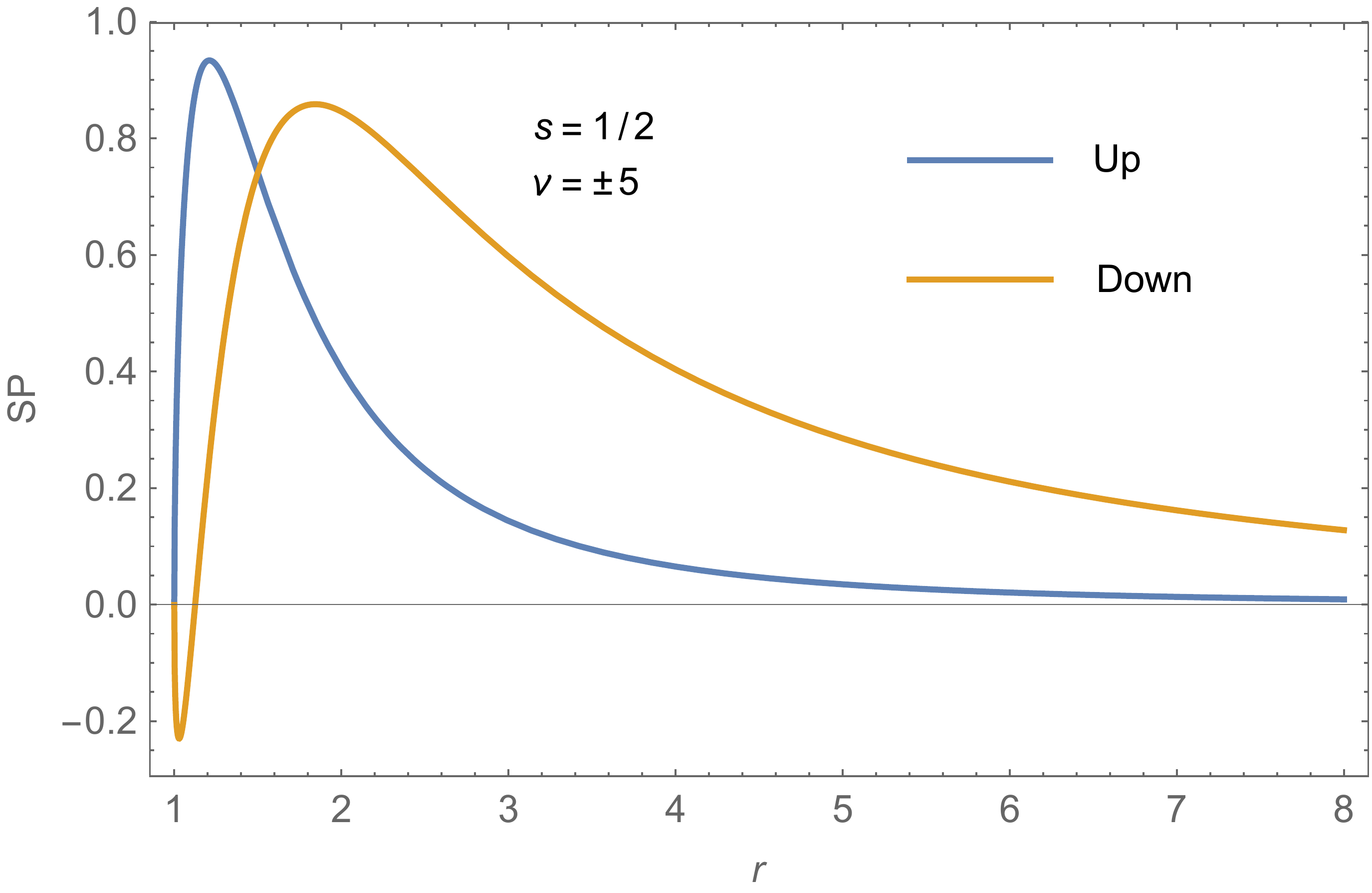}
  \caption{Plots of the effective potential \eqref{EPSSF} associated with the spinorial components (up and down) in Schwarzschild spacetime in units $2M =1$, where we take $\nu=\pm 5$.}\label{FSPSpinSM}
\end{figure}

The spinor field\index{Spinor field} case, namely $s = 1/2$, has a different form for the potential. For the massless
case, for instance, such a potential is given by \cite{Cho2003} (see Figure \ref{FSPSpinSM}):
\begin{equation}\label{EPSSF}
V^{\pm}_{s=1/2}(r) = f_{S} \left[\, \dfrac{|\nu|}{r^{2}}  \pm \dfrac{|\nu|}{\sqrt{f_{S}}} \dfrac{M}{r^{3}}  \mp \dfrac{|\nu| \sqrt{f_{S}}}{r^{2}} \,\right] \,,
\end{equation}
where $\nu$ are nonzero integers, $\nu = \pm 1, \pm 2, \ldots$, and the $\pm$ labels standing for the potential associated with the up ($+$) and down ($-$) components of the spinorial perturbation. 
This case was discussed first by Brill and Wheeler in Ref. \cite{Brill1957} and then extended by Page \cite{Page1976} and Unruh \cite{Unruh}. A detailed derivation for massive spin-$1/2$ field perturbations in generalized Nariai spacetimes\index{Generalized Nariai spacetimes} is presented in the coming chapters, see also \cite{Venancio2018}. For the spin-$3/2$ field perturbations see Ref. \cite{Torres1989}.

\begin{figure}[ht!]
  \centering
  \includegraphics[width=14cm]{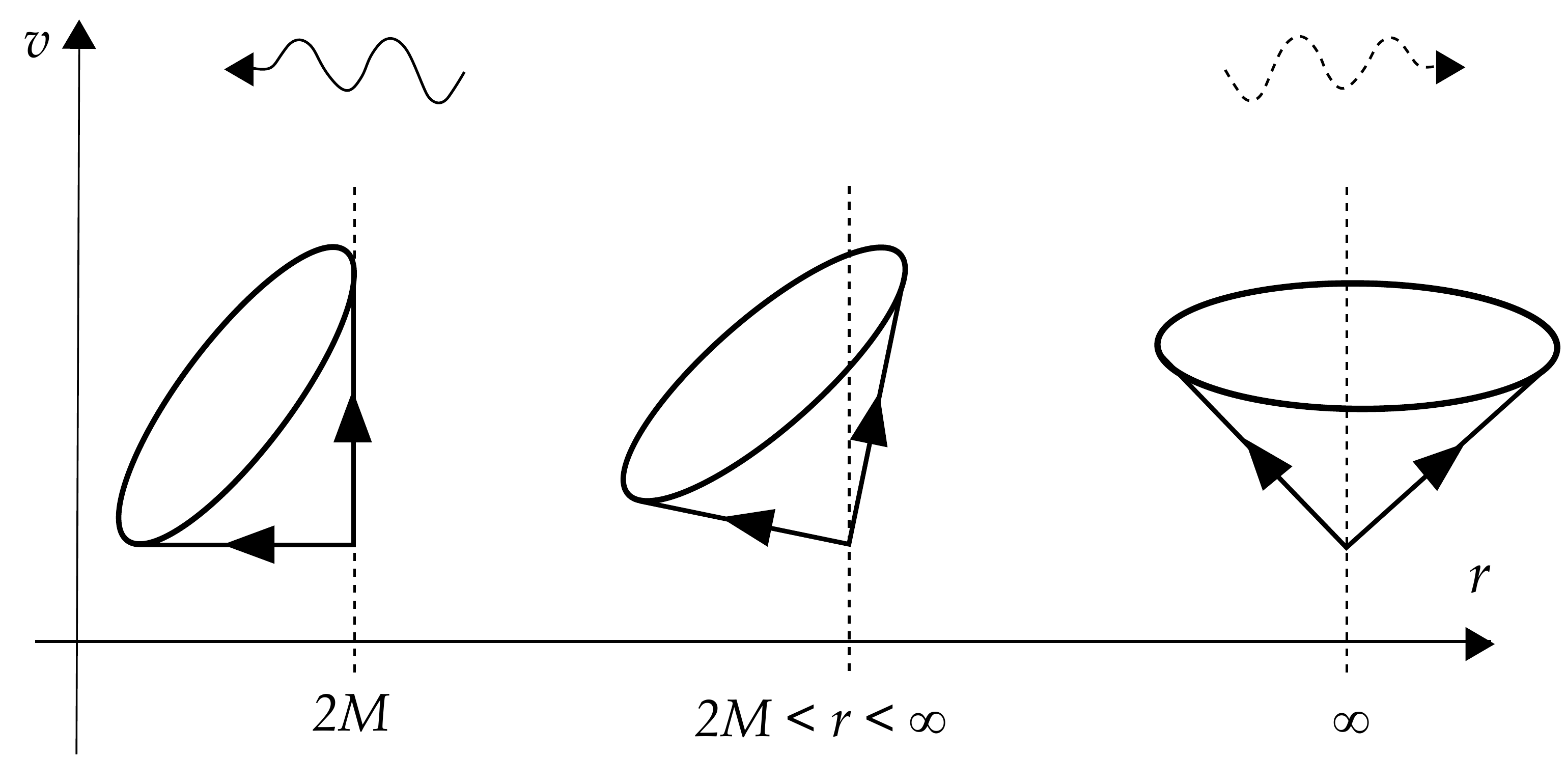}
  \caption{Illustration of the light cone structure in Schwarzschild spacetime, with $r = 2M$ being the event horizon\index{Event horizon}. The wavy arrows represent the natural boundary conditions.}\label{FBCSB}
\end{figure}

Once the potential was obtained, we should look for solutions satisfying appropriate boundary conditions\index{Boundary conditions}, the QNMs\index{Quasinormal Modes}. In order to understand what QNMs are, it is convenient to decompose the dependence of the field $Q$ in the coordinate $t$ in the Fourier basis\index{Fourier basis}, namely,
\begin{equation}\label{TDFQ}
Q(t, x) = e^{-i\omega t} H(x)  \,,
\end{equation}
with the final general solution for the field Q including a ``sum'' over all values of the Fourier frequencies $\omega$ with arbitrary Fourier coefficients. This is particularly convenient in backgrounds in which $\partial_{t}$ is a Killing vector, so that $t$ appears in the perturbation equation just through derivative operators $\partial_{t}$. Inserting this decomposition into Eq. \eqref{WEQ}, we end up with the following Schr\"{o}dinger-like differential equation\index{Schr\"{o}dinger-like differential equation} for the field $H$:
\begin{equation}\label{WEH}
\left[ \dfrac{d^{2}}{dx^{2}} + \omega^{2} - V(x) \right] H(x) = 0 \,,
\end{equation}
which is the ideal form to study QNMs in a way that parallels a normal mode analysis. Once we have an equation of the above form, it needs to be solved with the appropriate boundary conditions\index{Boundary conditions}. QNMs are precisely the solutions of the perturbation equations
%whose radial part can be transformed into a Schr\"{o}dinger-like equation \eqref{WEH}, 
satisfying specific boundary conditions\index{Boundary conditions}. It is worth pointing out that the boundary conditions which define a QNM solution depend directly on the background. For the Schwarzschild background, the boundaries are generally chosen to be the event horizon\index{Event horizon} and the infinity. In particular, the event horizon\index{Event horizon} is at $r = 2M$ where the coefficient of $dr^{2}$ in \eqref{SBM} blows up. Notice
that near the boundaries $r = 2M$ and $r = \infty$ the tortoise coordinate\index{Tortoise coordinate} \eqref{TCSB} behaves as
\begin{equation}
\left.x\right|_{r\rightarrow 2M} \rightarrow - \infty \quad \text{and} \quad  \left.x\right|_{r\rightarrow \infty} \rightarrow +\infty \,.
\end{equation}
%\begin{equation}
%\left.V(r)\right|_{r\rightarrow 2M} \rightarrow 0 \quad , \quad  \left.V(r)\right|_{r\rightarrow \infty} %\rightarrow 0\,.
%\end{equation}
In such boundaries, the real potential $V_{s}$ given in equation \eqref{EPSSF} satisfies
\begin{equation}\label{AFSBCP}
\left.V_{s}(x)\right|_{x\rightarrow +\infty} \rightarrow 0 \quad \text{and} \quad  \left.V_{s}(x)\right|_{x\rightarrow -\infty} \rightarrow 0\,,
\end{equation}
where $V_{s}(x)$ is implicitly defined as $V_{s}(x) = V_{s}[r(x)]$, with $r(x)$ being found by inverting equation \eqref{TCSB}. The above equation means that the function $H$ can be taken to be plane waves near the boundaries $x = \pm \infty$. 
%Since the time dependence of the field $Q$ is of the type \eqref{TDFQ}, 
Indeed, since the effective potential satisfies Eq. \eqref{AFSBCP}, such solutions can be
\begin{equation}
\left.H(x)\right|_{x\rightarrow -\infty} \simeq e^{\pm i\omega x} \quad \text{and} \quad  \left.H(x)\right|_{x\rightarrow +\infty} \simeq e^{\pm i\omega x}\,.
\end{equation}
In order to fix a sign in the above exponentials, we need to apply the appropriate boundary conditions\index{Boundary conditions}. Let us first recall the physical reasoning behind the boundary conditions for the QNMs\index{Quasinormal Modes} in Schwarzschild background. In order to guess the meaningful boundary conditions for the quasinormal modes in a given background, we should look
at its light cone structure. In Schwarzschild background, for instance, it is useful to introduce the coordinate $v=t+x$, so that the relation $dv = dt+\frac{1}{f_{S}}\,dr$ holds, where the identity $dx = \frac{1}{f_{S}}\,dr$ has been used. By analyzing the radial wave propagation, namely $d\theta = d
\phi = 0$, the Schwarzschild line element reduces to $g_{\mu\nu}^{S}dx^{\mu}dx^{\nu} = -f_{S}dv^{2} + 2dvdr$. Now, noting that the function $f_{S}> 0$ at $r > 2M$ and  $f_{S} < 0$ at $r < 2M$, the null rays of this spacetime, namely $g_{\mu\nu}^{S}dx^{\mu}dx^{\nu}= 0$, are given by
\begin{equation}
v = cte  \quad \text{and} \quad \dfrac{dv}{dr} = \dfrac{2}{f_{S}} \left\{\begin{matrix}
> 0 \quad \text{if} \quad r>2M \,,\\ 
< 0 \quad \text{if} \quad r<2M \,.
\end{matrix}\right.
\end{equation}
In particular, $\dfrac{dv}{dr}$ tends to infinity as it approaches the region $r = 2M$. Since the
propagation occur within (massive case) or on the light cones (massless case), it is impossible for an observer at the event horizon\index{Event horizon} ($r = 2M$) to increase its radial coordinate, it will inexorably fall towards a smaller value of r, as illustrated in Fig. \ref{FBCSB}. 
Therefore, it is natural to use as boundary conditions\index{Boundary conditions} at $r = 2M$ that the waves are ingoing, which is represented by an infalling wavy arrow in Fig. \ref{FBCSB}. In its turn, at the infinity, the usual boundary condition is that no wave comes from infinity, whereas some wave can arrive at infinity after scattering by the black hole \cite{Nollert1992,Choudhury2003}. Therefore, at infinity, it is natural to impose that waves are outgoing, as represented by a dashed wavy arrow in Fig. \ref{FBCSB}.
The perturbation field $H$ is then a QNM solution when assumed to be ingoing at the horizon and outgoing at infinity,
\begin{equation}\label{BCSB}
\left.H(x)\right|_{x\rightarrow -\infty} \simeq e^{- i\omega x} \quad \text{and} \quad  \left.H(x)\right|_{x\rightarrow +\infty} \simeq e^{+ i\omega x}\,.
\end{equation}
These boundary conditions\index{Boundary conditions} impose a non-trivial condition on $\omega$, the so-called QNFs \cite{ Kokkotas1999, Berti2009, Zhidenko2006,Venancio2018,Ortega2009}. Indeed, having defined the boundary conditions\index{Boundary conditions} to be used, we need to solve for the frequencies $\omega$ from differential equation \eqref{WEH} viewed as an eigenvalue problem:
\begin{equation}
\hat{D}H(x) = -\omega^{2}H(x)  \quad \text{where} \quad \hat{D} = \frac{d^{2}}{dx^{2}} - V(x) \,.
\end{equation}
Now, the problem boils down to finding the eigenvalues of the $\hat{D}$ operator. In order to satisfy the boundary conditions\index{Boundary conditions} \eqref{BCSB}, we may look for a solution assuming the following ansatz
\begin{equation}\label{AFFH}
H(x) = exp\left[i \int_{0}^{x} h(x)\,dx \right] \,,
\end{equation}
where the function $h(x)$ behaves as
\begin{equation}
\left.h(x)\right|_{x\rightarrow +\infty} \rightarrow +\omega \quad \text{and} \quad  \left.h(x)\right|_{x\rightarrow -\infty} \rightarrow -\omega\,.
\end{equation}
Then, plugging the ansatz \eqref{AFFH} into Eq. \eqref{WEH}, we are left with the nonlinear equation, called Riccati equation
\begin{equation}
i \dfrac{dh(x)}{dx} -h(x)^{2}+ \omega^{2} - V(x) = 0 \,.
\end{equation}
Then, in order to obtain QNMs one needs to integrate the Riccati equation numerically \cite{Schutz1985}. Chandrasekhar has shown that there are only discrete values of $\omega$ which allow solutions
of such an equation \cite{ChandrasekharBook}. There is, actually, a discrete infinity of $\omega$ as shown by Bachelot and Motet-Bachelot \cite{Bachelot1993}. Instead of normal modes, the corresponding eigenvalues of this problem are naturally complex quantities and come in pairs
\begin{equation}\label{QNFsG}
\omega = \pm \text{Re}(\omega) + i\,\text{Im}(\omega) \,,
\end{equation}
where the real part, $\text{Re}(\omega)$, stands for the oscillation frequency of the perturbation, namely $\text{Re}(\omega) = 2\pi/T$ 
%\begin{equation}
%\text{Re}(\omega) = \dfrac{2\pi}{T}
%\end{equation}
with $T$ being the period of the oscillation, while the imaginary part, $\text{Im}(\omega)$, gives the
characteristic timescale $\tau$ as $\text{Im}(\omega) = 1/\tau$
%\begin{equation} 
%\text{Im}(\omega) = \dfrac{1}{\tau} \,.
%\end{equation}
which indicates how rapidly the energy is leaked out in the form of gravitational radiation. Such QNFs are independent of the processes which give rise to oscillations, depending only on the potential parameters which in their turn carry information about both the field perturbation and the background, for example, the mass, electric charge, and angular momentum. Thus, QNMs modes are completely determined by theses parameters. 

For many relevant backgrounds, for instance, Schwarzschild and Kerr, it is not possible to calculate the values of their QNFs in exact form, we must use approximate or numerical methods. There are several numerical and semianalytical methods  for solving a Schr\"{o}dinger-like differential equation\index{Schr\"{o}dinger-like differential equation} of the form \eqref{WEH} and then obtain QNFs with high accuracy, among which are:\\
\\
$\bullet$\quad Mashhoon method \cite{Mashhoon1984} \\
$\bullet$\quad Shooting methods \cite{Detweiler1975}\\
$\bullet$\quad WKB method \cite{Will1985}\\
$\bullet$\quad Characteristic integration \cite{Gundlach1994}\\
$\bullet$\quad Continued fractions \cite{Leaver1985}\\
$\bullet$\quad Frobenius series \cite{Arfken1995}\\
$\bullet$\quad Confluent Heun's equation \cite{Fiziev2009}\\
\\

For instance, using continued fractions technique, Nollert found that spin-$s$ perturbations for $s = 0, 2$  propagating in the Schwarzchild background have QNFs given by \cite{Nollert1993}
\begin{equation}\label{QNFBN}
\omega = 0.0437123 M^{-1} - \dfrac{i}{4M} (2n+1) + \mathcal{O}[(n+1)^{-1}]  \quad \, (n \in \mathbb{N}\,, n \gg 1)\,.
\end{equation}
The index $n$ which labels the modes is called \textit{overtone index}. In particular, notice that $\omega$ is completely determined by only one parameter, the mass $M$, which is the signature of the Schwarzschild black hole itself.
%so that QNFs can be used to infer the properties of the Schwarzschild background.
%In general, for a given black hole, QNFs can be used as an efficient and accurate tool to infer their mass and angular momentum properties of a given background. 
In general, QNFs can be used as an efficient and accurate tool to infer the charges which define the geometry of a background in which a given perturbation is propagating, such as the mass, electric charge,
and angular momentum.

%Such a QNFs are independent of the processes which give rise to oscillations, depending only on the background parameters, for example, the mass, electric charge, and angular momentum. Thus, QNMs modes are completely determined by background's parameters \cite{Zhidenko2009, Birmingham2002,Chirenti2018}. The reason why this happens is that the process of reducing the perturbation equation to a Schr\"{o}dinger-like differential equation for different fields furnishes a potential depending on different parameters. These parameters, in turn, carry information from both the field perturbation and the background. So, since QNFs depend on the parameters of the potential, they will also depend on the parameters of the background.

%\section{Quasinormal Modes in the era of LIGO}

%QNMs and their spectrum are of great physical relevance, inasmuch as these are the modes that survive for a longer time when a background is perturbed and, therefore, these are the configurations that are generally measured by experiments \cite{Kokkotas1999,Berti2009,Nollert1999}. Therefore, this theme acquired even greater importance after the recent measurement of gravitational radiation.

\vspace{.5cm}
%\section{\uppercase{Quasinormal Modes and Background Stability}}
\section{Quasinormal Modes and Background Stability}
\vspace{.5cm}
%\section*{2.3\quad Quasinormal Modes and Background Stability}
%\addcontentsline{toc}{section}{2.3\qquad Quasinormal Modes and Background Stability}

We have argued that there is a discrete infinity of QNFs which come in pairs, that is, if $\omega  = \text{Re}(\omega) + i\,\text{Im}(\omega)$ is a QNF, then $\omega  = -\text{Re}(\omega) + i\,\text{Im}(\omega)$ also will be, and are usually counted by  their imaginary part. For instance, the fundamental frequency is labeled with the trivial overtone index $(n = 0)$, that is the frequency with the lowest imaginary part; the frequency with second lowest imaginary part is labelled with the overtone index $(n = 1)$ and so on. The presence of the imaginary part in the frequency is a relevant feature of perturbations in the presence of horizons. In particular, the sign of $\text{Im}(\omega)$ allows us to analyze the linear stability of a given background. Using \eqref{QNFsG}, the temporal decomposition of the field $Q$ in the Fourier basis\index{Fourier basis}, \eqref{TDFQ}, can be rewritten as
\begin{equation}\label{TDBS}
Q(t, x) = e^{-i [\text{Re}(\omega) + i \text{Im}(\omega)]t}H(x) = e^{\text{Im}(\omega)t} \left[\cos \text{Re}(\omega)t -i \sin\text{Re}(\omega)t \right] H(x)\,,
\end{equation}
from which one can conclude that the field $Q$ grows exponentially for $\text{Im}(\omega) > 0$. Thus,
this is an indication that there might be an instability. Indeed, multiplying \eqref{WEH} by
the complex conjugated field of $H$, here denoted by $H^{\star}$\label{CC}, and integrating the result we
obtain
\begin{equation}
\int_{-\infty}^{+\infty} \left[ H^{\star}(x)\, \dfrac{d^{2} H(x)}{dx^{2}} + (\omega^{2} - V(x)) \left|H(x)\right|^{2} \right] dx = 0\,.
\end{equation}
By integrating by parts the first term of the above expression, we end up with the expression
\begin{equation}
\left. H^{\star}(x)\, \dfrac{d H(x)}{dx} \right|_{-\infty}^{+\infty} + \int_{-\infty}^{+\infty} \left[(\omega^{2} - V(x)) \left|H(x)\right|^{2} - \left| \dfrac{d H(x)}{dx}\right|^{2}\, \right] dx = 0 \,.
\end{equation}
Now, we should apply the appropriate boundary condition in order to fix the sign of $\text{Im}(\omega)$. In asymptotically flat background, the potential $V$ is positive and satisfies Eq. \eqref{AFSBCP} and therefore the QNMs boundary conditions\index{Boundary conditions} are given by Eq. \eqref{BCSB}. Taking into account this latter boundary conditions\index{Boundary conditions} and the fact that the potential is real everywhere, the imaginary part of the previous expression leads to the following constraint:
\begin{equation}
\int_{-\infty}^{+\infty}  \left|H(x)\right|^{2} dx =  - \dfrac{\left|H(\infty)\right|^{2} + \left|H(\infty)\right|^{2}}{2\,\text{Im}(\omega)}  \quad \text{for} \quad \text{Re}(\omega) \neq 0 \,.
\end{equation}
Hence, being $\text{Re}(\omega) \neq 0$ then the imaginary part of $\omega$ has to be $\text{Im}(\omega) < 0$. This means that the asymptotically flat backgrounds are stable \cite{Vishveshwara1970PRD}.
In general, a background is said to be unstable if there is at least one growing mode in its spectrum, otherwise, it is stable, at least under the assumptions made in this analysis. In summary we have:
\begin{equation}
\begin{matrix}\label{ICAFS}
\text{Im}(\omega) < 0 &:&  \text{exponential\,\,damping (stable)} \,,\\
\text{Im}(\omega) > 0 &:&   \text{ exponential\,\,growth (unstable)} \,.
\end{matrix}
\end{equation}
Studying evolution of perturbations on the Schwarzschild background, Vishveshwara found that the equation \eqref{WEH} along with boundary conditions\index{Boundary conditions} \eqref{BCSB} cannot admit any solution with $\text{Im}(\omega) > 0$, so that the background is stable, according to \eqref{ICAFS}. Indeed, Wald furnished a rigorous proof that linear perturbations of the Schwarzschild background must remain uniformly bounded for all time \cite{Wald1979}. Therefore, QNMs are of great relevance for studying the stability of certain backgrounds. %From the numerical point of view, in order to prove the stability of a given background, we must show that the QNFs do not contain any growing mode, $\text{Im}(\omega) > 0$. 
This, however, requires an extremely complicated numerical proof, see \cite{CardosoThesis,Zhidenko2009} for more details. In general, it is not possible to calculate the values of their QNFs in exact form, There are a few exceptions, one of which we must present in the part II of this book.

\vspace{.5cm}
%\section{\uppercase{Quasinormal Modes and Black Hole Area Quantization}}
\section{Quasinormal Modes and Black Hole Area Quantization}
\vspace{.5cm}

%\section*{2.4\quad Quasinormal Modes and Black Hole Area Quantization}
%\addcontentsline{toc}{section}{2.4\qquad Quasinormal Modes and Black Hole Area Quantizationy}

When the perturbation propagation takes place on the background of black holes, which are gravitational configurations where the effects of gravity are extreme, quantum effects in their vicinity cannot be ignored. So, these objects set the ideal scene for testing the ideas of quantum gravity very similarly to the role developed by the Hydrogen atom the quantum mechanics. This, in its turn, means that interpretations of their oscillations can have a major role in understanding various puzzles in fundamental physics. That is the reason why there are several attempts to connect the QNMs with
the quantum spectrum of black hole excitations. In particular, it has been recently conjectured a connection between the real part of the QNFs with very large $n$ and the level spacing of the black hole area spectrum. Bekenstein and Mukhanov proposed a heuristic argument to the quantization of the black hole area according to which a quantum of area is given by \cite{Bekenstein1995}
\begin{equation}\label{BHAQ}
\delta A = \alpha l_{P}^{2} \,,
\end{equation}
where $l_{P}$, which is equal to $1$ in natural units, is the Planck length and $\alpha$ is an undefined
dimensionless constant. For instance, the geometry of the Schwarzschild black hole is completely determined by only one parameter, its mass. In the context of black hole thermodynamics, the horizon area $A$ is related to the mass $M$ by $A = 16\pi M^{2}$, from which we see immediately that a change $\delta M$ in the mass corresponds to a change $\delta A$ in black hole area given by
\begin{equation}\label{ACAQ}
\delta A = 32 \pi M \delta M \,.
\end{equation}
Bekenstein's argument suggests then that the quantization of the mass would lead to the quantization of the black hole area \cite{Bekenstein1995}. This looks like good and intuitive conjecture but completely non-trivial. Indeed, what is the correct $\delta M$ to be used? Up to now, there exist still no final answer to this question, and no confirmation that this is actually correct, reason why this is still just a conjecture.

Inspired by Bekenstein's ideas, Hod proposed to determine $\alpha$ via a version of Bohr's correspondence principle in which the QNFs with very large $n$ play a fundamental role \cite{Hod1998}. 
%At the time, the only available result of QNFs with very large $n$ was the numerical study by Nollert in Schwarzschild context, namely \eqref{QNFBN}. 
At the time, the only available data of QNFs with very large $n$ was the frequencies displayed in Eq. \eqref{QNFBN} obtained numerically by Nollert in Schwarzschild black hole context \cite{Nollert1993}. Realizing that $0.0437123 \sim \text{ln}3/(8\pi)$, Hod then conjectured that such frequencies can be written as
\begin{equation}\label{QNFBM}
\omega = \dfrac{\text{ln}3}{8\pi M} - \dfrac{i}{4M} (2n+1) + \mathcal{O}[(n+1)^{-1}]  \,.
\end{equation}
Interestingly, a few years later this result was analytically obtained by Motl in \cite{Motl2003} and later verified by Andersson using an independent analysis \cite{Andersson1993}. Notice in particular that in this limit, $\text{Re}(\omega)$ depends just on the black hole mass $M$ and is independent of the parameters $\ell, m$ and $n$, being thus the signature of the black hole itself. This crucial feature can be used to investigate a very interesting conjecture that links the QNMs to black hole thermodynamics. Based on Bohr's correspondence principle, namely the transition frequencies at large quantum numbers should equal classical oscillation frequencies, Hod postulated then that the energy difference between two subsequent modes is equal to the real part of their QNFs \cite{Hod1998}, that is
\begin{equation}\label{HCAQ}
\delta M = \text{Re}(\omega)  = \dfrac{\text{ln}3}{8\pi M} \,.
\end{equation}
Using this, the variation $\delta A$ can be written as
\begin{equation}\label{DBI}
\delta A = 4 \text{ln}3 l_{P}^{2} \,,
\end{equation}
from which we can identify the factor appearing in Eq. \eqref{BHAQ} as $\alpha = 4 \text{ln}3$.

From the results obtained by Hod, a few years later Dreyer was able to fix the so-called Barbero-Immirzi parameter, the only free parameter in Loop Quantum Gravity (LQG) \cite{Dreyer2003}; for a recent review, see \cite{Perez2017} and Refs. \cite{Ashtekar2004,Rovelli2004,Thiemann2008,Gambini2001}. Supposing that transitions of a quantum black hole are characterized by the appearance or disappearance of a puncture carrying the lowest allowed irreducible representation $j \in \{0, \frac{1}{2}, 1,\frac{3}{2}, \ldots \}$ of the gauge group SU(2), Dreyer found in complete agreement with the Bekenstein-Hawking result for the entropy that the count of black hole horizon states is dominated by configurations in which $j =j_{min} = 1$, fixing the value for the Barbero-Immirzi parameter. Explicitly, once the area of the black hole would change by an amount given by $A(j) = 8\pi \gamma l_{P}^{2} \sqrt{j(j + 1)}$ in LQG context, assuming $\delta A = A(j_{min} = 1)$ Dreyer concluded then from \eqref{DBI} that the factor $\gamma$ is given by $\gamma = \text{ln}3/(2\pi) \approx 0.124$. However, this result entails some technical problems. For instance, the above value is in disagreement with the approximate value coming from the entropy of large horizons obtained a few years later \cite{Meissner2004}. It is worth recalling that statistical mechanics describes the entropy of a system by the natural logarithm of the number of microscopic states realizing a given macroscopic state. Claiming that the procedure for state counting used in the literature contains an error, Dogamala and Lewandowski and then Gosh-Mitra provided the correct value for the Barbero-Immirzi parameter that is needed to obtain agreement with the Hawking-Bekenstein formula for large black holes. 
In particular, in the Dogamala-Lewandowski count of the number of microscopic states Barbero-Immirzi parameter has to be $\gamma \approx 0.238$ \cite{Dogamala2004}, while in the Gosh-Mitra count it has to be $\gamma \approx 0.274$ \cite{Ghosh2005}. Besides that, Barbero-Immirzi parameter has to be fixed in a way that it is independent from the black hole considered. For instance, computing the QNFs of Kerr black hole in the limit of large $n$ and then taking the limit when the rotational parameter an approaches zero, the result does not reduce to Eq. \eqref{HCAQ}. Actually, these asymptotic QNMs are not analytical at $a = 0$, they go as $a^{1/3}$ in the $a\rightarrow 0$ limit. This means that the large $n$ limit and the $a\rightarrow 0$ limit do not commute \cite{Keshet2007,Motl2003A}, so that asymptotic value of $\text{Re}(\omega)$, namely Eq. \eqref{HCAQ}, depends on the spin of the perturbation and is not an intrinsic property of the black hole. Due to these problems, the Hod's main assumption was later criticised by Maggiore in \cite{Maggiore2008}. As an alternative approach to remove some of the difficulties posed in Hod's conjecture, Maggiore suggested that instead of using $\text{Re}(\omega)$ we shall use the difference between the natural frequencies of subsequent modes, $\delta \omega_{0} = (\omega_{0})_{n} - (\omega_{0})_{n-1}$, of a damped harmonic oscillator. This gives a different value for $\delta M$. Indeed, by considering $\xi = \xi(t)$ as a solution of the equation
\begin{equation}
\ddot{\xi} + 2\beta \dot{\xi} + \omega_{0} \xi = 0 \,,
\end{equation}
where $\beta$ is the damping constant and $\omega_{0} $ the proper frequency of the harmonic oscillator,
we find that
\begin{equation}
\omega = \pm \sqrt{\omega_{0}- \beta^{2}} - i \beta \,
\end{equation}
are the two roots of the characteristic equation $\omega + 2i\beta\omega - \omega_{0} = 0$. Therefore, the field \eqref{TDBS} is reproduced by a damped harmonic oscillator, with the identifications
\begin{equation}
\text{Re}(\omega) = \pm \sqrt{\omega_{0}- \beta^{2}} \quad \text{and} \quad \text{Im}(\omega) = - \beta \,.
\end{equation}
Inverting these expressions, we find that $\omega_{0}$ can be written in terms of $\text{Re}(\omega)$ and $\text{Im}(\omega)$ as follows:
\begin{equation}
\omega_{0} = \sqrt{\text{Re}(\omega)^{2} + \text{Im}(\omega)^{2}} \,.
\end{equation}
For very large $n$, $\text{Im}(\omega) \gg \text{Re}(\omega)$ and if we consider a transition $n \rightarrow n-1$ we obtain from \eqref{QNFBM} that
\begin{equation}\label{MCBM}
\delta M = \delta \omega_{0} \approx \delta \text{Im}(\omega)  = \dfrac{1}{4M} \,.
\end{equation}
Using Eq. \eqref{ACAQ}, Maggiore concluded that
\begin{equation}
\delta A = 8 \pi l_{P}^{2} \,,
\end{equation}
from which we can identify the factor appearing in Eq. \eqref{BHAQ} as $\alpha = 8 \pi$. This is exactly
a quantum of area suggested by Bekenstein on the basis of a different reasoning. In contrast with what happens for $\text{Re}(\omega)$, the quantum of area obtained as of $\delta \omega_{0} \approx \delta \text{Im}(\omega)$ for very large $n$ looks like to be the most natural candidate as an intrinsic property of black holes. In particular, the large $n$ limit and the $a\rightarrow 0$ limit commute
and the value \eqref{MCBM} does not depend on the spin of the perturbation. While the highly-damped regime is not as simple for charged and rotating four-dimensional geometries as suggested by Hod's conjecture, Maggiore's suggestion can be extended also to Kerr black holes. Indeed, let us consider the case of an extremal Kerr black hole for which the horizon area $A$ is related to the mass $M$ by $A = 8\pi M^{2}$. It follows that a change $\delta M$ in the mass produces a change $\delta A$ in black hole area given by
\begin{equation}
\delta A = 16 \pi M \delta M \approx 16 \pi M \delta \text{Im}(\omega) \,.
\end{equation}
QNFs with $n$ very large for a Kerr black hole has been numerically found by Berti and collaborators in Ref. \cite{Berti2004}. In particular, they showed that for any $a$, the imaginary part
$\text{Im}(\omega) \gg \text{Re}(\omega)$ and is a monotonically increasing function of $a$, namely
\begin{equation}
\text{Im}(\omega) = \dfrac{1}{2} +  0.0438 a  - 0.0356 a^{2} \,,
\end{equation}
where $a$ is the dimensionless Kerr rotation parameter. In this case, the extremal case
corresponds to the values $M = 1/2$ and $a = 1/2$ \cite{Berti2004,Carneiro2020}. Assuming these latter values, we are led to
\begin{equation}
\text{Im}(\omega) = 1.026 \, \dfrac{1}{2} = 1.026 \, \dfrac{1}{4M} \,, 
\end{equation}
from which we see that this value is $2.6 \%$ above the Schwarzschild value. This implies
in the following approximated value for a quantum of area
\begin{equation}
\delta A \approx 4\pi l_{P}^{2} \,,
\end{equation}
a half of the Bekenstein's value. Up to now, it is not clear if Maggiore's approach can be extended in a consistent way to all geometries. All of this is still a conjecture but, if it can be proven to be true, it would provide a unique link between general relativity and quantum mechanics. It is worth mentioning that although Hod's conjecture cannot be generalized to the charged and rotating black holes in any simple way, being mostly regarded as a strange coincidence, 
%whether a relation between QNMs and the quantum behavior of black holes exists or not, 
Hod's conjecture was at the very least a crucial step towards our current scenario of quantum gravity.
%%%%%%%%%%%%%%%%%%%%%%%%%%%%%%%%%%%%%%%%%%%%%%%%%%%%%%%%%%%%%%%%%%%%%%%%%%%%%%%%%%%
%%%%%%%%%%%%%%%%%%%%%%%%%%% End: Chapter 1. Quasinormal Modes: An Introduction
%%%%%%%%%%%%%%%%%%%%%%%%%%%%%%%%%%%%%%%%%%%%%%%%%%%%%%%%%%%%%%%%%%%%%%%%%%%%%%%%%%%

%%%%%%%%%%%%%%%%%%%%%%%%%%%%%%%%%%%%%%%%%%%%%%%%%%%%%%%%%%%%%%%%%%%
%%%%%%%%%%%%%%%%%%%%%%%%%%% B: Part I
%%%%%%%%%%%%%%%%%%%%%%%%%%%%%%%%%%%%%%%%%%%%%%%%%%%%%%%%%%%%%%%%%%%
\part{On the Quasinormal Modes in Generalized Nariai Background}\label{Part-Orig}

%%%%%%%%%%%%%%%%%%%%%%%%%%%%%%%%%%%%%%%%%%%%%%%%%%%%%%%%%%%%%%%%%%%
%%%%%%%%%%%%%%%%%%%%%%%%%%% E: Part I
%%%%%%%%%%%%%%%%%%%%%%%%%%%%%%%%%%%%%%%%%%%%%%%%%%%%%%%%%%%%%%%%%%%

%%%%%%%%%%%%%%%%%%%%%%%%%%%%%%%%%%%%%%%%%%%%%%%%%%%%%%%%%%%%%%%%%%%%%%%%%%%%%%%%%%%
%%%%%%%%%%%%%%%%%%%%%%%%%%% B: Chapter 2. Field Perturbations: Spins 0 and 1
%%%%%%%%%%%%%%%%%%%%%%%%%%%%%%%%%%%%%%%%%%%%%%%%%%%%%%%%%%%%%%%%%%%%%%%%%%%%%%%%%%%

%\chapter{\uppercase{Field Perturbations: Spins $0$ and $1$}}\label{FPS01}
\chapter{Field Perturbations: Spins $0$ and $1$}\label{FPS01}
%\chapter*{3.1\quad Field Perturbations: Spins $0$ and $1$}\label{FPS01}
%\addcontentsline{toc}{chapter}{3.1\qquad Field Perturbations: Spins $0$ and $1$}

The study of scalar, spinorial and gauge fields (abelian and non-abelian) 
propagating in curved spacetimes plays a central role in the study of General relativity
and any other theory of gravity. The main reason is that besides the detection of
gravitational radiation and observation of the direct interaction between objects via
gravitation, the most natural and simple way to probe the gravitational field permeating
our spacetime is by letting other fields interact with it. In this chapter, we shall perform the integration of the Schr\"{o}dinger-like differential equation\index{Schr\"{o}dinger-like differential equation} for the Rose-Morse class of potential to study the dynamics of spin-$s$ field perturbations\index{Spin-$s$ field perturbations} for $s=0,1$ in generalized Nariai spacetime. In particular, we should analytically obtain the quasinormal spectrum associated to these fields.

\vspace{.5cm}
%\section{\uppercase{Generalized Nariai Spacetimes}}\label{GNS}
\section{Generalized Nariai Spacetimes}\label{GNS}
\vspace{.5cm}

Let us consider matter fields propagating in the background described in Ref. \cite{Batista2016}, a
higher-dimensional generalization of the Nariai spacetime. We take the point of view that, as well as being of interest in its own right, the generalized Nariai spacetime can provide insight into the propagation of waves on the generalized Schwarzschild spacetime. In generalized Nariai (GN)\label{GN} background, the metric in $D=2d$ dimensions is formed from the direct product of the two-dimensional de Sitter space $dS_{2}$ with $(d-1)$ spheres $S^{2}$ possessing different radii $R_{j}$, namely
\begin{equation}\label{nariai-metric}
g^{GN}_{\mu\nu}dx^{\mu}dx^{\nu} = -f(r)dt^{2} + \frac{1}{f(r)}dr^{2}+ \sum_{l=2}^{d}R_{l}^{2}\,d\Omega_{l}^{2} \,,
\end{equation}
%\begin{align}
%g^{GN}_{\mu\nu}dx^{\mu}dx^{\nu} \,=&\, -\left(1-\frac{r^2}{R_1^2}\right)dt^{2} + \left(1-\frac{r^2}%{R_1^2}\right)^{-1}dr^{2} \nonumber\\
%\,+&\, \sum_{l=2}^{d}R_{l}^{2}\left(d\theta_{l}^{2}+\text{sin}^{2}\theta_{l}\,d\phi_{l}^{2}\right)
%\end{align}
where $d\Omega_{l}^{2}$ is the line element of the $l$th two-sphere $S^{2}$, 
\begin{equation}\label{MTOS}
 d\Omega_{l}^{2} = \hat{g}_{a_{l}b_{l}}dx^{a_{l}}dx^{b_{l}} = d\theta_{l}^{2}\,+\, \text{sin}^{2}\theta_{l}\,d\phi_{l}^{2} \quad \forall \,\, a_{l},b_{l} \in \{\theta_{l}, \phi_{l}\}  \,,
\end{equation}
with the symmetric second order tensor $\hat{g}_{a_{l}b_{l}}$ being the metric on the $l$th two-sphere.
The function $f(r)$ has the following dependence on coordinate $r$:
\begin{equation}\label{functionf1}
f(r) = 1 - \dfrac{r^{2}}{R_{1}^{2}} \,,
\end{equation}
and the radius $R_{1}$ and $R_{l}$ are constants given by
\begin{equation}\label{radii}
    \begin{array}{ll}
      R_{1}&\,=\,\left[\Lambda \,-\,\frac{1}{2}Q_{1}^{2}\,+\,\dfrac{Q}{2(D-2)} \right]^{-1/2} \,, \\
R_{l}&\,=\,\left[\Lambda \,+\,\frac{1}{2}Q_{l}^{2}\,+\,\dfrac{Q}{2(D-2)} \right]^{-1/2} \,,
    \end{array}
 \end{equation}
with $Q_{1}$  and $Q_{l}$ being the electric and magnetic charges, respectively, while $Q$ is defined by	
\begin{equation}
Q \,\equiv\, Q_{1}^{2} \,-\, \sum_{l=2}^{d}Q_{l}^{2} \,.
\end{equation}
Generalized Nariai spacetime is locally a static solution of the equation \eqref{EE} in the presence of the electromagnetic gauge field
\begin{equation}\label{gauge-field}
\bl{\mathcal{A}}^{GN} \,=\,Q_{1}\, r\, dt \,+\, \sum_{l=2}^{d}Q_{l}R^{2}_{l} \,\text{cos}\theta_{l}\,d\phi_{l} \,.
\end{equation}

It is worth recalling that a spacetime is said to be spherically symmetric if there exists an action of $SO(3)$ by isometries whose orbits are spacelike two-dimensional spheres.  Clearly this is not the
case, since the angular part of the line element is the direct product of several two-spheres and therefore, the background has $SO(3)\times SO(3) \times \ldots \times SO(3)$, $d-1$ times,
whereas the usual $D$-dimensional Nariai (N)\label{N} background has a $SO(D-1)$ symmetry. For each of the $(d-1)$ spheres, there exists three independent Killing vectors that generate rotations, namely
\begin{equation}\label{KilllingV}
  \left\{
     \begin{array}{ll}
        \mathbf{K}_{1,l} = \sin\phi_l\, \partial_{\theta_l} + \cot\theta_l \cos\phi_l\,\partial_{\phi_l}\,,\\
       \mathbf{K}_{2,l} = \cos\phi_l\, \partial_{\theta_l} - \cot\theta_l \sin\phi_l\,\partial_{\phi_l} \,,\\
       \mathbf{K}_{3,l} = \partial_{\phi_l}  \,.
     \end{array}
   \right.
\end{equation}
In addition to these Killing vectors, $\mathbf{K}_t = \partial_t$ also generates an isometry. In particular, this Killing vector is light-like at the closed surfaces $r = \pm \Lambda^{-1/2}$, so that these are Killing horizons. The boundary conditions\index{Boundary conditions} of the quasinormal modes will be posed at these surfaces, as discussed in Ref. \cite{Venancio2018}. These surfaces in which the boundary conditions will be imposed, are the boundaries of the static region of the generalized Nariai spacetime \cite{Casals2009}. This is exactly the region covered by the static coordinates $\{t, r, \theta_2, \phi_2,\ldots,\theta_d, \phi_d\}$ and, therefore, the Killing horizons are well-characterized geometrically, that is, they are not arbitrary. This is particularly interesting in order to introduce QNMs, inasmuch as in static coordinates, the coefficients of the metric are
independents of the coordinate $t$, and therefore the background metric possesses the Killing vector field $\boldsymbol{K}_t = \partial_t$.  In this case, it is convenient to decompose the time dependence of the fields in this coordinate in the Fourier basis\index{Fourier basis}. Outside of the static region,  nothing in the arguments put forward stops us the Fourier basis\index{Fourier basis} to expand the field components. However, this is not the most suitable choice inasmuch as the notion of time is essential in order to introduce QNMs. For this reason, in this book, we will consider just the static region of the generalized Nariai spacetime.

Besides the continuous symmetries generated by Killing vectors \eqref{KilllingV}, there are also some discrete symmetries. For instance, we have seen in example $2$ that the line
element on $S^{2}$ is invariant under the transformation $(\theta, \phi) \rightarrow (\pi - \theta ,\phi + \pi)$, called parity transformation\index{Parity transformation} (spatial inversion). Here, however, the line element is invariant under
the parity\index{Parity transformation} transformation\label{PT} in each of the spheres. More precisely, the changes
\begin{equation}\label{parityT}
  \theta_l \,\rightarrow\, \pi - \theta_l \quad  \textrm{and} \quad \phi_l \,\rightarrow\,\phi_l + \pi
\end{equation}
%\begin{equation}
%  P_l \,\,:\,\, (\theta_l,\phi_l) \rightarrow (\pi - \theta_l, \phi_l + \pi) \,,
%\end{equation}
%\begin{equation}
%P_l^2 = 1 \,.
%\end{equation}
do not modify the line element (\ref{nariai-metric}). Denoting this transformation by $P_l$, it follows that $P_l^2$ is the identity transformation, so that the eigenvalues of this transformation are $\pm 1$. Objects unchanged under $P_l$ (eigenvalue $+$1) are said to have even parity, while those that change by a global sign (eigenvalue $-$1) are said to have odd parity. 

\begin{figure}[ht!]
  \centering
\includegraphics[width=14cm]{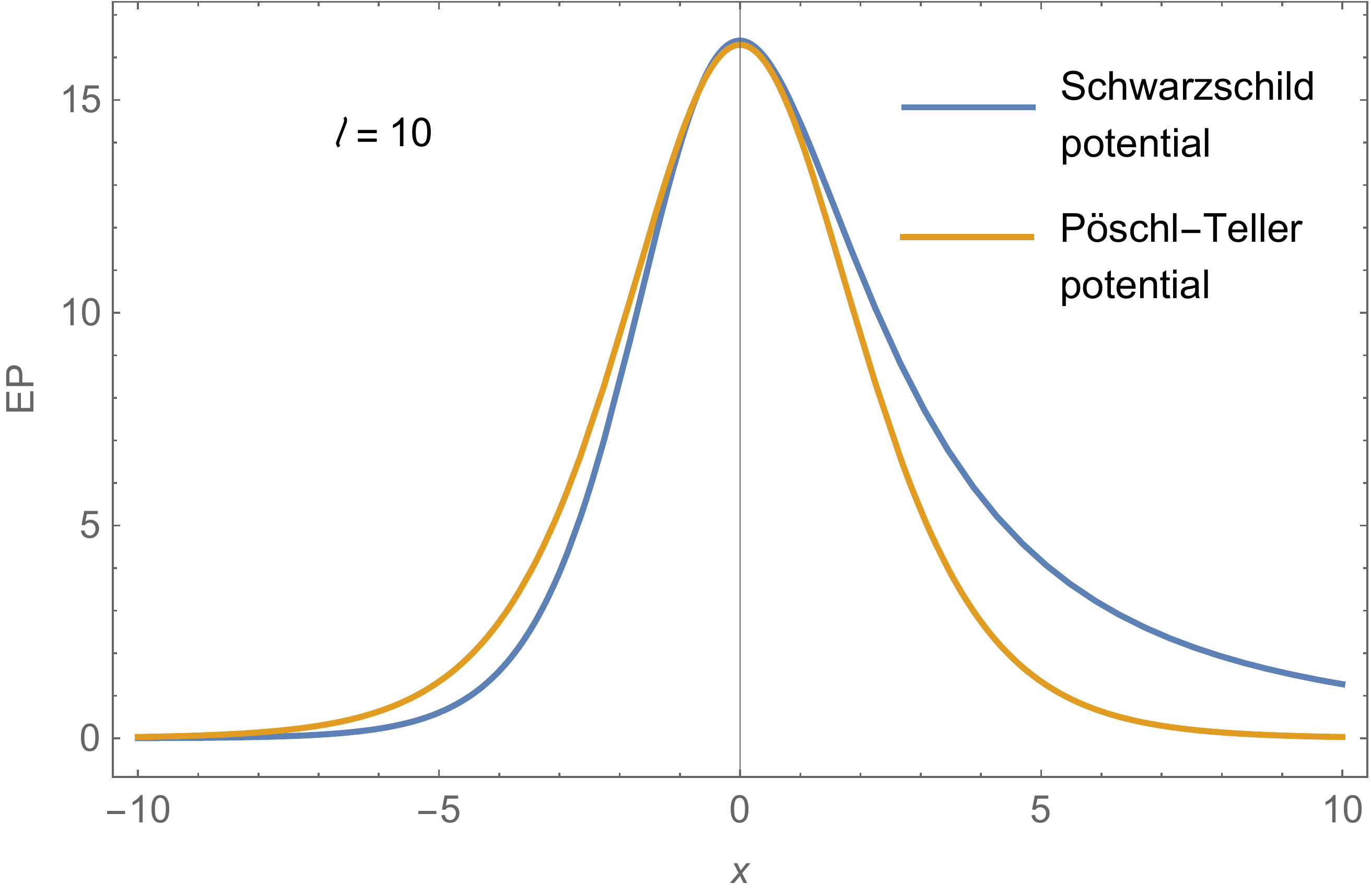}
  \caption{Comparison of effective potentials of perturbations with spin $s=0,1,2$ in Schwarzschild and Nariai spacetimes in four dimensions, in units $2M=1$. The constants $x_0=0, C=2/\sqrt{27}$ and $V_0=\ell (\ell +1)$. Here, we have chosen the integration constant in Schwarzschild tortoise coordinate\index{Tortoise coordinate} defined in \eqref{TCSB} for our convenience, so that the the peak of the potential barrier (at $r = 3M$) coincides with $x_0=0$, namely $x = r + 2M\,\text{ln}(r-2M) -(3M- 2M\,\text{ln}2)$.}\label{FS-PT}
\end{figure}

As we will see in what follows, one important property of studying perturbation equations in the
background considered here, the higher-dimensional generalization of the Nariai spacetime presented in Ref. \cite{Batista2016}, is that all equations turn out to be analytically integrable. Indeed, we are going to see that the problem of solving the perturbation equation for
the scalar field\index{Scalar field} (spin-$0$), the Dirac field\index{Dirac field} (spin-$1/2$) and the Maxwell field\index{Maxwell field} (spin-$1$) published in Ref. \cite{Venancio2018} as well as for the gravitational field\index{Gravitational field} (spin-$2$) published in Ref. \cite{Venancio2020} boils down to integrating a Schr\"{o}dinger-like equation whose effective potential $V(x)$ is contained in the Rosen-Morse\index{Rosen-Morse class of potentials} class of integrable potentials, as displayed in Table I of
the Ref. \cite{Dutt1988}. For this class of potentials, which has the well-known P\"{o}schl-Teller\label{PTPL} (PT) potential\index{P\"{o}schl-Teller potential} as a particular case, all the solutions of \eqref{WEH} are analytical. In particular, the P\"{o}schl-Teller\index{P\"{o}schl-Teller potential} potential was originally introduced as a potential for which the Schr\"{o}dinger equation is exactly solvable \cite{Poschl1933} and has the form
\begin{equation}\label{PTTI}
  V^{PT}(x) = \frac{C^{2} V_{0}}{\cosh^2\left[C\,(x-x_0)\right]} \,,
\end{equation}
where $C, V_0$ and $x_{0}$ are constants. As was said previously, we can prove that the generalized Nariai spacetime can provide insight into the propagation of waves on the generalized Schwarzschild spacetime. Notice that the P\"{o}schl-Teller\index{P\"{o}schl-Teller potential} potential is symmetric about $x_0$ and decays exponentially in the $x\rightarrow 0$ limit, whereas the Schwarzschild potential does not share these properties. In spite of this, the Schwarzschild potential has a single peak (see Fig. \ref{FS-PT}),  with a suitable choice of constants, the Po\"{o}schl-Teller potential can be made to fit the Schwarzschild potential in the vicinity of this peak. 
For instance, the Schr\"{o}dinger-like differential equation\index{Schr\"{o}dinger-like differential equation} for the massless scalar field\index{Scalar field}, the Dirac field\index{Dirac field}, the Maxwell field\index{Maxwell field} as well as for the gravitational field\index{Gravitational field} in four-dimensional Schwarzschild background can be made to fit the Schr\"{o}dinger-like differential equation\index{Schr\"{o}dinger-like differential equation} with the P\"{o}schl-Teller\index{P\"{o}schl-Teller potential} potential \eqref{PTTI} if we adopt the transformations
\begin{equation}
x_{S}\rightarrow \dfrac{x_{N}}{C}\,,\quad \quad  t_{S}\rightarrow \dfrac{t_{N}}{C}\,, \quad \quad \omega_{S} \rightarrow C\, \omega_{N}\,, \quad \text{where}\quad C=\dfrac{1}{\sqrt{27}M}\,,
\end{equation}
where it is worthwhile recalling that the labels S and N denote Schwarzschild and Nariai, respectively. 
Under this transformation, we can approximate these two spacetimes in four dimensions. 
%This is the closest analogy between the two spacetimes in four dimensions. 
In particular, this means that the Nariai spacetime can be taken as a model for exploring properties of the Schwarzschild spacetime, for example the QNM frequency spectrum, by using the exact solutions for the P\"{o}schl-Teller\index{P\"{o}schl-Teller potential} potential.
%In particular, this means that the P\"{o}schl-Teller potential can be used as a model for exploring properties of the Schwarzschild solution, for example the QNM frequency spectrum.
This is particularly useful, inasmuch as the effective potential for any field perturbation in Schwarzschild geometry is non-integrable. For these reasons, the next section is devoted to integrating the Rosen-Morse\index{Rosen-Morse class of potentials} class of integrable potentials.

\vspace{.5cm}
%\section{\uppercase{Integrating the Rosen-Morse Class of Integrable Potentials}}\label{IRMCIP}
\section{Integrating the Rosen-Morse Class of Integrable Potentials}\label{IRMCIP}
\vspace{.5cm}

Consider the problem of solving the Schr\"{o}dinger-like differential equation\index{Schr\"{o}dinger-like differential equation} \eqref{WEH}  with
the potential $V(x)$ given by
\begin{equation}\label{Potential_Generic}
  V(x) = \mathfrak{a} + \mathfrak{b} \tanh(\mathfrak{d}\,x) + \frac{\mathfrak{c}}{\cosh^2(\mathfrak{d}\, x)} \,,
\end{equation}
where $\mathfrak{a}$, $\mathfrak{b}$, $\mathfrak{c}$ and $\mathfrak{d}$ are constants with $\mathfrak{d} > 0$. These constants assume different values depending on the type of the field perturbation. Such $V(x)$ is contained in the Morse class of integrable potentials, with the case $\mathfrak{a}=\mathfrak{b}=0$ being the well-known P\"{o}schl-Teller potential\index{P\"{o}schl-Teller potential}, see \cite{Poschl1933}. In order to solve the latter ordinary differential equation, let us define a new independent variable defined by
\begin{equation}
  y = \frac{1}{2} + \frac{1}{2} \tanh(\mathfrak{d} \, x).
\end{equation}
Assuming that the domain of $x$ is the entire real line, we find that $y\in(0,1)$, with the boundaries $x = \pm\infty$ being given by $y = 0$ and $y=1$.  In particular, near the boundaries, the relation between the coordinates $x$ and $y$ assumes the simpler form
\begin{equation}\label{yr_boudary}
\left\{
  \begin{array}{ll}
   x \rightarrow -\infty \; \Rightarrow \; y \simeq e^{2\mathfrak{d}\, x} \,, \\
    x \rightarrow +\infty \; \Rightarrow \; (1-y) \simeq   e^{-2\mathfrak{d}\, x} \,.
  \end{array}
\right.
\end{equation}

Now, let us define the constant parameters $a$, $b$, and $c$ as follows
\begin{equation}\label{abc}
     \left\{  \begin{array}{ll}
       a = \dfrac{1}{2\mathfrak{d}}\left(\mathfrak{d}  +  \sqrt{\mathfrak{a} - \mathfrak{b} - \omega^2} -  \sqrt{\mathfrak{a} + \mathfrak{b} - \omega^2} + \sqrt{\mathfrak{d}^2 - 4\mathfrak{c}} \right) \,, \\
       \\
       b = \dfrac{1}{2\mathfrak{d}}\left( \mathfrak{d}  + \sqrt{\mathfrak{a} - \mathfrak{b} - \omega^2} -  \sqrt{\mathfrak{a} + \mathfrak{b} - \omega^2} - \sqrt{\mathfrak{d}^2 - 4 \mathfrak{c}}   \right)  \,,\\
       \\
       c =   \dfrac{1}{\mathfrak{d}}\sqrt{\mathfrak{a} - \mathfrak{b} - \omega^2} + 1 \,,
     \end{array}  \right.
\end{equation}
and, instead of $H(x)$, let us use the dependent variable $G(y)$ defined by
\begin{equation}\label{fieldH}
  H(x) = y^{(c-1)/2}\,( 1- y )^{\frac{1}{2}(a+b-c)} \,G(y) \,.
\end{equation}
Then, after some algebra, one can check that the function $G(y)$ obeys the equation
\begin{equation}\label{HeperGeomEq}
  y(1-y) \frac{d^2G}{dy^2} + \left[ c - y(a+b+1)\right]\frac{d G}{dy }  - a b \,G = 0 .
\end{equation}
This is the hypergeometric equation, whose general solution is given by
\begin{equation}\label{G(y)}
  G(y) = \alpha \,F(a,b,c;y) + \beta \,y^{(1-c)}\,F(1+a-c,1+b-c,2-c;y) \,,
\end{equation}
where $F$ is the hypergeometric function\label{HG} (usually denoted by $_2F_1$), while $\alpha$ and $\beta$ are arbitrary integration constants that can be fixed by the boundary conditions\index{Boundary conditions}. Summing up these results, we conclude, from Eqs. (\ref{fieldH}) and (\ref{G(y)}), that the solution for the function $H$ obeying Eq. (\ref{WEH}) is given by
\begin{align}\label{SolutionH}
  H &=\,\, ( 1 - y )^{\frac{1}{2}(a+b-c)} \left[\,\alpha\, y^{(c-1)/2}\, F(a,b,c;y) \right. \nonumber\\  
 &+\,\, \left. \beta \,y^{-(c-1)/2}\,F(1+a-c,1+b-c,2-c;y)  \, \right] \,.
\end{align}
There are two properties of the hypergeometric function which shall be needed in what follows. The first concerns the hypergeometric function at $y=0$, which is
\begin{equation}
F(a,b,c;0)=1 \,.
\end{equation}
Once this happens, it turns out that the latter way of writing the solution is particularly useful to apply the boundary conditions\index{Boundary conditions} at $y=0$, i.e. $x=-\infty$. Indeed, using Eq. (\ref{yr_boudary}), one can promptly verify that the following limit holds
\begin{equation}\label{SolutionH_y=0}
   \left. H\right|_{x\rightarrow -\infty} = \,\alpha\, e^{\mathfrak{d}(c-1)x}\,  
   + \beta \,e^{-\mathfrak{d}(c-1)x}   \,,
\end{equation}
which will be of relevance to impose the boundary conditions\index{Boundary conditions} at $x=-\infty$. The second
concerns the hypergeometric function at $y=1$, which can be evaluated using the identity
\begin{equation}
F(a,b,c;1)= \dfrac{\Gamma(c-a-b)\Gamma(c)}{\Gamma(c-a)\Gamma(c-b)} \,,
\end{equation}
where $\Gamma$ stands for the gamma function. Then, in order to apply the boundary conditions at $y=1$, i.e. $x=+\infty$, it is more useful to write the hypergeometric functions as functions of $(1-y)$, so that they become unit at the boundary. This can be done rewriting the hypergeometric functions appearing in Eq. (\ref{SolutionH}) by means of the following identity \cite{Abramowitz1972}:
\begin{align}
  F(a,b,c;y) \,=\,\,  F(a,b,c;1)\,F(a,b,a+b-c+1;1-y) \nonumber\\
 \,+\,\, F(c-a,c-b,c;1) \,(1-y)^{(c-a-b)}F(c-a,c-b,c-a-b+1;1-y)\,. \label{HyperGeomIdentity}
\end{align}
Doing so, and using Eq. (\ref{yr_boudary}) we eventually arrive at the following behavior  of the solution at
$x=+\infty$:
\begin{align}
  \left. H\right|_{x\rightarrow +\infty} \,\simeq & \,    e^{-\mathfrak{d}(a+b-c)x} \left[ \,\alpha\, \frac{\Gamma(c-a-b)\Gamma(c)}{ \Gamma(c-a) \Gamma(c-b)} +
\beta \,   \frac{\Gamma(c-a-b)\Gamma(2-c)}{ \Gamma(1-a) \Gamma(1-b)   } \,\right] \nonumber\\
& + e^{\mathfrak{d}(a+b-c)x}  \left[\,
  \alpha\, \frac{\Gamma(a+b-c)\Gamma(c)}{ \Gamma(a) \Gamma(b) }  + \beta \, \frac{\Gamma(a+b-c)\Gamma(2-c)}{ \Gamma(a-c+1) \Gamma(b-c+1) }    \,\right] \,. \label{SolutionH_y=1}
\end{align}

The exact expression for the constants $a, b$ and $c$ depends on the type of perturbation under study. In what follows, we must use the solution obtained in this section to investigate the QNMs of spin $0, 1/2, 1$ and $2$ fields in the background \eqref{nariai-metric}. We shall use as the boundaries of this space the horizons $r = \pm R_{1}$ and consider four types of boundary conditions\index{Boundary conditions}, as described in the following section. Since the spacetime considered here is the direct product of the de Sitter spacetime with several spheres, it is not asymptotically flat and, therefore, the issue of choosing suitable boundary conditions
can be troublesome. Indeed, the problem of which boundary conditions one should impose to compute well-defined QNMs in pure de Sitter space has been subject to several discussions in the literature \cite{Brady1999,Abdalla2002,Du2004,Myung2004}. Likewise, the problem of adopting suitable boundary conditions\index{Boundary conditions} for QNMs in anti-de Sitter spacetimes has also been addressed elsewhere \cite{Avis1978,Breitenlohner1982,Burgess}. In the upcoming section we intend to add to the existing discussion available in the literature.

\vspace{.5cm}
%\section{\uppercase{Boundary Conditions}}\label{Sec.Boundary Cond}
\section{Boundary Conditions}\label{Sec.Boundary Cond}
\vspace{.5cm}

Quasinormal modes are solutions of wave-like equations\index{Wave-like equation} satisfying specific boundary conditions\index{Boundary conditions}, generally forming a discrete set. Therefore, the boundary conditions for the fields are a central piece of information behind the quasinormal frequencies\index{Quasinormal frequencies} \cite{Berti2009,Kokkotas1999,Nollert1999,Natario2004}. The aim of the present section is to discuss the suitable boundary conditions for the quasinormal modes in the class of generalized Nariai spacetimes\index{Generalized Nariai spacetimes} considered in this book.

In order to motivate the boundary conditions considered in what follows, let us first recall the physical reasoning behind the boundary conditions\index{Boundary conditions} for the quasinormal modes in Schwarzschild spacetime. Looking at the light cone structure of Schwarzschild spacetime, shown in Fig. \ref{FBCSB}, we note that at the event horizon\index{Event horizon} ($r=2M$) it is impossible for an observer to increase its radial coordinate, it will inexorably fall towards smaller values of $r$. Therefore, it is natural to use as boundary conditions at $r=2M$ that the waves are ingoing, which is represented by an infalling wavy arrow in Fig. \ref{FBCSB}. In its turn, at the infinity, the usual boundary condition is that no wave comes from infinity, whereas some waves can arrive at infinity after being scattered by the black hole \cite{Nollert1992,Choudhury2003}. Therefore, at infinity, it is natural to impose that waves are outgoing, as represented by a dashed wavy arrow in Fig. \ref{FBCSB}.
%\begin{figure}[ht!]
%  \centering
%  \includegraphics[width=8cm]{Cones_artigo_3}
%  \caption{Illustration of the light cone structure in Schwarzschild spacetime, with $r=2M$ being the event horizon. The wavy arrows represent the natural boundary conditions.}\label{FigCones_3}
%\end{figure}

\begin{figure}[ht!]
  \centering
 (a) \includegraphics[width=14cm]{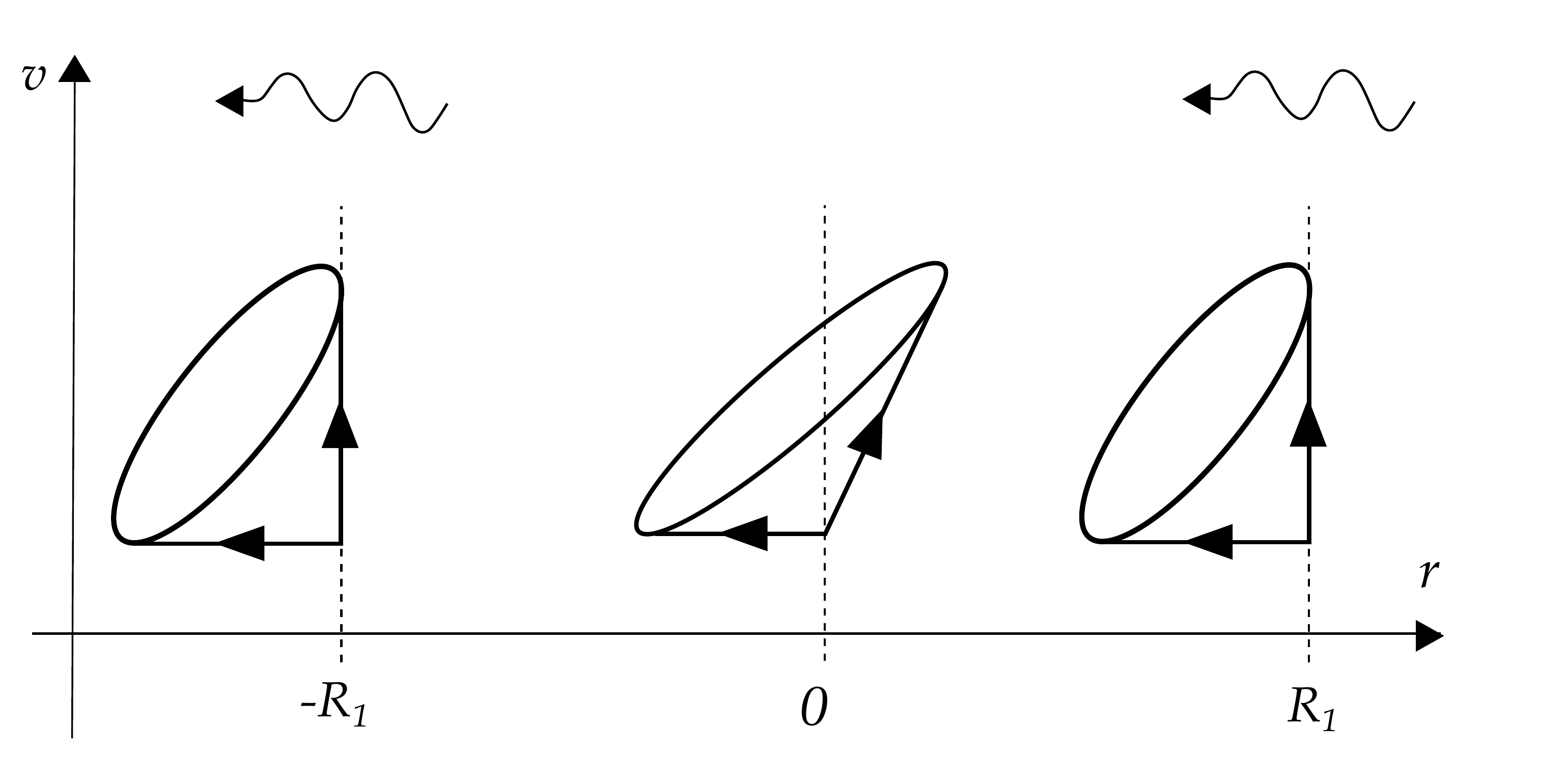}
 (b)\includegraphics[width=14cm]{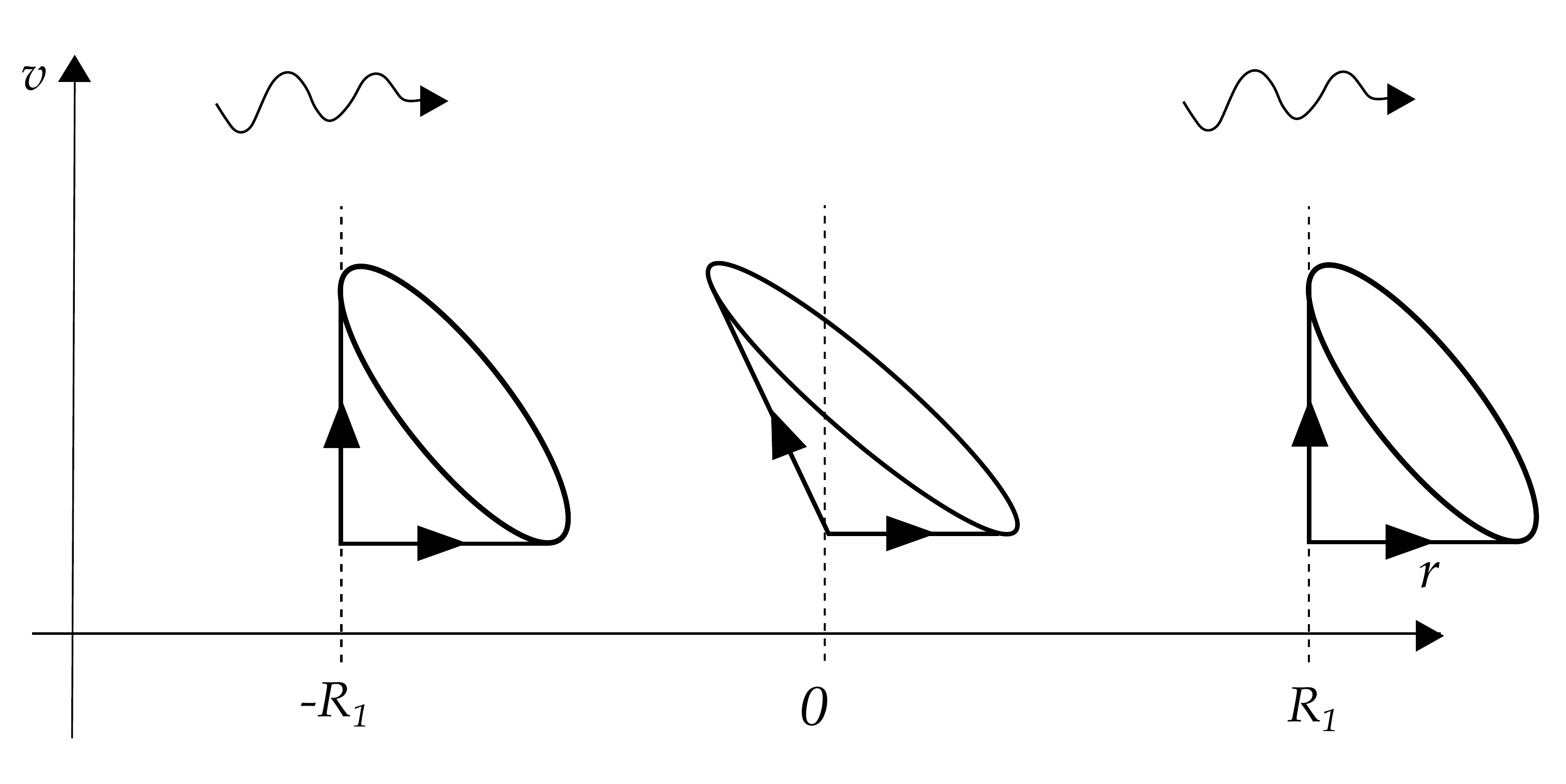}
  \caption{Illustration of the light cone structure of generalized Nariai spacetime. The wavy arrows represent the natural boundary conditions\index{Boundary conditions}. In part (a) it is assumed that the time-like vector field $\partial_v$ points to the future, while in part (b) it is assumed that $\partial_v$ points to the past. }\label{FigCones_4}
\end{figure}

Analogously, in order to guess the meaningful boundary conditions\index{Boundary conditions} for the quasinormal modes in generalized Nariai spacetime, we should look at its light cone structure. Such spacetime is the direct product of the two-dimensional de Sitter spacetime, $dS_2$, with several spheres. In the case of radial wave propagation, namely  $d\theta_l = d\phi_l = 0$, the line element is given by the $dS_2$ one,
\begin{equation}\label{dS2_LineElem}
  ds^2 = -\,f\, dt^2 + \frac{1}{f}\, dr^2 \quad \textrm{where} \quad f = 1 -  \frac{r^2}{R_1^2} \;,
\end{equation}
with $R_1$ being a positive constant. Actually, the latter line element represents just a patch of the whole $dS_2$ spacetime, which is rigorously defined as the surface
\begin{equation}\label{Surface_dS}
  - T^2 + X^2 + Y^2 = R_1^2
\end{equation}
immersed into the three-dimentional flat space with Lorentzian line element $ds^2 = - dT^2 + dX^2 + dY^2$. A parametrization of this surface is given by the coordinates $\{t,r\}$ defined by
\begin{equation}
  \left\{
     \begin{array}{ll}
       T = \sqrt{R_1^2 - r^2} \,\sinh(t/R_1) \,,\\
       X = \sqrt{R_1^2 - r^2} \,\cosh(t/R_1) \,,\\
       Y = r \,.
     \end{array}
   \right.
\end{equation}
In terms of the parameters $\{t,r\}$, the metric of the surface (\ref{Surface_dS}) is the one given in Eq. (\ref{dS2_LineElem}). In order for the coordinates $T$ and $X$ to be real, we must have $r\in [-R_1,R_1]$. Thus, in particular, we find that $-R_1\leq Y \leq R_1$ and $X\geq 0$, so that the surface that defines $dS_2$ is not fully covered by the coordinate system $\{t,r\}$. In addition, note that we should not ignore the negative values of $r$, since this part of the domain of $r$ describes a portion of $dS_2$ that is different from the one covered by $r>0$ \cite{Anninos2012, Spradlin2001}. This is an important point that differs from what happens in higher-dimensional de Sitter spacetimes \footnote{For instance, $dS_3$ is the surface $- T^2 + X^2 + Y^2 + Z^2 = R^2$ immersed into the flat space $ds^2 = - dT^2 + dX^2 + dY^2 +  dZ^2$. The coordinates $\{t,r,\phi\}$ defined by $T = \sqrt{R_1^2 - r^2}\sinh(t/R_1)$, $X = \sqrt{R_1^2 - r^2}\cosh(t/R_1)$, $Y=r\cos(\phi)$ and $Z=r \sin(\phi)$ cover part of $dS_3$. In this case, note that if we adopt the domain $\phi\in[0,2\pi]$ we just need to consider the positive branch of $r$.}. Aiming at studying the light cones in $dS_2$, it is useful to introduce the coordinate $v$ defined by the relation $dv = dt + \frac{1}{f}dr$, in terms of which the line element reads
\begin{equation}
  ds^2 = - f dv^2 + 2 dv dr \,.
\end{equation}
In particular, since $f> 0$ in the domain $r\in (-R_1,R_1)$, we see that $\partial_v$ is a time-like vector field, so that $v$ can be pictured as a time coordinate. We can assume that this coordinate increases as time passes by, namely that $\partial_v$ points to the future. The null rays of this spacetime are given by
\begin{equation}
  ds^2 = 0 \; \Rightarrow \; \left\{
                               \begin{array}{ll}
                                 dv = 0 \,, \\
                                 dv = (2/f)dr \,.
                               \end{array}
                             \right.
\end{equation}
The first light ray, defined by $dv=0$, is tangent to the vector field $\partial_r$. Since the inner product of $\partial_r$ and $\partial_v$ is positive, it follows that the light-like vector field $\partial_r$ points to the past or, in other words, $-\partial_r$ points to the future. Thus, as times passes by, this light ray must decrease its radial coordinate, as illustrated by the horizontal arrows in the line cones in part (a) of Fig. \ref{FigCones_4}.
The second light ray, given by $dv = (2/f)dr$, is tangent to $2\partial_v + f\partial_r$, which is a null vector field pointing to the future. Since the coefficient in front of $\partial_v$ is positive, it follows that, as time passes by, this light ray increases its coordinate $v$, just as illustrated by the arrows in the line cones in part (a) of Fig. \ref{FigCones_4}. Also, note that at the boundaries $r=\pm R_1$ we have $f=0$, so that the second light ray points in the direction of $\partial_v$. Then, analyzing the light cone structure shown in part (a) of Fig. \ref{FigCones_4}, we can see that an observer cannot increase its radial coordinate when it is at the boundaries $r=\pm R_1$. This suggests that the natural boundary condition for the waves in this spacetime is that they are infalling at both boundaries, as depicted by the wavy arrows. The latter conclusion was based on the arbitrary assumption that $\partial_v$ is oriented to the future. Have we had considered that $\partial_v$ pointed to the past, we would have found the light cone structure depicted in part (b) of Fig. \ref{FigCones_4}. In the latter case, the natural boundary condition is that the waves should be outgoing at both boundaries, as illustrated by the wavy arrows. Due to the symmetry $t\rightarrow -t$ and $r\rightarrow -r$ of the line element (\ref{dS2_LineElem}), it follows that both choices of time orientation for $\partial_v$ are equally valid, there is no preferred choice.

\begin{figure}[ht!]
  \centering
  \includegraphics[width=14cm]{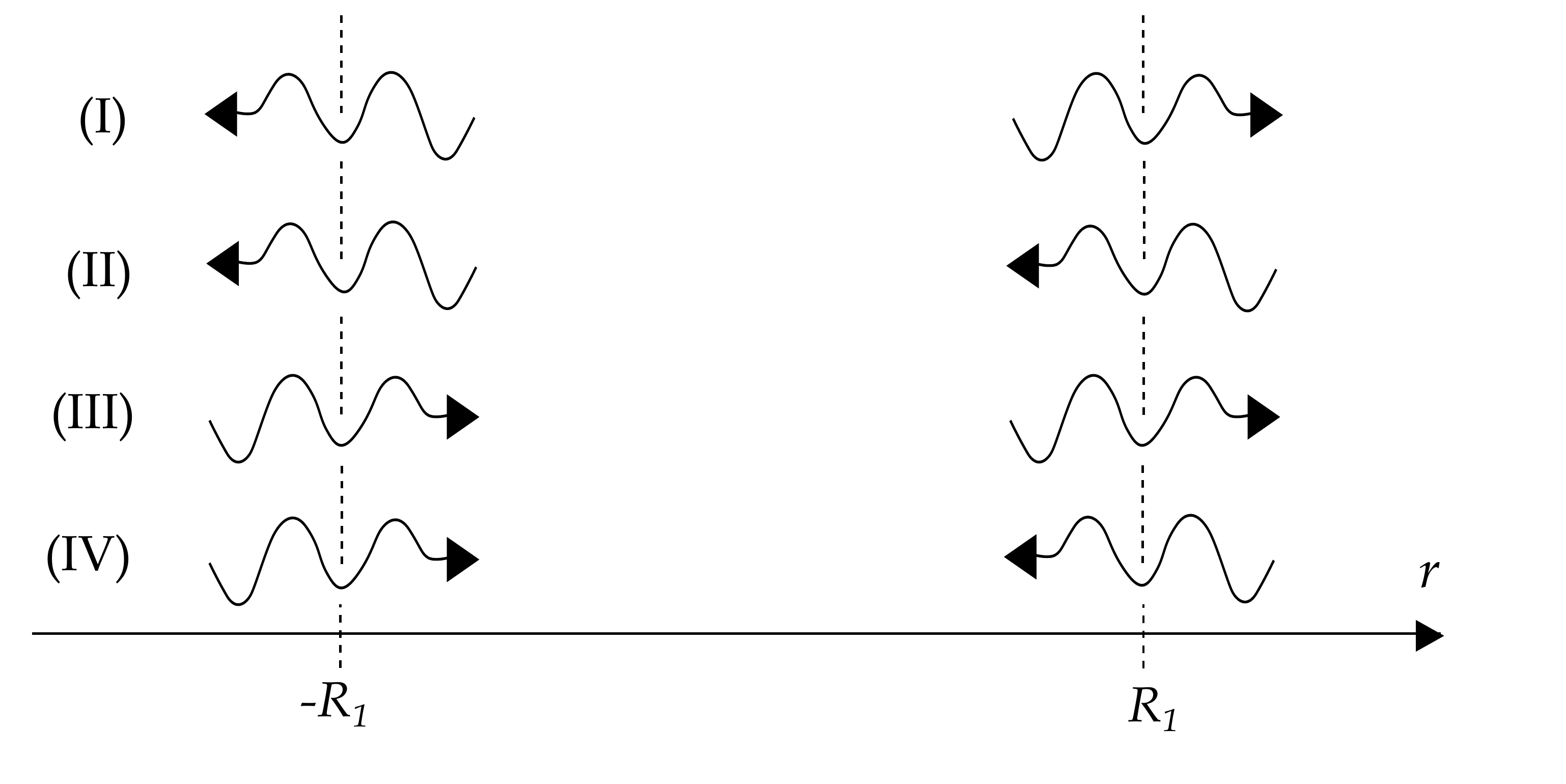}
  \caption{Types of boundary conditions considered in this article. Mathematically, wavy arrows pointing to the right represent $e^{-i\omega(t-x)}$, while wavy arrows pointing to the left represent $e^{-i\omega(t+x)}$.}\label{FigBoundCond}
\end{figure}

Thus, we can say that the natural boundary condition for the waves is that either the waves are infalling at both boundaries or the waves are outgoing at both boundaries. Nevertheless, as we shall see in the sequel,  it turns out that these boundary conditions\index{Boundary conditions}, although physically motivated, do not lead to quasinormal modes. On the other hand, when we impose that the wave is infalling at one boundary and outgoing at the other, we find what we are looking for: a discrete set of quasinormal modes. Therefore, in order for our calculations to be more complete, in the following sections we will consider four different types of boundary conditions\index{Boundary conditions}, the ones defined in Fig. \ref{FigBoundCond}.
In this figure, wavy arrows pointing to the right represent waves moving toward higher values of $r$, mathematically represented by $e^{-i\omega(t-x)}$, while wavy arrows pointing to the left represent waves moving toward lower values of $r$, mathematically represented by $e^{-i\omega(t+x)}$, where the coordinate $x$ will be defined below. As argued in the previous paragraph, boundary conditions\index{Boundary conditions} (II) and (III) are the ones physically motivated, although they will not lead to quasinormal modes. In contrast, we will see that conditions (I) and (IV) are associated to quasinormal modes. In spite of this, the physical reason why the boundary conditions (I) and (IV) will be taken into account is because they are the analogous boundary conditions to the \textit{ingoing at horizon} and \textit{outgoing at infinity} solutions that are causally appropriate in the Schwarzschild case. 

%%%%%%%%%%%%%%%%%%%%%%%%%%%%%%%%%%%%%%%%%%%%%%%%%%%%%%%%%%%%%%%%%%%%%%%%%%%%%%%%%%%%%%%%%%%%%%%%%%%%%%%%%%%%%%%%%%%%%%%%%%%%%%%%%%%%%%%%%%%%
%%%%%%%%%%%%%%%%%%%%%%%%%%%%%%%%%%%%%%%%%%%%%%%%%%%%%%%%%%%%%%%%%%%%%%%%%%%%%%%%%%%%%%%%%%%%%%%%%%%%%%%%%%%%%%%%%%%%%%%%%%%%%%%%%%%%%%%%%%%%
%%%%%%%%%%%%%%%%%%%%%%%%%%%%%%%%%%%%%%%%%%%%%%%%%%%%%%%%%%%%%%%%%%%%%%%%%%%%%%%%%%%%%%%%%%%%%%%%%%%%%%%%%%%%%%%%%%%%%%%%%%%%%%%%%%%%%%%%%%%%
%%%%%%%%%%%%%%%%%%%%%%%%%%%%%%%%%%%%%%%%%%%%%%%%%%%%%%%%%%%%%%%%%%%%%%%%%%%%%%%%%%%%%%%%%%%%%%%%%%%%%%%%%%%%%%%%%%%%%%%%%%%%%%%%%%%%%%%%%%%%

\vspace{.5cm}
%\section{\uppercase{Spin-$0$ Field Perturbations}}
\section{Spin-$0$ Field Perturbations}
\vspace{.5cm}
%\section*{3.4\quad Spin-$0$ Field Perturbations}
%\addcontentsline{toc}{section}{3.4\qquad Spin-$0$ Field Perturbations}

With the integration of the equation \eqref{WEH} for the generic potential \eqref{Potential_Generic} at hand, we are ready to move on and study the perturbation of several matter fields. Let us start with
the perturbations in a spin-$0$ field, a scalar field\index{Scalar field} $\boldsymbol{\Phi}$ of mass $\mu$. It is worth pointing out that the study of quasinormal modes associated to a scalar perturbation around several
backgrounds is a subject of active investigation. In particular, on the background of black holes, massive scalar perturbation has been shown to be quite different from that of the massless one in many aspects. For instance, it presents the so-called superradiant instability which does not appear in the massless case \cite{Bekenstein1998}, it may also have infinitely many long-living modes known as quasi-resonances \cite{Konoplya2005,Ohashi2004}. Finally, at asymptotically late times the massive fields show universal behavior independent of the spin of the field \cite{Konoplya2007} and it can be interpreted as a self-interacting scalar field\index{Scalar field} under the regime of small perturbations \cite{Hod1998}. Yet the scalar perturbation with mass has been investigated only in very few studies as to its quasinormal spectrum \cite{Konoplya2002,Xue2002,Simone1992}. So, it is important that we investigate the quasinormal modes associated to a massive scalar perturbation around the generalized Nariai background.

%Now, with the integration of the general equation (\ref{WEH}) at hand, we are ready to move on and study the perturbation of several matter fields. Let us start with the perturbations in a scalar field $\Phi$ of mass $\mu$. For the study of the quasinormal modes of scalar fields in other backgrounds, see \cite{Scalar1,Scalar2,Scalar3,Scalar4}.

The equation obeyed by the scalar field\index{Scalar field} while it propagates in the background (\ref{nariai-metric}) is the Klein-Gordon equation given by
\begin{equation}\label{KGE}
\frac{1}{\sqrt{\left | g^{GN} \right |}} \,\partial_{\mu} \left( g^{GN\,\mu\nu}\sqrt{ \left | g^{GN} \right |} \,\partial_{\nu} \right)\boldsymbol{\Phi} \,=\,\mu^{2}\boldsymbol{\Phi} \,,
\end{equation}
where $g_{\mu\nu}^{GN}$ is the metric in generalized Nariai background, namely \eqref{nariai-metric}. In order to accomplish the integrability of this equation, it is useful to introduce the tortoise coordinate\index{Tortoise coordinate} $x$ defined by the equation
\begin{equation}\label{TC}
dx=\frac{1}{f(r)}\,dr \;\;\Rightarrow \;\; x=R_{1} \, \text{arctanh}\left(\frac{r}{R_{1}}\right) \,.
\end{equation}
In particular, note that the tortoise coordinate\index{Tortoise coordinate} maps the domain between two horizons, $r\in(-R_1,R_1)$, into the interval $x \in (-\infty, \infty)$. In terms of this coordinate, the line element is written as
\begin{equation}\label{FF1}
ds^{2}= f\,(-dt^{2} + dx^{2}) + \sum_{l=2}^{d}R_{l}^{2}\,d\Omega_{l}^{2} \,.
\end{equation}
where
\begin{equation*}
 f =  f(x) = \textrm{sech}^2(x/R_{1}) \,.
\end{equation*}
Thus, writing  \eqref{KGE} in these coordinates, we eventually arrive at the following field equation
\begin{equation}\label{KGE1}
\left[ \dfrac{1}{f}\,\left(\partial_{x}^{2}-\partial_{t}^{2}\right) +\sum_{l=2}^{d}\frac{\Delta_{l}}{R_{l}^2} -\mu^{2} \right ]\boldsymbol{\Phi}(x) =0 \,,
\end{equation}
where
\begin{equation}
\Delta_{l} \equiv \frac{1}{\text{sin}\theta_{l}} \,\partial_{\theta_{l}}(\text{sin}\theta_{l}\,\partial_{\theta_{l}}) \,+\,\frac{1}{\text{sin}^{2}\theta_{l}} \,\partial^{2}_{\phi_{l}} \,,
\end{equation}
is the Laplace-Beltrami operator on the unit sphere.  The eigenfunctions of $\Delta_{l}$ are the well-known scalar spherical harmonics\index{Scalar spherical harmonics}, $Y_{\ell_{l},m_{l}}(\theta_{l}, \phi_{l})$, with eigenvalues determined by the equation
\begin{equation}
\Delta_{l}Y_{\ell_{l},m_{l}}(\theta_{l}, \phi_{l})\,=\,-\ell_l(\ell_l + 1)\,Y_{\ell_{l},m_{l}}(\theta_{l}, \phi_{l}) \,,
\end{equation}
with $\ell_l$ and $m_l$ being integers satisfying $|m_l|\leq \ell_l$ and $\ell_l \geq 0$ in order to ensure that the $Y_{\ell_{l},m_{l}}$ is regular at the points $\theta_{l} = 0$ and $\theta_{l} = \pi$ , where our coordinate system breaks down.  The index $\ell_l$ labels the irreducible
representations of the $SO(3)$ isometry subgroup associated with the spherical parts of the line element, while $m_l$ labels the $(2\ell_l+1)$ elements of the basis of the irreducible representation $\ell_l$. We have seen that whenever the background has spherical symmetry, it is useful to expand the angular dependence of a scalar field\index{Scalar field} in terms of scalar spherical harmonics\index{Scalar spherical harmonics} which are the objects with the same nature with respect to the action of the isometry subgroup $SO(3)$, see example 1. The symmetry of the generalized Nariai background, however, is a product of spherical symmetries, namely $SO(3) \times SO(3) \times \ldots SO(3)$. In such a case, the Klein-Gordon equation for a scalar field\index{Scalar field} is separable by the decomposition
%Thus, it is fruitful to expand the scalar field in terms of the spherical harmonics as follows
\begin{equation}\label{ExpansionSacalar}
\boldsymbol{\Phi} = \sum_{\ell, m}  e^{-i\omega t} \phi^{\omega}_{\ell,m}(x)\,\mathcal{Y}_{\ell,m} ,
\end{equation}
where
\begin{equation}\label{YLM}
  \mathcal{Y}_{\ell,m} = \prod_{l=2}^d \, Y_{\ell_l,m_l}(\theta_l,\phi_l)  \,.
\end{equation}
Here and in the rest of this book, for notational simplicity, we usually omit the ``sum'' over frequency $\omega$ in the Fourier transform. The general solution for the field $\boldsymbol{\Phi}$ must, then, include a ``sum'' over all values of the Fourier frequency $\omega$ with arbitrary Fourier coefficients. In Eq. (\ref{ExpansionSacalar}) we have taken into account the fact that $t$ is a cyclic coordinate of the metric, so that it is useful to decompose the temporal dependence of the field $\Phi$ in the Fourier basis\index{Fourier basis}. The sum over the collective index $\{\ell, m \}$ means that we are summing over all values of the set $\{\ell_{2}, m_{2}, \ell_{3}, m_{3}, \ldots , \ell_{d}, m_{d} \}$.

Then, by inserting the expression \eqref{ExpansionSacalar} into the filed equation, we are lead to the following ordinary differential equation for the components $\phi^{\omega}_{\ell,m}$:
\begin{equation}\label{WE}
\left[ \frac{d^{2}}{dx^{2}} \,+\, \omega^{2}\,-\,V_{s=0}(x)  \right ]\phi^{\omega}_{\ell,m} \,=\,0 \,,
\end{equation}
where the potential $V_{s=0}$ is the one studied in the previous section, see Eq. (\ref{Potential_Generic}), with the parameters $\mathfrak{a}$, $\mathfrak{b}$, $\mathfrak{c}$, and $\mathfrak{d}$ given by:
\begin{equation}
  \mathfrak{a} = 0 \quad , \quad \mathfrak{b} = 0 \quad , \quad \mathfrak{c} = \mu^{2} +\sum_{l=2}^d \dfrac{\ell_{l}(\ell_l + 1)}{R_l^2} \quad , \quad \mathfrak{d} = \dfrac{1}{R_1}\,.
\end{equation}
%\begin{equation}
%V(x) =  \frac{\mu^{2} +\sum_{j=2}^d \ell_{j}(\ell_j \,+\, 1)R_{j}^{-2} }{ \text{cosh}^{2}( x/R_1)}  ,
%\end{equation}
%\begin{equation}
%V(x) =  \frac{1 }{ \text{cosh}^{2}( x/R_1)} \left[ \mu^{2} +\sum_{j=2}^d\frac{\ell_{j}(\ell_j \,+\, 1)}{R_{j}^{2}} \right] ,
%\end{equation}
Inserting these parameters into Eq. (\ref{abc}), we find that the constants appearing in the hypergeometric equation are given by
\begin{equation}\label{abcScalar}
      \begin{array}{ll}
      a = \dfrac{1}{2} + i R_{1}\sqrt{\mu^2 + \sum_{l=2}^d  \dfrac{\ell_{l}(\ell_l + 1)}{R_l^2} - \dfrac{1}{4R_{1}^{2}} }  ,   \\
      \\
      b = \dfrac{1}{2} - i R_{1}\sqrt{\mu^2 + \sum_{l=2}^d  \dfrac{\ell_{l}(\ell_l + 1)}{R_l^2} - \dfrac{1}{4R_{1}^{2}} } ,  \\
      \\
      c = 1+ i\,R_1\,\omega \,.
    \end{array}
  \end{equation}
In particular, the following relations hold:
\begin{equation}
\mathfrak{d}(c-1) = i\omega \quad \text{and} \quad \mathfrak{d}(a+b-c) = -i\omega\,.
\end{equation}

%\subsection{Scalar Quasinormal Modes}
\vspace{.5cm}
\subsection{Scalar Quasinormal Modes}
\vspace{.5cm}

Now, we are ready to impose the boundary conditions\index{Boundary conditions}. 

\subsubsection{Boundary Condition (I)}

In order to investigate QNMs solutions we must impose the appropriate boundary conditions. Let us start with the boundary condition (I), described in Fig. \ref{FigBoundCond}. In this case, the field is assumed to move to decreasing $x$ at the boundary $x = -\infty$  while at the boundary $x = +\infty$ it should move towards increasing values of $x$. Since the time dependence of the mode $\phi^{\omega}_{\ell,m}$ is of the type $e^{-i\omega t}$, this means that  $\left.e^{-i\omega t}\phi^{\omega}_{\ell,m}\right|_{x\rightarrow -\infty}$ should behave as $e^{-i\omega(t+ x)}$ which is a plane wave propagating to the left (negative $x$-direction), while $\left.e^{-i\omega t}\phi^{\omega}_{\ell,m}\right|_{x\rightarrow +\infty}$ should go as $e^{-i\omega(t-x)}$ which is a plane wave propagating to the right (positive $x$-direction). Notice that, in the case considered in this section, Eq. (\ref{SolutionH_y=0}) translates to
\begin{equation}\label{Scalar-Iy=0}
   \left.e^{-i\omega t}\phi^{\omega}_{\ell,m}\right|_{x\rightarrow -\infty} = \,\alpha\, e^{-i\omega(t-x)} + \beta \,e^{-i\omega(t+x)}  \,.
\end{equation}
For the boundary condition (I), the condition for QNMs near the boundary $x = -\infty$ is therefore
\begin{equation}
\alpha = 0 \,.
\end{equation}
Next we study the behavior of the scalar mode near the boundary $x = \infty$. In this
region, assuming this latter requirement to hold, Eqs. (\ref{SolutionH_y=1}) and (\ref{abcScalar}) immediately yield
\begin{align}\label{Scalar-Iy=1}
  \left. \phi^{\omega}_{\ell,m}\right|_{x\rightarrow +\infty}  &\simeq\,   \beta \,   \frac{\Gamma(c-a-b)\Gamma(2-c)}{ \Gamma(1-a) \Gamma(1-b)}
 \,e^{-i\omega(t-x)}  \nonumber \\
&+\, \beta  \, \frac{\Gamma(a+b-c)\Gamma(2-c)}{ \Gamma(a-c+1) \Gamma(b-c+1) }\, e^{-i\omega(t+x)}   \,.
\end{align}
One can ensure the boundary condition (I) by requiring that the coefficient multiplying
$e^{-i\omega(t+x)}  $ should vanish and that the coefficient multiplying $e^{-i\omega(t-x)}  $ is nonvanishing. Since $\beta$ cannot be zero, otherwise the mode would vanish identically, the combination of the gamma functions has to be such that
\begin{equation}
\frac{\Gamma(c-a-b)\Gamma(2-c)}{ \Gamma(1-a) \Gamma(1-b)} \neq 0 \quad \text{and} \quad \frac{\Gamma(a+b-c)\Gamma(2-c)}{ \Gamma(a-c+1) \Gamma(b-c+1) } = 0 \,.
\end{equation}
Now, once the gamma function has no zeros, the way to achieve this is by letting the gamma
functions at the denominator to diverge, $\Gamma(a-c+1) = \infty$ or $\Gamma(b-c+1) = \infty$. Since
the gamma function diverges only at non-positive integers, we are led to the following constraint:
\begin{equation}
  a-c+1 = -n \quad \textrm{or} \quad b-c+1 = -n \quad \textrm{where} \quad   n\in\{0,1,2,\ldots\}.
\end{equation}
Therefore, assuming the latter constraints and using Eq. \eqref{abcScalar}, one eventually obtains
that
\begin{equation}
  \omega_{\text{I}} = \pm\sqrt{\mu^2  + \sum_{l=2}^d  \frac{\ell_{l}(\ell_l + 1)}{R_l^2}   - \frac{1}{4R_1^2} } + \frac{i}{2R_1} (2n+1),
\end{equation}
with $n$ being any non-negative integer, called \textit{overtone index}. Here, we have started to employ the notation $\omega_{\text{I}}$ for the frequencies when the boundary condition is (I), $\omega_{\text{II}}$ for the boundary condition (II) and so on. The important point to note is that these frequencies are the only ones compatible with the boundary condition (I). They are the so-called frequencies of the quasinormal modes. Note the presence of the imaginary part in the frequency, which accounts for a damping of the field, a feature of perturbations in the presence of horizons. Moreover, note that the square root could also lead to an imaginary part of the frequency, in which case the perturbation mode would be solely damped, with no characteristic oscillation. For instance, in the case of a massless field the frequency of the spherically symmetric mode ($\ell_l=0$) will be purely imaginary.

\subsubsection{Boundary Condition (II)}

Now, let us investigate the boundary condition (II). In this case the mode $e^{-i\omega t}\phi^{\omega}_{\ell,m}$ should behave as $e^{-i\omega (t+x)}$ at both boundaries $x = \pm\infty$, as depicted in Fig. \ref{FigBoundCond}. Thus, since the behavior at $x = -\infty$ is the same as at boundary condition (I), it follows that Eq. (\ref{Scalar-Iy=0}) remains valid for the boundary condition (II). The only difference is that in Eq. (\ref{Scalar-Iy=1}) we should eliminate the term $e^{-i\omega (t-x)}$, which is possible only if either $\Gamma(1-a)$ or $\Gamma(1-b)$ diverge. Since $a$ and $b$ do not depend on the frequency $\omega$, see Eq. (\ref{abcScalar}), it follows that the constraints $1-a=-n$ and $1-b=-n$, with $n$ a non-negative integer, would represent restriction on parameters that are already fixed, like the mass $\mu$ and the radii $R_l$ that describe the background. Therefore, we conclude that, generally, we have no solution for the perturbation when the boundary condition (II) is assumed.

\subsubsection{Boundary Condition (III)}

For the boundary condition (III), the mode $e^{-i\omega t}\phi^{\omega}_{\ell,m}$ should behave as $e^{-i\omega (t-x)}$ at both boundaries $x = \pm\infty$.
Therefore, at  Eq. (\ref{Scalar-Iy=0}) we should set $\beta=0$, in which case we are left with the following form at $x = +\infty$:
\begin{align}
  \left. e^{-i\omega t}\phi^{\omega}_{\ell,m}\right|_{x\rightarrow \infty} & \simeq      \alpha \,
  \frac{\Gamma(c-a-b)\Gamma(c)}{ \Gamma(c-a) \Gamma(c-b)   } \, e^{-i\omega (t-x)}  \nonumber\\
& +  \alpha\,
   \frac{\Gamma(a+b-c)\Gamma(c)}{ \Gamma(a) \Gamma(b) } \, e^{-i\omega (t+x)}  . \label{SolutionH_y=1III}
\end{align}
In order to eliminate the term  $e^{-i\omega (t+x)}$, we need to set $a = -n$ or $b=-n$, with $n\in\{0,1,2,\cdots\}$. Just as in the case of boundary condition (II), this constraint cannot be satisfied in general. Thus, we have no quasinormal modes obeying the boundary condition (III).

\subsubsection{Boundary Condition (IV)}

Finally, for the boundary condition (IV), the field must behave as $e^{-i\omega (t-x)}$ at $x = -\infty$ while at  $x = +\infty$ it should go as
$e^{-i\omega (t+x)}$. Hence, at Eq. (\ref{SolutionH_y=1III}) we should get rid of the term $e^{-i\omega (t-x)}$, which can be accomplished by setting
$c-a=-n$ or $c-b=-n$, with $n$ being a non-negative integer. The latter constraints along with Eq. (\ref{abcScalar}) lead to the following quasinormal frequencies\index{Quasinormal frequencies}:
\begin{equation}
  \omega_{\text{IV}} = \pm\sqrt{\mu^2  + \sum_{j=2}^d  \frac{\ell_{j}(\ell_j + 1)}{R_j^2}   - \frac{1}{4R_1^2} } - \frac{i}{2R_1} (2n+1).
\end{equation}
This spectrum is almost equal to the one found for the boundary condition (I), the only difference being the sign of the imaginary part. Thus, while for the boundary condition (IV) the modes dwindle for $t\rightarrow-\infty$ and diverge for $t\rightarrow+\infty$, for the boundary condition (I) it is the other way around.

%Actually, should we have chosen another branch of the square root while computing the coefficients $c$ at Eq. (\ref{abcScalar}) from the general expression %at Eq. (\ref{abc}), we would have found that the spectrum associated to the boundary conditions I and IV would have been interchanged.

\vspace{.5cm}
%\section{\uppercase{Spin-$1$ Field Perturbation}}
\section{Spin-$1$ Field Perturbation}
\vspace{.5cm}

In this section we shall consider the perturbations on the Maxwell field\index{Maxwell field} $\bl{\mathcal{A}}$, a massless spin-1 field. In this case we shall assume that the electromagnetic charges of the background are zero, namely $Q_1=Q_l=0$, so that we have a vanishing Maxwell field\index{Maxwell field} in the background, $\bl{\mathcal{A}}^{GN}=0$. This is important to validate the separability\index{Separability} of the perturbations in the background metric and the matter fields, as discussed in chapter \ref{ChapPTP}. In particular this means that the radii $R_1$ and $R_l$ are all equal in such a case, see Eq. (\ref{radii}). For the calculation of quasinormal modes of spin-$1$ fields in other backgrounds, see \cite{Cardoso2001,Koutsoumbas2006,Konoplya2006,Ortega2008}.

%\subsection{Ansatz for the Separation of Maxwell's Equation}
\vspace{.5cm}
\subsection{Ansatz for the Separation of Maxwell's Equation}
\vspace{.5cm}

%\begin{equation}
%\boldsymbol{\mathcal{A}} = \mathcal{A}_t\,dt + \mathcal{A}_x\,dx + \sum_{l=2}^{d}\mathcal{A}_{a_l}dx^{a^l} \,.
%\end{equation}

As we have proved in the example 2, Maxwell's equation for a spin-$1$ gauge field $\boldsymbol{A} = A_{\mu} dx^{\mu}$ in a four-dimensional background with spherical symmetry is separable by the decomposition
\begin{align}
\boldsymbol{A}(t, r, \theta_{l},\phi_{l}) &=\,\,  \sum_{\ell_{l} , m_{l}} \left[ \left(G^{+}_{0, \ell_{l} m_{l}} \,dt + G^{+}_{1, \ell_{l} m_{l}}\,dr \right) Y_{\ell_{l},m_{l}} \right.\nonumber\\
&+\,\, \left. G^{+}_{\ell_{l}, m_{l}} \,V_{\ell,m}^{+} + G^{-}_{\ell_l, m_l} \,V_{\ell_l,m_l}^{-} \,\right] \,.
\end{align}
Here, $\{\theta_{l}, \phi_{l}\}$ denote the angular variables in the unit two-dimensional sphere whose
metric tensor is $\hat{g}_{a_{l}b_{l}}$ as defined in Eq. \eqref{MTOS}, with the indices $a_{l}b_{l}$ running through $\{\theta_{l}, \phi_{l}\}$. Under this decomposition, we can rewrite each component of $\nabla^{\mu}\nabla_{[\mu}A_{\nu]}$ as products of a function depending on the variables $\{t, r\}$ and a non-vanishing function depending on the angular variables $\{\theta_{l}, \phi_{l}\}$, so that $\nabla^{\mu}\nabla_{[\mu}A_{\nu]} = 0$ depends just on the coordinates $\{t, r\}$. This makes it clear the fundamental importance of using the angular basis\index{Angular basis} $\{Y_{\ell_l,m_l},V_{\ell_l,m_l}^{+},V_{\ell_l,m_l}^{-}\}$ whenever the background has spherical symmetry, where $Y_{\ell_l,m_l}$ are scalar spherical harmonics\index{Scalar spherical harmonics} and $V_{\ell_l,m_l}^{\pm}$ are vector spherical harmonics\index{Vector spherical harmonics} as defined in example 2, namely
\begin{align}\label{B1F4D}
V_{\ell_l,m_l}^{+} &=\,\ \partial_{\theta_l} Y_{\ell_l,m_l}d\theta_l +  \partial_{\phi_l} Y_{\ell_l,m_l}d\phi_l \,,\nonumber\\
\\
V_{\ell_l,m_l}^{-} &=\,\, \dfrac{1}{\sin\theta_l }\, \partial_{\phi_l} Y_{\ell_l,m_l} \, d\theta_l -\sin\theta_l \, \partial_{\theta_l} Y_{\ell_l,m_l} \, d\phi_l \nonumber \,.
\end{align}
Notice that, since the scalar spherical harmonics\index{Scalar spherical harmonics} are a basis for the functions in the
sphere, nothing in the arguments put forward stops us from using them to expand the
components $A_{\mu}$ just as we did for the scalar field\index{Scalar field}, namely
\begin{equation}
A_{\mu} =  \sum_{\ell_{l} , m_{l}}  G_{\mu, \ell_{l} m_{l}}(t,r) Y_{\ell_l,m_l}(\theta_l,\phi_l) \,.
\end{equation}
However, this is not the most suitable choice inasmuch as each type of field should be expanded in terms of an angular basis\index{Angular basis} that has the same nature. Indeed, while the components $A_{t}$ and $A_{r}$ transform as scalar fields\index{Scalar field} with respect to the action of the isometry subgroup $SO(3)$, the components $A_{\theta_l}$ and $A_{\phi_l}$ transform as the components of a covector in the two-sphere. Therefore, the most natural way to expand  $A_{\theta_l}$ and $A_{\phi_l}$ is using a basis of $1$-forms in the sphere, namely \eqref{B1F4D}. In particular, this latter basis of $1$-forms acquires an elegant form in terms of covariant derivatives in the two-sphere, denoted here by $\hat{\nabla}_{a_l}$. Starting from the spherical harmonic $Y_{\ell_l,m_l}$, which are scalar fields\index{Scalar field} in the two-sphere, and taking covariant derivatives, we find that the $1$-forms \eqref{B1F4D} can be elegantly expressed as:
\begin{equation}
V_{\ell_l,m_l}^{\pm} =  V_{a_l}^{\pm}dx^{a_l} \,,
\end{equation}
where the components $V_{a_l}^{\pm} = V_{a_l}^{\pm}(\theta_l,\phi_l)$ are defined as
\begin{equation}\label{B1FCD}
V_{a_l}^{+} = \hat{\nabla}_{a_l}Y_{\ell_l,m_l} \quad \text{and} \quad V_{a_l}^{-} = \epsilon_{a_lc_l}\hat{\nabla}^{c_l}Y_{\ell_l,m_l} \,,
\end{equation}
with $\epsilon_{a_l b_l}$ being the volume form in the $l$th two-sphere.

We are ready to expand, in a natural way, the spin-$1$ field perturbations $\boldsymbol{\mathcal{A}} = \mathcal{A}_{\mu} dx^{\mu}$ in the generalized Nariai background whose symmetry is a product of $(d-1)$ spherical symmetries. So, a lot of the formulas thus established for the spherically symmetric
in four dimensions case still remain valid by formally replacing $Y_{\ell_l,m_l}$ by the product of
scalar spherical harmonics defined in Eq. \eqref{YLM}, namely $\mathcal{Y}_{\ell m} = \prod_{l=2}^{d} Y_{\ell_l,m_l}$. For instance, $\mathcal{A}_{t}$ and $\mathcal{A}_{x}$ are scalars with respect to the $(d-1)$ spheres and should be expanded in terms of product of scalar spherical harmonics\index{Scalar spherical harmonics}  just as we did for the scalar field\index{Scalar field}, namely Eq. \eqref{ExpansionSacalar}
\begin{align}
\mathcal{A}_{t} &=\, \sum_{\ell, m} e^{-i\omega t} A^{+}_{0,\ell m} \mathcal{Y}_{\ell m} \,,\nonumber\\
\mathcal{A}_{x} &=\, \sum_{\ell, m} e^{-i\omega t} A^{+}_{1,\ell m} \mathcal{Y}_{\ell m} \,,
\end{align}
where $A^{+}_{0,\ell m}, A^{+}_{1,\ell m}$ are arbitrary functions of the tortoise coordinate\index{Tortoise coordinate} $x$, Eq. \eqref{TC}, and the sum over $\{\ell,m\}$ means the sum over all possible values of the set $\{\ell_2,m_2,\ell_3,m_3,\\ \ldots,\ell_d,m_d \}$. Beside this, we would say that the objects $V_{\ell_l,m_l}^{\pm}$ are $1$-forms with respect to rotations in the $l$th sphere. However, strictly speaking, $V_{\ell_l,m_l}^{\pm}$ must depend on all indices $\{\ell_2,m_2,\ell_3,m_3, \ldots, \ell_d,m_d \}$ and not just on $\{\ell_2,m_2\}$ as suggested by the notation $V_{\ell_l,m_l}^{\pm}$. So in order to be consistent with this requirement, a natural generalization for arbitrary $d$ is provided by
\begin{equation}\label{B1FDD}
\mathcal{V}_{l,\ell m}^{\pm} =  \mathcal{V}_{a_l}^{\pm}dx^{a_l} \,,
\end{equation}
with $\mathcal{V}_{a_l}^{\pm}$ being defined in terms of $V_{a_l}^{\pm}$ as follows:
\begin{equation}
\mathcal{V}_{a_l}^{\pm} = V_{a_l}^{\pm}(\theta_l,\phi_l) \prod_{n=2,n\neq l}^{d} Y_{\ell_n,m_n}(\theta_n,\phi_n)  \,,
\end{equation}
where $V_{a_l}^{\pm}$ have been defined in Eq. \eqref{B1FCD}. It follows, then, that
\begin{align}\label{ABMF}
\mathcal{A}_{a_l}dx^{a_l} \,=\, \sum_{\ell, m} e^{-i\omega t} \left( A^{+}_{l,\ell m} \mathcal{V}_{l,\ell m}^{+} + A^{-}_{l,\ell m} \mathcal{V}_{l,\ell m}^{-} \right) \,,
\end{align}
where $ A^{\pm}_{l,\ell m}$ are generic functions of $x$. One can easily check that the $1$-forms $\mathcal{V}_{l,\ell,m}^{\pm}$ defined in Eq. \eqref{B1FDD} have the same form as those for the $1$-forms $V_{\ell_{l},m_{l}}^{\pm}$ defined in Eq. \eqref{B1F4D} just replacing $\ell_l,m_l$ by $l, \ell m$ and $Y_{\ell_l,m_l}$ by $\mathcal{Y}_{\ell m}$.

With these objects and taking into account the symmetries of the background considered here, namely Eq. \eqref{FF1}, a suitable ansatz for the gauge field in order to separate the field equation is provided by
\begin{align}
\boldsymbol{\mathcal{A}} = \sum_{\ell,m} e^{-i\omega t} \left[ \left(A^{+}_{0,\ell m} \,dt + A^{+}_{1,\ell m}\,dx \right) \mathcal{Y}_{\ell m} + \sum_{l=2}^{d} \left( A^{+}_{l,\ell m} \,\mathcal{V}_{l,\ell m}^{+} + A^{-}_{l,\ell m} \,\mathcal{V}_{l,\ell m}^{-}\right) \right] \,.
\end{align}

%\begin{align}
%\boldsymbol{\mathcal{A}} \,=&\, \sum_{\ell,m} e^{-i\omega t} \left[ \left(A^{+}_{0,\ell m} \,dt + A^{+}_{1,\ell m}\,dx \right) \mathcal{Y}_{\ell m} \right. \nonumber\\
%\,+&\, \left.\sum_{l=2}^{d} \left( A^{+}_{l,\ell m} \,\mathcal{V}_{l,\ell m}^{+} + A^{-}_{l,\ell m} \,\mathcal{V}_{l,\ell m}^{-}\right) \right] \,.
%\end{align}

The final general solution for the field $\boldsymbol{\mathcal{A}} $ must then include a ``sum'' over all values of the Fourier frequencies $\omega$ with arbitrary Fourier coefficients. This is the most natural way of writing the degrees of freedom of the Maxwell field\index{Maxwell field} which comes from the fact that it is a spin-$1$ field and, therefore, the spherical symmetries should show up in terms of vector spherical harmonics\index{Vector spherical harmonics}.

Notice that we can use the freedom of choosing a gauge to simplify the general form of the Maxwell perturbations. Indeed, the above expression can be rewritten in the following form
\begin{align}
\boldsymbol{\mathcal{A}} = \sum_{\ell,m} e^{-i\omega t} \left[\left(A^{+}_{0,\ell m} \,dt + \tilde{A}^{+}_{1,\ell m}\,dx \right) \mathcal{Y}_{\ell m} + \sum_{l=2}^{d} A^{-}_{l,\ell m} \,\mathcal{V}_{l,\ell m}^{-} \right. \nonumber\\
\,+\, \left. d\left(\sum_{l=2}^{d} A^{+}_{l,\ell m}\mathcal{Y}_{\ell m} \right) \right] \,,
\end{align}
where
\begin{equation}
\tilde{A}^{+}_{1,\ell m}  = A^{+}_{1,\ell m} - \partial_{x} \sum_{l=2}^{d} A^{+}_{1,\ell m}  \,. 
\end{equation}
In a $U(1)$ gauge field theory, the last term does not represent a relevant degree of
freedom inasmuch as it is an exact differential and therefore we can eliminate it by
means of a gauge transformation\index{Gauge transformation}. Doing so and dropping the tilde, we can say that the
most natural ansatz for a $1$-form gauge field in generalized Nariai background, which is
a problem with a product of $(d-1)$ spherical symmetries is:
\begin{align}
\boldsymbol{\mathcal{A}} = \sum_{\ell,m} e^{-i\omega t} \left[\left(A^{+}_{0,\ell m} \,dt + A^{+}_{1,\ell m}\,dx \right) \mathcal{Y}_{\ell m} + \sum_{l=2}^{d} A^{-}_{l,\ell m} \,\mathcal{V}_{l,\ell m}^{-}  \right] \,.
\end{align}
At this point, it is worth recalling that for a given $\ell_l$, the object $V_{\ell_l,m_l}^{+}$
should transform as $(-1)^{\ell_l}$, while $V_{\ell_l,m_l}^{-}$ should go as $(-1)^{\ell_l+1}$ under a parity transformation\index{Parity transformation} in the $l$th two-sphere, namely Eq. \eqref{parityT}, which is a consequence from the fact that $Y_{\ell_l,m_l} \overset{P_l\,}{\rightarrow } (-1)^{\ell_l} Y_{\ell_l,m_l}$.  However, the symmetry of the generalized Nariai background is a product of spherical symmetries. So, under a parity transformation\index{Parity transformation} in each of the spheres, we find that the scalar $\mathcal{Y}_{\ell m} = \prod_{l=2}^{d} Y_{\ell_l,m_l}$ transforms as
\begin{equation}
\mathcal{Y}_{\ell m} \,\overset{\text{parity}}{\longrightarrow }\, (-1)^{\ell_2 + \ell_3 + \ldots + \ell_d} \mathcal{Y}_{\ell m} \,,
\end{equation}
which implies that $\mathcal{V}_{l,\ell m}^{\pm} \overset{\text{parity}}{\longrightarrow }\, \pm (-1)^{\ell_2 + \ell_3 + \ldots + \ell_d} \,\mathcal{V}_{l,\ell m}^{\pm}$. Objects that transform in the same way as $\mathcal{Y}_{\ell m}$ under a parity transformation\index{Parity transformation} are said to be even, while objects that gain an extra minus sign compared to $\mathcal{Y}_{\ell m}$ are said to be odd. In particular, we say that $\mathcal{V}_{l,\ell m}^{+}$ has even parity, $(-1)^{\ell_2 + \ell_3 + \ldots + \ell_d}$, while $\mathcal{V}_{l,\ell m}^{-}$ has odd parity, $(-1)^{\ell_2 + \ell_3 + \ldots + \ell_d \,+\,1}$. It follows that we can write $\boldsymbol{\mathcal{A}}$ as
\begin{equation}\label{PTGF}
\boldsymbol{\mathcal{A}}  = \boldsymbol{\mathcal{A}}^{+} + \boldsymbol{\mathcal{A}}^{-} \,,
\end{equation}
where the objects $\boldsymbol{\mathcal{A}}^{\pm}$ defined by
\begin{align}
\boldsymbol{\mathcal{A}}^{+}  &=\,\, \sum_{\ell,m} e^{-i\omega t} \left(A^{+}_{0,\ell m} \,dt + A^{+}_{1,\ell m}\,dx \right) \mathcal{Y}_{\ell m}  \,,\nonumber\\
\\
\boldsymbol{\mathcal{A}}^{-}  &=\,\,  \sum_{\ell,m}\sum_{l=2}^{d} A^{-}_{l,\ell m} \,\mathcal{V}_{l,\ell m}^{-} \,,\nonumber
\end{align}
transform as $\boldsymbol{\mathcal{A}}^{\pm} \overset{\text{parity}}{\longrightarrow }\, \pm (-1)^{\ell_2 + \ell_3 + \ldots + \ell_d} \,\boldsymbol{\mathcal{A}}^{\pm}$. In particular, this means that $\boldsymbol{\mathcal{A}}^{+}$ is an even field with parity $(-1)^{\ell_2 + \ell_3 + \ldots + \ell_d} $, while $\boldsymbol{\mathcal{A}}^{-} $ is an odd field with parity $(-1)^{\ell_2 + \ell_3 + \ldots + \ell_d \,+\,1} $.

\vspace{.5cm}
%\subsection{Maxwell Quasinormal Modes}
\subsection{Maxwell Quasinormal Modes}
\vspace{.5cm}

Spin-$1$ field perturbations are governed by Maxwell's source-free equations
\begin{equation}\label{Maxwell-equation}
\nabla_{\mu}\mathcal{F}^{\mu\nu}\,=\,0 \,, \quad \text{with} \quad \mathcal{F}_{\mu\nu}\,=\,\partial_{\mu}\mathcal{A}_{\nu}\,-\,\partial_{\nu}\mathcal{A}_{\mu}\,,
\end{equation}
where $\mathcal{F}^{\mu\nu}$ is the Maxwell tensor and $\mathcal{A}^{\mu}$ are the components of the gauge field \eqref{PTGF}. Now, since the generalized Nariai background metric does not change when a parity transformation\index{Parity transformation} is applied in each of the spheres, we expect that the perturbation equations will not mix the $\boldsymbol{\mathcal{A}}^{+}$ and $\boldsymbol{\mathcal{A}}^{-}$ parts since these have different parities, namely $(-1)^{\ell_2 + \ell_3 + \ldots + \ell_d}$ and $(-1)^{\ell_2 + \ell_3 + \ldots + \ell_d \,+\,1}$, and the background is invariant under parity transformation\index{Parity transformation}. 
%So, we can without loss of generality separate the perturbation into its $\boldsymbol{\mathcal{A}}^{+}$ and $\boldsymbol{\mathcal{A}}^{-}$ parts and study them separately.
Thus, in order to find the general solution one can first ignore the part $\boldsymbol{\mathcal{A}}^{-}$ and integrate for $\boldsymbol{\mathcal{A}}^{+}$; then, set $\boldsymbol{\mathcal{A}}^{+}$ to zero and find $\boldsymbol{\mathcal{A}}^{-}$. This separation represents no loss of generality.

\subsubsection{Odd Perturbation}

By an odd perturbation\index{Odd perturbation} we mean the most general perturbation for a given set of
spherical harmonic indices $\{\ell_2,m_2,\ell_3,m_3,\ldots,\ell_d,m_d \}$ and parity $(-1)^{\ell_2 + \ell_3 + \ldots + \ell_d +1}$, namely
\begin{equation}
\boldsymbol{\mathcal{A}}^{-}  \,=\,  \sum_{\ell,m}\sum_{l=2}^{d} e^{-i\omega t} A^{-}_{l,\ell m} \,\mathcal{V}_{l,\ell m}^{-} \,.
\end{equation}
Inserting this ansatz into Maxwell's source-free equation, we end up with the following differential equations obeyed by the components $A^{-}_{l,\ell m}$:
\begin{align}
E_{\theta_l}^{-} &\equiv\,\, \left[  \dfrac{d}{dx} + \omega^{2} - V_{s=1}(x) \right]A^{-}_{l,\ell m}(x) = 0 \,,\\
E_{\phi_l}^{-} &\equiv\,\, E_{\theta_l}^{-}  = 0 \,, 
\end{align}
\begin{align}
E_{\theta_l}^{-} &\equiv\,\, \left[  \dfrac{d}{dx} + \omega^{2} - V(x) \right]A^{-}_{l,\ell m}(x) = 0 \,,\\
E_{\phi_l}^{-} &\equiv\,\, E_{\theta_l}^{-}  = 0 \,, 
\end{align}
where the potential $V_{s=1}$ is the one studied in the previous section, see \eqref{Potential_Generic}, with the parameters $\mathfrak{a}, \mathfrak{b}, \mathfrak{c}$, and $\mathfrak{d}$ given by:
\begin{equation}
  \mathfrak{a} = 0 \quad , \quad \mathfrak{b} = 0 \quad , \quad \mathfrak{c} = \sum_{l=2}^d \dfrac{\ell_{l}(\ell_l + 1)}{R_l^2} \quad , \quad \mathfrak{d} = \dfrac{1}{R_1}\,.
\end{equation}
Then, assuming that $E_{\theta_l}^{-} = 0$, which implies that $E_{\phi_l}^{-} = 0$, it follows directly that $A^{-}_{l,\ell m}$ obeys the same equation as that for the scalar field\index{Scalar field} mode $\phi_{\ell m}^{\omega}$ when the scalar field has vanishing mass ($\mu=0$). Thus, the quasinormal spectrum associated to this component of the Maxwell field\index{Maxwell field} must be the same as the one for the massless scalar field\index{Scalar field}. In particular, this means that for the boundary conditions\index{Boundary conditions} (II) and (III) we have no QNMs, while for the boundary conditions\index{Boundary conditions} (I) and (IV) the allowed frequencies must have the form
\begin{align}\label{QNFS1F}
  \omega_{\text{I}} &=\,\, \pm\sqrt{\sum_{l=2}^d  \frac{\ell_{l}(\ell_l + 1)}{R_l^2}   - \frac{1}{4R_1^2} } + \frac{i}{2R_1} (2n+1) \,,\nonumber\\
  \\ 
   \omega_{\text{IV}} &=\,\, \pm\sqrt{\sum_{l=2}^d  \frac{\ell_{l}(\ell_l + 1)}{R_l^2}   - \frac{1}{4R_1^2} } - \frac{i}{2R_1} (2n+1) \,.\nonumber
\end{align}

\subsubsection{Even Perturbation}

By an even perturbation\index{Even perturbation} we mean the most general perturbation for a given set of
spherical harmonic indices $\{\ell_2,m_2,\ell_3,m_3,\ldots,\ell_d,m_d \}$ and parity $(-1)^{\ell_2 + \ell_3 + \ldots + \ell_d }$, namely
\begin{equation}
\boldsymbol{\mathcal{A}}^{+}  = \sum_{\ell,m} e^{-i\omega t} \left(A^{+}_{0,\ell m} \,dt + A^{+}_{1,\ell m}\,dx \right) \mathcal{Y}_{\ell m} \,.
\end{equation}
Inserting this field perturbation into the Maxwell equation, we are eventually led to the
following equations obeyed by the components $A^{+}_{0,\ell m}$ and $A^{+}_{1,\ell m}$:
\begin{align}
E_{t}^{+} &\equiv\,\, \dfrac{d}{dx} \left[\dfrac{1}{f}\left(\dfrac{d}{dx} A^{+}_{0, \ell m} + i\omega A^{+}_{1, \ell m} \right ) \right ] - \sum_{l=2}^{d}\dfrac{\ell_l (\ell_l +1)}{R_l^2}A^{+}_{0, \ell m} = 0 \,, \\
E_{x}^{+} &\equiv\,\, i\omega\left[\dfrac{1}{f}\left(\dfrac{d}{dx} A^{+}_{0, \ell m} + i\omega A^{+}_{1, \ell m} \right ) \right ] + \sum_{l=2}^{d}\dfrac{\ell_l (\ell_l +1)}{R_l^2}A^{+}_{1, \ell m} = 0  \,, \\
E_{\theta_l}^{+} &\equiv\,\, \dfrac{d}{dx} E_{x}^{+} -i\omega E_{t}^{+} = 0 \,,\\
E_{\phi_l}^{+} &\equiv\,\, E_{\theta_l}^{+}  = 0 \,. 
\end{align}
In order to solve this set of equations, it is useful to use the function $\breve{A}_{\ell m} = \breve{A}_{\ell m}(x)$ defined by
\begin{equation}
\breve{A}_{\ell m} = \dfrac{1}{f}\left(\dfrac{d}{dx} A^{+}_{0, \ell m} + i\omega A^{+}_{1, \ell m} \right ) \,,
\end{equation}
instead of the degrees of freedom $A^{+}_{0, \ell m}$ and $A^{+}_{1, \ell m}$ inasmuch as $\breve{A}_{\ell m}$ is the field that satisfies a Schr\"{o}dinger-like differential equation\index{Schr\"{o}dinger-like differential equation}. Indeed, it follows immediately from the relation
\begin{equation}\label{SLDES1F}
\dfrac{d}{dx} E_{t}^{+} + i\omega E_{x}^{+} = \left[  \dfrac{d}{dx} + \omega^{2} - V_{s=1}(x) \right]\breve{A}_{\ell m}(x) = 0  \,,
\end{equation}
which is a consequence of the components $E_{t}^{+} = 0$ and $E_{t}^{+} = 0$ of Maxwell's field
equation, that $\breve{A}_{\ell m}$ obeys the same Schr\"{o}dinger equation as $A^{-}_{l,\ell m}$ with the same effective pontential and, therefore, has the same spectrum, namely Eq. \eqref{QNFS1F}. Now, assuming that $\breve{A}_{\ell m}$ is a solution of Eq. \eqref{SLDES1F}, the identities $E_{t}^{+} = 0$ and $E_{t}^{+} = 0$ lead to the fact that $A^{+}_{0, \ell m}$ and $A^{+}_{1, \ell m}$ are related to $\breve{A}_{\ell m}$ by the following equation:
\begin{align}
A_{0, \ell m}^{+} &=\,\, \left( \sum_{l=2}^{d}\dfrac{\ell_l(\ell_l +1)}{R_{l}^{2}} \right)^{-1} \dfrac{d}{dx} \, \breve{A}_{\ell m} \,,\\
A_{1, \ell m}^{+} &=\,\,  -\left( \sum_{l=2}^{d}\dfrac{\ell_l(\ell_l +1)}{R_{l}^{2}} \right)^{-1} i\omega  \breve{A}_{\ell m} \,.
\end{align}
Thus, the components $A_{0, \ell m}^{+}$ and $A_{1, \ell m}^{+}$ of the gauge field must have the same spectrum of $\breve{A}_{\ell m}$. Indeed, using that $\breve{A}_{\ell m}$ satisfies \eqref{SLDES1F}, we can easily check that $A_{0, \ell m}^{+}$ and $A_{1, \ell m}^{+}$ obey \eqref{SLDES1F}, but with a source, namely,
\begin{equation}
\left[  \dfrac{d}{dx} + \omega^{2} - V_{s=1}(x) \right] A_{i, \ell m}^{+}(x) \,\propto\, \delta_{i0}  \,\breve{A}_{\ell m}(t, x) \dfrac{d}{dx} V_{s=1}(x)\quad (i = 0,1) \,,
\end{equation}
where $\delta_{i0}$ is the Kronecker delta. The general solution for a linear differential equation with a source is given by the general solution for the homogeneous part of the equation plus a particular solution that depends linearly on the source. Now, once the $d V_{s=1}/dx$ goes to zero at the boundaries $x = \pm \infty$, it follows that near these boundaries $A_{i, \ell m}^{+}$  satisfies the same Schr\"{o}dinger-like differential equation\index{Schr\"{o}dinger-like differential equation} as $\breve{A}_{\ell m}$ and, therefore, yield the same spectrum. Summing up, we have obtained that all the degrees of freedom of the Maxwell field\index{Maxwell field} have the same spectrum, given by Eq. \eqref{QNFS1F}.

%%%%%%%%%%%%%%%%%%%%%%%%%%%%%%%%%%%%%%%%%%%%%%%%%%%%%%%%%%%%%%%%%%%%%%%%%%%%%%%%%%%
%%%%%%%%%%%%%%%%%%%%%%%%%%% E: Chapter 2. Field Perturbations: Spins 0 and 1
%%%%%%%%%%%%%%%%%%%%%%%%%%%%%%%%%%%%%%%%%%%%%%%%%%%%%%%%%%%%%%%%%%%%%%%%%%%%%%%%%%%

%%%%%%%%%%%%%%%%%%%%%%%%%%%%%%%%%%%%%%%%%%%%%%%%%%%%%%%%%%%%%%%%%%%%%%%%%%%%%%%%%%%
%%%%%%%%%%%%%%%%%%%%%%%%%%% B: Chapter 3. Spin-2 Field Perturbations
%%%%%%%%%%%%%%%%%%%%%%%%%%%%%%%%%%%%%%%%%%%%%%%%%%%%%%%%%%%%%%%%%%%%%%%%%%%%%%%%%%%

%\chapter{\uppercase{Spin-$2$ Field Perturbation}}\label{S2FP}

\chapter{Spin-$2$ Field Perturbation}\label{S2FP}

It is well known that linear differential equations are widely used for describing a lot
of phenomena related to perturbation propagation of different kind of information in different branches of the sciences. In physics, more precisely in General Relativity, the great triumph of the perturbation theory involves its application in gravitational waves, which are probably one of the most relevant predictions of Einstein's General Relativity theory. This theme acquired even greater importance after the recent measurement of gravitational radiation. This makes spin-$2$ field perturbations one of the most important perturbation types among several types of field perturbations. In particular,  the detection of their quasinormal modes in gravitational wave experiments allows precise measurements of the charges of the gravitational background, such as the mass and spin. In this chapter, we analytically obtain the quasinormal spectrum for the spin-$2$ field perturbations in generalized Nariai spacetime. A key step in order to attain this result is to use a suitable basis for the angular functions depending on the rank of the tensorial degree of freedom that one needs to describe. Here we define such a basis, which is a generalization of the tensor spherical harmonics\index{Tensor spherical harmonics} that is suited for spaces that are the product of several spaces of constant curvature.

\vspace{.5cm}
%\section{\uppercase{Field Equation for the Spin-$2$ Perturbation}}\label{Sec.Problem}
\section{Field Equation for the Spin-$2$ Perturbation}\label{Sec.Problem}
\vspace{.5cm}

In this chapter we shall consider the perturbations on the gravitational field\index{Gravitational field}, a massless
spin-$2$ field, in the generalized Nariai Background, $g_{\mu\nu}^{GN}$. Here we shall assume that the electromagnetic charges of the background are zero, namely $Q_1 = Q_l = 0$, so that we have a vanishing Maxwell field\index{Maxwell field} in the background, $\bl{\mathcal{A}}^{GN}=0$ and hence the gravitational perturbation decouples from the electromagnetic perturbation. In this case, the field equations reduce to Einstein's vacuum equation with a cosmological constant $\Lambda$
\begin{equation}\label{FES2}
  \mathcal{R}_{\mu\nu} = \Lambda \,g_{\mu\nu}   \,.
\end{equation}
It is worth pointing out that while in previous chapters Einstein's vacuum equation was not assumed to
hold, so that the spheres of the generalized Nariai background could have different radii, depending on
the electromagnetic charges of the background, here we have assumed vanishing charges, so that the gravitational perturbation decouples from the electromagnetic perturbation. Otherwise, we would have to consider the gravitational and electromagnetic perturbations simultaneously, since the electromagnetic perturbation field would be a source for the gravitational perturbation. 

%Gravitational perturbations can be written in the linear approximation in the form
Let us perform a small perturbation $h_{\mu\nu}$  in $g_{\mu\nu}^{GN}$ such that the perturbed metric can be taken as the sum of unperturbed background metric and the perturbation,
\begin{equation}\label{PerturbPhi}
g_{\mu\nu}\,=\,g^{GN}_{\mu\nu}\,+\,h_{\mu\nu} \,,
\end{equation}
where $h_{\mu\nu}$ is assumed to be ``small''. By small we mean that plugging the above equation into Eq. \eqref{FES2},  the terms of order $\sim h_{\mu\nu}^{2}$ and higher can be neglected in the first order approximation. Linearizing then Einstein's vacuum equation around $g^{GN}_{\mu\nu}$ we end up with the following equation for $h_{\mu\nu}$:
\begin{align}
\delta\left(  \mathcal{R}_{\mu\nu} - \Lambda \,g_{\mu\nu}\right) = 0 \quad\Rightarrow\quad  2\nabla^{\sigma}\nabla_{(\mu}h_{\nu)\sigma}\,-\,\Box h_{\mu\nu} \,-\, \nabla_{\mu}\nabla_{\nu}\,h - \,2\Lambda\,h_{\mu\nu}   &= 0\,, \label{Eqh}
\end{align}
where $\nabla_\mu$ is the Levi-Civita covariant derivative with respect to unperturbed background $g^{GN}_{\mu\nu}$, and $\Box = \nabla^\mu\nabla_\mu$, with the indices being raised using the inverse background metric, $g^{GN\,\mu\nu}$. The background spacetime considered here is the direct product of the de Sitter space $dS_{2}$ with $(d-1)$ spheres $S^{2}$ possessing three independent Killing vectors $\mathbf{K}_{p,l}\,(p=1,2,3)$ that generate rotations, namely Eq. \eqref{KilllingV}. In particular, this means that
\begin{equation}
 \pounds_{\mathbf{K}_{p,l}} \, g_{\mu\nu}^{GN} =0 \,,
 \end{equation}
where the operator $\pounds_{\mathbf{K}_{p,l}}$\label{LD} is the Lie derivative along $\mathbf{K}_{p,l}$. Now, since the Levi-Civita covariant derivative depends only on $g_{\mu\nu}^{GN}$ it follows that $\pounds_{\mathbf{K}_{p,l}}\nabla_\mu = \nabla_\mu \pounds_{\mathbf{K}_{p,l}}$. 
%once the $\mathbf{K}_{I,\,l}$ are Killing vector fields of the background metric. 
Hence, the operator that acts on $h_{\mu\nu}$ in Eq. (\ref{Eqh}) commutes with $\pounds_{\mathbf{K}_{p,l}}$. 
%Indeed, since $\mathbf{K}_{I,\,l}$ are Killing vector fields of the background metric, it follows that the action of $\mathcal{L}_{\mathbf{K}_{I,\,l}}$ on $g_{\mu\nu}^{GN}$ yields zero. Since the Levi-Civita covariant derivative depends only on the background metric it follows that $\mathcal{L}_{\mathbf{K}_{I,j}}\nabla_\mu = \nabla_\mu \mathcal{L}_{\mathbf{K}_{I,j}}$. 
Thus, since $\pounds_{\mathbf{K}_{p,l}}$ generates infinitesimal rotations in the $l$th sphere, it turns out that if $h_{\mu\nu}$ is a solution of Eq. (\ref{Eqh}) its rotated version will also be a solution. This humble assertion has an important practical consequence, namely when we expand $h_{\mu\nu}$ in terms of irreducible representations of $SO(3)$ we just need to consider the elements of the representation basis with $m_l=0$, where $m_l$ is the eigenvalue with respect to $\mathbf{K}_{3,\,l}$. The other possible values for $m_l$ can be attained by applying the ladder operators, which are just linear combinations of rotations generated by $\mathbf{K}_{1,l}$ and $\mathbf{K}_{2,\,l}$. This leads to great simplification in the calculations. We shall return to this point when we introduce the basis used to expand the components of $h_{\mu\nu}$.

When the electromagnetic charges of the background vanish, the radii $R_1$ and $R_l$ are all equal in such a case and given by
\begin{equation}
R_1 = R_l = \Lambda^{-1/2} \,.
\end{equation}
In particular, the surfaces $r = \pm R_{1}$ are event horizons\index{Event horizon} in which the boundary conditions\index{Boundary conditions} of the quasinormal modes will be posed, as discussed in the section \ref{Sec.Boundary Cond}, see also \cite{Venancio2020}.

%It is worth noting that the Killing vectors $\mathbf{K}_t = \partial_t$ is light-like at Killing horizons $r = \pm \Lambda^{-1/2}$. The boundary conditions of the quasinormal modes will be posed at these surfaces, as discussed in the section \ref{Sec.Boundary Cond}, see also \cite{Venancio2020}, and the domain of interest will be $r\in ( -\Lambda^{-1/2}, \Lambda^{-1/2})$. 
%In such domain one can use the tortoise coordinate  $x$ defined by $r = \Lambda^{-1/2} \tanh(x$%%\Lambda^{1/2})$, in terms of which the line element becomes
%\begin{equation}\label{nariai-metric2}
%g_{\mu\nu}dx^{\mu}dx^{\nu} = f\,\left( -\,dt^{2} + dx^{2}\right) + \frac{1}{\Lambda}\sum_{j=2}^{d}\,d%\Omega_{j}^{2} \,,%
%\end{equation}
%where
%\begin{equation*}
% f =  f(x) = \textrm{sech}^2(x\sqrt{\Lambda}) \,.
%\end{equation*}

\vspace{.5cm}
%\section{\uppercase{Anzatz for the Separation of the Linearized Einstein Field Equation}}
\section{Anzatz for the Separation of the Linearized Einstein Field Equation}
\vspace{.5cm}

As we have proved in example 3, the linearized Einstein field equation for a spin-$2$ field perturbation $\boldsymbol{h} = h_{\mu\nu} dx^{\mu}dx^{\nu}$ in a four-dimensional background with spherical symmetry is separable by the decomposition
%\begin{equation}
%\boldsymbol{h} = \boldsymbol{h}^{+} + \boldsymbol{h}^{-} \,,
%\end{equation}
%where $\boldsymbol{h}^{\pm}$ parts are given by
\begin{align}\label{DFH}
\boldsymbol{h} &=\,\, \sum_{\ell_l. m_l} e^{i\omega t}\left[ \left(H_{tt}\, dt^2 + H_{x x} \,dr^2 + 2H_{tx} \, dtdx \right) Y_{\ell_l,m_l} + \left( H^{+}_{t} dt  +  H^{+}_{x} dx \right) V^{+}_{\ell_l,m_l} \right.\nonumber\\
\\
&+\,\,\left.\left(H^{-}_{t}dt  +  H^{-}_{x}dx \right)\,V^{-}_{\ell_l,m_l}  +  H^{-} \,T^{-}_{\ell_l,m_l}  +H^{\oplus}\,T^{\oplus}_{\ell_l,m_l} + H^{+}\,T^{+}_{\ell_l,m_l} \right]\,,\nonumber
%\\
%\boldsymbol{h}^{-} &=\,\,e^{-i\omega t}\left[ \left(H^{-}_{t}dt  +  H^{-}_{x}dx \right)\,V^{-}%_{\ell_l,m_l}  +  H^{-} \,T^{-}_{\ell_l,m_l} \right] \,,\nonumber
\end{align}
where the basis $\{Y_{\ell_l,m_l},V^{\pm}_{\ell_l,m_l},T^{\oplus}_{\ell_l,m_l},T^{\pm}_{\ell_l,m_l}\}$ has been defined in chapter \ref{ChapPTP}. For instance, the objects $V^{\pm}_{\ell_l,m_l}$ stand for $V^{\pm}_{a_l}dx^{a_l}$ where
\begin{equation}\label{V1}
V_{a_l}^{+} = \hat{\nabla}_{a_l}Y_{\ell_l,m_l} \quad \text{and} \quad V_{a_l}^{-} = \epsilon_{a_lc_l}\hat{\nabla}^{c_l}Y_{\ell_l,m_l} \,,
\end{equation}
and the objects $T^{\oplus,\pm}_{\ell_l,m_l}$ stands for $T^{\oplus,\pm}_{a_lb_l}dx^{a_l}dx^{b_l}$ where
\begin{align}
  T^\oplus_{a_lb_l} &= Y_{\ell_l,m_l}\, \hat{g}_{a_lb_l} \,,\nonumber \\
  T^+_{a_lb_l} &= \hat{\nabla}_{a_l} \hat{\nabla}_{b_l} Y_{\ell_l,m_l} \,,\label{T2} \\
  T^-_{a_lb_l} &= \hat{\epsilon}_{a_lc_l}  \hat{\nabla}_{b_l} \hat{\nabla}^{c_l} Y_{\ell_l,m_l} +
\hat{\epsilon}_{b_lc_l}    \hat{\nabla}_{a_l} \hat{\nabla}^{c_l} Y_{\ell_l,m_l}\,.\nonumber
\end{align}
The ten $H$'s are functions only of $x$ and they account for the ten degrees of freedom associated to $h_{\mu\nu}$ in four dimensions. $V^{+}_{\ell_l,m_l}, T^\oplus_{\ell_l,m_l}$ and $T^+_{\ell_l,m_l}$ have even parity, namely transform in the same way as the scalar $Y_{\ell_l}^{m_l}$ under a parity transformation\index{Parity transformation}, given by Eq. \eqref{parityT}, while $V_{\ell_l,m_l}^{-}, T^-_{\ell_l,m_l}$ have odd parity. With this ansatz for $h_{\mu\nu}$, it is much easier to integrate Eq. (\ref{Eqh}) than using just the scalar spherical harmonics\index{Scalar spherical harmonics} to expand the angular part of the field.

With Eq. \eqref{DFH} at hand, we are ready to expand, in a natural way, the gravitational perturbation $\mathbf{h} = h_{\mu\nu}dx^\mu dx^\nu$ in the generalized Nariai spacetime for arbitrary $d$. 
Indeed, let us decompose the perturbation $\mathbf{h} = h_{\mu\nu}dx^{\mu}dx^{\nu}$ as 
\begin{align}\label{GES2P}
\mathbf{h} &=\,\, h_{tt}\,dt^{2} +  2 h_{tx}\,dt dx +  h_{xx}\,dx^{2} + \sum_{l=2}^{d} h_{ta_l}dx^{a_l} dt + h_{xa_l}dx^{a_l} dx \nonumber\\
 &+\,\, \sum_{l=2}^{d} h_{a_lb_l}\,dx^{a_l}dx^{b_l} + \sum_{l=2}^{d}\sum_{n>l}^{d} h_{a_lb_n}\,dx^{a_l}dx^{b_n} \quad \text{where} \quad a_l,b_l \in \{\theta_l, \phi_l \} \,.
\end{align}

\begin{align}
\mathbf{h} &=\,\, h_{tt}\,dt^{2} +  2 h_{tx}\,dt dx +  h_{xx}\,dx^{2} \nonumber\\
&+\,\, \sum_{l=2}^{d} h_{ta_l}dx^{a_l} dt + h_{xa_l}dx^{a_l} dx \nonumber\\
 &+\,\, \sum_{l=2}^{d} h_{a_lb_l}\,dx^{a_l}dx^{b_l} + \sum_{l=2}^{d}\sum_{n>l}^{d} h_{a_lb_n}\,dx^{a_l}dx^{b_n} \,.
\end{align}

Given the spherical symmetry of the background in each of the spheres, we now decompose and classify the gravitational perturbations according to the $SO(3)$ isometry subgroup associated
to each spherical part of the line element. The components $h_{tt}$,  $h_{xx}$, and $h_{tx}$ are scalars with respect to the $(d-1)$ spheres and, therefore, their angular dependence should be given by the product of scalar spherical harmonics $\mathcal{Y}_{\ell m}$ just as we did for the scalar field\index{Scalar field}, namely Eq. \eqref{ExpansionSacalar}
\begin{align}\label{ASFS2P}
h_{tt} &=\, \sum_{\ell, m} e^{-i\omega t} H^{+}_{tt} \,\mathcal{Y}_{\ell m} \,,\nonumber\\
h_{tx} &=\, \sum_{\ell, m} e^{-i\omega t} H^{+}_{tx} \,\mathcal{Y}_{\ell m} \,,\\
h_{xx} &=\, \sum_{\ell, m} e^{-i\omega t} H^{+}_{xx} \,\mathcal{Y}_{\ell m} \,,\nonumber
\end{align}
where $H^{+}_{tt}, H^{+}_{tx}$ and $H^{+}_{xx}$ are generic functions of the tortoise coordinate\index{Tortoise coordinate} $x$, namely Eq. \eqref{TC}. Strictly speaking, all $2d(2d+1)/2$ functions $H$'s  which account for the $2d(2d+1)/2$ degrees of freedom associated to $h_{\mu\nu}$ in $D=2d$ dimensions depend on spherical harmonic indices $\{\ell, m\}$. From now on, such indices will be omitted in the fields for notational simplicity.
%\begin{equation}
%  \mathcal{Y}= Y_{\ell_2}^{m_2}(\theta_2,\phi_2)\, Y_{\ell_3}^{m_3}(\theta_3,\phi_3)\,\cdots \, Y_{\ell_d}^{m_d}(\theta_d,\phi_d)\,.
%\end{equation}
%When we perform a parity transformation in each of the spheres this scalar transforms as
%\begin{equation*}
%  \mathcal{Y} \xrightarrow{\textrm{parity}} (-1)^{\ell_2 + \cdots + \ell_d} \,\mathcal{Y} \,.
%\end{equation*}
%Objects that transform in the same way as $\mathcal{Y}$ under a parity transformation will be said to have even parity, while objects that gains an extra minus sign compared to $\mathcal{Y}$ will be said to have odd parity.

In their turn, $h_{ta_l}$ and $h_{xa_l}$ behave as the components of a $1$-form with respect to rotations in the $l$th sphere, but behave as scalars with respect to rotations in the other $(d-2)$ spheres. Thus, a suitable basis for the angular dependence would be $\{\mathcal{V}_{l,\ell,m}^{\pm}\}$ just as we did for the Maxwell field\index{Maxwell field}, namely Eq. \eqref{ABMF}
 \begin{align}\label{AMFS2P}
h_{ta_l}dx^{a_l} &=\, \sum_{\ell, m} e^{-i\omega t} \left( H^{+}_{tl}\, \mathcal{V}_{l,\ell m}^{+} + H^{-}_{tl} \,\mathcal{V}_{l,\ell m}^{-} \right) \,, \nonumber\\
\\
h_{xa_l}dx^{a_l} &=\, \sum_{\ell, m} e^{-i\omega t} \left( H^{+}_{xl} \,\mathcal{V}_{l,\ell m}^{+} + H^{-}_{xl} \,\mathcal{V}_{l,\ell m}^{-} \right) \,,\nonumber
\end{align}
where $ H^{\pm}_{tl,\ell m}$ and $H^{\pm}_{xl,\ell m}$ are generic functions of $x$.

In an analogous fashion, $h_{a_lb_l}$ behaves as a symmetric rank two tensor with respect to rotations in the $l$th sphere and as a scalar with respect to rotations in the $n$th sphere when $n\neq l$. Thus, a suitable basis for the angular dependence of this part is
\begin{equation}
  \mathcal{T}^{\oplus,\pm}_{a_l b_l} = T^{\oplus,\pm}_{a_lb_l}(\theta_l,\phi_l)  \prod_{n=2,n\neq l }^d\,  Y_{\ell_n}^{m_n}(\theta_n,\phi_n)\,,
\end{equation}
where $T^{\oplus}_{a_lb_l}$ and $T^{\pm}_{a_jb_j}$ have been defined in Eq. (\ref{T2}).
The corresponding tensors are, then, defined by
\begin{equation}
  \mathcal{T}^{\oplus,\pm}_{l, \ell m} = \mathcal{T}^{\oplus,\pm}_{\theta_l \theta_l} d\theta_l^2 \,+\,
 2\mathcal{T}^{\oplus,\pm}_{\theta_l \phi_l} d\theta_l d\phi_l
\,+\, \mathcal{T}^{\oplus,\pm}_{\phi_l \phi_l} d\phi_l^2 \,,
\end{equation}
where $\mathcal{T}^{\oplus}_{l, \ell m}$ and $\mathcal{T}^{+}_{l, \ell m}$ have even parity, while $\mathcal{T}^{-}_{l, \ell m}$ has odd parity. Any symmetric rank two tensor satisfying the above properties under rotation can be written as a linear combination of these tensors 
\begin{align}\label{ATFS2P}
h_{a_lb_l}dx^{a_l}dx^{b_l} =\, \sum_{\ell, m} e^{-i\omega t} \left(H^{\oplus}_{l}\,  \mathcal{T}^{\oplus}_{l, \ell m} +  H^{+}_{l}\,  \mathcal{T}^{+}_{l, \ell m} + H^{-}_{l}\,  \mathcal{T}^{-}_{l, \ell m}\right) \,,
\end{align}
with $H^{\oplus,\pm}_{l}$ being arbitrary functions of $x$.

A more tricky type of component is $h_{a_nb_l}$ with $n\neq l$, which behaves as the components of a $1$-form under rotations in the $n$th and $l$th
spheres, while it behaves as scalars with respect to rotations in the other spheres. We need a basis for the angular dependence that has this property. On top of that, we would like the basis elements to have a definite parity. A way to fulfill these constraints is defining
\begin{align}
   \mathcal{W}^{+}_{a_n b_l} &= V^{+}_{a_n}(\theta_n,\phi_n)  V^{+}_{b_l}(\theta_l,\phi_l)  \prod_{k\neq n,l}   Y_{\ell_k, m_k}(\theta_k,\phi_k) \,,\\
\mathcal{W}^{\oplus}_{a_n b_l} &= V^{-}_{a_n}(\theta_n,\phi_n)  V^{-}_{b_l}(\theta_l,\phi_l)
 \prod_{k\neq n,l }   Y_{\ell_k,m_k}(\theta_k,\phi_k) \,, \\
 \mathcal{W}^{-}_{a_n b_l} &= V^{+}_{a_n}(\theta_n,\phi_n)  V^{-}_{b_l}(\theta_l,\phi_l)
\prod_{k\neq n,l}   Y_{\ell_k,m_k}(\theta_k,\phi_k) \,,\\
\mathcal{W}^{\ominus}_{a_n b_l} &= V^{-}_{a_n}(\theta_n,\phi_n)  V^{+}_{b_l}(\theta_l,\phi_l)
\prod_{k\neq n,l}   Y_{\ell_k,m_k}(\theta_k,\phi_k) \,.
\end{align}
Using  these components we can define the symmetric rank two tensors $\mathcal{W}_{nl}^{+}$,  $\mathcal{W}_{nl}^{\oplus}$, $\mathcal{W}_{nl}^{-}$, and $\mathcal{W}_{nl}^{\ominus}$ in the natural way. For instance,
\begin{align}
  \mathcal{W}_{nl,\ell m}^{+} =\, & \mathcal{W}^{+}_{\theta_n \theta_l} d\theta_n d\theta_l \,+\,
  \mathcal{W}^{+}_{\theta_n \phi_l} d\theta_n d\phi_l \\
&\,+\, \mathcal{W}^{+}_{\phi_n \theta_l} d\phi_n d\theta_l \,+\, \mathcal{W}^{+}_{\phi_n \phi_l} d\phi_n d\phi_l   \,,
\end{align}
and analogously for the other three tensors, so that
\begin{align}\label{AMF2S2P}
h_{a_lb_n}dx^{a_l}dx^{b_n} =\, \sum_{\ell, m} e^{-i\omega t} \left( H^{+}_{ln}\,  \mathcal{W}^{+}_{ln,\ell m} + H^{\oplus}_{ln}\,  \mathcal{W}^{\oplus}_{ln,\ell m} + H^{-}_{ln}\,  \mathcal{W}^{-}_{ln,\ell m} + H^{\ominus} _{ln}\,  \mathcal{W}^{\ominus}_{ln,\ell m}\right) \,,
\end{align}
with $H^{\pm, \oplus}_{ln}$ being arbitrary functions of $x$. $\mathcal{W}^{+}_{ln,\ell m}$ and $\mathcal{W}^{\oplus}_{ln,\ell m}$ have positive parity, as they transform in the same way as $\mathcal{Y}_{\ell m}$ under a parity transformation\index{Parity transformation}, while $\mathcal{W}^{-}_{ln,\ell m}$ and $\mathcal{W}^{\ominus}_{ln, \ell m}$ have negative parity. Note that the odd parity modes come from the product of modes with opposite parities, whereas the positive parity modes arise from the product of elements with the same parity. The preceding steps used to find a suitable basis for the angular dependence should not be underestimated. Indeed, the perturbation equation for $h_{\mu\nu}$ is quite involved and can lead to an unbearable entanglement between the components of $h_{\mu\nu}$  if a natural basis is not adopted.

Thus, inserting the expansions \eqref{ASFS2P}, \eqref{AMFS2P}, \eqref{ATFS2P}  and \eqref{AMF2S2P} into Eq. \eqref{GES2P}, we conclude that a suitable way to expand the gravitational perturbation $\mathbf{h}=h_{\mu\nu}dx^\mu dx^\nu$ is as follows:
\begin{align}
  \mathbf{h}&= \sum_{\ell m} e^{-i\omega t}\Big[   (H_{tt}^{+} dt^2 + 2 H_{tx}^{+} dtdx + H_{xx}^{+} dx^2) \mathcal{Y}_{\ell m}   \\
& + \sum_{l=2}^d  \left( H_{tl}^+ \mathcal{V}_{l,\ell m}^{+} +  H_{tl}^- \mathcal{V}_{l,\ell m}^{-} \right) dt +
\left( H_{xl}^+ \mathcal{V}_{l,\ell m}^{+} +  H_{xl}^- \mathcal{V}_{l,\ell m}^{-} \right) dx\\
& + \sum_{l=2}^d  H_l^+ \,\mathcal{T}_{l,\ell m}^{+} +  H_l^\oplus \,\mathcal{T}_{l,\ell m}^{\oplus} + H_l^- \,\mathcal{T}_{l,\ell m}^{-}  \\
& + \sum_{l=2}^d \sum_{n>l}^d  H_{ln}^+  \mathcal{W}_{ln,\ell m}^{+} +  H_{ln}^\oplus  \mathcal{W}_{ln,\ell m}^{\oplus}
+ H_{ln}^-  \mathcal{W}_{ln,\ell m}^{-} + H_{ln}^\ominus  \mathcal{W}_{ln,\ell m}^{\ominus} \Big],
\end{align}
where the $H$'s are all functions of the coordinate $x$. Counting the number of independent functions, we have three coming from the first line of the right hand side of the previous equation, namely from $H_{tt}^{+}$,  $H_{tx}^{+}$, and $H_{xx}^{+}$; in the second line there are $(d-1)$ functions  $H_{tl}^+$ and, analogously, more $3(d-1)$ components stemming from $H_{tl}^-$, $H_{xl}^+$, and $H_{xl}^-$; in the third line we have $3(d-1)$ independent functions; finally, in the fourth line we should recall that $n>l$, so that there are $4\frac{(d-1)(d-2)}{2}$ functions. Summing the number of these functions, we have:
\begin{equation}
  3 + 4(d-1) + 3(d-1) + 2(d-1)(d-2) = \frac{2d(2d+1)}{2},
\end{equation}
which is exactly the number of independent components of $h_{\mu\nu}$ in $2d$ dimensions, as it should be. This proves that no possible degree of freedom of the perturbation field is being neglected.

Once we have made an appropriate expansion for $h_{\mu\nu}$, we are ready to start the integration process of Eq. (\ref{Eqh}). In order to do so, we can take advantage of the spherical symmetries and only consider the cases $m_l = 0$, for all $l$, so that no $\phi_l$ dependence will show up. Thus, any derivative of the type $\partial_{\phi_l}$ will not contribute, including those appearing in the definition of our basis. For instance, $\mathcal{V}_{\phi_l}^{+}$ and $\mathcal{V}_{\theta_l}^{-}$ are automatically zero in such a case. As explained before, this will represent no important loss of generality, since the other solutions can be generated by applying rotations to the ones with $m_l=0$. Moreover, in this work we are only interested in the frequencies of the quasinormal modes, which are invariant under the rotations in the spheres, so that we do not even need to bother on generating solutions with nonzero values of $m_l$.

\vspace{.5cm}
%\subsection{Gauge Transformation}
\subsection{Gauge Transformation}
\vspace{.5cm}

While working with QNMs, a source of simplification in the calculations performed arise from the gauge freedom in choosing the elements $h_{\mu\nu}$ coming from freedom in choice of the coordinate system. Indeed, if we perform the change in the coordinates
\begin{equation}\label{CoordTransf}
  x^\mu \,\mapsto\, \tilde{x}^\mu = x^\mu +  \zeta^\mu \,,
\end{equation}
where $\zeta^\mu = \zeta^\mu(x)$ is infinitesimal, it follows that the components of the metric in the new coordinate system are given by
\begin{equation*}
  g_{\mu\nu} \,\mapsto\,  \tilde{g}_{\mu\nu} = g_{\mu\nu} + \nabla_\mu \zeta_\nu +  \nabla_\nu \zeta_\mu \,.
\end{equation*}
Thus, performing the perturbation (\ref{PerturbPhi}) in the metric  followed by the infinitesimal coordinate transformation (\ref{CoordTransf}) is equivalent, to first order in the infinitesimal parameters, to performing just a metric perturbation with the perturbation field being
\begin{equation}
  \tilde{h}_{\mu\nu} = h_{\mu\nu} + \nabla_\mu \zeta_\nu +  \nabla_\nu \zeta_\mu \,.
\end{equation}
Since physics is insensitive to coordinate transformations, it follows that the transformation
\begin{equation}\label{GaugetTransf}
  h_{\mu\nu}\,\mapsto\,  h_{\mu\nu} + \nabla_\mu \zeta_\nu +  \nabla_\nu \zeta_\mu \,.
\end{equation}
is just a gauge transformation\index{Gauge transformation}, namely it does not lead to changes in the physical results. In particular, these transformations do not change the quasinormal spectrum of the gravitational perturbation. In what follows we will perform a wise choice for the vector field $\zeta^\mu$ in order to eliminate some degrees of freedom of the perturbation field.

\vspace{.5cm}
%\section{\uppercase{Gravitational Quasinormal Modes}}
\section{Gravitational Quasinormal Modes}
\vspace{.5cm}

When a parity transformation\index{Parity transformation} \eqref{parityT} is applied in each of the spheres, we can split $\mathbf{h}$ into a sum of two distinct classes of perturbation under this latter transformation as follows:
\begin{equation}
 \mathbf{h} =  \mathbf{h}^{+} +  \mathbf{h}^{-} \,,
\end{equation}
where the $ \mathbf{h}^{\pm}$ parts given by
\begin{align}
  \mathbf{h}^{+} &=\, \sum_{\ell m}e^{-i\omega t}\Big[   (H_{tt}^+ dt^2 + 2 H_{tx}^+ dtdx + H_{xx}^+ dx^2) \mathcal{Y}_{\ell m}   \nonumber \\
& + \sum_{l=2}^d  \left( H_{tl}^+ \mathcal{V}_{l,\ell m}^{+} dt + H_{xl}^+ \mathcal{V}_{l,\ell m}^{+} dx + H_l^+ \,\mathcal{T}_{l,\ell m}^{+} +  H_l^\oplus \,\mathcal{T}_{l,\ell m}^{\oplus} \right) \nonumber \\
& + \sum_{l=2}^d \sum_{n>l}^d  H_{ln}^+ \,\mathcal{W}_{ln,\ell m }^{+} +  H_{ln}^\oplus \,\mathcal{W}_{ln,\ell m}^{\oplus}\Big] \,,\nonumber\\
\\
\mathbf{h}^{-} &=\,  \sum_{\ell m} \sum_{l=2}^d e^{-i\omega t} \left[ \left(  H_{tl}^- \,dt + H_{xl}^- \,dx \right) \mathcal{V}_{l,\ell m}^{-} \,+\, H_l^- \,\mathcal{T}_{l,\ell m}^{-}  \right. \nonumber\\
&+\, \sum_{n>l}^d \left.\left( H_{ln}^- \,\mathcal{W}_{ln,\ell m}^{-} + H_{ln}^\ominus \,\mathcal{W}_{ln,\ell m}^{\ominus} \right) \right]\,,
\end{align}
transform as $\mathbf{h}^{\pm} \overset{\text{parity}}{\longrightarrow }\, \pm (-1)^{\ell_2 + \ell_3 + \ldots + \ell_d} \,\mathbf{h}^{\pm}$. In particular, this means that $\mathbf{h}^{+}$ is an even perturbation\index{Even perturbation} with parity $(-1)^{\ell_2 + \ell_3 + \ldots + \ell_d} $, while $\mathbf{h}^{-}$ is an odd perturbation\index{Odd perturbation} with parity $(-1)^{\ell_2 + \ell_3 + \ldots + \ell_d \,+\,1} $. We can take advantage of this fact inasmuch as the field equation for $h_{\mu\nu}$ do not mix components with opposite parities. Thus, in what follows we will separate the integration of the perturbation equation in the odd degrees of freedom, which will be tackled in the next section, and the even degrees of freedom, which will be considered in section \ref{Sec.Even}.

\vspace{.5cm}
%\subsection{Odd Perturbations}\label{Sec.Odd}
\subsection{Odd Perturbations}\label{Sec.Odd}
\vspace{.5cm}

By an odd perturbation\index{Odd perturbation} we mean the most general perturbation for a given set of
spherical harmonic indices $\{\ell_2,m_2,\ell_3,m_3,\ldots,\ell_d,m_d\}$ and parity $(-1)^{\ell_2 + \ell_3 + \ldots + \ell_d \,+\,1}$, namely
\begin{eqnarray}\label{hodd}
\mathbf{h}^{-} &=&\, \sum_{\ell m} \sum_{l=2}^d e^{-i\omega t} \left[ \left(  H_{tl}^- \,dt + H_{xl}^- \,dx \right) \mathcal{V}_{l,\ell m}^{-} \,+\, H_l^- \,\mathcal{T}_{l,\ell m}^{-}  \right. \nonumber\\
 &+&\,\sum_{n>l}^d  \left.\left( H_{ln}^- \,\mathcal{W}_{ln,\ell m}^{-} + H_{ln}^\ominus \,\mathcal{W}_{ln,\ell m}^{\ominus} \right) \right]\,.\nonumber
\end{eqnarray}
However, one can eliminate some degrees of freedom by means of a gauge transformation\index{Gauge transformation}. Indeed, performing the transformation (\ref{GaugetTransf}) with $\zeta_\mu $ given by
\begin{equation}
  \zeta_\mu dx^\mu = - e^{-i\omega t}\sum_{l=2}^d H_l^-\,\mathcal{V}_{l,\ell m}^{-} \,,
\end{equation}
it follows that the transformed field $\tilde{h}_{\mu\nu}$ is such that it has the same form as depicted in the expansion \eqref{hodd} but with the fields $H^-(x)$ transformed to $\tilde{H}^-(x)$ where
\begin{equation}
\left\{
  \begin{array}{ll}
     \tilde{H}_{tl}^- =  H_{tl}^- + i\omega H_l^- \,, \\
    \tilde{H}_{xl}^- =  H_{xl}^- - \frac{d}{dx} H_l^- \,,\\
    \tilde{H}_{l}^- = 0 \,,\\
     \tilde{H}_{ln}^- =  H_{ln}^- -  H_n^- \,,\\
     \tilde{H}_{ln}^\ominus =  H_{ln}^\ominus -   H_l^- \,.
  \end{array}
\right.
\end{equation}
Thus, we see that the components  $H_{l}^-$ of the ansatz \eqref{hodd} can be eliminated by a gauge transformation\index{Gauge transformation}, while the other components just get redefined. Thus, in what follows we can ignore the degrees of freedom $\tilde{H}_{l}^-$ and consider that the gravitational perturbation is given by
\begin{eqnarray}\label{hodd2}
  \mathbf{h}^{-} &=& \sum_{\ell m} \sum_{l=2}^d e^{-i\omega t} \left[ \left(  H_{tl}^- \,dt + H_{xl}^- \,dx \right) \mathcal{V}_{l,\ell m}^{-} \right. \nonumber\\
&+&\sum_{n>l}^d  \left.\left( H_{ln}^- \,\mathcal{W}_{ln,\ell m}^{-} + H_{ln}^\ominus \,\mathcal{W}_{ln,\ell m}^{\ominus} \right) \right]\,.\nonumber
\end{eqnarray}

Now, inserting this perturbation into the field equation (\ref{Eqh}), we are eventually led to the following equations:
\begin{equation}\label{Emenos}
  \left.
     \begin{array}{ll}
      E^-_{t\phi_l}\,\equiv & \, \dfrac{d}{dx}\left[ \dfrac{1}{f}\left( \dfrac{d}{dx} H^-_{tj} + i \omega  H^-_{xl} \right) \right] -
\Lambda  \,(\kappa -2  ) \,H^-_{tl} - i\omega \Lambda \sum_{n\neq l} \kappa_n  ( H_{ln}^\ominus + H_{nl}^-)  = 0 \,,  \\
\\
E^-_{x\phi_j} \,\equiv & \,  \dfrac{i\omega}{f}\left( \dfrac{d}{dx} H^-_{tl} + i \omega  H^-_{xl} \right) -
\Lambda \, (\kappa - 2) \,H^-_{xl} +   \Lambda \sum_{n\neq j} \kappa_n \dfrac{d}{dx}\left( H_{ln}^\ominus + H_{nl}^-\right)  = 0  \,, \\
\\
E^-_{\theta_l\phi_l} \,\equiv & \,  \dfrac{1}{f}\left( \dfrac{d}{dx} H^-_{xl} + i \omega  H^-_{tl} \right) -
  \Lambda \sum_{n\neq l} \kappa_n \left(  H_{ln}^\ominus + H_{nl}^-\right)  = 0  \,,  \\
\\
  E^-_{\theta_l\phi_n} \,\equiv & \, \dfrac{d^2}{dx^2}\left(  H_{nl}^\ominus + H_{ln}^-\right) +
  \left[\omega^2 - f\,\Lambda\, (\kappa - 2) \right]\left(  H_{nl}^\ominus + H_{ln}^-\right)
  -f\,E^-_{\theta_n\phi_n} = 0  \,.
     \end{array}
   \right.
\end{equation}
In the left hand side of these equations, the objects $E^-_{\mu\nu}$ are just to stress that the equation $E^-_{\mu\nu}=0$ comes from imposing the component $\mu\nu$ of Eq. (\ref{Eqh}) to hold. The components that do not appear above, like $E^-_{tt}$ are identically vanishing. In the last line of Eq. (\ref{Emenos}) it is being assumed that $n\neq l$.  Above, we have also used the definitions
\begin{equation}
 \kappa_l = \ell_l(\ell_l + 1)  \quad \text{and} \quad  \kappa =\sum_{l=2}^d\,\kappa_l\,.
\end{equation}
Thus, the above equations comprise all the restrictions associated to the odd perturbation\index{Odd perturbation} equation obeyed by $h_{\mu\nu}$.

In order to attain Eq. (\ref{Emenos}), we have assumed that the spherical harmonics\index{Scalar spherical harmonics} $Y_{\ell_l,m_l}(\theta_l,\phi_l)$ have $m_l=0$, which is justified by the spherical symmetry, as explained before. So, we have used $Y_{\ell_l,m_l} = Y_{\ell_l}(\theta_l)$ where $Y_{\ell_l}(\theta_l)$ obeys the following differential equation:
\begin{equation}
  \dfrac{1}{\sin\theta_l}\dfrac{d}{d\theta_l}\left( \sin\theta_l\,\dfrac{d}{d\theta_l} Y_{\ell_l} \right) + \kappa_l Y_{\ell_l} = 0\,.
\end{equation}

In the equations displayed in (\ref{Emenos}), the fields $H_{ln}^\ominus$ and $H_{ln}^-$ appear only by means of the combination $(H_{nl}^\ominus + H_{ln}^-)$. Note, however, that we are always assuming that $n\neq l$, so that either $n>l$ or $n<l$. These fields were defined through Eq. (\ref{hodd}), where it is always assumed that the second index is greater than the first. Thus, the fields  $H_{ln}^\ominus$ and $H_{ln}^-$ with $l>n$ are not defined. Hence, the convention in Eq. (\ref{Emenos}) is that these undefined fields are zero. So, what might appear as two fields in the sum $(H_{ln}^\ominus + H_{nl}^-)$ is, actually, just one field. Indeed, if $n>l$ it follows that $H_{nl}^-$ vanishes, so that $(H_{ln}^\ominus + H_{nl}^-) = H_{ln}^\ominus$, while if $l>n$ we have $(H_{ln}^\ominus + H_{nl}^-) = H_{nl}^-$. Summing up, in Eq. (\ref{Emenos}) we have
\begin{equation}
  (H_{ln}^\ominus + H_{nl}^-) \,=\, \left\{
                                      \begin{array}{ll}
                                        H_{nl}^- \;,\;\; \textrm{if } l> n\\
                                        \;\\
                                        H_{ln}^\ominus \;,\;\; \textrm{if } n>l
                                      \end{array}
                                    \right.\,.
\end{equation}

Assuming that $E^-_{\theta_l\phi_l}$ vanishes, in accordance with the third equation in (\ref{Emenos}), it follows from the last line in (\ref{Emenos}) that the fields $H_{ln}^-$ and $ H_{nl}^\ominus$ both obey a Schr\"{o}dinger-like differential equation\index{Schr\"{o}dinger-like differential equation} where the effective potential is the well-known P\"{o}schl-Teller potential\index{P\"{o}schl-Teller potential}, namely is the one studied in the previous chapter, see \eqref{Potential_Generic}, with the parameters $\mathfrak{a}, \mathfrak{b}, \mathfrak{c}$, and $\mathfrak{d}$ given by:
\begin{equation}\label{PoschlTeller}
  \mathfrak{a} = 0 \quad , \quad \mathfrak{b} = 0 \quad , \quad \mathfrak{c} = \Lambda (\kappa -2) \quad , \quad \mathfrak{d} = \sqrt{\Lambda}\,.
\end{equation}
%\begin{equation}\label{PoschlTeller}
%  \dfrac{d^2}{dx^2}H + \left[ \omega^2 - \frac{\Lambda\, (\kappa - 2) }{\cosh^2(\sqrt{\Lambda}\,x )}%\right] H = 0 \,.%
%\end{equation}
Such a Schr\"{o}dinger-like differential equation\index{Schr\"{o}dinger-like differential equation} with the above potential is the well-known P\"{o}schl-Teller\index{P\"{o}schl-Teller equation} equation that can be integrated analytically \cite{Poschl1933}. In order to find the spectrum of frequencies of these components, we need to aplly the appropriate boundary conditions\index{Boundary conditions}. However, the above equation is the same equation obeyed by the scalar field\index{Scalar field} mode $\phi_{\ell m}^{\omega}$ when the scalar field has vanishing mass ($\mu=0$) and when the constraint $R_1 = R_l = \Lambda^{-1/2}$ holds. Thus, the quasinormal spectrum associated to this component of the gravitational field\index{Gravitational field} must be the same as the massless scalar field one. In particular, assuming that the boundary condition for the perturbation field is as depicted in Fig. \ref{FigBoundCond}, it follows that the boundary conditions\index{Boundary conditions} (II) and (III) lead to no QNMs, while for the boundary conditions (I) and (IV) the spectrum of allowed
frequencies is
\begin{align}\label{QNFS2F}
  \omega_{\text{I}} &=\,\, \sqrt{\Lambda}\left[\, \sqrt{\kappa -\dfrac{9}{4} } \,-\, i \,(n+\dfrac{1}{2}) \,\right]\,,\nonumber\\
  \\ 
   \omega_{\text{IV}} &=\,\,\sqrt{\Lambda}\left[\, \sqrt{\kappa -\dfrac{9}{4} } \,+\, i \,(n+\dfrac{1}{2}) \,\right] \,,\nonumber
\end{align}
%This is the well-known P\"{o}schl-Teller equation, that can be integrated analytically. In particular, assuming that the boundary condition for the perturbation field is as depicted in Fig. \ref{FigCones1}, which is the appropriate boundary condition for quasinormal modes, it follows that the spectrum of allowed frequencies is
%\begin{equation}\label{Spectrum}
%  \omega = \sqrt{\Lambda}\left[\, \sqrt{\kappa -9/4 } \,+\, i \,(n+1/2) \,\right] \,,
%\end{equation}
where $n\in\{0,1,2,\cdots \}$. For more details on the calculation of the spectrum and on the choice of boundary condition, the reader is referred to Refs. \cite{Venancio2018,Venancio2020}.
%\begin{figure}[ht!]
 % \centering
 % \includegraphics[width=8cm]{Cones_artigo3}
 % \caption{The wavy arrows depicts the direction of the perturbation field at the boundaries, while the cones are the local light cones. This boundary conditions is the appropriate one to attain a quasinormal mode at the generalized Nariai spacetime, see Ref. \cite{VenancioBatista2}. }\label{FigCones1}
%\end{figure}
Thus, summing up, we have just proved that the spectrum of the degrees of freedom $H_{ln}^-$ and $ H_{ln}^\ominus$  is the one given in Eq. (\ref{QNFS2F}). It remains to check whether $H^-_{tl}$ and $H^-_{xl}$ have the same spectrum. Defining the field
\begin{equation}
  \breve{H}^-_l =\frac{1}{f}\left( \frac{d}{dx} H^-_{tl} + i \omega  H^-_{xl} \right) \,,
\end{equation}
it follows immediately from the equation
\begin{equation}
\frac{d}{dx}E_{t\phi_l} + i\omega E_{x\phi_l} = 0
\end{equation}
that $\breve{H}^-_l$ also obeys the P\"{o}schl-Teller\index{P\"{o}schl-Teller equation} equation (\ref{PoschlTeller}) and, therefore, have the same spectrum of the fields $H_{ln}^-$ and $ H_{ln}^\ominus$, namely (\ref{QNFS2F}). Then, by means of the equations $E^-_{t\phi_l}=0$ and $E^-_{x\phi_l}=0$ we can write the fields $H^-_{tl}$ and $H^-_{xl}$ in terms of the fields that obey the P\"{o}schl-Teller\index{P\"{o}schl-Teller equation} equation. More precisely, we have:
\begin{equation}
     \begin{array}{ll}
       H^-_{tl} &= \dfrac{1}{\Lambda(\kappa - 2)} \dfrac{d}{dx} \breve{H}^-_l  - \dfrac{i\omega}{\kappa - 2}\sum_{n\neq l} \kappa_n ( H_{ln}^\ominus + H_{nl}^-)\,,\\
\\
       H^-_{xl} &=  \dfrac{1}{\kappa - 2}\sum_{n\neq l} \kappa_n \dfrac{d}{dx}( H_{ln}^\ominus + H_{nl}^-) -\dfrac{i\omega}{\Lambda(\kappa - 2)} \breve{H}^-_l\,.
     \end{array}
\end{equation}
So, $H^-_{tl}$ and $H^-_{xl}$  must have the same spectrum of $\breve{H}^-_l$, $H^-_{ln}$, and $H^\ominus_{ln}$, namely (\ref{QNFS2F}). Indeed, since fields $\breve{H}^-_l$, $H^-_{ln}$, and $H^\ominus_{ln}$ obey the boundary condition depicted in Fig. \ref{FigBoundCond}, it follows that near the boundaries $x\rightarrow \pm \infty$ ($r\rightarrow \pm 1/\sqrt{\Lambda}$) the behavior of these fields is $e^{\pm i \omega x}$. Thus, linear combinations of these fields and their derivatives will also obey the same boundary conditions\index{Boundary conditions}. Another way to understand why $H^-_{tl}$ and $H^-_{xl}$ have the spectrum (\ref{QNFS2F}) is by applying the differential operator that acts on $H$ in Eq. (\ref{PoschlTeller}) to the above expressions for $H^-_{tl}$ and $H^-_{xl}$. Doing so, we can check that $H^-_{tl}$ and $H^-_{xl}$ obey the P\"{o}schl-Teller\index{P\"{o}schl-Teller equation} equation with a source, namely
\begin{equation}
   \left[ \frac{d^2}{dx^2}+ \omega^2 - \frac{\Lambda\, (\kappa - 2) }{\cosh^2(\sqrt{\Lambda}\,x )}\right] H^-_{tl} =
F_l   \frac{df}{dx} \,,
\end{equation}
where $F_l=F_l(x)$ is some field obeying the P\"{o}schl-Teller equation\index{P\"{o}schl-Teller equation} and likewise for  $H^-_{xl}$. The general solution for a linear differential equation with a source is given by the general solution for the homogeneous part of the equation, which in the latter case is the P\"{o}schl-Teller equation\index{P\"{o}schl-Teller equation}, plus a particular solution that depends linearly on the source. In the case of interest, the source goes to zero exponentially at the boundaries, due to the term $df/dx$. Hence, near the boundaries $H^-_{tl}$ and $H^-_{xl}$ obey the P\"{o}schl-Teller equation and, therefore, yield the same spectrum (\ref{QNFS2F}).

So far, we have imposed and solved the equations $ E^-_{t\phi_l}=0$, $ E^-_{x\phi_l}=0$, and $ E^-_{\theta_l\phi_n}=0$, whereas we have just assumed $ E^-_{\theta_l\phi_l}=0$ to be true, without really solving it. However, inserting the latter expressions for $H^-_{tl}$ and $H^-_{xl}$ in the third line of Eq. (\ref{Emenos}) it follows that $ E^-_{\theta_l\phi_l}=0$ whenever $H_{ln}^\ominus$ and $H_{ln}^-$ obey the P\"{o}schl-Teller equation\index{P\"{o}schl-Teller equation} (\ref{PoschlTeller}), so that the constraint $ E^-_{\theta_l\phi_l}=0$ is already guaranteed to hold once the other equations in (\ref{Emenos}) are solved.  In conclusion, all degrees of freedom of the odd perturbation\index{Odd perturbation} have the spectrum (\ref{QNFS2F}).

\vspace{.5cm}
%\subsection{Even Perturbations}\label{Sec.Even}
\subsection{Even Perturbations}\label{Sec.Even}
\vspace{.5cm}

By an even perturbation\index{Even perturbation} we mean the most general perturbation for a given set of
spherical harmonic indices $\{\ell_2,m_2,\ell_3,m_3,\ldots,\ell_d,m_d\}$ and parity $(-1)^{\ell_2 + \ell_3 + \ldots + \ell_d}$, namely
\begin{align}
  \mathbf{h}&= \sum_{\ell m}e^{-i\omega t}\Big[   (H_{tt} dt^2 + 2 H_{tx} dtdx + H_{xx} dx^2) \mathcal{Y}_{\ell m}   \nonumber \\
& + \sum_{l=2}^d  \left( H_{tl}^+ \mathcal{V}_{l,\ell m}^{+} dt + H_{xl}^+ \mathcal{V}_{l,\ell m}^{+} dx + H_l^+ \,\mathcal{T}_{l,\ell m}^{+} +  H_l^\oplus \,\mathcal{T}_{l,\ell m}^{\oplus} \right) \nonumber \\
& + \sum_{l=2}^d \sum_{n>l}^d  H_{ln}^+ \,\mathcal{W}_{ln,\ell m}^{+} +  H_{ln}^\oplus \,\mathcal{W}_{ln,\ell m}^{\oplus}\Big]\,, \label{heven1}
\end{align}
Then, performing a gauge transformation\index{Gauge transformation} (\ref{GaugetTransf}) with
\begin{equation*}
  \zeta_\mu dx^\mu =\frac{ e^{-i\omega t}}{2} \left[ A \,\mathcal{Y}dt + B \,\mathcal{Y} dx - \sum_{l=2}^d H_{l}^{+}\,\mathcal{V}_{l,\ell m}^{+} \right] \,,
\end{equation*}
where $A = A(x)$ and $B = B(x)$ are functions of the coordinate $x$ given by
\begin{eqnarray}
A & = & -i \omega\,H_{l}^{+}\,-\,2 H_{t2}^{+}\,,\\
B & = & \frac{d}{dx}\,H_{l}^{+}\,-\,2 H_{x2}^{+} \,,
\end{eqnarray}
it follows that the transformed perturbation field $\tilde{h}_{\mu\nu}$ is such that it admits an expansion just as depicted in Eq. (\ref{heven1}) but with the fields $H(x)$ transformed to $\tilde{H}(x)$ where
\begin{eqnarray}
\tilde{H}_{tt} &=& H_{tt}\,-\,\frac{f'}{2f}\left(\frac{d}{dx}H_{2}^{+}\,-\,2H_{x2}^{+} \right ) \,-\,\omega^{2}H_{2}^{+}\,+\,2i\omega H_{x2}^{+} \,,\nonumber\\
\tilde{H}_{xx} &=& H_{tt}\,-\,\frac{f'}{2f}\left(\frac{d}{dx}H_{2}^{+}\,-\,2H_{x2}^{+} \right ) \,+\,\frac{d^{2}}{dx^{2}}H_{2}^{+}\,-\,2\frac{d}{dx}H_{x2}^{+} \,,\nonumber\\
\tilde{H}_{tx}&=& H_{tx}\,-\,\frac{f'}{2f}\left(\frac{d}{dx}H_{2}^{+}\,-\,2H_{x2}^{+} \right ) \,-\,i\omega \frac{d}{dx}H_{2}^{+}\,-\,2\frac{d}{dx}H_{t2}^{+}\,+\,i\omega H_{x2}^{+} \,,\nonumber\\
\tilde{H}_{tl}^{+} &=& H_{tl}^{+}\,-\,H_{t2}^{+}\,-\,\frac{i\omega}{2}\left(H_{2}^{+}\,-\,H_{l}^{+}\right )  \quad \forall \, l\neq 2 \,,\nonumber\\
\tilde{H}_{xl}^{+} &=& H_{xl}^{+}\,-\,H_{x2}^{+}\,+\,\frac{1}{2}\frac{d}{dx}\left(H_{2}^{+}\,-\,H_{l}^{+}\right ) \quad \forall \, l\neq 2 \,,\nonumber\\
\tilde{H}_{ln}^{\oplus} &=& H_{ln}^{\oplus}\,-\,\frac{1}{2}\left(H_{l}^{+}\,+\,H_{n}^{+} \right )\nonumber\\
\tilde{H}_{t2}^{+} &=& 0 \,,\nonumber\\
\tilde{H}_{x2}^{+} &=& 0 \,,\nonumber\\
\tilde{H}_{l}^{+} &=& 0 \,.
\end{eqnarray}
Hence, without loss of generality, we can set
%Thus, we can ignore the $(d+1)$ degrees of freedom from
\begin{equation}
\tilde{H}_{t2}^{+} \,=\, 0 \quad , \quad \tilde{H}_{x2}^{+} \,=\, 0 \quad , \quad \tilde{H}_{l}^{+} \,=\, 0 \,,
\end{equation}
since the these $(d+1)$ degrees of freedom can be eliminated by a gauge transformation\index{Gauge transformation}, whereas the other $\tilde{H}$'s are just equal to the previous $H$'s added by functions of $x$. Thus, assuming this gauge choice and dropping the tildes, we can assume that the perturbation field has the form
\begin{align}
  \mathbf{h}&= \sum_{\ell m}e^{-i\omega t}\Big[   (H_{tt} dt^2 + 2 H_{tx} dtdx + H_{xx} dx^2) \mathcal{Y}_{\ell m} + \sum_{l=3}^d  \left( H_{tl}^+ \mathcal{V}_{l,\ell m}^{+} dt + H_{xl}^+ \mathcal{V}_{l,\ell m}^{+} dx \right) \nonumber \\
& + \sum_{l=2}^d \sum_{n>l}^d \left(H_{ln}^+ \,\mathcal{W}_{ln,\ell m}^{+} +  H_{ln}^\oplus \,\mathcal{W}_{ln,\ell m}^{\oplus}\right) + \sum_{l=2}^d H_{l}^\oplus \,\mathcal{T}_{l,\ell m}^{\oplus}\Big]\,, \label{heven2}
\end{align}

Now, inserting this ansatz into the field equation \eqref{Eqh}, we find the following differential equations obeyed by the fields $H$'s:
\begin{align}\label{Emais}
E^+_{tt} &\equiv\, \frac{d^2}{dx^2} H_{tt} - 2i\omega\, \frac{d}{dx}H_{tx} + \Lambda\,(2 - \kappa f)\,H_{tt} - (\omega^2 + 2\Lambda)\,H_{xx} \nonumber\\
&+\, 2\Lambda f \sum_{l=2}^{d}\left[\,i \omega \left(\kappa_{l}\,H_{tl}^{+} - i\omega H_l^{\oplus} \right) - \frac{f'}{2f}\left(\kappa_{l} H_{xl}^{+} + \frac{d}{dx}H_l^{\oplus}\right ) \right] \nonumber\\
&+\,\frac{f'}{2f}\left(\frac{d}{dx} H_{xx} -3\,\partial_{x}H_{tt} + 2i\omega\, H_{tx} \right ) \,=\,0 \,, \nonumber\\
E^+_{tx} &\equiv\,  \sum_{l}\left[ \kappa_l f \frac{d}{dx}\left( \frac{1}{f}H_{tl}^+ \right) + i\omega \kappa_l H_{xl}^+
- \kappa_l\, H_{tx} + 2i\omega \sqrt{f}\frac{d}{dx}\left( \frac{1}{\sqrt{f}}H_l^{\oplus} \right)   \right] = 0  \,, \nonumber\\
E^+_{xx} &\equiv\, \frac{d^2}{dx^2} H_{tt} - 2i\omega\, \frac{d}{dx} H_{tx} - \left[\omega^{2}+ \Lambda\,(2 - \kappa f)\right]H_{xx} +  2\Lambda\,H_{tt} \nonumber\\
&-\,2\Lambda f \sum_{l=2}^{d}\left[\kappa_{l}\,\frac{d}{dx} H_{xl}^{+} + \frac{d^2}{dx^2} H_l^{\oplus} - \frac{f'}{2f}\left(\kappa_{l} H_{xl}^{+} + \frac{d}{dx} H_l^{\oplus}\right ) \right] \nonumber\\
&+\,  \frac{f'}{2f}\left(\frac{d}{dx} H_{xx} -3\,\frac{d}{dx}H_{tt} + 2i\omega\, H_{tx} \right ) \,=\,0 \,,\nonumber\\
E^+_{t\theta_l} &\equiv\, \frac{d}{dx}\left[\frac{1}{f}\left(\frac{d}{dx}H_{tl}^+ - i \omega H_{xl}^+  \right) \right] + i\omega \left[\frac{1}{f}\left(\frac{d}{dx}H_{xl}^{+} - i \omega H_{tl}^+  \right) \right]  
- \Lambda (\kappa-2) H_{tl}^{+} +\nonumber\\
&-\, \frac{1}{f}\,\frac{d}{dx} H_{tx} + \frac{i \omega}{2f}\,(H_{tt} + H_{xx}) + \Lambda \sum_{n=2}^{d}\left(i\omega H_n^{\oplus} + \kappa_{n} H_{tn}^{+}  \right) - i \omega  E^{I}_{\phi_{l}\phi_{l}} = 0 \,,\nonumber\\
E^+_{x\theta_l} &\equiv\, i\omega \left[\frac{1}{f}\left(\frac{d}{dx}H_{tl}^+ - i \omega H_{xl}^+  \right) \right] + \frac{d}{dx} \left[\frac{1}{f}\left(\frac{d}{dx}H_{xl}^{+} - i \omega H_{tl}^+  \right) \right]  
- \Lambda (\kappa-2) H_{xl}^{+} +\nonumber\\
&+\, \frac{i\omega}{f}\,H_{tx} -\frac{1}{2f}\,\frac{d}{dx}\,(H_{tt} + H_{xx}) + \Lambda \sum_{n=2}^{d}\left(\frac{d}{dx}H_n^{\oplus} + \kappa_{n} H_{xn}^{+}  \right) - \frac{d}{dx}  E^{I}_{\phi_{l}\phi_{l}} = 0 \,,\nonumber
\end{align}
\begin{align}
E^+_{\theta_l\theta_n} &\equiv\, \frac{d^2}{dx^2} (H_{ln}^+ + H_{nl}^+ ) + \left[ \omega^2 - \Lambda f (\kappa-2) \right] (H_{ln}^+ + H_{nl}^+ ) 
-  E^{I}_{\phi_l\phi_l} - E^{I}_{\phi_n\phi_n} = 0 \,,\nonumber\\
E^+_{\phi_l\phi_n} &\equiv\,  \frac{d^2}{dx^2} (H_{ln}^\oplus  + H_{nl}^\oplus ) +
 \left[ \omega^2 - \Lambda f (\kappa-2) \right] (H_{ln}^\oplus + H_{nl}^\oplus ) = 0 \,,\nonumber\\
E^{I}_{\phi_l\phi_l} &\equiv\,   \frac{1}{f}\left(\frac{d}{dx} H_{xl}^+ - i \omega  H_{tl}^{+} \right) + \Lambda H_{l}^\oplus - 
 \Lambda  \sum_{n=2}^{d} \left[  H_{n}^{\oplus}  + \kappa_{n} (H_{ln}^+  + H_{nl}^+ )  \right] \nonumber\\
&+\, \frac{1}{2f}(H_{tt} - H_{xx}) = 0 \,,\nonumber \\
E^{II}_{\phi_l\phi_l} &\equiv\,  \frac{d^2}{dx^2} H_{l}^{\oplus} + \left[ \omega^2 - \Lambda f ( \kappa -2 ) \right] H_{l}^\oplus  = 0 \,, \nonumber\\
E^{I}_{\theta_l\theta_l} &\equiv\, E^{I}_{\phi_l\phi_l} = 0  \,,    \nonumber\\
E^{II}_{\theta_l\theta_l} &\equiv\, E^{II}_{\phi_l\phi_l} + 2 \kappa_l \,E^{I}_{\phi_l\phi_l}  = 0 \,,
\end{align}
where $f'$ stands for $df/dx$. The great advantage of using the angular basis\index{Angular basis} $\{\mathcal{Y}_{\ell m}, \mathcal{V}_{l,\ell m}^{+},\\
\mathcal{T}_{l,\ell m}^{\oplus},\cdots \}$, instead of just using the scalar spherical harmonics\index{Scalar spherical harmonics} $\mathcal{Y}_{\ell m}$ is that when we compute the components of the perturbation equation (\ref{Eqh}) the angular dependence automatically factors as a global multiplicative term, so that we end up with equations that depend just on the coordinate $x$, as we have seen in the odd perturbation\index{Odd perturbation} in the previous section and as we just saw in the above equations for the even perturbations\index{Even perturbation}. Nevertheless, in components $\phi_l\phi_l$ and $\theta_l\theta_l$ of the even perturbation\index{Even perturbation} equation the angular functions do not factor out automatically, rather we face an equation of the following type
\begin{equation}\label{LI}
 P(x)\,Y_{\ell_l, 0}(\theta_l) + Q(x)\,\cot\theta_l \frac{d}{d\theta_l}Y_{\ell_l,0}(\theta_l) = 0 \,. 
\end{equation}
However, in general, the spherical harmonic $Y_{\ell_l,0}$ is linearly independent from $\cot\theta_l \frac{d}{d\theta_l}Y_{\ell_l,0}$, so that the latter equation implies both $P(x) =0$ and $Q(x)=0$. This is the reason why the equations that stem from the components $\phi_l\phi_l$ and $\theta_l\theta_l$ are split in two separate constraints, which are denoted in Eq. (\ref{Emais}) by $E^{I}_{\phi_l\phi_l}$, $E^{II}_{\phi_l\phi_l}$, $E^{I}_{\theta_l\theta_l}$, and $E^{II}_{\theta_l\theta_l}$. The only case in which we cannot conclude that $P(x)$ and $Q(x)$ are both zero in Eq. (\ref{LI}) is when the two angular functions are linearly dependent, namely when
\begin{equation}
  \alpha \,  Y_{\ell_l,0}(\theta_l) + \beta\, \cot\theta_l \frac{d}{d\theta_l}Y_{\ell_l,0}(\theta_l) = 0 
\end{equation}
for some constants $\alpha$ and $\beta$. Integrating this constraint, we conclude that the linear dependence happens only if
\begin{equation}
  Y_{\ell_l,0}(\theta_l) = c\, (\cos\theta_l)^{\alpha/\beta} \,, 
\end{equation}
where $c$ is some constant. This is true only for $\ell_l=0$, in which case $\alpha/\beta = 0$, and for $\ell_l=1$, in which case $\alpha/\beta = 1$. Thus, for any $\ell_l > 1$ we can promptly assume that, in Eq. (\ref{LI}), $P(x)$ and $Q(x)$ are independently zero.

From the equations $E_{\theta_{l}\theta_{n}}^{+} = 0, E_{\phi_{l}\phi_{n}}^{+}= 0, E_{\phi_{l}\phi_{l}}^{II} = 0$, we have that the fields $H_{ln}^{+}, H_{ln}^{\oplus}$ and $H_{l}^{\oplus}$ obey the
P\"oschl-Teller equation, namely Eq. \eqref{PoschlTeller}. In particular, assuming the QNMs
boundary condition, it follows that the spectrum of allowed frequencies is give by Eq. \eqref{QNFS2F}.
%\begin{equation}\label{PoschlTeller}
 % \frac{d^2}{dx^2}H + \left[ \omega^2 - \frac{\Lambda\, (\kappa - 2) }{\cosh^2(\sqrt{\Lambda}\,x )}\right] H = 0 \,.
%\end{equation}
Now, defining the fields $V_{ln}^{I} = V_{[ln]}^{I}$ and $V_{ln}^{II} = V_{[ln]}^{II}$ as
\begin{eqnarray}\label{HIAII}
V_{ln}^{I} &:=& \frac{1}{f}\left( \frac{d}{dx} H^{+}_{tl} + i \omega  H^{+}_{xl} - \frac{d}{dx} H^{+}_{tn} - i \omega  H^{+}_{xn}\right) \,,\nonumber\\
V_{ln}^{II} &:=& \frac{1}{f}\left( \frac{d}{dx} H^{+}_{xl} + i \omega  H^{+}_{tl} - \frac{d}{dx} H^{+}_{xn} - i \omega  H^{+}_{tn}\right) \,,
\end{eqnarray}
it follows immediately from the identities
\begin{align}
\frac{d}{dx}\left( E_{t\theta_{l}}^{+} - E_{t\theta_{n}}^{+} \right) + i\omega \left( E_{x\theta_{l}}^{+} - E_{x\theta_{n}}^{+} \right)  &=\, \frac{d^2}{dx^2}V_{ln}^{I} + \left[ \omega^2 - (\kappa - 2)\,f\right] V_{ln}^{I} = 0 \,,\nonumber\\
\\
\frac{d}{dx}\left( E_{x\theta_{l}}^{+} - E_{x\theta_{n}}^{+} \right) + i\omega \left( E_{t\theta_{l}}^{+} - E_{t\theta_{n}}^{+} \right) &=\,  \frac{d^2}{dx^2}V_{ln}^{II} + \left[ \omega^2 - \Lambda\, (\kappa - 2)\,f \right] V_{ln}^{II} = 0 \,,\nonumber
\end{align}
which is a consequence of the fact that the components $E_{t\theta_{l}}^{+} = 0$ and $E_{x\theta_{l}}^{+} = 0$, that $V_{ln}^{I}$ and $V_{ln}^{II}$ also obey the P\"oschl-Teller equation and thus, the spectrum associated to these degrees of freedom is given by Eq. \eqref{QNFS2F}. 

It is worth recalling that we use a gauge transformation\index{Gauge transformation} to eliminate some degrees of freedom of the perturbation, in particular $H_{t2}^{+} = H_{x2}^{+} = 0$. So, from the identities $E_{t\theta_{l}} - E_{t\theta_{2}} = 0$ and $E_{x\theta_{l}} - E_{x\theta_{2}} = 0$ and assuming that $E^{I}_{\phi_l\phi_l} =0$, it follows that 
\begin{eqnarray}
H_{tl}^{+} &=&\frac{1}{\Lambda (\kappa - 2)} \left(\frac{d}{dx} V_{2l}^{I} - i\omega V_{2l}^{II}   \right) \,,\nonumber\\
H_{xl}^{+} &=& \frac{1}{\Lambda (\kappa - 2)} \left(\frac{d}{dx} V_{2l}^{II} - i\omega V_{2l}^{I}   \right)\,,
\end{eqnarray}
and thus the spectrum associated to these fields must be the same as that for $V_{ln}^{I}$ and $V_{ln}^{II}$, namely, Eq. \eqref{QNFS2F}.

It remains to check whether $H_{tt}^+, H_{tx}^+$ and $H_{xx}^+$ have the same spectrum. From $E_{tx}^{+} = 0$, we have directly that 
\begin{equation}
H_{tx}^+\,=\,\frac{1}{\kappa}\sum_{l}\left[ \kappa_l f \frac{d}{dx}\left( \frac{1}{f}H_{tl}^+ \right) - i\omega \kappa_l H_{xl}^+ - 2i\omega \sqrt{f}\frac{d}{dx}\left( \frac{1}{\sqrt{f}}H_l^{\oplus} \right)   \right] \,,
\end{equation}
and therefore, yield the same spectrum of the fields $H_{tl}^+, H_{xl}^+$ and $H_{l}^{\oplus}$ that, in its turn, it is the same spectrum of the fields that obey the P\"oschl-Teller equation. Finally, from the identities $E^{I}_{\phi_l\phi_l} =0$ and $E^{+}_{t\theta_l} =0$, we find that the fields $H_{tt}^+ - H_{xx}^+$ and $H_{tt}^+ + H_{xx}^+$ are related to $V_{ln}^{II}, H_{l}^{\oplus}, H_{ln}^{+}, H_{tl}^{+}, H_{tx}^+$ by the following equations:
\begin{align}
H_{tt}^+ - H_{xx}^+ &=\, V_{2l}^{II} -2\Lambda f H_{l}^\oplus + 
 \Lambda f  \sum_n \left[ H_{n}^{\oplus}  + \kappa_{n} (H_{ln}^+  + H_{nl}^+ )  \right] \,,\nonumber\\
H_{tt}^+ + H_{xx}^+ &=\, \frac{2i}{\omega }\left[\frac{d}{dx} H_{tx} - \Lambda f \sum_{n}\left(\kappa_{n} H_{tn}^{+} -i\omega H_{n}^{\oplus}\right)  \right]\,.
\end{align}
Thus, from these equations for $H_{tt}^+ \pm H_{xx}^+$ we conclude that these fields are written in terms of fields that we already proved that have the spectrum \eqref{QNFS2F}. This finishes the proof
that in the generalized Nariai spacetime all degrees of freedom of the gravitational perturbation, scalar, vectorial, and tensorial, even and odd, have the same spectrum of quasinormal modes. This differs, for example, from what happens in other higher-dimensional spacetimes like Schwarzschild and (anti) de Sitter \cite{ Dreyer2003, Hughston1973, Frolov2008}, in which different parts of the gravitational perturbation have different spectra. This isospectral property of the higher-dimensional  Nariai spacetime considered here proves that the existence of different spectra to different degrees of freedom of the gravitational field\index{Gravitational field} is much more related to the symmetries of the spacetime than to the tensorial nature of the degree of freedom of the perturbation or to the dimension of the background. In particular, when $D=4$ we have $\kappa = \ell (\ell +1)$ and the spectrum \eqref{QNFS2F} when the  assumed boundary condition is (I) can be written as
\begin{equation}
\frac{\omega}{\sqrt{\Lambda}} = \pm \sqrt{(\ell + 2)(\ell - 1) -\frac{1}{4} } \,-\, i \left(n+\frac{1}{2}\right) \,.
\end{equation}
This result coincides with the spectrum of frequencies in $D=4$ shown by Cardoso in Ref. \cite{Cardoso2003PRD} in which an exact expression for the quasinormal modes of gravitational perturbations of a near extremal Schwarzschild-de Sitter black hole in four dimensions was obtained.  It is well known that the extremal limit of the Schwarzschild-de Sitter solution, when the black hole horizon coalesces with the cosmological horizon, yield the Nariai spacetime \cite{Ginsparg1983,Bousso1996,Cardoso2004}. Nevertheless, when $D = 4$, our analytical results are in disagreement with the quasinormal frequencies\index{Quasinormal frequencies} for the tensorial degrees of freedom of the gravitational perturbation displayed in Ref. \cite{Vanzo2004,Ortega2009}. We believe that this difference might have come from a typo in Ref. \cite{Vanzo2004} that was replicated in Ref. \cite{Ortega2009}. 

In order to obtain the spectrum of the even part of the gravitational perturbation it was not necessary to use all field equations displayed in Eq. \eqref{Emais}. More precisely, we have not solved $E^{+}_{tt} = 0, E^{+}_{xx} =0$ and $E^{I}_{\phi_n\phi_n} = 0$ for $n>2$. Therefore, it is prudential to check if these remaining equations are consistent with the solutions of the ones that we have used. After some algebra, we have checked that this consistency holds indeed. Thus, once we assume that $H^{+}_{ln}, H^{\oplus}_{ln},H^{\oplus}_{l}, V^{I}_{ln}$ and $V^{II}_{ln}$ obey the P\"{o}schl-Teller
equation\index{P\"{o}schl-Teller equation} \eqref{PoschlTeller}, and that $H^{+}_{tl}, H^{+}_{xl}, H^{+}_{tx},H^{+}_{tt}$ and $H^{+}_{xx}$ are given by the expressions displayed above, it follows that the remaining components of Einstein's equation are automatically satisfied.

%%%%%%%%%%%%%%%%%%%%%%%%%%%%%%%%%%%%%%%%%%%%%%%%%%%%%%%%%%%%%%%%%%%%%%%%%%%%%%%%%%%%%%
%%%%%%%%%%%%%%%%%%%%%%%%%%%%%%%%%%%%%%%%%%%%%%%%%%%%%%%%%%%%%%%%%%%%%%%%%%%%%%%%%%%%%%
%%%%%%%%%%% B: Chapter 5: Spin-1/2 Field Perturbations
%%%%%%%%%%%%%%%%%%%%%%%%%%%%%%%%%%%%%%%%%%%%%%%%%%%%%%%%%%%%%%%%%%%%%%%%%%%%%%%%%%%%%%
%%%%%%%%%%%%%%%%%%%%%%%%%%%%%%%%%%%%%%%%%%%%%%%%%%%%%%%%%%%%%%%%%%%%%%%%%%%%%%%%%%%%%%
%%%%%%%%%%%%%%%%%%%%%%%%%%%%%%%%%%%%%%%%%%%%%%%%%%%%%%%%%%%%%%%%%%%%%%%%%%%%%%%%%%%%%%

%\chapter{\uppercase{Spin-$1/2$ Field Perturbation}}\label{S12FP}
\chapter{Spin-$1/2$ Field Perturbation}\label{S12FP}

%\chapter{\LARGE{5\quad \uppercase{Spin-$1/2$ Field Perturbation}}}\label{S12FP}

%\chapter{Spin-$1/2$ Field Perturbation}\label{S12FP}

In this chapter, we consider the perturbations in a spin-$1/2$ field, a Dirac field\index{Dirac field} of mass $\mu$. As a first step, we shall attain the separation of the Dirac equation in the generalized Nariai background and its reduction into a set of Schr\"{o}dinger-like differential equations\index{Schr\"{o}dinger-like differential equation} with a particular effective potential, the Rosen-Morse\index{Rosen-Morse class of potentials} class of potential. It is worth pointing out that although the study of quasinormal modes has a long history, in the context of general relativity it started with a stability problem that concerned the evolutions of spin-$2$ perturbations \cite{Regge1957}. Inasmuch as spin-$2$ perturbations comprise effectively components transforming as fields of spins $0,1$, and $2$, quasinormal mode frequencies are mainly obtained for fields with these spins. So we hope that our investigation of the quasinormal modes associated to a massive Dirac perturbation around the generalized Nariai spacetime can serve to fill a part of this shortfall. For previous works on quasinormal modes of Dirac fields\index{Dirac field} in other backgrounds, see \cite{Cho2003,Ortega2006,Becar2014}.

\vspace{.5cm}
%\section{\uppercase{Clifford Algebra and Spinors}}
\section{Clifford Algebra and Spinors}
\vspace{.5cm}

There are several ways to define Clifford algebras\index{Clifford algebra} and spinors. Let us here present one of them, for more details, see \cite{VenancioBook,BennBook,LounestoBook}. We choose a simplified approach just to achieve our intent, which is to motivate the natural ansatz for a spin-$1/2$ field perturbation, a spinorial field $\hat{\bl{\Psi}}$ satisfying the Dirac equation.

Let us work out with the Dirac equation in the $l$th two-dimensional unit sphere $S^2$ with coordinates $\{\theta_l,\phi_l \}$ , whose line element was given in Eq. \eqref{nariai-metric}, namely
\begin{equation}
    d \Omega^{2}_{l} = d \theta^{2}_{l} + \sin^2 \theta_{l} d \phi^{2}_{l}\,.
\end{equation}
At each point of $S^2$, the orthonormal frame field given by

\begin{equation}
    \hat{\bl{e}}_{1} = \partial_{\theta_{l}} \quad \text{and} \quad\hat{\bl{e}}_{2} = \frac{1}{\sin \theta_{l}} \partial_{\phi_{l}} \,,
\end{equation}
spans a two-dimensional vector space, denoted here by $\mathbb{V}$\label{TVS} . By an orthonormal frame we mean that the components of the metric $\hat{\bl{g}}$ with respect to this frame field are given by
\begin{equation}
    \hat{\bl{g}} (\hat{\bl{e}}_{\mathfrak{m}},\hat{\bl{e}}_{\mathfrak{n}}) = \delta_{\mathfrak{mn}}\quad \forall \,\mathfrak{m},\mathfrak{n} \in \{1,2\}\,.
\end{equation}
In this frame, any vector field $\bl{V} \in \mathbb{V}$ can be written as the linear expression

\begin{equation}
    \bl{V} = V^{1} \hat{\bl{e}}_{1} + V^{2} \hat{\bl{e}}_{2}  = V^{\mathfrak{m}} \hat{\bl{e}}_{\mathfrak{m}} \,.
\end{equation}
Let us now introduce the Clifford algebra\index{Clifford algebra}, a special kind of algebra defined on vector spaces endowed with a metric. In order to perform this, let us write the quadratic form $\hat{\bl{g}} (\bl{V},\bl{V}) = \hat{\bl{g}} (\hat{\bl{e}}_{\mathfrak{m}},\hat{\bl{e}}_{\mathfrak{n}})V^{\mathfrak{m}} V^{\mathfrak{n}}$ as the square of $\bl{V}$, namely $\bl{V} \bl{V} = \hat{\bl{g}} (\bl{V},\bl{V})$. This defines the so-called Clifford product\index{Clifford product} which has been denoted by juxtaposition. As result, assuming the distributive property of multiplication, the elements of the frame $\{ \hat{\bl{e}}_{\mathfrak{m}} \}$ must obey the following algebra
\begin{equation}\label{CAD}
    \hat{\bl{e}}_{\mathfrak{m}} \hat{\bl{e}}_{\mathfrak{n}} + \hat{\bl{e}}_{\mathfrak{n}} \hat{\bl{e}}_{\mathfrak{m}} = 2 \,\hat{\bl{g}} (\hat{\bl{e}}_{\mathfrak{m}},\hat{\bl{e}}_{\mathfrak{n}})\,,
\end{equation}
which is just the very definition of Clifford algebra\index{Clifford algebra} of the vector space $\mathbb{V}$ endowed with a metric $\hat{\bl{g}}$, denoted by $\mathcal{C} \ell (\mathbb{V}, \hat{\bl{g}})$\label{CA}. 
%The Eq. \eqref{CAD} defines the Clifford product which has been denoted by juxtaposition. 
It is worth noting that, inasmuch as the Clifford product\index{Clifford product} of two vectors $\hat{\bl{e}}_{\mathfrak{m}}$ and $\hat{\bl{e}}_{\mathfrak{n}}$ is defined to be such that its symmetric part gives the metric components, it is defined only up to a product which is skew-symmetric on vectors.
Indeed, we can without loss of generality write it as
\begin{equation}
    \hat{\bl{e}}_{\mathfrak{m}} \hat{\bl{e}}_{\mathfrak{n}} = \hat{\bl{g}} (\hat{\bl{e}}_{\mathfrak{m}},\hat{\bl{e}}_{\mathfrak{n}}) + \hat{\bl{e}}_{\mathfrak{m}} \wedge \hat{\bl{e}}_{\mathfrak{n}}\,,
\end{equation}
where exterior product\label{EP} of two vectors $\hat{\bl{e}}_{\mathfrak{m}} \wedge \hat{\bl{e}}_{\mathfrak{n}} \equiv - \hat{\bl{e}}_{\mathfrak{n}} \wedge \hat{\bl{e}}_{\mathfrak{m}}$ is defined by the following relation:
\begin{equation}\label{EPD}
    \hat{\bl{e}}_{\mathfrak{m}} \hat{\bl{e}}_{\mathfrak{n}} - \hat{\bl{e}}_{\mathfrak{n}} \hat{\bl{e}}_{\mathfrak{m}} \equiv 2 \,\hat{\bl{e}}_{\mathfrak{m}} \wedge \hat{\bl{e}}_{\mathfrak{n}}\,,
\end{equation}
that in its turn defines the exterior product as the totally antisymmetric part of the Clifford product\index{Clifford product} of $\hat{\bl{e}}_{\mathfrak{m}}$ and $\hat{\bl{e}}_{\mathfrak{n}}$.

Thus, in two dimensions the set $\{ 1,\hat{\bl{e}}_{\mathfrak{m}},\hat{\bl{e}}_{\mathfrak{m}} \wedge \hat{\bl{e}}_{\mathfrak{n}} \}$ contains $2^2 = 4$ elements and forms a basis for $\mathcal{C}\ell (\mathbb{V},\hat{\bl{g}})$, so that a general element $\bl{\mathcal{C}} \in C \ell (\mathbb{V},\hat{\bl{g}})$ can always be put in the form:

\begin{equation}
   \bl{\mathcal{C}} = S + V^{\mathfrak{m}} \hat{\bl{e}}_{\mathfrak{m}} + B^{\mathfrak{mn}} \,\hat{\bl{e}}_{\mathfrak{m}} \wedge \hat{\bl{e}}_{\mathfrak{n}}\,,
\end{equation}
where the term in the sum denoted by $S$ in the above equation transforms like scalar under rotation on the two-sphere, while $V^{\mathfrak{m}}$ transform like the components of a vector and finally the elements $B^{\mathfrak{mn}} = - B^{\mathfrak{nm}}$ transform like the components of a skew-symmetric second order tensor. It is worth mentioning that in higher dimensions, we must consider antisymmetric products of a larger order. For instance, in $3$ dimensions the set $\{1, \hat{\boldsymbol{e}}_{\mathfrak{m}}, \hat{\boldsymbol{e}}_{\mathfrak{m}}\wedge\hat{\boldsymbol{e}}_{\mathfrak{n}}, \hat{\boldsymbol{e}}_{\mathfrak{m}}\wedge\hat{\boldsymbol{e}}_{\mathfrak{n}}\wedge\hat{\boldsymbol{e}}_{\mathfrak{p}}\}$, which contains $2^{3}=8$ elements, furnish a basis for the Clifford algebra\index{Clifford algebra}, with each of the three objects $\hat{\boldsymbol{e}}_{\mathfrak{m}}\wedge\hat{\boldsymbol{e}}_{\mathfrak{n}}$  written as Eq. \eqref{EPD} for each choice of $\mathfrak{m}\neq \mathfrak{n}=1, 2, 3$ and a single highest order element $\hat{\boldsymbol{e}}_{\mathfrak{m}}\wedge\hat{\boldsymbol{e}}_{\mathfrak{n}}\wedge\hat{\boldsymbol{e}}_{\mathfrak{p}}$
%$\hat{\boldsymbol{e}}_{\mathfrak{m}}\wedge\hat{\boldsymbol{e}}_{\mathfrak{n}}\wedge\hat{\boldsymbol{e}}%_{\mathfrak{p}}$ defined by
%\begin{align}
%\hat{\boldsymbol{e}}_{\mathfrak{m}}\wedge\hat{\boldsymbol{e}}_{\mathfrak{n}}\wedge\hat{\boldsymbol{e}}%_{\mathfrak{p}}&=\,\dfrac{1}{3!}\left(\hat{\boldsymbol{e}}_{\mathfrak{m}}\hat{\boldsymbol{e}}%_{\mathfrak{n}}\hat{\boldsymbol{e}}_{\mathfrak{p}} + \hat{\boldsymbol{e}}_{\mathfrak{n}}\hat{\boldsymbol{e}}_{\mathfrak{p}}\hat{\boldsymbol{e}}_{\mathfrak{m}} + \hat{\boldsymbol{e}}%_{\mathfrak{p}}\hat{\boldsymbol{e}}_{\mathfrak{m}}\hat{\boldsymbol{e}}_{\mathfrak{n}} \right. \nonumber\\
%&-\,\left.\hat{\boldsymbol{e}}_{\mathfrak{n}}\hat{\boldsymbol{e}}_{\mathfrak{m}}\hat{\boldsymbol{e}}_{\mathfrak{p}} - \hat{\boldsymbol{e}}_{\mathfrak{p}}\hat{\boldsymbol{e}}_{\mathfrak{n}}\hat{\boldsymbol{e}}_{\mathfrak{m}} - \hat{\boldsymbol{e}}_{\mathfrak{m}}\hat{\boldsymbol{e}}_{\mathfrak{p}}\hat{\boldsymbol{e}}_{\mathfrak{n}}\right) \nonumber\,,
%\end{align}
defined as the totally antisymmetric part of the Clifford product\index{Clifford product}. In general, in $D$ dimensions, a basis for Clifford algebra\index{Clifford algebra} contains $2^{D}$ elements \cite{VenancioBook,BennBook}.

Once the Clifford algebra\index{Clifford algebra} $\mathcal{C}\ell (\mathbb{V},\hat{\bl{g}})$ has been defined, we will use it to define the so-called spinors. Spinors can be defined as the elements of a vector space, denoted here by $\mathbb{S} \subset \mathcal{C}\ell (\mathbb{V},\hat{\bl{g}})$\label{SP}, on which a linear and faithful representation for the Clifford algebra\index{Clifford algebra} acts, the so-called spinorial representation\index{Spinorial representation}. This means that if $\{ \boldsymbol{\xi }_{A} \}$ is an arbitrary spinor frame for $\mathbb{S}$ with the index $A$ running over $\{+,-\}$, we can choose it conveniently so that the Clifford action of the frame $\{ \hat{\bl{e}}_{\mathfrak{m}} \}$ on the spinors $\{ \boldsymbol{\xi }_{A} \}$ is a constant in a given patch of $S^2$.
\begin{equation}\label{CAOSCP}
    \hat{\bl{e}}_{\mathfrak{m}} \bl{\xi}_{A} = {(\sigma_{\mathfrak{m}})^{B}}_{A} \bl{\xi}_{B}\,,
\end{equation}
where the constant matrices ${(\sigma_{\mathfrak{m}})^{B}}_{A}$ are the known Dirac matrices\index{Dirac matrices}. 
%One can easily prove that this action is correct. 
Multiplying \eqref{CAOSCP} by $\hat{\bl{e}}_{\mathfrak{n}}$ and then adding the result to $\hat{\bl{e}}_{\mathfrak{m}} \hat{\bl{e}}_{\mathfrak{n}} \bl{\xi}_{A}$, one obtains from \eqref{CAD} that these matrices satisfy the following anticommutation relation
\begin{equation}
    {(\sigma_{\mathfrak{m}})^{A}}_{B} {(\sigma_{\mathfrak{n}})^{B}}_{C} + {(\sigma_{\mathfrak{n}})^{A}}_{B} {(\sigma_{\mathfrak{m}})^{B}}_{C} = \delta_{\mathfrak{mn}} \,\delta^{A}_{\phantom{A}C} \,,
\end{equation}
which is the definition of the Clifford algebra\index{Clifford algebra}. 

Note that $\mathbb{S} \subset \mathcal{C}\ell (\mathbb{V},\hat{\bl{g}})$ has been introduced to provide a representation of the Clifford algebra\index{Clifford algebra} $\mathcal{C}\ell (\mathbb{V},\hat{\bl{g}})$, since $\mathbb{S}$ is a vector space and, by definition, $\mathcal{C}\ell (\mathbb{V},\hat{\bl{g}})$ maps $\mathbb{S}$ into $\mathbb{S}$. As a vector space, the space $\mathbb{S}$ is called spinor space and its elements are called spinor fields\index{Spinor field}.  In particular, note also that the spinor space is a subalgebra of $\mathcal{C}\ell (\mathbb{V},\hat{\bl{g}})$ under Clifford product\index{Clifford product}. In two dimensions, a spinor frame $\{\bl{\xi}_{A}\}$ for the spinor space $\mathbb{S}$ can be spanned by the following elements of $\mathcal{C}\ell (\mathbb{V},\hat{\bl{g}})$
\begin{equation}
    \bl{\xi}_{A} = \bl{m}_{A} (1 + \bl{m}_{- A})\,,
\end{equation}
%which are also elements of $\mathcal{C}\ell (\mathbb{V},\hat{\bl{g}})$, 
where the complex vectors $\bl{m}_{A}$ defined as
\begin{equation}\label{NVD}
    \bl{m}_{A} = \hat{\bl{e}}_{1} + i A \hat{\bl{e}}_{2}\,,
\end{equation}
span a null frame on sphere. Indeed, using Eq. \eqref{CAD}, it is straightforward to prove that $\bl{m}_{(A}\bl{m}_{B)} = \hat{\bl{g}} (\bl{m}_{A},\bl{m}_{B})$ with the metric components given by
%Indeed, it is straightforward to prove that such vectors satisfy the Clifford algebra with the metric components given by
\begin{equation}
\hat{\bl{g}} (\bl{m}_{A},\bl{m}_{B}) = \left\{\begin{matrix}
0 & \textrm{  for } A = B \,,\\ 
1 & \textrm{  for } A \neq B \,.
\end{matrix}\right. 
\end{equation}
Under the action of $\hat{\bl{e}}_{\mathfrak{m}}$, the spinors $\boldsymbol{\xi}_{A}$ satisfy concisely the relations
\begin{equation}\label{CAOS}
    \hat{\bl{e}}_{1}\bl{\xi}_{A} = \bl{\xi}_{- A} \quad \text{and}\quad \hat{\bl{e}}_{2}\bl{\xi}_{A} = i A \bl{\xi}_{- A} \quad \text{for} \quad A \in \{+,-\}\,,
\end{equation}
that is, the action of $\mathbb{V}$ on $\mathbb{S}$ yields elements on $\mathbb{S}$. In particular, this means that $\mathbb{S}$ is invariant by the action of $\mathcal{C}\ell(\mathbb{V},\hat{\bl{g}})$. Thus, if $\Psi^{A} = \Psi^{A} (\theta_{l}, \psi_{l})$ are the components of $\hat{\boldsymbol{\Psi}}$ with respect to the frame $\{ \boldsymbol{\xi}_{A} \}$, then the spinor space $\mathbb{S}$ has dimension $2^{1}$ and defined by
\begin{equation}
    \mathbb{S} = \left \{ \hat{\boldsymbol{\Psi}} \in \mathcal{C}\ell(\mathbb{V}, \hat{\boldsymbol{g}}) \,\,|\,\, \hat{\boldsymbol{\Psi}} = 
    \Psi^{A} \boldsymbol{\xi}_{A} \,\,\forall\,\, \Psi^{A} \in \mathbb{C} \right \}\,.
\end{equation}
The elements $\hat{\boldsymbol{\Psi}}$ are the known spinor\index{Spinor field} fields\label{CF}. In higher dimensions, it can be proved that if $D = 2d$ is the dimension of the vector space, then the dimension of the spinor space is $2^d$; for thorough reviews, see Refs. \cite{VenancioBook,BennBook,CartanBook}.

The fact that $\mathbb{S}$ is invariant by the action of $\mathcal{C} \ell (\mathbb{V},\hat{\bl{g}})$ implies that the algebra $\mathcal{C}\ell(\mathbb{V},\hat{\bl{g}})$ can be faithfully represented by $2 \times 2$ matrices. In order to see this explicitly, we only need to act the elements that span $\mathbb{V}$ which are $\hat{\bl{e}}_{1}$ and $\hat{\bl{e}}_{2}$ on the a general element of $\mathbb{S}$, namely
\begin{equation}\label{CAOGS}
    \hat{\bl{e}}_{1} \hat{\boldsymbol{\Psi}} =  \Psi^{+} \boldsymbol{\xi}_{-} + \Psi^{-} \boldsymbol{\xi}_{+} \quad \text{and} \quad \hat{\bl{e}}_{2} \hat{\boldsymbol{\Psi}} =  i \Psi^{+} \boldsymbol{\xi}_{-} - i \Psi^{-} \boldsymbol{\xi}_{+}\\,,
\end{equation}
where Eq. \eqref{CAOS} has been used. In particular, this enables us to find explicitly the spinor representation of the vectors $\hat{\bl{e}}_{1}$ and $\hat{\bl{e}}_{2}$. Indeed, Eq. \eqref{CAOGS} implies the following spinor representation for theses vectors of the basis:
\begin{equation}
\begin{matrix}
 \sigma_{1} = \begin{bmatrix}
 0 & 1\\ 
 1 & 0 
\end{bmatrix} \quad , \quad 

\sigma_{2} = \begin{bmatrix}
 0 & -i\\ 
 i & 0 
\end{bmatrix}\,,
\end{matrix}
\end{equation}
where 
%$\sigma_{3} = - i \sigma_{1} \sigma_{2}$ and 
the spinorial indices have been omitted for simplicity. 
Hence, in two dimensions, 
%the Dirac matrices ${(\sigma_{\mathfrak{m}})^{A}}_{B}$ represent faithfully the Clifford algebra by the Pauli matrices. In this case,
the spinor frame $\{ \boldsymbol{\xi}_{A} \}$ for $\mathbb{S}$ can be represented by the following column vectors on which these constant matrices act
\begin{equation}
  \xi_{+} = \left[
              \begin{array}{c}
                1 \\
                0 \\
              \end{array}
            \right] \quad \text{and} \quad
 \xi_{-} = \left[
              \begin{array}{c}
                0 \\
                1 \\
              \end{array}
            \right] \,.
\end{equation}
Indeed, notice that the action of the matrices ${(\sigma_{\mathfrak{m}})^{A}}_{B}$ on the above column vectors can be summarized quite concisely as
\begin{equation}\label{CAOSMF}
    \sigma_{1}\,\xi_{A} = \xi_{- A}  \quad \textrm{and} \quad   \sigma_{2}\,\xi_{A} = i A \,\xi_{- A} \,,
\end{equation}
which is just the very matrix representation of \eqref{CAOS}. Thus, if $V^{\mathfrak{m}}$ are the components of the vector field\index{Vector field} $\bl{V}$ expanded in the orthonormal frame field $\{\hat{\bl{e}}_{\mathfrak{m}}\}$, we have

\begin{equation}
  \bl{V} = V^{\mathfrak{m}} \hat{\bl{e}}_{\mathfrak{m}} \quad\Longleftrightarrow\quad {V^{A}}_{B} = V^{\mathfrak{m}} {(\sigma_{\mathfrak{m}})^{A}}_{B}\,,
\end{equation}
where ${V^{A}}_{B}$ is the spinorial representation\index{Spinorial representation} of the vector field\index{Vector field} $\bl{V}$ in two dimensions. In
higher dimensions, it can be proved that if $D = 2d$ is the dimension of the vector space, then the Dirac matrices\index{Dirac matrices} represent faithfully the Clifford algebra\index{Clifford algebra} by $2^{d} \times 2^{d}$ matrices.

Clifford algebra provides a very clear and compact method for performing rotations, which is considerably more powerful than working with the vector representation of the rotation group, which is the usual approach. Indeed, let $\bl{n} \in \mathbb{V}$ be a non-null vector, $\bl{n}^{2} = \hat{\bl{g}} (\bl{n},\bl{n}) \neq 0$. Theses elements are invertible in $\mathcal{C} \ell (\mathbb{V},\hat{\bl{g}})$,

\begin{equation}
  \bl{n}^{-1} = \frac{\bl{n}}{\hat{\bl{g}} (\bl{n},\bl{n})}\,.
\end{equation}
In particular, note that $\bl{n}^{-1} = \bl{n}$ when $\bl{n}$ is a normalized vector, namely $\hat{\bl{g}} (\bl{n},\bl{n}) = 1$. Now, let us consider two normalized vectors $ \bl{n}_{1}$ and $ \bl{n}_{2}$ given by

\begin{equation}
  \bl{n}_{1} = \cos \left ( \frac{\varphi_{1}}{2} \right ) \hat{\bl{e}}_{1} + \sin \left ( \frac{\varphi_{1}}{2} \right ) \hat{\bl{e}}_{2} \quad \textrm{  and  }  \quad \bl{n}_{2} = \cos \left ( \frac{\varphi_{2}}{2} \right ) \hat{\bl{e}}_{1} + \sin \left ( \frac{\varphi_{2}}{2} \right ) \hat{\bl{e}}_{2}\,.
\end{equation}
Then, we can construct the following element\label{RO}
\begin{equation}\label{OR}
  \bl{R}_{\zeta} := \bl{n}_{1} \bl{n}_{2} = \cos \left ( \frac{\zeta}{2} \right ) + \sin \left ( \frac{\zeta}{2} \right ) \hat{\bl{e}}_{1} \wedge \hat{\bl{e}}_{2}\,,
\end{equation}
labeled by a single real parameter $\zeta = \varphi_{2} - \varphi_{1} \in [0, 2 \pi]$. So, to each $\bl{R}_{\zeta}$ given by \eqref{OR}, we can define a rotation of $\zeta$ on the plane generated by $\{\hat{\bl{e}}_{1} ,\hat{\bl{e}}_{2} \}$ as follows:
\begin{equation}\label{ROV}
  \hat{\bl{e}}_{\mathfrak{m}} \overset{\bl{R}_{\zeta}\,}{\longrightarrow} \hat{\bl{e}}'_{\mathfrak{m}} = \bl{R}_{\zeta} \hat{\bl{e}}_{\mathfrak{m}} \bl{R}^{-1}_{\zeta}\,.
\end{equation}
Since, by definition, a rotation is a linear transformation that preserve the metric, we should check that this expression for the rotation has the desired property of leaving the metric invariant. Using \eqref{CAD}, a simple proof is given by:

\begin{align}
\hat{\bl{g}} (\bl{R}_{\zeta} \hat{\bl{e}}_{\mathfrak{m}} \bl{R}_{\zeta}^{-1},\bl{R}_{\zeta} \hat{\bl{e}}_{\mathfrak{n}} \bl{R}_{\zeta}^{-1})  &=\,\, \dfrac{(\bl{R}_{\zeta} \hat{\bl{e}}_{\mathfrak{m}} \bl{R}_{\zeta}^{-1})(\bl{R}_{\zeta} \hat{\bl{e}}_{\mathfrak{n}} \bl{R}_{\zeta}^{-1}) + (\bl{R}_{\zeta} \hat{\bl{e}}_{\mathfrak{n}} \bl{R}_{\zeta}^{-1})(\bl{R}_{\zeta} \hat{\bl{e}}_{\mathfrak{m}} \bl{R}_{\zeta}^{-1})}{2} \nonumber\\
  &=\,\, \bl{R}_{\zeta} \left ( \dfrac{\hat{\bl{e}}_{\mathfrak{m}} \hat{\bl{e}}_{\mathfrak{n}} + \hat{\bl{e}}_{\mathfrak{n}} \hat{\bl{e}}_{\mathfrak{m}}}{2}  \right ) \bl{R}_{\zeta}^{-1} \\
 &=\,\, \hat{\bl{g}} (\hat{\bl{e}}_{\mathfrak{m}}, \hat{\bl{e}}_{\mathfrak{n}}) \,,\nonumber
\end{align}
which is, by definition, a rotation. It is not so hard to prove the following relations:
\begin{equation}
\begin{matrix}
\bl{R}_{\zeta} \hat{\bl{e}}_{1}\bl{R}_{\zeta}^{-1} &= \bl{n}_{2} \bl{n}_{1} \,\hat{\bl{e}}_{1}\, \bl{n}_{1} \bl{n}_{2} = \cos \zeta \,\hat{\bl{e}}_{1} - \sin \,\zeta \hat{\bl{e}}_{2}, \\ 
\bl{R}_{\zeta} \hat{\bl{e}}_{2} \bl{R}_{\zeta}^{-1} &= \bl{n}_{2} \bl{n}_{1} \,\hat{\bl{e}}_{2}\, \bl{n}_{1} \bl{n}_{2} = \sin \zeta \,\hat{\bl{e}}_{1} + \cos \zeta \,\hat{\bl{e}}_{2}\,. 
\end{matrix}
\end{equation}
This is clearly a rotation on the plane $\hat{\bl{e}}_{1} \wedge \hat{\bl{e}}_{2}$, where $\zeta$ is the angle of rotation. Now, inasmuch as the composition of rotations is also a rotation, the set of all elements $\bl{R}_{\zeta}$ form a group under Clifford product\index{Clifford product}. Denoted by $\textrm{\textit{SPin}}(\mathbb{V})$, this is called spin group\label{SG}

\begin{equation}
    \textrm{\textit{SPin}}(\mathbb{V}) = \{ \bl{R} \in \mathcal{C} \ell (\mathbb{V},\hat{\bl{g}}) \,\,|\,\, \bl{R} = \bl{R}_{\zeta_{p}} \bl{R}_{\zeta_{p-1}} \ldots \bl{R}_{\zeta_{1}} \}\,.
\end{equation}
Indeed, noting that

\begin{align}
    \bl{R}_{\zeta_{1}} \bl{R}_{\zeta_{2}} &=\,\, \left[ \cos \left( \dfrac{\zeta_{1}}{2} \right) + \sin \left( \dfrac{\zeta_{1}}{2} \right) \hat{\bl{e}}_{1} \wedge \hat{\bl{e}}_{2} \right] \left[ \cos \left( \dfrac{\zeta_{2}}{2} \right) + \sin \left( \dfrac{\zeta_{2}}{2} \right) \hat{\bl{e}}_{1} \wedge \hat{\bl{e}}_{2} \right] \nonumber\\
    &=\,\, \cos \left( \dfrac{\zeta_{1} + \zeta_{2}}{2}  \right) + \sin \left( \dfrac{\zeta_{1} + \zeta_{2}}{2}  \right) \hat{\bl{e}}_{1} \wedge \hat{\bl{e}}_{2} = \bl{R}_{\zeta_{1} + \zeta_{2}} \,,
\end{align}
we see that there is an element $e:=\bl{R}_{\zeta=0} \in SPin(\mathbb{V})$, called the identity element, such that $e \bl{R}_{\zeta} = \bl{R}_{\zeta} \,\forall\, \bl{R}_{\zeta} \in SPin(\mathbb{V})$; there is an element $\bl{R}_{\zeta}^{-1}:=\bl{R}_{-\zeta} \in SPin(\mathbb{V})$, called
the inverse of $\bl{R}_{\zeta} \in SPin(\mathbb{V})$, such that $\bl{R}_{\zeta}\bl{R}_{\zeta}^{-1} = e$; finally, the product is associative, namely $\bl{R}_{\zeta_{1}}(\bl{R}_{\zeta_{2}} \bl{R}_{\zeta_{3}}) = (\bl{R}_{\zeta_{1}}\bl{R}_{\zeta_{2}})\bl{R}_{\zeta_{3}} \,\forall\, \bl{R}_{\zeta_{1}},\bl{R}_{\zeta_{2}}, \bl{R}_{\zeta_{3}} \in SPin(\mathbb{V})$.

While a vector $\hat{\boldsymbol{e}}_{\mathfrak{m}}$ transform under rotations as \eqref{ROV}, the spinors $ \bl{\xi}_{A}$ transform as
follows:
\begin{equation}\label{ROS}
 \bl{\xi}_{A} \overset{\bl{R}_{\zeta}\,}{\longrightarrow} \bl{\xi}'_{A} = \bl{R}_{\zeta}\,\bl{\xi}_{A}\,.
\end{equation}
Indeed, it is simple matter to prove that the action of $\bl{R}_{\zeta}$ on spinor frame $\{\bl{\xi}_{A}\}$ is given by\footnote{This transformation must preserve an inner product defined on the spinor space $\mathbb{V}$. Such a product
is defined as $\left<\bl{\xi}_{A},\bl{\xi}_{B}\right> = \widetilde{\bl{\xi}_{A}}\bl{\xi}_{B}$, where the operation $\widetilde{\,\,}$ reverses the order of vectors in any product,
$\widetilde{\hat{\boldsymbol{e}}_{\mathfrak{m}} \wedge \hat{\boldsymbol{e}}_{\mathfrak{n}}} = - \hat{\boldsymbol{e}}_{\mathfrak{n}} \wedge \hat{\boldsymbol{e}}_{\mathfrak{m}}$. In particular, this means that $\bl{R}_{\zeta}^{-1} = \widetilde{\bl{R}_{\zeta}}$ . Using this, under the action of $\bl{R}_{\zeta}$, the following relation holds: $\left<\bl{R}_{\zeta}\,\bl{\xi}_{A},\bl{R}_{\zeta}\,\bl{\xi}_{B}\right> = \left<\bl{\xi}_{A},\bl{\xi}_{B}\right>$ , as should be. Hence, this inner product on the spinor space is invariant under the action of the spin group.}
\begin{equation}\label{AROSF}
\bl{R}_{\zeta}\,\bl{\xi}_{+} = e^{i\zeta/2}\,\bl{\xi}_{+} \quad\text{and}\quad \bl{R}_{\zeta}\,\bl{\xi}_{-} = e^{-i\zeta/2}\,\bl{\xi}_{-} \,. 
\end{equation}
%which is a rotation of $\zeta/2$ on the spinor space $\mathbb{S}$. 
Notice in particular that, when a rotation of  $\zeta = 2\pi$ is applied, vectors remain unchanged under the action of $\bl{R}_{2\pi} \in SPin(\mathbb{V})$, while the spinors are multiplied by $-1$ when $\bl{R}_{2\pi}$ acts on the spinor space $\mathbb{S}$.

\vspace{.5cm}
%\section{\uppercase{Ansatz for the Separation of Dirac Equation}}
\section{Ansatz for the Separation of Dirac Equation}
\vspace{.5cm}

In the previous chapters, we started to address the problem recalling that in order to integrate the field equation for spin-$s$ field perturbations\index{Spin-$s$ field perturbations} for $s = 0, 1, 2$, their angular dependence should be expanded in terms of an angular basis\index{Angular basis} that has the same nature. For spin-$0$ perturbations, those that transform as scalar fields\index{Scalar field} under rotations on the sphere, is convenient to expand their angular dependence in the basis $\{Y_{\ell_l m_l}\}$ inasmuch as the scalar spherical harmonics\index{Scalar spherical harmonics} $Y_{\ell_l m_l}$ are a basis for the functions in the sphere. For the spin-$1$ field perturbations, however, which have fields transforming as scalar fields\index{Scalar field} under rotation on the sphere and fields transforming as the components of $1$-forms with respect to the sphere, a suitable basis is then provided by $\{Y_{\ell_l m_l}, V^{\pm}_{\ell_l m_l}\}$. In the same vein, spin-$2$ field perturbations have components transforming as scalar fields, components transforming as the components of $1$-forms with respect to the sphere, and fields transforming as the components of a rank two symmetric tensor in the sphere. In this case, a suitable basis is given by $\{Y_{\ell_l m_l}, V^{\pm}_{\ell_l m_l}, T^{\oplus}_{\ell_l m_l},T^{\pm}_{\ell_l m_l}\}$. 
%Besides that, the elements of theses bases can be generated from the scalar spherical harmonics by repeated applications of the covariant derivative operator as it was shown in the previous chapters, namely
Besides that, the elements of these bases can be generated from the scalar spherical harmonics\index{Scalar spherical harmonics} by applications of the covariant derivative operator as it was shown in the previous chapters.
%\begin{equation}
%Y_{\ell_{l} m_l} \overset{\hat{\nabla}_{\mathfrak{m}}\,}{\longrightarrow }\,V^{\pm}_{\ell m}\,\overset{\hat{\nabla}_{\mathfrak{m}}\,}{\longrightarrow }\,T^{\oplus}_{\ell m},T^{\pm}_{\ell m}\,.
%\end{equation}
Is it possible to extend this procedure to spin-$1/2$ perturbations? Spin-$1/2$ perturbations
possess components that transform as the components of a spinor field\index{Spinor field} under rotation on the sphere. So, as a first step, we must extend the covariant derivative operator to be able to act on spinor fields\index{Spinor field}.

The covariant derivatives  $\hat{\nabla}_{\mathfrak{m}}$ of the frame vector fields\index{Vector field} $\hat{\bl{e}}_\mathfrak{m}$ determine components of the spin connection\label{CSC} $\hat{\omega}_{\mathfrak{mn}}^{\phantom{mn}p}$ by means of the following relation:
\begin{equation}
\hat{\nabla}_{\mathfrak{m}}\,\hat{\bl{e}}_\mathfrak{n} = \hat{\omega}_{\mathfrak{mn}}^{\phantom{mn}\mathfrak{p}}\,\hat{\bl{e}}_\mathfrak{p} \,.
\end{equation}
In particular, since the metric $\hat{\bl{g}}$ is a covariantly constant tensor, it follows that the
coefficients of the spin connection with all indices down $\hat{\omega}_{\mathfrak{mnp}} = \hat{\omega}_{\mathfrak{mn}}^{\phantom{mn}\mathfrak{q}}\delta_{\mathfrak{qp}}$ are antisymmetric
in their two last indices, $\hat{\omega}_{\mathfrak{mnp}} = -\hat{\omega}_{\mathfrak{mpn}}$. Indeed, the only components of the spin connection that are potentially nonvanishing are
\begin{equation}\label{NSPCS2}
\hat{\omega}_{212} = - \hat{\omega}_{221} = \text{cot}\theta_l \,.
\end{equation}
Notice that, however, the indices of the spin connection are raised and lowered with $\delta_{\mathfrak{mn}}$ and $\delta^{\mathfrak{mn}}$, respectively, so that frame indices can be raised and lowered unpunished. In particular, $ \hat{\omega}_{\mathfrak{m}}^{\phantom{m}\mathfrak{np}} =  \hat{\omega}_{\mathfrak{m}}^{\phantom{m}\mathfrak{[np]}}$, where indices inside the square brackets are antisymmetrized.
Then, one can show that the covariant derivative $\nabla_{\mathfrak{m}}$ must have the following action in
the spinor frame\footnote{Formally, in order to obtain Eq. \eqref{CDAOSF}, it is imposed for the covariant derivative to satisfy the Leibniz rule with respect to the Clifford action $\hat{\nabla}_{\mathfrak{m}}(\boldsymbol{V}\hat{\boldsymbol{\Psi}}) = (\hat{\nabla}_{\mathfrak{m}}\boldsymbol{V}) \hat{\boldsymbol{\Psi}} + \boldsymbol{V}(\hat{\nabla}_{\mathfrak{m}} \hat{\boldsymbol{\Psi}})\,\forall \,\boldsymbol{V} \in \mathbb{V} \,,\hat{\boldsymbol{\Psi}} \in \mathbb{S}$ and also to be compatible with the inner product on the spinor space, $\hat{\nabla}_{\mathfrak{m}}\left<\hat{\boldsymbol{\Psi}}, \hat{\boldsymbol{\Phi}}\right> = \left< \hat{\nabla}_{\mathfrak{m}}\hat{\boldsymbol{\Psi}}, \hat{\boldsymbol{\Phi}}\right> + \left<\hat{\boldsymbol{\Psi}}, \hat{\nabla}_{\mathfrak{m}}\hat{\boldsymbol{\Phi}}\right>\,\forall \,\hat{\boldsymbol{\Psi}},\hat{\boldsymbol{\Phi}} \in \mathbb{S}
 $.}
\begin{eqnarray}\label{CDAOSF}
\hat{\nabla}_{\mathfrak{m}}\,\boldsymbol{\xi}_{A} \,=\, (\hat{\Omega}_{\mathfrak{m}})^{B}_{\phantom{\beta}A}\boldsymbol{\xi}_{B}  \quad \text{where} \quad \hat{\Omega}_{\mathfrak{m}} = -\dfrac{1}{4}\,\hat{\omega}_{\mathfrak{m}}^{\phantom{m}\mathfrak{np}} \,\hat{\boldsymbol{e}}_{\mathfrak{n}}\hat{\boldsymbol{e}}_{\mathfrak{p}} \,.
\end{eqnarray}
The covariant derivative of a spinorial field $\hat{\boldsymbol{\Psi}} \in \mathbb{S}$ is, then, given by
\begin{equation}
\hat{\nabla}_{\mathfrak{m}}\hat{\boldsymbol{\Psi}} = \left(\partial_{\mathfrak{m}}  -\dfrac{1}{4}\,\hat{\omega}_{\mathfrak{m}}^{\phantom{m}\mathfrak{np}} \,\hat{\boldsymbol{e}}_{\mathfrak{n}}\hat{\boldsymbol{e}}_{\mathfrak{p}}  \right)\hat{\boldsymbol{\Psi}} \,,
\end{equation}
with $\partial_{\mathfrak{m}}$ denoting the partial derivative along the vector field\index{Vector field} $\hat{\boldsymbol{e}}_{\mathfrak{m}}$.

Now that we know how to act covariant derivatives on spinor fields\index{Spinor field}, we could use this
covariant derivative to build a class of functions defined on the sphere from its action on
scalar spherical harmonics\index{Scalar spherical harmonics}, exactly how we did in the previous chapters. And then, we
could use this class of functions as a basis in terms of which the spinor components will
be expanded. Note, however, that the covariant derivative operator $\nabla_{\mathfrak{m}}$ carries a vector index coming from $\hat{\boldsymbol{e}}_{\mathfrak{m}}$, meaning that its spinorial equivalent carries two spinorial indices, namely $\nabla^{A}_{\phantom{A}B}$, while the spinor components have only one spinorial index, namely $\Psi^{A}$. We would then need to find a way to obtain a covariant derivative with only one spinor index. In order to circumvent this limitation, instead of using the covariant derivative along $\hat{\boldsymbol{e}}_{\mathfrak{m}}$, it is more convenient for our purposes using the covariant along the null vector $\boldsymbol{m}_{A}$ which under the action of $\bl{R}_{\zeta} \in SPin(\mathbb{V})$ transforms as
\begin{equation}\label{TNVM}
 \boldsymbol{m}_{A} \overset{\bl{R}_{\zeta}\,}{\longrightarrow} \boldsymbol{m}'_{A} = e^{iA\zeta}\,\boldsymbol{m}_{A}\,.
\end{equation}
Using that the action of the vectors $\hat{\boldsymbol{e}}_{\mathfrak{m}}$ on the spinors $\bl{\xi}_{A}$ satisfy Eq. \eqref{CAOS} along with spin coefficients \eqref{NSPCS2}, we find by projecting the covariant the along the null vector $\boldsymbol{m}_{A}$, namely $\hat{\nabla}_{A} \hat{\bl{\Psi}} = (\nabla_{1} + iA\nabla_{2}\hat{\bl{\Psi}})$, that
\begin{equation}
\hat{\nabla}_{A} \hat{\bl{\Psi}} = \left(\hat{\nabla}_{A} \Psi^{B}\right) \bl{\xi}_{B} \,,
\end{equation}
with
%\begin{equation}\label{Comp1}
%\hat{\nabla}_{A} \Psi^{B} = \left(\partial_{\theta_l} + \dfrac{iA}{\sin\theta_l}\,\partial_{\phi_l} + %\dfrac{AB}{2} \cot\theta_l \right)\Psi^{B} \,,%
%\end{equation}
\begin{align}\label{Comp1}
\hat{\nabla}_{A} \Psi^{B} &=\, \left(\partial_{\theta_l} + \dfrac{iA}{\sin\theta_l}\,\partial_{\phi_l}  + \frac{AB}{2}\, \cot\theta_l \right)\Psi^{B} \nonumber\\
&=\, (\sin\theta_l)^{A s_l}\,\boldsymbol{m}_{A}\,(\sin\theta_l)^{-A s_l}\Psi^{B} \,,
\end{align}
where the null vectors $\bl{m}_{A}$ defined in Eq. \eqref{NVD} can be seen as differential operators that act on the space of the functions over $S^{2}$ and we have introduced the parameter $s_l = - B/2$. Under the action of the operator $\nabla_{A}$, the components $\Psi^{B}$ satisfying Eq. \eqref{Comp1} are said to have spin weight $s_l = - B/2$. This is an alternative way of characterizing spin weight quantities, very similar to what was done by Newman and Penrose\footnote{The differential operators $\eth$ and $\overline{\eth}$ acting on a quantity $Q$ of spin weight $s_l$ can be written in a more compact form if we denote the operator $\eth$ by $\eth_{+}$ and $\overline{\eth}$ by $\eth_{-}$. In a particular $\{\theta_l,\phi_l\}$ coordinate system, the latter operators are defined to satisfy:
\begin{align}\label{Comp2}
\eth_{A}Q &=\, \left[(\sin\theta_l)^{A s_l}\,\boldsymbol{m}_{A}\,(\sin\theta_l)^{-A s_l}\right ]Q\nonumber\\
&=\,\left(\partial_{\theta_l} + \dfrac{iA}{\sin\theta_l}\,\partial_{\phi_l} - A \,s_l \cot\theta_l \right)Q \,,\nonumber
\end{align}
where $s_l$ is so-called spin weight of the quantity $Q$ \cite{Goldberg1967}. By comparing equations \eqref{Comp1} with the above equation, it is immediate to conclude that the spinor
component $\Psi^{A}$ has spin weight $s_l = - A/2$.} by introducing first-order differential operators $\eth$ and $\overline{\eth}$ \cite{Goldberg1967,Campbell1971,Boyle2016}. 
%being the so-called spin weight of the quantity $\Psi^{B}$. 
In general, a quantity $Q$ defined on $S^{2}$ is said to have spin weight $s_l$ if, under the transformation \eqref{TNVM}, it transforms into  \cite{Goldberg1967,Campbell1971,Boyle2016}
\begin{equation}
Q \rightarrow Q' = e^{i s_l \zeta_l} \,Q \,.
\end{equation}
We should then check that the components $\Psi^{A}$ have in fact the expected spin weight under the transformation \eqref{TNVM}. In order to check this, let us use Eq. \eqref{AROSF} from which it follows   that
\begin{equation}
 \hat{\boldsymbol{\Psi}} \overset{\bl{R}_{\zeta}\,}{\longrightarrow} \hat{\boldsymbol{\Psi}}' =  \Psi'^{+}\boldsymbol{\xi}'_{+}  +  \Psi'^{-}\boldsymbol{\xi}'_{-}\,.
\end{equation}
where $\Psi^{A}$ is given by
\begin{equation}
\Psi'^{+} = e^{-i\zeta/2} \Psi^{+} \quad \text{and} \quad \Psi'^{-} = e^{i\zeta/2} \Psi^{-} \,,
\end{equation}
so that the spinor component $\Psi^{+}$ has in fact spin weight $s_l=-1/2$, while the spinor component $\Psi^{-}$ has spin weight $s_l=+1/2$ which is in perfect accordance with Eq. \eqref{Comp1}.

The great utility of using the covariant derivative operator $\hat{\nabla}_{A}$ instead of $\nabla_{\mathfrak{m}}$ is that the effect of $\hat{\nabla}_{A}$ on the so-called spin-$
\frac{A}{2}$ spherical harmonics\index{Spin-1/2 spherical harmonics}, usally denoted by $\,_{\frac{A}{2}}Y_{j_l,m_l}(\theta_l,\phi_l) $, is well known and any quantity with spin weight $s_l =-A/2$ can be expanded in a series in $\,_{\frac{A}{2}}Y_{j_l,m_l}(\theta_l,\phi_l)$. For a given set of half-integer parameters $\{j_l, m_l,s_l\}$,  spin-$s_l$ spherical harmonics\index{Spin-1/2 spherical harmonics} can be defined by means of the following equation \cite{Goldberg1967,Campbell1971,Boyle2016}\label{SSSH}:
\begin{align}\label{EOAOSH}
\hat{\nabla}_{A}\left(_{s_l}Y_{j_l,m_l}\right) &=\, A \sqrt{j_l(j_l+1) - s_l(s_l+A)} \,_{s_l+A}Y_{j_l,m_l} \quad  \text{for}\quad  j_l \in \{1/2,3/2,5/2 \ldots \} \,,
\end{align}
from which we see that if $_{s_l}Y_{j_l,m_l}$ has spin weight $s_l$, then $\hat{\nabla}_{A}\left(_{s_l}Y_{j_l,m_l}\right)$ has spin weight $s_l +A$. Notice that $_{s_l}Y_{j_l,m_l}$ are not defined for $|s_l| > j_l$. Indeed, it is straightforward to see that $\hat{\nabla}_{A}$ annihilates $_{j_l}Y_{j_l,m_l}$
for $A = +1$ while $\hat{\nabla}_{A}$ annihilates $_{-j_l}Y_{j_l,m_l}$
for $A = -1$. Spin-$s_l$ spherical harmonics\index{Spin-1/2 spherical harmonics} form a complete orthogonal set for each value of $s_l$ satisfying the condition $|s_l| \leq j_l$, that is, they define a basis in terms of which any function with spin weight $s_l$ can be expanded in a series in $_{s_l}Y_{j_l,m_l}$.

Taking into account that the spinor components $\Psi^{A}$ of a spinor field\index{Spinor field} $\hat{\bl{\Psi}}$ possess
components with half-integer spin weight $s_l=-A/2$, it is handy to expand them in terms of spin-$\frac{A}{2}$ spherical harmonics\index{Spin-1/2 spherical harmonics} $_{\frac{A}{2}}Y_{\ell_l,m_l}$, so that the natural ansatz for spinor fields\index{Spinor field}
on the two-sphere is given by
\begin{equation}
\hat{\bl{\Psi}}(\theta_l,\phi_l) = \sum_{A} \Psi_{A}(\theta_l,\phi_l) \bl{\xi}_{A} \quad \text{with}\quad  \Psi_{A}(\theta_l,\phi_l) = \sum_{j_l,m_l}c_{j_l,m_l}\,{ _{-\frac{A}{2}}Y_{j_l,m_l}(\theta_l,\phi_l)}\,,
\end{equation}
%where for the spinor index $A$ no summation convention has been employed, reason why the sum over $A$ has been explicited. 
Now, a Dirac spinor is a eigenfunction of the Dirac operator $\hat{\bl{D}}_l = \delta^{\mathfrak{mn}}\hat{\boldsymbol{e}}_{\mathfrak{m}}\hat{\nabla}_{\mathfrak{n}} $ with eigenvalue $\lambda_l$ , namely
\begin{equation}
\hat{\bl{D}}_l \hat{\bl{\Psi}} = \lambda_{l} \, \hat{\bl{\Psi}} \,.
\end{equation}
So, since that the action of the vectors $\hat{\boldsymbol{e}}_{\mathfrak{m}}$ on the spinors $\bl{\xi}_{A}$ satisfy Eq. \eqref{CAOS}, it follows that the above Dirac equation in terms of $ _{\frac{A}{2}}Y_{j_l,m_l}(\theta_l,\phi_l)$ is written as:
\begin{equation}\label{DETSSH}
\hat{\nabla}_{A}\left( _{-\frac{A}{2}}Y_{j_l,m_l}\right) = \lambda_l\, { _{\frac{A}{2}}Y_{j_l,m_l}} \,.
\end{equation}
In order to obtain the above expression, we have changed the index $A$ to $-A$ in the sum.
Once that the sum over $A$ runs over all values of the set $\{+,-\}$, which comprise the
same list of the values of $-A$, the final result remains unchanged. Using Eq. \eqref{EOAOSH}, we conclude that the Dirac equation \eqref{DETSSH} admits regular analytical solutions only when the eigenvalues are nonzero integers \cite{Camporesi1996,Trautman1993}
\begin{equation}\label{EVDEOS}
\lambda_l = A \left|j_l + \dfrac{1}{2}\right|  = \pm 1, \pm 2, \pm 3, \ldots \quad \text{for} \quad j_l \in \{ 1/2, 3/2,5/2 \ldots\} \,.
\end{equation}

%*******************
%(sobre mudar a base)
%*******************

Now, the ansatz for the spinor field\index{Spinor field} can be naturally generalized to higher dimensions.
In order to perform this, the first step consists of introducing a suitable orthonormal
frame of vector fields\index{Vector field}. In the problem considered in the present book, the background
is the direct product of $dS_2$ with two-spheres, so that we have spherical symmetry in
each of these two-spheres. A suitable orthonormal frame ${\bl{e}_\alpha}$ for this space, with
$\alpha = 1, 2, \ldots, D=2d$, is then given by
\begin{eqnarray}
\boldsymbol{e}_{1} \,=\,-i\,\text{cosh}(x/R_{1})\,\partial_{t}  & , &  \boldsymbol{e}_{l} \,=\,\frac{1}{R_{l}\,\text{sin}\,\theta_{l}}\,\partial_{\phi_{l}} \,,\nonumber\\
\boldsymbol{e}_{\tilde{1}}\,=\,\text{cosh}(x/R_{1})\,\partial_{x}  & , &  \boldsymbol{e}_{\tilde{l}} \,=\,\frac{1}{R_{l}}\,\partial_{\theta_{l}} \,,
\end{eqnarray}
where the index $l$ ranges from $2$ to $d$. Since $\{\boldsymbol{e}_{\alpha}\}$ is orthonormal, the components of the metric $\bl{g}$ in this frame are constants. With this notation, we have in particular that 
\begin{equation}
\boldsymbol{g}(\boldsymbol{e}_{\alpha}, \boldsymbol{e}_{\beta}) \,=\,\delta_{\alpha \beta} \quad \leftrightarrow \quad \left\{\begin{matrix}
\boldsymbol{g}(\boldsymbol{e}_{a}, \boldsymbol{e}_{b})  & = & \delta_{ab} \,, \\ 
\boldsymbol{g}(\boldsymbol{e}_{a}, \boldsymbol{e}_{\tilde{b}})  & = & 0 \,,\\ 
\boldsymbol{g}(\boldsymbol{e}_{\tilde{a}}, \boldsymbol{e}_{\tilde{b}})  & = & \delta_{\tilde{a}\tilde{b}} \,,
\end{matrix}\right. 
\end{equation} 
where $a$ and $\tilde{a}$ are indices that range from $1$ to $d$. The index $a$, for instance, is only a label for the first $d$ vector fields\index{Vector field} of the orthonormal frame $\{\boldsymbol{e}_{a} \}$, while the index $\tilde{a}$ is a label for the remaining $d$ vectors of the frame $\{\boldsymbol{e}_{\tilde{a}} \}$. These vector fields\index{Vector field} can be represented  by Dirac matrices\index{Dirac matrices} ${\Gamma_{\alpha}}$ which in $D = 2d$ dimensions represent faithfully the Clifford algebra\index{Clifford algebra} by $2^d \times 2^d$ matrices obeying the following relation:
\begin{equation}\label{CAMF}
\Gamma_{\alpha}\,\Gamma_{\beta} + \Gamma_{\beta}\,\Gamma_{\alpha} = 2\,\boldsymbol{g}(\boldsymbol{e}_{\alpha}, \boldsymbol{e}_{\beta})\,\mathbb{I}_{d} \,,
\end{equation}
with $\mathbb{I}_{d}$ standing for the $2^{d} \times 2^{d}$ identity matrix\label{Metric}. 
%Indeed, since in two dimensions the vector fields $\hat{\bl{e}}_{1}$ and $\hat{\bl{e}}_{1}$ admit a matrix representation by Pauli matrices $\sigma_1,\sigma_2$ and $\sigma_3=-i\sigma_1 \sigma_2$, 
In order to accomplish the separability\index{Separability} of the Dirac equation, it is necessary to use a suitable representation for the Dirac matrices\index{Dirac matrices}. In what follows, the $2 \times 2$ identity
matrix will be denoted by $\mathbb{I}$, while the usual notation for the Pauli matrices is going to be adopted: 
\begin{equation}
 \sigma_1\,=\,\left[
           \begin{array}{cc}
             0 & 1 \\
             1 & 0 \\
           \end{array}
         \right] \quad , \quad\sigma_2\,=\,\left[
                                 \begin{array}{cc}
                                   0 & -i \\
                                   i & 0 \\
                                 \end{array}
                               \right]  \quad , \quad \sigma_3\,=\, \left[
                                 \begin{array}{cc}
                                   1 & 0 \\
                                   0 & -1 \\
                                 \end{array}
                               \right]\,.
 \end{equation} 
Using this notation, a convenient representation for the Dirac matrices\index{Dirac matrices} is the following:
\begin{align}\label{dirac.matrices.rep}
\Gamma_{a} &=\, \underbrace{\sigma_3\otimes\ldots\otimes\sigma_3}_{(a-1)\,\,\text{times}}\otimes\,\sigma_1 \otimes  \underbrace{\mathbb{I}\otimes\ldots \otimes \mathbb{I} }_{(d-a) \,\,\text{times}	} \,,\nonumber\\
\Gamma_{\tilde{a}} &=\, \underbrace{\sigma_3\otimes\ldots\otimes\sigma_3}_{(a-1)\,\,\text{times}}\otimes\,\sigma_2 \otimes  \underbrace{\mathbb{I}\otimes\ldots \otimes \mathbb{I} }_{(d-a) \,\,\text{times}	}
  \,.
\end{align}
%where $\mathbb{I}$ stands for the $2 \times 2$ identity matrix. 
Indeed, we can easily check that the Clifford algebra\index{Clifford algebra} given in equation \eqref{CAMF} is properly satisfied by the above matrices\footnote{In $D = 2d + 1$, besides the $2d$ Dirac matrices\index{Dirac matrices} $\Gamma_{a}$ and $\Gamma_{\tilde{a}}$ we need to add one further matrix, which will be denoted by $\Gamma_{d+1}$ given by $\Gamma_{d+1} = \underbrace{\sigma_{3}\otimes \sigma_{3} \ldots \otimes \sigma_{3}}_{d \,\, \text{times}}$.}. In this case, spinorial fields are represented by the column vectors with $d$ components on which these matrices act.  So, since the spinors $\{\bl{\xi}_{A}\}$
can be represented by the column vectors $\{\xi_{A}\}$, if we introduce a spinor index $A_{a}$ which can take the values $``+1''$ and $``-1''$, then a basis in $D = 2d$ dimensions for the spinor space is spanned by the direct product of the elements $\xi_{A_a}$ d times, namely $\xi_{A_1} \otimes \xi_{A_2} \otimes \ldots \xi_{A_d}$. Once the base is defined, any spinor field\index{Spinor field} can be represented on this basis as
\begin{equation}\label{SpinorFieldD}
\Psi \,=\,\sum_{A}\Psi_{A_{1}A_{2}\dots A_{d}}\,{\xi}_{A_{1}}\otimes {\xi}_{A_{2}} \otimes \dots \otimes {\xi}_{A_{d}}\,.
\end{equation} 

Since each of the indices $A_{a}$ can take just two values, it follows the sum over $\{A\} \equiv \{A_{1}, A_{2}, \ldots, A_{d}\}$ comprises $2^{d}$ terms, which is exactly the number of degrees of freedom of a spinorial field in $D = 2d$ dimensions. The components $\Psi_{A_{1}A_{2}\dots A_{d}}$ transform as the components of a spin-$1/2$ field under rotation, therefore, their angular dependence should be given by the product of spin-$A/2$ spherical harmonics\index{Spin-1/2 spherical harmonics} $ _{\frac{A}{2}}Y_{\ell_l,m_l}$, something very similar to what we did for the scalar, Maxwell and gravitational fields\index{Gravitational field}. Thus, the ansatz for the spinor components which is in agreement with the symmetries of the background is 
\begin{equation}\label{AFSDE}
\Psi_{A_{1}A_{2}\dots A_{d}} = \sum_{j,m} \Psi_{A_1}^{jm}(t,x)\,  { _{\frac{\breve{A}}{2}}\mathcal{Y}_{jm}} \,,
\end{equation}
where
\begin{equation}
_{\frac{\breve{A}}{2}}\mathcal{Y}_{jm} = \prod_{l=2}^{d} { _{\frac{A_l}{2}}Y_{j_l,m_l}(\theta_l,\phi_l)} \,.
\end{equation}
The sum over the collective index $\{j, m\}$ means that we are summing over all values of
the set $\{j_2, m_2,j_3,m_3,\ldots j_d,m_d\}$ while the collective index $\{\breve{A}\}$ means all values of the set $\{A_2,A_3,\ldots A_d\}$.

\vspace{.5cm}
%\section{\uppercase{Separability of Dirac's Equation}}
\section{Separability of Dirac's Equation}
\vspace{.5cm}
%\begin{equation}
%\gamma^{\alpha}\nabla_{\alpha} \boldsymbol{\Psi} = \mu \boldsymbol{\Psi}
%\end{equation}

All that was have seen above are necessary tools to attack our problem of separating the
Dirac equation in generalized Nariai background. A spin-1/2 field $\Psi$, as defined in Eq.
\eqref{SpinorFieldD}, with electric charge $q$ and mass $\mu$ propagating in such background is a spinor
field obeying the following version of the Dirac equation:
\begin{equation}
\not{D}\Psi = \mu \Psi \,,
\end{equation}
with the operator $\not{D}$ being the Dirac operator minimally coupled to the components of
the background gauge field $\mathcal{A}^{GN}_{\alpha}$, namely
\begin{equation}\label{DiracEMF}
\not{D} = \Gamma^\alpha(\nabla_\alpha-iq\mathcal{A}_\alpha^{GN})\,,
\end{equation}
where the operator $\nabla_\alpha$ stand for the spinor covariant derivative whose action on the a
spinor field\index{Spinor field} $\Psi$ is represented by
\begin{equation}
\nabla_\alpha \Psi =\left(\partial_\alpha-\frac{1}{4}\,\omega_\alpha^{\phantom{\alpha}\beta\gamma}\,\Gamma_\beta\Gamma_\gamma\right)\Psi \,,
\end{equation}
with $\partial_\alpha$ denoting the partial derivative along the vector field\index{Vector field} $\mathbf{e}_\alpha$. 

Remember that, the components of the spin connection $\omega_\alpha^{\phantom{\alpha}\beta\gamma}$ which satisfy the anti-symmetry property $\omega_{\alpha\beta\gamma}=-\omega_{\alpha\gamma\beta}$, are determined from the action of the covariant derivative operator $\nabla_\alpha$ on the frame of vector fields\index{Vector field} $\bl{e}_\alpha$, namely $\nabla_\alpha\bl{e}_\beta=\omega_\alpha^{\phantom{\alpha}\beta\gamma}\bl{e}_\gamma$. By
doing this, we find that the only components of the spin connection that are potentially
nonvanishing are
\begin{align}
\omega_{1\tilde{1}1} &= -\,\omega_{11\tilde{1}}=-\dfrac{1}{R_1}\,\sinh(x/R_1)\,,\label{omega1}\\
\omega_{l\tilde{l}l}&= -\,\omega_{ll\tilde{l}}=\dfrac{1}{R_l}\,\text{cot}\theta_l \label{omega2}\,.
\end{align}
To solve the Dirac equation, we need to separate the degrees of freedom of the field which can be quite challenging in general. However, introducing the dual frame of $1$-forms $\{\bl{E}^\alpha\}$, defined to be such that its action on $\bl{e}_\alpha$ is $\bl{E}^\alpha(\bl{e}_\beta)=\delta_\beta^\alpha$, namely
\begin{align}
\bl{E}^1 &=\,\dfrac{i}{\cosh(x/R_1)}\,dt\quad ,\quad \bl{E}^l = R_l \sin\theta_l d\phi_l,\nonumber\\
\bl{E}^{\tilde{1}} &=\, \dfrac{1}{\cosh(x/R_1)}\,dx \quad,\quad \bl{E}^{\tilde{l}} = R_l\,d\theta_l\,,
\end{align}
we see that the line element and the background gauge field $\mathcal{A}_\mu^{GN}$ can be written as:
\begin{align}
g_{\mu\nu}^{GN}dx^\mu dx^\nu &=\,\sum\limits_{a=1}^d\left(\bl{E}^a\bl{E}^a+\bl{E}^{\tilde{a}}\bl{E}^{\tilde{a}}\right)\\
\mathcal{A}_\mu^{GN}dx^\mu &=\,\sum\limits_{a=1}^d\left(\mathcal{A}_a^{GN}\bl{E}^a+\mathcal{A}_{\tilde{a}}^{GN}\bl{E}^{\tilde{a}}\right)
\end{align}
where the components $\mathcal{A}_a^{GN}$ and $\mathcal{A}_{\tilde{a}}^{GN}$ in the considered frame are given by
\begin{equation}\label{NVGaugField}
\mathcal{A}_1^{GN}=-iQ_1R_1\sinh(x/R_1)\quad,\quad\mathcal{A}_j^{GN}=Q_lR_l\cot\theta_l\quad,\quad\mathcal{A}_{\tilde{a}}^{GN}=0 \,.
\end{equation}
We should note that the background fields $g^{GN}_{\mu\nu}$, $\mathcal{A}_\mu^{GN}$ belong exactly the class of solutions $\mu$ studied in Ref. \cite{Venancio2017}. Indeed, our main goal in this reference is to show that the Dirac equation minimally coupled to a background gauge field is separable in backgrounds that are the direct product of bidimensional spaces. So, let us now present the key points of this latter separation.

The spinor basis in terms of the elements $\xi_{A_a}$ introduced previously is very convenient, since the action of the Dirac matrices\index{Dirac matrices} on the spinor fields\index{Spinor field} can be easily comptuted. Indeed, using the equations \eqref{dirac.matrices.rep} and \eqref{SpinorFieldD} along with the fact that the action of the Pauli matrices on the column vectors $\xi_{A_a}$ satisfy the relations
\begin{equation}
\sigma_1\xi_{A_a}=\xi_{-A_a}\quad,\quad\sigma_2\xi_{A_a}=iA_a\xi_{-A_a}\quad,\quad\sigma_3\xi_{A_a}=A_a\xi_{A_a} \,,
\end{equation}
we eventually arrive at the following equation
\begin{align}\label{action1}
\Gamma_a\Psi &=\,\sum_A(A_1A_2...A_{a-1})\Psi_{A_1A_2...A_d}\xi_{A_1}\,\otimes\xi_{A_2}\otimes\ldots\otimes\xi_{A_{a-1}}\otimes\xi_{-A_a}\otimes\xi_{A_{a+1}}\otimes\nonumber\\
&\ldots\, \otimes\xi_{A_d}=\sum_A(A_1A_2\ldots A_a)A_a\Psi_{A_1A_2\ldots A_{a-1}(-A)A_{a+1}\ldots A_d}\xi_{A_1}\,\otimes\xi_{A_2}\otimes\nonumber\\
&\ldots\, \otimes\xi_{A_{a-1}}\otimes\xi_{A_a}\otimes\xi_{A_{a+1}}\otimes\ldots\otimes\xi_{A_d},
\end{align}
where from the first to the second line we have changed the index $A_a$ to $-A_a$, which does not change the final result, since we are summing over all values of $A_a$, which comprise the same list of the values of $-A_a$. Moreover, we have used that $(A_a)^2=1$. Analogously, we have:
\begin{align}\label{action2}
\Gamma_{\tilde{a}}\Psi&=\sum_A(A_1A_2...A_{a-1})(iA_a)\Psi_{A_1A_2...A_d}\,\xi_{A_1}\otimes\xi_{A_2}\otimes...\otimes\xi_{A_{a-1}}\otimes\xi_{-A_a}\otimes\xi_{A_{a+1}}\otimes\nonumber\\
&...\ \otimes\xi_{A_d}=-i\sum_A(A_1A_2...A_a)\Psi_{A_1A_2...A_{a-1}(-A_a)A_{a+1}...A_d}\,\xi_{A_1}\otimes\xi_{A_2}\otimes\nonumber\\
&..\ \otimes\xi_{A_{a-1}}\otimes\xi_{A_a}\otimes\xi_{A_{a+1}}\otimes...\otimes\xi_{A_d}\,.
\end{align}
In order to accomplish the separation of the general equation \eqref{DiracEMF}, let us assume the decomposition of the spinor components defined in Eq. \eqref{AFSDE}. From this important decomposition which is crucial in order to attain the integrability of the Dirac equation and with Eqs. \eqref{action1} and \eqref{action2} in hand, after some careful algebra, one can show that the component $\Psi_{A_1}^{jm}$ obeys the following differential equation (the reader is invited to demonstrate the two equations below or consult more details in Ref. \cite{Venancio2017}):
\begin{align}\label{Separa1}
\left[\partial_{\tilde{1}}+\frac{\omega_{1\tilde{1}1}}{2}-iq\mathcal{A}_{\tilde{1}}^{GN} -\,iA_1\left(\partial_1+\frac{\omega_{\tilde{1}1\tilde{1}}}{2}-iq\mathcal{A}_1^{GN}\right)\right]\Psi_{A_1}^{jm} \nonumber\\
=\,(c_1-iA_1\mu)\Psi_{-A_1}^{jm}&\,,
\end{align}
where the parameter $c_1$ appearing in the latter equation is part of a set of $(d-1)$
separation constants, namely $\{c_1,c_2,..., c_{d-1}\}$, determined by the angular part of the Dirac equation, in which each of the angular components $ _{\frac{A_l}{2}}Y_{j_l,m_l}$ satisfies the following differential equation:
\begin{align}\label{Separa2}
\left[\partial_{\tilde{l}}+\frac{\omega_{l\tilde{l}l}}{2}-iq\mathcal{A}_{\tilde{l}}^{GN}-iA_l\left(\partial_l+\frac{\omega_{\tilde{l}l\tilde{l}}}{2}-iq\mathcal{A}_{l}^{GN}\right)\right]{ _{\frac{A_l}{2}}Y_{j_l,m_l}} \nonumber\\
=(c_l + A_1 c_{l-1}){ _{-\frac{A_l}{2}}Y_{j_l,m_l}}&\,,
\end{align}
where $\omega_{\alpha\beta\gamma}$ and $\mathcal{A}_\alpha^{GN}$ are the components of the spin connection and the background gauge field, respectively, thus achieving the separability\index{Separability} that we were looking for.

Since spin-$s_l$ spherical harmonics\index{Spin-1/2 spherical harmonics} satisfy regularity requirements on the sphere, namely at the points $\theta_l=0$ and $\theta_l=\pi$, where our coordinate system breaks down, the separation constants $\{c_1,c_2,...,c_{d-1} \}$ should only take discrete values. In particular, we will see that these separation constants are exactly the eigenvalues of the Dirac operator on sphere under certain conditions. Since we have been able to separate the generalized Dirac equation into the Eqs. \eqref{Separa1} and \eqref{Separa2}, we now shall investigate a little further these equations. In order to study the QNMs, we need to transform the first order differential equation \eqref{Separa1} into a Schr\"{o}dinger differential equation\index{Schr\"{o}dinger-like differential equation} for $\Psi_{A_1}^{jm}$ which, in turn, carries information of the angular part $\ _{\frac{A_l}{2}}Y_{j_l,m_l}$ through the separation constant $c_1$. So, before proceeding we first need to explicitly determine this separation constant.

\vspace{.5cm}
%\subsection{The Angular Part of the Dirac Equation}
\subsection{The Angular Part of the Dirac Equation}
\vspace{.5cm}

Let us work out the angular part of the Dirac equation, namely, the equation for $_{\frac{A_l}{2}}Y_{j_l,m_l}$. The frame considered here, the only angular components of $\omega_{\alpha\beta\gamma}$, $\mathcal{A}_\alpha^{GN}$ that can be nonvanishing according to the Eqs. \eqref{omega2} and \eqref{NVGaugField} are
\begin{equation}
\omega_{l\tilde{l}l}=-\omega_{l\tilde{l}l}=\frac{1}{R_l}\cot\theta_l\quad \text{and}\quad\mathcal{A}_l^{GN}=Q_lR_l\cot\theta_l,.
\end{equation}
Then, by inserting theses expressions into Eq. \eqref{Separa2} we are left with the following differential equation:
\begin{equation}\label{Separa3}
\left(\nabla_{A_l}+A_lqQ_lR_l^2\cot\theta_l\right){_{-\frac{A_l}{2}}Y_{j_l,m_l}} = R_l(A_l\,c_{l-1}-c_l)_{\frac{A_l}{2}}Y_{j_l,m_l} \,.
\end{equation}
In order to write this equation in a more convenient form, instead of using the $(d-1)$
separation constants $\{c_1,c_2,...,c_{d-1}\}$, let us introduce the parameters $\{\lambda_2,\lambda_3,...,\lambda_d\}$ defined by
\begin{equation}
\frac{\lambda_l}{R_l}\equiv \sqrt{c_{l-1}^2-c_l^2}\quad\text{and}\quad c_d=0.
\end{equation}
So, inverting these expressions, we can prove that the parameters $\{c_l\}$ are related to the parameters $\{\lambda_l\}$ by the following identity
\begin{equation}\label{ConstanteS}
c_{l-1}=\sqrt{\frac{\lambda_l^2}{R_l^2}+\frac{\lambda_{l+1}^2}{R_{l+1}^2}+\ldots+\frac{\lambda_d^2}{R_d^2}}\,.
\end{equation}
%An advantage of this in order to simplify the equation for the field ${_{\frac{A_l}{2}}Y_{j_l,m_l}}$ is that 
Next, if we introduce the parameters
\begin{equation}
\varphi_l=\text{arctanh}(c_l/c_{l-1})\,,
\end{equation}
it is a simple matter to prove the identities
\begin{equation}
c_l=\frac{\lambda_l}{R_l}\sinh\varphi_l\quad\text{and}\quad c_{l-1}=\frac{\lambda_l}{R_l}\cosh\varphi_l\,,
\end{equation}
so that, assuming the latter relations, the multiplicative factor of ${_{\frac{A_l}{2}}Y_{j_l,m_l}}$ on the right-hand side of Eq. \eqref{Separa3} can be written in terms of $\lambda_l$ and $\varphi_l$ as follows:
\begin{equation}
R_l\left(A_lc_{l-1}-c_l\right)=A_l\lambda_le^{-A_l\varphi_l}\,.
\end{equation}
Inserting the above expression into Eq. \eqref{Separa3}, we are left with the following differential
equation:
\begin{equation}\label{Separa20}
\left(\nabla_{A_l}+A_lqQ_lR_l^2\cot\theta_l\right)\left(e^{A_l\varphi_l/2}{_{-\frac{A_l}{2}}Y_{j_l,m_l}}\right)=A_l\lambda_l\left(e^{-A_l\varphi_l/2}{_{\frac{A_l}{2}}Y_{j_l,m_l}}\right)\,.
\end{equation}
In addition to this change of parameters, if we perform yet a field redefinition, we can obtain a differential equation that is independent of the parameters $\varphi_l$. Indeed, performing the field redefinition
\begin{equation}
{_{\frac{A_l}{2}}\mathbb{Y}_{j_l,m_l}} := e^{-A_l\varphi_l/2}\,{_{\frac{A_l}{2}}Y_{j_l,m_l}}\,,
\end{equation}
Eq. \eqref{Separa20} for ${_{\frac{A_l}{2}}Y_{j_l,m_l}}$ acquires the following form:
\begin{equation}\label{Separa4}
\left(\nabla_{A_l}+A_lqQ_lR_l^2\cot\theta_l\right){_{-\frac{A_l}{2}}\mathbb{Y}_{j_l,m_l}}=A_l\lambda_l\,{_{\frac{A_l}{2}}\mathbb{Y}_{j_l,m_l}}\,.
\end{equation}
Notice that, inasmuch as $e^{-A_l\varphi_l/2}$ is just a constant multiplicative factor, ${_{\frac{A_l}{2}}\mathbb{Y}_{j_l,m_l}}$ possess the same properties satisfied by
${_{\frac{A_l}{2}}Y_{j_l,m_l}}$. In particular, ${_{\frac{A_l}{2}}\mathbb{Y}_{j_l,m_l}}$ is also a spin-$A/2$ spherical harmonic and therefore, obeys Eq. \eqref{EOAOSH}.

The great advantage of using $\{\lambda_l,{_{\frac{A_l}{2}}\mathbb{Y}_{j_l,m_l}}\}$ instead of $\{c_l,{_{\frac{A_l}{2}}{Y}_{j_l,m_l}}\}$ shows up when the magnetic charges of the background vanish, $Q_l=0$. In this case, the equation for ${_{\frac{A_l}{2}}{\mathbb{Y}}_{j_l,m_l}}$ reduces to
\begin{equation}
\nabla_{A_l}\left({_{-\frac{A_l}{2}}\mathbb{Y}_{j_l,m_l}}\right)=A_l\lambda_l\,{_{\frac{A_l}{2}}\mathbb{Y}_{j_l,m_l}} \,,
\end{equation}
which is exactly the Dirac equation at the $l$th two-dimensional unit sphere whose eigenvalues are known, namely \eqref{EVDEOS}. Indeed, spin-$1/2$ spherical harmonics\index{Spin-1/2 spherical harmonics} admit regular analytical function on sphere only when the eigenvalues $\lambda_l$ are nonzero integers [125, 126]
\begin{equation}
\lambda_l=\pm1,\pm2,\pm3,\ldots .
\end{equation}

It is worth stressing that the parameter $c_1$ is the only separation constant showing up in the equation for $\Psi_{A_1}^{jm}$. In particular, according to Eq. \eqref{ConstanteS}, this separation constant is related to the eigenvalues of the Dirac equation by 
\begin{equation}\label{ctec1}
c_1=\sqrt{\sum_{l=2}^d\frac{\lambda_l^2}{R_l^2}}\,.
\end{equation}

Since the case $Q_l=0$ in Eq. \eqref{Separa4} has a known solution, as described above, it follows that we can look for solutions for the case $Q_l\neq 0$ by means of perturbation methods, with $Q_l$ being
the perturbation parameter. Indeed, in the celebrated paper \cite{Press1973}, a similar path has been taken by Press and Teukolsky in order to find the solutions and their eigenvalues for the angular part of the equations of motion for fields with arbitrary spin on Kerr spacetime, in which case the angular momentum of the black hole was the order parameter. In this respect, see also Ref. \cite{Chakrabarti1984}.

%\subsection{\textit{The Radial Part of the Dirac Equation}}

%\subsection{\textit{The Radial Part of the Dirac Equation}}

Concerning the differential equation for the fields $\Psi^{jm}_{A_1}$, the only nonzero radial compo-
nents of the spin coefficients and the background electromagnetic field in the considered
frame are given by
\begin{align}
  \omega_{1\tilde{1}1} =  -\,\omega_{11\tilde{1}}= - \frac{1}{R_{1}}\,\text{sinh}(x/R_1) \quad \text{and} \quad \mathcal{A}^{GN}_{1} =  -i Q_{1}R_{1}\,\text{sinh}(x/R_1) \,.
\end{align}
Inasmuch as the coefficients in the equation for $\Psi^{jm}_{A_1}$ do not depend on the coordinates $t$ which stems from the fact that $\partial_t$ is a Killing vector fields\index{Vector field} of our metric, we
can expand the time dependence of $\Psi^{jm}_{A_1}$ in the Fourier basis\index{Fourier basis},
\begin{equation}
\Psi^{jm}_{A_1}(t, x)\,=\, e^{-i\omega t}\,\psi^{\omega}_{A_1,jm}(x) \,.
\end{equation}
Here, we are omitting the integral over all values of the Fourier frequencies $\omega$ for notational simplicity. 
%While ? j are related to angular momentum of the eld, ? can be interpreted as related to the energy. 
It follows that the field equation \eqref{Separa1} yields
\begin{align}
 \bigg[    \frac{d}{dx}  + i A_{1} \omega   + \left( i  A_{1} q  Q_{1}R_{1} -  \frac{1}{2R_{1}}  \right) \text{tanh}(x/R_1)  \bigg] \psi^{\omega}_{A_1,jm}    \nonumber \\
   = \frac{(c_1 -iA_{1}\,\mu )}{\cosh(x/R_1)}\,\psi^{\omega}_{-A_1,jm} &\,.  \nonumber
\end{align}
Notice that these first order differential equations are coupled, namely, the spinor component $\psi^{\omega}_{+,jm}$ is a source for component $\psi^{\omega}_{-,jm}$ and vice versa. Eliminating, for instance, $\psi^{\omega}_{-,jm}$, give us  a second order differential equation for component  $\psi^{\omega}_{+,jm}$.  By doing this, after some algebra, we
are left with the following Schr\"{o}dinger-like differential equation\index{Schr\"{o}dinger-like differential equation} for the field $\psi^{\omega}_{A_1,jm}$
\begin{equation}\label{WED}
\left[  \frac{d^{2}}{dx^{2}} \,+\,\omega^{2} \,-\, V_{s=1/2}(x)
 \right ]\psi^{\omega}_{A_1,jm}\,=\,0 \,,
\end{equation}
where the potential $V_{s=1/2}(x)$ is the one considered in Eq. (\ref{Potential_Generic}) with the parameters $\mathfrak{a}$, $\mathfrak{b},\mathfrak{c}$ and $\mathfrak{d}$ given by
\begin{align}
  \mathfrak{a} &\,=\,\frac{1}{4 R_1^2} -q\,Q_{1} (i A_1 +q\,Q_{1} \, R_1^2 )\,,  \nonumber \\
       \mathfrak{b} &\,=\, -\frac{\omega}{R_1}\, (iA_{1} + 2\,q \,Q_{1} \, R_1^2) \,,  \label{ABC_Spin}\\
       \mathfrak{c} &\,=\,\mu^{2} + \sum_{l=2}^d\frac{\lambda_l^2}{R_l^2} +  \frac{1}{4 R_1^2} + q^{2}\,Q^{2}_{1}  R_1^2 \,,\nonumber\\
       \mathfrak{d} &\,=\,\frac{1}{R_1} \nonumber\,.
\end{align}
The expression for constant separation $c_1$ has been used, namely Eq. \eqref{ctec1}.  These are
known as potentials of Rosen-Morse\index{Rosen-Morse class of potentials} type, which are generalizations of the P\"{o}schl-Teller potential\index{P\"{o}schl-Teller potential} \cite{Dutt1988, Poschl1933}. It is straightforward see that this potential satisfies the following
properties:
\begin{equation}\label{Prop1SF}
\left. V_{s=1/2}\right|_{x\rightarrow+\infty} \rightarrow \mathfrak{a} + \mathfrak{b} \quad \text{and} \quad \left. V_{s=1/2}\right|_{x\rightarrow-\infty} \rightarrow \mathfrak{a} - \mathfrak{b} \,.
\end{equation}
In general, the potential function $V_{s=1/2}$ is assumed to be regular at $r=0$ ($x=0$), in
particular it can be equal to a constant different from zero. In our case, we find that
\begin{equation}\label{Prop2SF}
\left. V_{s=1/2}\right|_{x\rightarrow 0} \rightarrow \mathfrak{a} + \mathfrak{c} \,,
\end{equation}
which clearly is regular. So, we point out that for this potential both limits \eqref{Prop1SF} and
\eqref{Prop2SF} are finite and thus there is no reason to demand for a regular solution in these
points. Notice that the potential above is complex, whereas in most problems of QNMs the potentials turn out to be real. Although it is possible to make field redefinitions in order to make the potential real, we shall not do it here; see Ref. \cite{Guven1977}. Moreover, inasmuch as the potential does not vanish at $x\rightarrow \pm \infty$, the solution at the boundaries is not of the plane-wave type, see Eq. \eqref{SCF}.

\vspace{.5cm}
%\section{\uppercase{Spin-1/2 Quasinormal Modes}}
\section{Spin-1/2 Quasinormal Modes}
\vspace{.5cm}

By plugging Eq. \eqref{ABC_Spin} into Eq. \eqref{abc}, we find that the constants appearing in the hypergeometric equation can be written as
\begin{align}
a &= i R_1 \sqrt{\mu^2 +\sum_{l=2}^d\frac{\lambda_l^2}{R_l^2}+ q^2 Q_1^2 R_1^2} + (1+A_1)\left( \frac{1}{4} - i \omega \frac{R_1}{2}\right) -i (1-A_1)\frac{q Q_1 R_1^2}{2} \,,
\nonumber\\
b &= -i R_1 \sqrt{\mu^2 +\sum_{l=2}^d\frac{\lambda_l^2}{R_l^2}+ q^2 Q_1^2 R_1^2 } + (1+A_1)\left( \frac{1}{4} - i \omega \frac{R_1}{2}\right) -i (1-A_1)\frac{q Q_1 R_1^2}{2} \,, \label{abcSpinor}\nonumber\\
c &=  \frac{1}{2} + i A_1 \left(q Q_1 R_1^2 - \omega R_1 \right) \,.
\end{align}

%
%
%Erros: \mathcal{A}^0_{1}, acima da Eq. (42) - mudei sinal; correto: \mathcal{A}^0_{1} =- i Q_{1}R_{1}\text{sinh}(r_{\star}/R_1)
%Erros: sinais em frente a (1-s_1) nas expressções para a e b na Eq. (44)
%
%

Now, with Eq. (\ref{abcSpinor}) at hand, we are ready to impose the boundary conditions\index{Boundary conditions} in order to investigate the quasi-normal modes. Since all we need, for this end, are the asymptotic behavior obtained in Eqs. (\ref{SolutionH_y=0}) and (\ref{SolutionH_y=1}), and since they depend just on the exponents $\mathfrak{d}(c-1)$ and $\mathfrak{d}(a+b-c)$, it is useful to write the explicit expressions for these combinations:
\begin{align}\label{ExponentsSpinor}
       \mathfrak{d} (c-1) &=\, - i A_1 \omega + i A_1 q Q_1 R_1 - \dfrac{1}{2R_1} \,, \nonumber\\
       \\ 
\mathfrak{d} (a+b-c) &=\, - i  \omega - i  q Q_1 R_1 + \dfrac{A_1}{2R_1}  \,. \nonumber
\end{align}
Now we are ready to impose the boundary conditions\index{Boundary conditions}. Without loss of generality, we can consider that the spin $A_{1}$ is already chosen and fixed at $A_{1} = +$ or $s_{1} = -$ since the QNFs should not depend on choice of $A_{1} = \pm$. Let us impose, for instance, the boundary conditions (IV) for the component $A_{1} = +$ of the spinorial field. In this case, using the identity \eqref{ExponentsSpinor} along with the equation (\ref{SolutionH_y=0}), we eventually arrive at the following behavior of the
solution at $x \rightarrow -\infty$:
\begin{equation}\label{SCF}
\left.e^{-i\omega t}\psi^{\omega}_{+,jm}\right|_{x\rightarrow-\infty} \,=\, \alpha \,e^{-i \omega (t + x)}\,e^{\left(i q Q_1 R_1 - \frac{1}{2R_1} \right)x}\,+\, \beta\,e^{- i \omega (t - x)}\,e^{- \left(i q Q_1 R_1 - \frac{1}{2R_1} \right)x}\,,
\end{equation} 
which is not a solution of the plane-wave type, as expected, inasmuch as the potential
does not vanish at this point. For the boundary condition (IV), Fig. \ref{FigBoundCond}  tells us that
$e^{-i\omega t}\psi^{\omega}_{+,jm}$ must have a dependence of the type $e^{-i \omega (t - x)}$ at $x \rightarrow -\infty$, while it must
goes as $e^{-i \omega (t + x)}$ at $x \rightarrow +\infty$. Thus, from Eq. \eqref{SCF}, we conclude that we must set $\alpha = 0$. Then, inserting $\alpha = 0$ into (\ref{SolutionH_y=1}), we end up with the following behavior of the solution at $x \rightarrow +\infty$:
\begin{eqnarray}\label{SCFI}
 \left.e^{-i\omega t}\psi^{\omega}_{+,jm}\right|_{x \rightarrow +\infty} \,\simeq \,\beta \,\left[\frac{\Gamma(c-a-b)\Gamma(2-c)}{ \Gamma(1-a) \Gamma(1-b)   } \,\right] e^{-i \omega (t + x)}\,e^{\left(i q Q_1 R_1 - \frac{1}{2R_1} \right)x}  \nonumber\\
\,+\, \beta\,\left[\frac{\Gamma(a+b-c)\Gamma(2-c)}{ \Gamma(a-c+1) \Gamma(b-c+1) }    \,\right] e^{-i \omega (t - x)}\,e^{\left(- i q Q_1 R_1 + \frac{1}{2R_1} \right)x} \,. 
\end{eqnarray}
The boundary condition (IV) imposes that the coefficient multiplying $e^{-i \omega (t - x)}$ should
vanish. Since $\beta$ cannot be zero, as otherwise the mode would vanish identically, we need the combination of the gamma functions to be zero. Now, once the gamma function has no zeros, the way to achieve this is to let the gamma functions in the denominator diverge, $\Gamma(a-c+1) = \infty$ or $\Gamma(b-c+1) = \infty$. Since the gamma functions diverge only at nonpositive integers, we are led to the following  constraint:
\begin{equation}
a-c+1=-n  \quad \text{or} \quad b-c+1=-n \,,
\end{equation}
with $n$ being a non-negative integer. These imply that the frequencies must be given by
\begin{equation}\label{SpectrumSpinor1}
  \omega_{\text{IV}} =  \pm\,\sqrt{\mu^2 +\sum_{l=2}^d\frac{\lambda_l^2}{R_l^2}+ q^2 Q_1^2 R_1^2} + \frac{i}{2R_1}\left(2n+1\right) \,,
\end{equation}
with $n \in \{0,1,2,\ldots\}$. It is worth recalling that $\lambda_l$ are the eigenvalues of the Dirac
equation in the $l$th two-sphere. 
%which are the QNFs of the Dirac field propagating in $D$-dimensional generalized Nariai spacetimes. The real part of a QNF is associated with the oscillation frequency, while the imaginary part is related to its decay rate.
%At this point, it is worth recalling that $L$ is a separation constant of the Dirac equation that is related to the angular mode of the field.
Likewise, imposing the boundary condition (IV) to the component $A_1= -$ of the spinorial field, we  find that we must set $\beta = 0$ at Eq. (\ref{SolutionH_y=0}) and then $c-a = -n$ or $c-b = -n$, with $n$ being a non-negative integer. This, in turn, lead to the same spectrum obtained for the component  $A_1= +$, namely \eqref{SpectrumSpinor1}.

Analogously, imposing the boundary conditions\index{Boundary conditions} (II) and (III) for the spinorial field, we find that no quasinormal mode exists in these cases, just as happens with the scalar and Maxwell's fields. On the other hand, imposing the boundary condition (I), we find that the quasinormal frequencies\index{Quasinormal frequencies} are given by
\begin{equation}\label{SpectrumSpinor2}
  \omega_{\text{I}} =  \pm\,\sqrt{\mu^2 +\sum_{l=2}^d\frac{\lambda_l^2}{R_l^2}+ q^2 Q_1^2 R_1^2} - \frac{i}{2R_1}\left(2n+1\right) \,,
\end{equation}
where $n\in\{0,1,2,\ldots\}$.

\vspace{.5cm}
%\subsection{Analyzing the regularity of the solution}
\subsection{Analyzing the regularity of the solution}
\vspace{.5cm}

For the sake of notational simplicity, let us here omit indices $jm$ of $\Psi^{jm}_{A_1}$. We have seen that the solution for the spinor component $\Psi_{A_1}$ is not exactly a plane wave at the boundaries, which is a consequence of the fact that the potential $V_{s=1/2}(x)$ does not vanish at $x \rightarrow\pm \infty$. Indeed, computing the asymptotic form of the time-dependent fields $\Psi_{A_1}$ when the assumed  boundary condition is (I), with the spectrum given by (\ref{SpectrumSpinor1}), we find that
\begin{align}
 \left. \Psi_{+} \right|_{x\rightarrow -\infty}  &= \alpha \,e^{-i\omega t} e^{\mathfrak{d}(c-1)x}  \nonumber\\
   &= \alpha \,e^{\mp i \sqrt{\mu^2 +\sum_{l}\lambda_l^2/R_l^2+ q^2 Q_1^2 R_1^2} (t-x)}\,
e^{-(n+\frac{1}{2})\frac{t}{R_1}}   \, e^{  -( n +1 - i  q Q_1 R_1^2 )\frac{x}{R_1}  } \,,\\
 \left. \Psi_{+}\right|_{x\rightarrow +\infty} &= \,
\alpha  \frac{\Gamma(a+b-c)\Gamma(2-c)}{ \Gamma(a-c+1) \Gamma(b-c+1) } e^{-i\omega t} e^{-\mathfrak{d}(a+b-c)x}
\nonumber\\
   &= \alpha \frac{\Gamma(a+b-c)\Gamma(2-c)}{ \Gamma(a-c+1) \Gamma(b-c+1) } \,e^{\mp i \sqrt{\mu^2 +\sum_{l}\lambda_l^2/R_l^2+ q^2 Q_1^2 R_1^2} (t+x)}\nonumber\\
   &\times e^{-(n+\frac{1}{2})\frac{t}{R_1}}   \, e^{  ( n + i  q Q_1 R_1^2 )\frac{x}{R_1}  }  \,,
  \label{Psi-Asymptotic} \\
\left. \Psi_{-} \right|_{x\rightarrow -\infty}  &= \beta \,e^{-i\omega t} e^{-\mathfrak{d}(c-1)x}
\nonumber\\
   &= \beta \,e^{\mp i \sqrt{\mu^2 +\sum_{l}\lambda_l^2/R_l^2+ q^2 Q_1^2 R_1^2} (t-x)}\,
e^{-(n+\frac{1}{2})\frac{t}{R_1}}   \, e^{  -( n-i  q Q_1 R_1^2 )\frac{x}{R_1}  } \,,\\
 \left. \Psi_{-}\right|_{x\rightarrow +\infty} &= \,
  \beta\, \frac{\Gamma(a+b-c)\Gamma(c)}{ \Gamma(a) \Gamma(b) }  e^{-i\omega t} e^{\mathfrak{d}(a+b-c)x}
\nonumber\\
   &= \beta \, \frac{\Gamma(a+b-c)\Gamma(c)}{ \Gamma(a) \Gamma(b) } \,e^{\mp i \sqrt{\mu^2 +\sum_{l}\lambda_l^2/R_l^2+ q^2 Q_1^2 R_1^2} (t+x)}\,\nonumber\\
&\times e^{-(n+\frac{1}{2})\frac{t}{R_1}}   \, e^{  ( n  - i  q Q_1 R_1^2 )\frac{x}{R_1}  }  \,.
\end{align}
Looking at these asymptotic forms, two features stand out: (i) the solutions do not represent progressive waves moving to the right or left, as we should demand from the boundary condition; (ii) since $n$ is real and positive, both fields $\Psi_{\pm}$ diverge  exponentially at the boundaries. It seems that something is wrong. Nevertheless, this impression comes from the fact that we are looking at the fields themselves instead of analyzing the conserved current that describes the flux of Dirac particles.

%
%
%Erros: \mathcal{A}^0_{1}, acima da Eq. (42) - mudei sinal; correto: \mathcal{A}^0_{1} =- i Q_{1}R_{1}\text{sinh}(r_{\star}/R_1)
%Erros: sinais em frente a (1-s_1) nas expressções para a e b na Eq. (44)
%Erros: Sinal de + i  q Q_1 R_1^2  em  \Psi^{+} \right|_{r_\star\rightarrow -\infty} e \Psi^{-} \right|_{x\rightarrow -\infty}
%

The conserved current associated to the Dirac field\index{Dirac field} interacting with the background electromagnetic field is $J_\alpha = \bar{\Psi}\Gamma_\alpha \Psi$, where $\bar{\Psi}$ stands for  the adjoint of $\Psi$, which for the representation adopted here is given by
\begin{equation}
  \bar{\Psi} =  \Psi^{\dagger}\, (\sigma_3\otimes\sigma_3\otimes \cdots \otimes\sigma_3 ) \,,
\end{equation}
see also Ref. \cite{VenancioBook}. In particular, the current along the radial direction is given by:
\begin{align}
  J_{\tilde{1}} &= (e_{\tilde{1}})_\alpha \, J^\alpha \nonumber\\
&= \sum_{A}\sum_{B}\left[\Psi_{A_1}(t,x)\xi_{A_1}\right]^{\dagger}\sigma_2\sigma_3\Psi_{B_1}(t,x)\xi_{B_1} \times (\textrm{Angular Part}).
\end{align}
Thus, using that $\sigma_2 \xi_{A_1}=iA_1\xi_{-A_1}, \sigma_3 \xi_{A_1}= A_1\xi_{A_1}$ along with $\xi^{\dagger}_{A_1}\xi_{B_1} = \delta_{A_1B_1}$ and ignoring the multiplicative factor coming from angular dependence, it follows that the radial current is given by
\begin{equation}\label{J1-v2}
  J_{\tilde{1}} = \text{Re}(\Psi_{+}\Psi^{\star}_{-}) \,,
\end{equation}
where $\star$ stands for complex conjugation and $\text{Re}(\cdots)$ takes the real part of its argument. Then, inserting the asymptotic forms (\ref{Psi-Asymptotic}) into Eq. (\ref{J1-v2}), lead us to the following asymptotic behavior for the current when the boundary condition is (I):
\begin{equation}\label{J1Asymptotic}
      \begin{array}{ll}
     \left. J_{\tilde{1}} \right|_{x\rightarrow +\infty}  & \sim\, e^{-(2n+1)\frac{(t-x)}{R_1}} \,, \\
\left. J_{\tilde{1}} \right|_{x\rightarrow -\infty}    &\sim \,  e^{-(2n+1)\frac{(t+x)}{R_1}} \,.
    \end{array}
\end{equation}
\begin{equation}
      \begin{array}{ll}
     \left. J_{\tilde{1}} \right|_{x\rightarrow +\infty}  & \sim\, e^{-(2n+1)\frac{(t-x)}{R_1}} = e^{-(2n+1)\frac{u}{R_1}} \,, \\
\left. J_{\tilde{1}} \right|_{x\rightarrow -\infty}    &\sim \,  e^{-(2n+1)\frac{(t+x)}{R_1}} = e^{-(2n+1)\frac{v}{R_1}} \,.
    \end{array}
\end{equation}
Thus, since the dependence of $J^{\tilde{1}}$ on the coordinates $t$ and $x$ occur just through combinations $(t-x)$ and $(t+x)$, it follows that $J_{\tilde{1}}$ becomes a progressive wave at boundaries. In particular, at $x\rightarrow +\infty$ the flux of particles is in the direction of increasing $x$, while at $x\rightarrow -\infty$ the flux of particles is in the direction of decreasing $x$, which is in perfect accordance with the boundary condition (I).

\begin{figure}%[ht!]
  \centering
 \includegraphics[width=12cm]{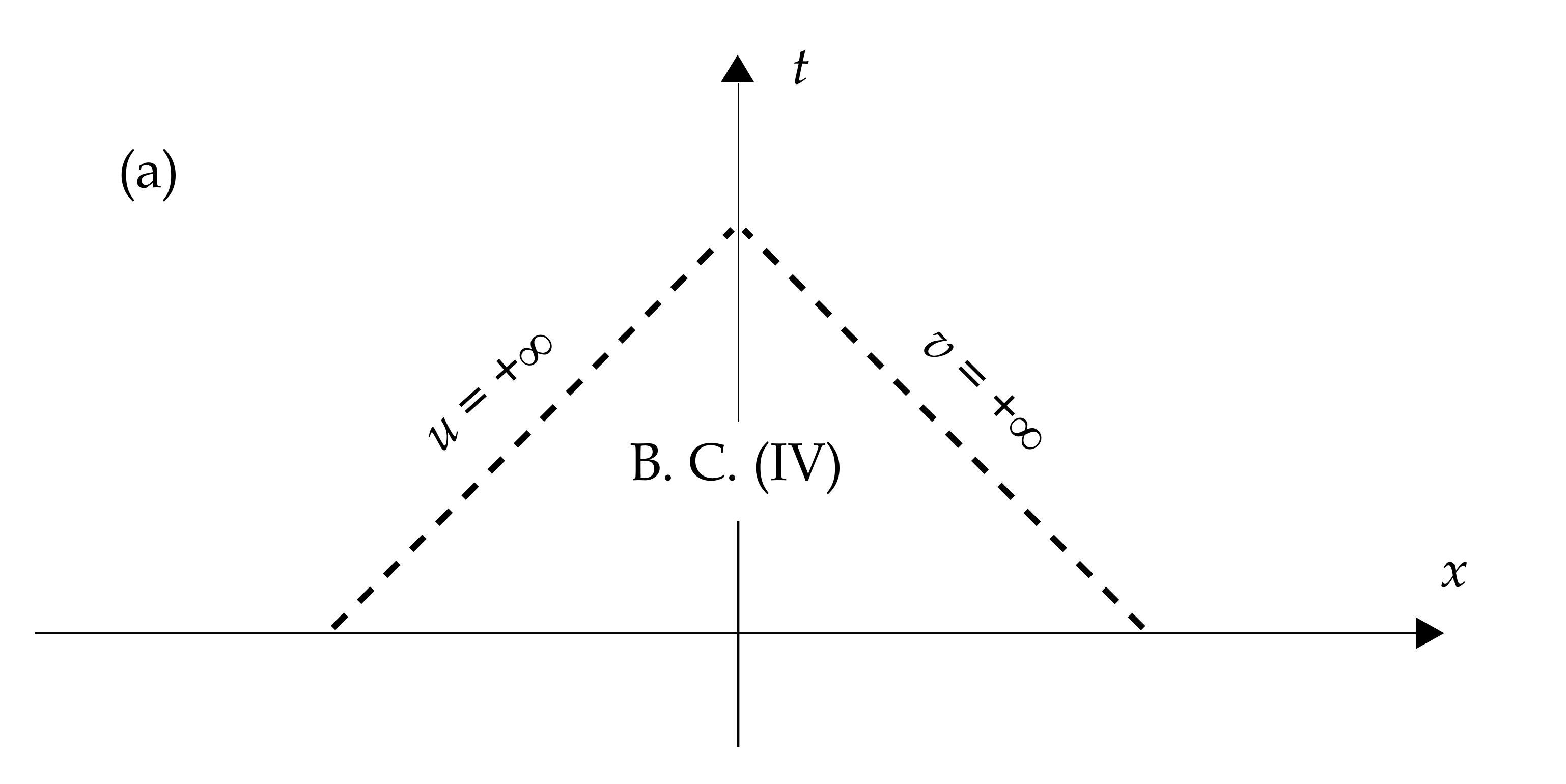}
 \includegraphics[width=12cm]{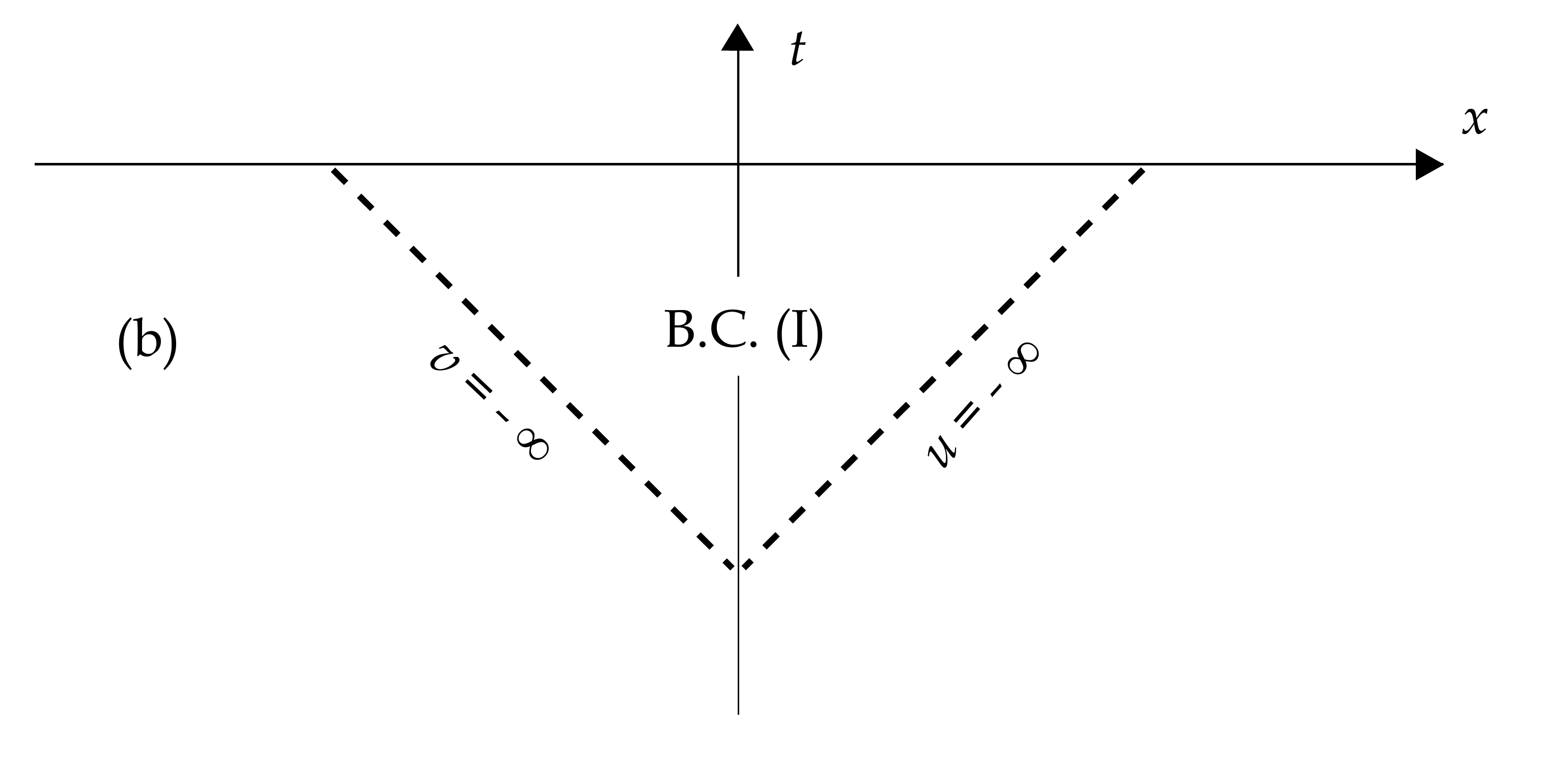}
  \caption{The dashed lines denote the region where the current is ill-defined. Part (a) corresponds to boundary condition (IV), B.C.(IV), while part (b) is corresponds to boundary condition (I),B.C.(I).}\label{FigBC}
\end{figure}

From the asymptotic behavior shown in Eq. (\ref{J1Asymptotic}), one could conclude that the current $J_{\tilde{1}}$ diverges exponentially at the boundaries. However, this can be circumvented for arbitrarily large negative times. Indeed, defining the null coordinates
\begin{equation}
  u = t-x \quad \text{and} \quad  v = t+x \,,
\end{equation}
we see, from Eq. (\ref{J1Asymptotic}), that for $x\rightarrow +\infty$ the current $J^{\tilde{1}}$  is ill-defined at $u\rightarrow -\infty$, but well-defined elsewhere. On the other hand, for $x\rightarrow +\infty$ the current diverges at $v\rightarrow - \infty$, while it is well-defined in other regions of the spacetime. This means that, for the boundary condition (I), the current is ill-defined at the past null infinity, but it is non-divergent elsewhere, as depicted in part (b) of Fig. \ref{FigBC}.

Analogously, computing the asymptotic form of the current $J_{\tilde{1}}$ for the solution corresponding to the boundary condition (IV), namely
\begin{align}
 \left. \Psi_{+} \right|_{x\rightarrow -\infty}  &= \beta \,e^{-i\omega t} e^{-\mathfrak{d}(c-1)x}  \nonumber\\
   &= \beta \,e^{\mp i \sqrt{\mu^2 +\sum_{l}\lambda_l^2/R_l^2+ q^2 Q_1^2 R_1^2} (t-x)}\,
e^{ (n+\frac{1}{2})\frac{t}{R_1}}   \, e^{  -( n + i  q Q_1 R_1^2 )\frac{x}{R_1}  } \,,\\
 \left. \Psi_{+}\right|_{x\rightarrow +\infty} &= \,
\beta  \frac{\Gamma(a+b-c)\Gamma(2-c)}{ \Gamma(a-c+1) \Gamma(b-c+1) } e^{-i\omega t} e^{\mathfrak{d}(a+b-c)x}
\nonumber\\
   &= \beta \frac{\Gamma(a+b-c)\Gamma(2-c)}{ \Gamma(a-c+1) \Gamma(b-c+1) } \,e^{\mp i \sqrt{\mu^2 +\sum_{l}\lambda_l^2/R_l^2+ q^2 Q_1^2 R_1^2} (t+x)}\nonumber\\
 &\times e^{ (n+\frac{1}{2})\frac{t}{R_1}}   \, e^{  ( n+1 - i  q Q_1 R_1^2 )\frac{x}{R_1}  }  \,,
  \label{Psi-Asymptotic2} \\
\left. \Psi_{-} \right|_{x\rightarrow -\infty}  &= \alpha \,e^{-i\omega t} e^{\mathfrak{d}(c-1)x}
\nonumber\\
   &= \alpha \,e^{\mp i \sqrt{\mu^2 +\sum_{l}\lambda_l^2/R_l^2+ q^2 Q_1^2 R_1^2} (t-x)}\,
e^{ (n+\frac{1}{2})\frac{t}{R_1}}   \, e^{  -( n+ 1 + i  q Q_1 R_1^2 )\frac{x}{R_1}  } \,,\\
 \left. \Psi_{-}\right|_{x\rightarrow +\infty} &= \,
  \alpha\, \frac{\Gamma(a+b-c)\Gamma(c)}{ \Gamma(a) \Gamma(b) }  e^{-i\omega t} e^{\mathfrak{d}(a+b-c)x}
\nonumber\\
   &= \alpha \, \frac{\Gamma(a+b-c)\Gamma(c)}{ \Gamma(a) \Gamma(b) } \,e^{\mp i \sqrt{\mu^2 +\sum_{l}\lambda_l^2/R_l^2+ q^2 Q_1^2 R_1^2} (t+x)}\nonumber\\
&\times e^{ (n+\frac{1}{2})\frac{t}{R_1}}   \, e^{  ( n  - i  q Q_1 R_1^2 )\frac{x}{R_1}  }  \,.
\end{align}
we find that
\begin{equation}\label{J1AsymptoticIV}
      \begin{array}{ll}
     \left. J_{\tilde{1}} \right|_{x\rightarrow -\infty}  & \sim\, e^{(2n+1)\frac{(t+x)}{R_1}} = e^{(2n+1)\frac{v}{R_1}}  \,, \\
\left. J_{\tilde{1}} \right|_{x\rightarrow +\infty}    &\sim \,  e^{(2n+1)\frac{(t-x)}{R_1}} = e^{(2n+1)\frac{u}{R_1}} \,.
    \end{array}
\end{equation}
Thus, for the boundary condition (IV), the current is divergent for $u\rightarrow +\infty$ and $v\rightarrow +\infty$. In other words, the current is ill-defined at the future null infinity, but it is well-defined elsewhere. The part (a) of Fig. \ref{FigBC} shows the region where the current is divergent for the boundary condition (IV).

The very same behavior is found for the current of the scalar field\index{Scalar field}, whose conserved current is defined by
\begin{equation}
  \mathcal{J}_\mu = \text{Im}(\Phi \,\partial_\mu  \Phi^{\star}) \,.
\end{equation}
Indeed, computing this current at the boundaries using the asymptotic form of the scalar field\index{Scalar field} obeying the boundary condition (I), namely Eqs. \eqref{Scalar-Iy=0} and \eqref{Scalar-Iy=1}, leads us to the following
current in the radial direction:
\begin{equation}
      \begin{array}{ll}
     \left.  \mathcal{J}_{x} \right|_{x\rightarrow +\infty}  & \sim\, e^{-(2n+1)\frac{(t-x)}{R_1}} = e^{-(2n+1)\frac{u}{R_1}}  \,, \\
\left. \mathcal{J}_{x} \right|_{x\rightarrow -\infty}    &\sim \,  e^{-(2n+1)\frac{(t+x)}{R_1}} = e^{-(2n+1)\frac{v}{R_1}} \,.
    \end{array}
\end{equation}
This is the same behavior of the spinorial current for the boundary condition (I), see Eq. (\ref{J1Asymptotic}). Likewise, when the adopted boundary condition is (IV), the component $\mathcal{J}_{x}$ at the boundaries using the asymptotic forms Eqs. \eqref{Scalar-Iy=0} and \eqref{SolutionH_y=1III} has the following asymptotic behaviours 

namely 
\begin{equation}
      \begin{array}{ll}
     \left.  \mathcal{J}_{x} \right|_{x\rightarrow +\infty}  & \sim\, e^{(2n+1)\frac{(t+x)}{R_1}} = e^{(2n+1)\frac{v}{R_1}}  \,, \\
\left. \mathcal{J}_{x} \right|_{x\rightarrow -\infty}    &\sim \,  e^{(2n+1)\frac{(t-x)}{R_1}} = e^{(2n+1)\frac{u}{R_1}} \,,
    \end{array}
\end{equation}
which is the same asymptotic form of current $J_{\tilde{1}}$ in Eq. (\ref{J1AsymptoticIV}).

The fact that there exists regions of the spacetime where the physical current is ill-defined should not come as a surprise. Indeed, the reason why the solutions for the boundary conditions (I) and (IV) are called quasinormal modes instead of normal modes is that the spectrum of allowed frequencies has also an imaginary part. Therefore, the time dependence of the fields, $e^{-i\omega t}$, blows up at $t\rightarrow \infty$ when $\text{Im}(\omega) > 0$, whereas it diverges at $t\rightarrow -\infty$ for $\text{Im}(\omega) < 0$. Generally, the QNMs are thought as states that do not exist at all times, rather they are excitations that occur at a particular time interval. In particular, they do not form a complete basis for the space of solutions of considered field equation \cite{Nollert1999}.

With all these results at hand, we complete the results obtained in previous chapters
in which the quasinormal spectrum for fields with spins $0, 1$ and $2$ have been explicitly
calculated. For the convenience of the reader, all the obtained QNFs are summarized below in table \ref{Summary}

\begin{table}[!htbp]
\begin{center}
\begin{tabular}{c|cccc}
  \hline
  \hline
  \textbf{Spin-$s$ Field} &   \qquad \qquad \text{$\omega_{\text{I}}$} & \text{$\omega_{\text{II}}$} & \text{$\omega_{\text{III}}$} & \text{$\omega_{\text{IV}}$}  \\ \hline
 $s=0$ & $\begin{matrix}
\pm \left[\mu^2  + \sum_{l=2}^d  \frac{\kappa_l}{R_l^2} - \frac{1}{4R_1^2}\right]^{1/2} \\
-\frac{i}{2R_1} (2n+1)
\end{matrix}$ & $\times$ &$\times$&$\begin{matrix}
\pm \left[\mu^2  + \sum_{l=2}^d  \frac{\kappa_l}{R_l^2} - \frac{1}{4R_1^2}\right]^{1/2} \\
+\frac{i}{2R_1} (2n+1)
\end{matrix}$\\

 $s=1/2$ &  $\begin{matrix}
\pm \left[\mu^2  + \sum_{l=2}^d  \frac{\lambda_l^2}{R_l^2} +q^2Q_1^2R_1^2\right]\\
-\frac{i}{2R_1} (2n+1)
\end{matrix}$ & $\times$ &$\times$&$\begin{matrix}
\pm \left[\mu^2  + \sum_{l=2}^d  \frac{\lambda_l^2}{R_l^2} +q^2Q_1^2R_1^2\right]\\
+\frac{i}{2R_1} (2n+1)
\end{matrix}$\\

 $s=1$ &  $\begin{matrix}
\pm \left[\sum_{l=2}^d  \frac{\kappa_l}{R_l^2} - \frac{1}{4R_1^2}\right]^{1/2} \\
-\frac{i}{2R_1} (2n+1)
\end{matrix}$ & $\times$ &$\times$&$\begin{matrix}
\pm \left[\sum_{l=2}^d  \frac{\kappa_l}{R_l^2} - \frac{1}{4R_1^2}\right]^{1/2} \\
+\frac{i}{2R_1} (2n+1)
\end{matrix}$\\

 $s=2$ &  $\begin{matrix}
\pm \left[\sum_{l=2}^d  \frac{\kappa_l}{R_l^2} - \frac{9}{4R_1^2}\right]^{1/2} \\
-\frac{i}{2R_1} (2n+1)
\end{matrix}$ & $\times$&$\times$&$\begin{matrix}
\pm \left[\sum_{l=2}^d  \frac{\kappa_l}{R_l^2} - \frac{9}{4R_1^2}\right]^{1/2} \\
+\frac{i}{2R_1} (2n+1)
\end{matrix}$\\
  \hline
  \hline
\end{tabular}
\caption{ Allowed frequencies for the spin-$s$ fields for each value $s$ of the spin considered
here and for each one of the four boundary conditions\index{Boundary conditions} described in Fig. \ref{FigBoundCond}. The
subindex I in $\omega_{\text{I}}$ , for instance, stands for the frequencies when the boundary condition
(I) is assumed and so on and $\times$ indicates the absence of QNFs.}\label{Summary}
\end{center}
\end{table}

%\chapter*{\center{\uppercase{Conclusions and Perspectives}}}
%\chapter{\uppercase{Conclusions and Perspectives}}
\chapter{Conclusions and Perspectives}
%\addcontentsline{toc}{chapter}{Conclusions and Perspectives}

In this book we have investigated spin-$s$ field perturbations\index{Spin-$s$ field perturbations} for $s = 0, 1/2, 1$ and $2$ propagating in a higher-dimensional generalization of the charged Nariai spacetime with dimension $D = 2d$. One interesting feature of this background is that the perturbations can also be analytically integrated. It has been shown that they all obey a Schr\"{o}dinger-like equation with an effective potential contained in the Rosen-Morse\index{Rosen-Morse class of potentials} class of integrable potentials, with the solution given in terms of hypergeometric functions, as shown in section \ref{IRMCIP}. This is a valuable property, since even the effective potential associated to the humble Schwarzschild background is non-integrable, in spite of being separable.  
We have also investigated the QNMs associated with these fields by choosing the natural boundary conditions\index{Boundary conditions} in the background considered here. From the causal point of view, we have seen that the natural boundary conditions to be imposed on the perturbations are (II) and (III) of Fig. \ref{FigBoundCond}, but they do not lead to QNMs. 
On the other hand, the ad hoc boundary conditions (I) and (IV) of Fig. \ref{FigBoundCond} do allow QNMs with the corresponding QNFs summarized in table \ref{Summary}. 
Analyzing this table, it is interesting noting that the imaginary parts of the QNFs, are generally the same for the four types of fields and they do not depend on any detail of the perturbation, rather they only hinge on the charges of the gravitational background, through the dependence on $R_{1}$.  Differently, the real parts of the QNFs depend on the mass of the field and on the angular mode of the perturbations. 
Another fact worth pointing out is that while the fermionic fields always have a real part on their QNFs spectra, meaning that they always oscillate, the bosonic fields can have purely imaginary QNMs frequencies. Indeed, due to the negative factor $`` -1/(4R_1^2)"$ inside the square root appearing in the scalar spectrum ($s = 0$), it follows that for small enough $R_1$, along with small enough mass and angular momentum, the argument of the square root can be negative, so that this term becomes imaginary.
The same argument remains valid for the Maxwell field\index{Maxwell field} ($s=1$) for which both scalar and
covector degrees of freedom have the same spectrum as displayed in table \ref{Summary}. Regarding
the gravitational field\index{Gravitational field} ($s = 2$), all degrees of freedom of the perturbation have the same
spectrum as displayed in table \ref{Summary}. This differs, for example, from what happens in
other higher-dimensional spacetimes like Schwarzschild and (anti) de Sitter \cite{ Dreyer2003, Hughston1973, Frolov2008}, in which different parts of the gravitational perturbation have different spectra. Thus, the isospectral property of the higher-dimensional the Nariai spacetime considered here
proves that the existence of different spectra to different degrees of freedom of the gravitational field\index{Gravitational field} is much more related to the symmetries of the spacetime than to the tensorial nature of the degree of freedom of the perturbation or to the dimension of the background. Here the background has $SO(3) \times SO(3) \times \ldots \times SO(3)$ symmetry, $d-1$ times, whereas the Schwarzschild black hole has a $SO(2d-1)$ symmetry.

The separability\index{Separability} of the degrees of freedom of perturbations with spin $s=0,1/2,1,2$ has been attained in chapters \ref{FPS01}, \ref{S2FP} and \ref{S12FP} thanks to construction a suitable angular basis\index{Angular basis}. In its turn, this angular basis constructed here can also be used to separate the degrees of freedom of  spin-$s$ field perturbations\index{Spin-$s$ field perturbations} propagating on other backgrounds with the symmetry $SO(3) \times SO(3) \times \ldots \times SO(3)$. In particular, the higher-dimensional black hole presented in Ref. \cite{Batista2016} can certainly be handled with the technique introduced here. The same idea can also be applied to any spacetime that is the direct product of several spaces of constant curvature. The QNF spectrum associated with each of the perturbation types have been analytically calculated. With all these spectra at hand, we can write down a unique formula that works for all of these cases whenever electromagnetic charges of the background are zero, namely, $Q_1 = Q_l =0$, so that all radii of the spheres of the generalized Nariai spacetime are equal in such a case: 
%\begin{equation}
%\omega = \sqrt{\Lambda}\left[ \pm \sqrt{\mu^{2} + \sum_{l=2}^{d}(\ell_l+s)(\ell_l -s +1) - \left(s-%\frac{1}{2}\right )^{2}} + i\epsilon\,(n + \frac{1}{2})\right ] \,,
%\end{equation}
\begin{equation}
\omega = \sqrt{\Lambda}\left[ \pm \sqrt{\mu^{2} + \sum_{l=2}^{d}\nu_{s, l} - \left(s-\frac{1}{2}\right )^{2}} + i\epsilon\,\left(n + \dfrac{1}{2}\right)\right ] \,,
\end{equation}
where $\epsilon = -1$ stands for the QNFs when the boundary condition (I) is assumed and $\epsilon = 1$ when the boundary condition is (IV). The parameter $s$ is the spin of the perturbation and $\mu$ its mass  while $\nu_{s, l}$ is a positive constant related with angular momentum eigenvalue. For instance, for gravitational perturbation that is a massless spin-2 field
($\mu$ = 0, $s$ = 2) this constant has to be $\nu_{s=2,l}= \ell_l(\ell_l + 1)$ which in its turn is the same for
the scalar field\index{Scalar field} ($s = 0$) and for the Maxwell field\index{Maxwell field} ($s = 1$), since these are all bosonic fields. On the other hand, for spinor perturbation that is a massive spin-$1/2$
field ($s = 1/2$) we must have $\nu_{s=1/2,l}=\lambda_l^2$ where $\lambda_l \in \{\pm 1, \pm 2, \pm 3, \ldots\}$ are the eigenvalues of the Dirac operator on the unit sphere. It is worth pointing out that while in chapter \ref{FPS01} Einstein's vacuum equation was not assumed to hold, so that the spheres of the generalized Nariai spacetime could have different radii, depending on the
electromagnetic charges of the background, in chapter \ref{S2FP} we have assumed vanishing charges, so that the gravitational perturbation decouples from the electromagnetic perturbation.  Otherwise, we would have to consider the gravitational and electromagnetic perturbations simultaneously, since the electromagnetic perturbation field would be a source for the gravitational perturbation, as discussed above in section \ref{GNS}. Note also that we have not analyzed the perturbations for the Proca field and for massive gravitational field\index{Gravitational field}, $i.e$., for spin one and two the above formula has been checked only for the case of vanishing mass, $\mu = 0$. However, it is natural to expect that the above formula for the spectrum will also hold for these cases not considered yet. 

%The exactly solvable systems are usually limits of more realistic systems and allow us to study in details some properties of a physical process and test some methods which can be used to analyze more complicated systems. Thus they are powerful tools in many research lines. Therefore we expect that the exactly computed QNFs for $D$-dimensional generalized Nariai spacetime may play an important role in future research. 

In view of the results obtained in this book, an interesting application is the investigation of superradiance phenomena for the spin-$1/2$ field. Although bosonic fields like scalar, electromagnetic, and gravitational fields\index{Gravitational field} can exhibit superradiant behavior in four-dimensional Kerr spacetime \cite{Rosa2017}, curiously, this is not the case for the Dirac field\index{Dirac field} \cite{Guven1977}. Thus, it would be interesting to investigate whether an analogous thing happens in the background considered here.

The next natural step, once we have integrated the perturbations of spin-$s$ fields for $s=0,1/2,1$ and $2$ in the background considered here, as well as studied their boundary conditions, is to consider the perturbations in a spin-$3/2$ field, a fermionic field satisfying the  Rarita-Schwinger equation, propagating in the higher-dimensional generalization of Nariai spacetime. Research on the latter problem is still ongoing and shall be considered in a future work.

Another interesting application is to extend the work that we have done on the generalized Nariai spacetimes\index{Generalized Nariai spacetimes} to the case of other spaces that are the product of several spaces of constant curvature. An interesting topic of research would be to investigate spin-$s$ field perturbation for $s=0,1/2,1,3/2,2$ propagating in the higher-dimensional generalization of Schwarzschild spacetime, which is a static black hole whose horizon topology is $\mathbb{R} \times S^{2}\times \ldots S^{2}$. One interesting feature of this black hole is that, in addition to the electric charge, it has a magnetic charge, differently from the higher-dimensional generalization of the Reissner-Nordstrom solution \cite{Tangherlini1963}, which only has electric charge. Thus, in spite of the static character of the black hole to be considered in a future work, the physics involved can be quite rich. We would like then to see, for instance, if QNFs are the same for fields with spins $0, 1/2, 3/2$ and $2$ in this black hole, as is the case of the extremal $4$-dimensional Reissner-Nordtr\"{o}m black hole \cite{Cho2006}.

%%%%%%%%%%%%%%%%%%%%%%%%%%%%%%%%%%%%%%%%%%%%%%%%%%%%%%%%%%%%%%%%%%%%%%%%%%%%%%%%%%%%%%%%%%%%%%%%%%%%%%%%%%%%%%%%%%%%%%%%%%%%%%%%%%%%%%%%%%%%
%%%%%%%%%%%%%%%%%%%%%%%%%%%%%%%%%%%%%%%%%%%%%%%%%%%%%%%%%%%%%%%%%%%%%%%%%%%%%%%%%%%%%%%%%%%%%%%%%%%%%%%%%%%%%%%%%%%%%%%%%%%%%%%%%%%%%%%%%%%%
%%%%%%%%%%%%%%%%%%%%%%%%%%%%%%%%%%%%%%%%%%%%%%%%%%%%%%%%%%%%%%%%%%%%%%%%%%%%%%%%%%%%%%%%%%%%%%%%%%%%%%%%%%%%%%%%%%%%%%%%%%%%%%%%%%%%%%%%%%%%
%%%%%%%%%%%%%%%%%%%%%%%%%%%%%%%%%%%%%%%%%%%%%%%%%%%%%%%%%%%%%%%%%%%%%%%%%%%%%%%%%%%%%%%%%%%%%%%%%%%%%%%%%%%%%%%%%%%%%%%%%%%%%%%%%%%%%%%%%%%%
%%%%%%%%%%%%%%%%%%%%%%%%%%%%%%%%%%%%%%%%%%%%%%%%%%%%%%%%%%%%%%%%%%%%%%%%%%%%%%%%%%%%%%%%%%%%%%%%%%%%%%%%%%%%%%%%%%%%%%%%%%%%%%%%%%%%%%%%%%%%

%%%%%%%%%%%%%%%%%%%%%%%%%%%%%%%%%%%%%%%%%%%%%%%%%%%%%%%%%%%%%%%%%%%%%%%%%%%%%%
%%%%%%%%%%%%%%%%%%%%%%%%%%% Bibliography
%%%%%%%%%%%%%%%%%%%%%%%%%%%%%%%%%%%%%%%%%%%%%%%%%%%%%%%%%%%%%%%%%%%%%%%%%%%%%%

%%%%%%%%%%%%%%%%%%%%%%%%%%%%%%%%%%%%%%%%%%%%%%%%%%%%%%%%%%%%%%%%%%%%%%%%%%%%%%
%%%%%%%%%%%%%%%%%%%%%%%%%%% End: Bibliography
%%%%%%%%%%%%%%%%%%%%%%%%%%%%%%%%%%%%%%%%%%%%%%%%%%%%%%%%%%%%%%%%%%%%%%%%%%%%%%

%\makeindex
\newpage
\printindex
%\contentsline {chapter}{\numberline { }Index}{}

\end{document}